\documentclass[11pt,a4paper,fleqn]{book}

\usepackage{amstext,amsfonts,amsmath,theorem,amssymb}

\usepackage{graphicx}

\usepackage[cp1252]{inputenc}

\allowdisplaybreaks  

\usepackage[vcentermath]{youngtab} 

\usepackage{geometry}
\usepackage[a4,center,noinfo]{crop}

\usepackage{sectsty}  
\chapternumberfont{\huge \centering}
\chaptertitlefont{\LARGE \centering}
\sectionfont{\normalsize}
\subsectionfont{\it \normalsize}
\subsubsectionfont{\it \normalsize}

\makeatletter
\usepackage{calc}
\newlength{\topspace}
\def\@makechapterhead#1{%
 \vspace*{\topspace}
 {\parindent \z@ \center \normalfont
   \ifnum \c@secnumdepth >\m@ne
     \if@mainmatter
       \huge\bfseries \@chapapp\space \thechapter 
       \par\nobreak
       \vskip 0.5cm 
     \fi
   \fi
   \interlinepenalty\@M
   \LARGE \bfseries #1\par\nobreak  
   \vskip 3cm 
 }%
}%
\makeatother

\usepackage{fancyhdr}

 \fancypagestyle{plain}{
\fancyhead{}

\fancyfoot[C]{ \thepage} }

\makeatletter
\def\cleardoublepage{\clearpage\if@twoside \ifodd\c@page\else%
    \hbox{}%
    \thispagestyle{empty}
    \newpage%
    \if@twocolumn\hbox{}\newpage\fi\fi\fi}
\makeatother

\newcommand{\fft}[2]{{\frac{#1}{#2}}}
\newcommand{\ft}[2]{{\textstyle\frac{#1}{#2}}}
\newcommand{\R}{{\mathbb R}}

\newcommand{\Hol}{{\mathrm{Hol}}}
\newcommand{\hol}{{\mathrm{hol}}}
\newcommand{\SO}{{\mathrm{SO}}}

\newcommand{\D}{{\mathcal D}}
\newcommand{\Rm}{{\mathcal R}}
\newcommand{\Cm}{{\mathcal C}}

\newcommand{\n}{\tilde{n}}

\usepackage{verbatim}
\makeatletter \@addtoreset{equation}{section}
\@addtoreset{table}{section}

\renewcommand{\theequation}{\thesection.\arabic{equation}}
\makeatother

       {\theorembodyfont{\rmfamily}
}

       {\theorembodyfont{\rmfamily}
}

{\end{list}}

\begin{document}
\newcounter{bean}



%
%
%
%
%
%
%
%
%
%
%

\pagenumbering{roman}

\setcounter{page}{1}

\thispagestyle{empty}

\begin{center}


\begin{huge}
\textbf{\mbox{SYMMETRY AND HOLONOMY} \\[.8cm] IN M THEORY}
\end{huge}


\vspace*{2cm}

\begin{Large}
{\bf OSCAR VARELA}
\end{Large}

\vspace*{0.5cm}

\begin{footnotesize}
Departament de F\'\i{}sica Te\`orica and IFIC \\
Universitat de Val\`encia \\
46100 Burjassot (Val\`encia), Spain

\vspace*{0.3cm}

oscar.varela@ific.uv.es

\end{footnotesize}

\vspace*{7cm}

\textbf{TESIS DOCTORAL}

\vspace*{0.3cm}

Submitted: 26th Abril 2006

Defended: 7th July 2006

\end{center}

\vspace*{1cm}



\begin{minipage}[c]{0.5\linewidth}
\centering \hspace{-1cm} \includegraphics[scale=0.02]{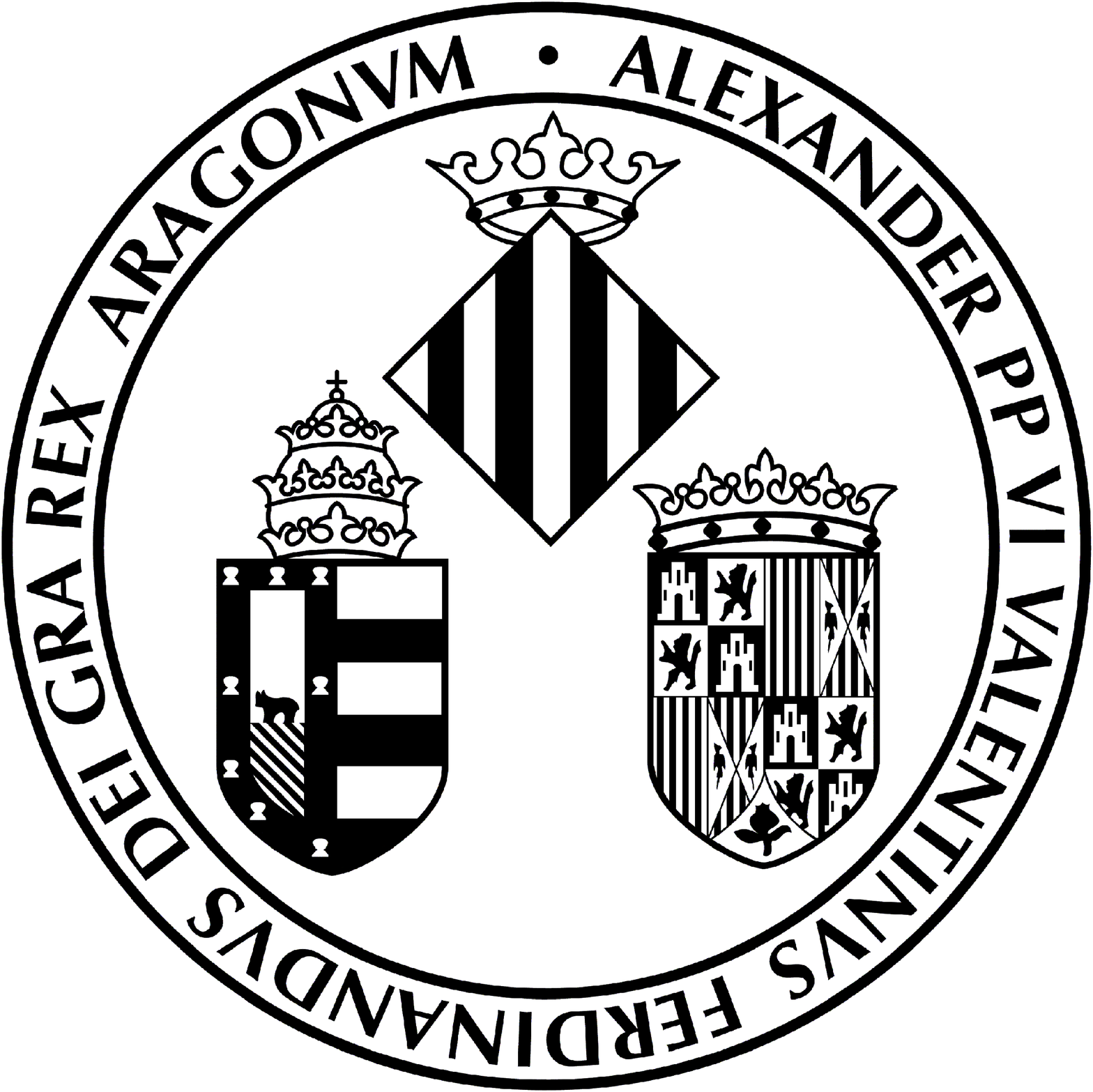}
\end{minipage}\hspace{-3.2cm}
\begin{minipage}[c]{0.5\linewidth}
\begin{flushleft}
\begin{small}
\centering \includegraphics[scale=0.06]{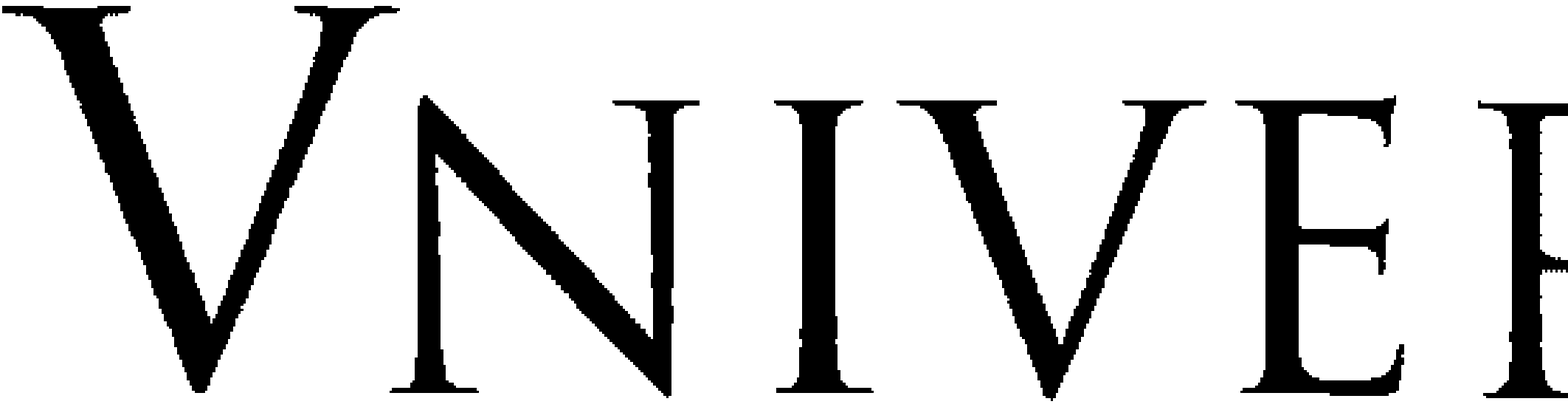} 
\hspace{0.1em} Departament de F\'\i sica Te\`orica \\
\hspace{1.6em} Institut de F\'\i sica Corpuscular
\end{small}
\end{flushleft}
\end{minipage}


\newpage{\pagestyle{empty}\cleardoublepage} 

\newpage{\pagestyle{empty}\cleardoublepage}

\newpage{\pagestyle{empty}\cleardoublepage} 

\newpage{\pagestyle{empty}\cleardoublepage}

\thispagestyle{empty}

\vspace*{2cm}

\noindent  D.~Jos\'e A.~de Azc\'arraga Feliu, Catedr\'atico de
F\'\i sica Te\'orica de la Universidad de Valencia,

\vspace{2cm}

\noindent  certifica:

\vspace{2cm}

\noindent que la presente memoria, {\it Symmetry and holonomy in M
Theory} ha sido realizada bajo su direcci\'on en el Departamento
de F\'\i sica Te\'orica de la Universidad de Valencia por \'Oscar
Varela Rizo y constituye su Tesis Doctoral.

\vspace{1.5cm}

\begin{flushright} Valencia, abril de 2006 \\

\vspace{4cm}

Jos\'e A.~de Azc\'arraga Feliu

\end{flushright}

\newpage{\pagestyle{empty}\cleardoublepage} 

\newpage{\pagestyle{empty}\cleardoublepage}

\thispagestyle{empty}

\vspace*{4cm}

\begin{flushright}

\begin{large}

{\it A Majo}

\end{large}

\end{flushright}

\newpage{\pagestyle{empty}\cleardoublepage} 

\newpage{\pagestyle{empty}\cleardoublepage}

\thispagestyle{empty}

\vspace*{2cm}

{\small {\it \noindent Les principes de la Mecanique
sont d\'ej\`a si solidement \'etablis, \\
qu'on auroit grand tort, si l'on vouloit encore douter de leur
verit\'e.} \footnote{The principles of Mechanics have already been
so solidly established that it would be a great mistake to still
question their truth.}

\vspace*{.5cm}

\noindent L. Euler, {\it Reflexions sur l'espace et le tems}
(1748)}

\vspace*{2cm}

{\small {\it \noindent The necessity to depart from classical
ideas when one wishes to account for the ultimate structure of
matter may be seen, not only from experimentally established
facts, but also from general philosophical grounds.}

\vspace*{.5cm}

\noindent P.A.M.~Dirac, {\it The principles of quantum mechanics}
(1930)}

\newpage{\pagestyle{empty}\cleardoublepage} 

\newpage{\pagestyle{empty}\cleardoublepage}

\thispagestyle{empty}


\tableofcontents




\chapter*{{\it Prefacio}}

\fancyhf{} \fancyhead[LE,RO]{\nouppercase{\it\thepage}}
\fancyhead[RE,LO]{\nouppercase{\it Prefacio}}

\addcontentsline{toc}{chapter}{{\it Prefacio}}

\vskip 1.4cm

A pesar de carecer actualmente de formulación dinámica, es posible
ob\-te\-ner gran cantidad de información sobre la Teoría M, la
teoría que se postula como unificadora de todas las interacciones,
a partir de sus sectores perturbativo y de baja energía. En las
regiones perturbativas adecuadas, la Teoría M adopta la apariencia
de la Teoría de Cuerdas. Al considerar su límite de baja energía,
surge la supergravedad en once dimensiones. Precisamente, esta
Tesis Doctoral, basada en las referencias
\cite{BAPV05}--\cite{30/32}, discute algunos aspectos de la Teoría
M desde el punto de vista de la supergravedad once-dimensional. En
el capítulo \ref{chapter1} se esboza una visión de conjunto de la
Tesis, y se argumenta la pertinencia del análisis de supergravedad
para el estudio de cuestiones relativas a la Teoría M. Es en el
capítulo \ref{chapter2} donde realmente comienza la discusión.

Esta Tesis ha sido realizada en su mayor parte en el Departamento
de F\'\i sica Te\'orica y el Instituto de F\'\i sica Corpuscular
de la Universidad de Valencia, con ayuda de una beca predoctoral
de la Generalitat Valenciana. Reciban mi agradecimiento todas
estas instituciones. Quisiera mostrar mi gratitud hacia mi
director de tesis, Jos\'e A.~de Azc\'arraga, por haber aprendido
de él tantas cosas, y no sólo sobre Física. Particular mención ha
de hacerse también de Igor Bandos, con quien he disfrutado tantas
conversaciones y de quien he recibido tanta ayuda. Estoy sumamente
agradecido a ambos por su estímulo y apoyo. Quisiera expresar mi
reconocimiento a Jos\'e M.~Izquierdo por conversaciones y
colaboraciones y a Dmitri Sorokin por sus comentarios. Los
agradecimientos han de hacerse extensivos a Mois\'es Pic\'on por
conversaciones y colaboraciones, a Miguel Nebot por conversaciones
y a Luis J.~Boya por una interesante discusión cuando esta Tesis
estaba ya siendo finalizada.

Parte del contenido de esta Tesis fue llevado a cabo fuera de
Valencia. Quisiera agradecer a Michael Duff su amable hospitalidad
durante mi visita al Michigan Center for Theoretical Physics y a
James Liu, Alex Batrachenko y Steve Wen por fértiles discusiones y
colaboraciones. Asimismo, estoy muy agradecido a Kellogg Stelle y
Jerome Gauntlett por su a\-ma\-ble hospitalidad durante mi visita
a Imperial College, y a  J.~Gauntlett, Daniel Waldram y Eoin
O'Colgain por tan fructíferas discusiones y colaboraciones. Pasé
un tiempo entrañable tanto en Ann Arbor como en Londres, y se lo
debo a todos ellos.

Por último, quisiera agradecer el apoyo de mi familia, en especial
el de mis padres Santiago y Antonia, y el amor, apoyo y afecto que
recibo cada día de Majo Rodríguez.

\vspace*{0.5cm}

\noindent O.V.

\vspace*{0.2cm}

\noindent Valencia, abril de 2006

\chapter*{{\it Preface}}

\fancyhf{} \fancyhead[LE,RO]{\nouppercase{\it\thepage}}
\fancyhead[RE,LO]{\nouppercase{\it Preface}}

\addcontentsline{toc}{chapter}{{\it Preface}}

\vskip 1.4cm

Despite its current lack of a dynamical formulation, a great deal
of information about M Theory, the conjectured theory unifying all
interactions, can be retrieved from its perturbative and low
energy corners. In the suitable perturbative regions, M Theory
adopts the ten-dimensional guise of String Theory. When its low
energy limit is considered, eleven-dimensional supergravity
arises. As a matter of fact, this PhD Thesis, based on references
\cite{BAPV05}--\cite{30/32}, is devoted to the discussion of some
topics about M Theory from the eleven-dimensional supergravity
point of view. A general overview is sketched in chapter
\ref{chapter1}, where the relevance of supergravity in order to
study M-theoretical issues is discussed, and the contents of the
Thesis outlined. It is, however, chapter \ref{chapter2} that
really starts the discussion.

This Thesis has been mostly made at the Departamento de F\'\i sica
Te\'orica and the Instituto de F\'\i sica Corpuscular of the
Universidad de Valencia, with the help of a Valencian Government
PhD fellowship. All these institutions are gratefully
acknowledged. I am indebted to my supervisor, Jos\'e A.~de
Azc\'arraga, from whom so much I have learned throughout all these
years, and not only about Physics. Particular mention should also
be made of Igor Bandos, with whom I have also enjoyed many
discussions and from whom I have received so much help. I am
extremely grateful to both of them for their encouragement and
support. Discussions and collaborations with Jos\'e M.~Izquierdo
are also gratefully acknowledged, as well as comments by Dmitri
Sorokin. Kind acknowledgements are extended to Mois\'es Pic\'on
for discussions and collaborations, to Miguel Nebot for
conversations and to Luis J.~Boya for a useful discussion in the
last stages of the writing up of this Thesis.

Some of the work contained in this Thesis was carried out outside
Valencia. I would like to thank Michael Duff for kind hospitality
during my visit to the Michigan Center for Theoretical Physics and
James Liu, Alex Batrachenko and Steve Wen for fertile discussions
and collaborations. Likewise, I am grateful to Kellogg Stelle and
Jerome Gauntlett for their kind hospitality during my visit to
Imperial College, and to J.~Gauntlett, Daniel Waldram and Eoin
O'Colgain for fruitful discussions and collaborations. I had a
wonderful time both in Ann Arbor and in London, and I owe it to
all of them.

Last, I should like to thank the support from my family,
especially from my parents Santiago and Antonia, and the love,
support and affection I receive every day from Majo Rodr\'\i guez.

\vspace*{0.5cm}

\noindent O.V.

\vspace*{0.2cm}

\noindent Valencia, April 2006


\chapter{Introduction} \label{chapter1}

\fancyhf{} \fancyhead[LE,RO]{\nouppercase{\it\thepage}}
\fancyhead[LO]{\it\rightmark}
\fancyhead[RE]{\nouppercase{\it\leftmark}}

\pagenumbering{arabic}

Amazing as it is the advance experienced in the last decades by
our understanding of the fundamental processes and laws that rule
the physical world, many important questions remain unsolved. Two
major developments, namely, General Relativity  and Quantum
Mechanics, contributed to shape the 20th century Physics. The
former, culminating the framework of  Classical Physics, is a
generalization of Special Relativity, the theory that revised the
Galilean and Newtonian notions of space and time and placed them
on an equal footing in a continuum spacetime. General Relativity
provides a geometrical description of gravity and the framework in
which the current cosmological models are formulated. Quantum
Mechanics, on the other hand, applies to physical phenomena
occurring (mostly) at the subatomic level, and is crucial in the
description of the rest of fundamental interactions. The
replacement of the original Galilean character of Quantum
Mechanics to make it consistent with Special Relativity led to the
development of Quantum Field Theory. Causality arguments could
then be invoked to insist on a local, rather than a global,
realization of some of the symmetries. The resulting Yang-Mills,
or gauge, theories describe all fundamental interactions
(electromagnetic, weak and strong forces) but gravity and,
together with a prescribed matter content, are the key components
of the Standard Model of particle physics.

\section{The road to M Theory}

The search for a unified description of different phenomena has
been historically a guiding principle for the progress of Physics.
From this point of view, it seems natural to look for a theory
that combines all four fundamental interactions within the same
descriptive scheme. A more solid argument for the unification of
fundamental interactions, going beyond aesthetical grounds, is
provided by the fact that the coupling constants of the
fundamental interactions, including gravity, seem to converge at
very high energies: at the grand unification scale of about
$10^{16}$ GeV. Even at low energies, though, the three
interactions of the Standard Model admit a description with the
same Yang-Mills gauge theory language. The reason why gravity does
not fit in this scheme is more serious than it may seem at first
sight: General Relativity is a classical theory, and no consistent
results are obtained when the usual prescriptions to account for
quantum effects are imposed; in other words, General Relativity is
a non-renormalizable theory. And yet, the energy scale at which
quantum gravity effects would be significant, the so-called Plank
scale, is $10^{19}$ GeV, relatively close to the grand unification
scale. This could be interpreted as a hint that a unifying theory
describing all four fundamental interactions does indeed exist.

Research on gravitation and high energy physics has traditionally
followed separate road maps, although some discoveries have been
fruitfully applied to both. That has been the case of
supersymmetry \cite{GolLik72,VolAk72,WeZu74} (see
\cite{FayFer77,Sohn85} for reviews and \cite{collsusy} for a
collection of reprints), a symmetry between bosons and fermions
based on the concept of Lie superalgebra, a structure containing
generators of both bosonic and fermionic character and thus
including both commutators and anticommutators. Soon after its
discovery, it was noticed that theories in which supersymmetry was
realized locally automatically contained gravity. Roughly, the
argument goes as follows: the anticommutator of two supersymmetry
generators is a translation; {\it locally} realized supersymmetry
therefore produces a {\it local} translation, to be identified
with a diffeomorphism or local coordinate transformation;
invariance under local supersymmetry implies, thus, invariance
under diffeomorphisms and hence gravity. Such theories of local
supersymmetry were consequently called supergravities (see
\cite{vNPhysRep,vProeyen03,Ortin04,Cd'AF91}). The first and
simplest supergravity theory to be constructed was its
four-dimensional ($D=4$) version with only one supercharge
($N=1$), and it was consequently called $D=4$, $N=1$, or simple,
supergravity \cite{FvNF76,DZ76} (see \cite{vNPhysRep} for a
review).

In the seventies, supergravity was regarded as a promising
candidate for a quantum theory of gravity, since fermionic
contributions to the gravitational perturbative expansions were
expected to cancel the divergent scattering amplitudes. Simple
supergravity turned out not to completely fulfil this prospect
since, although successfully proved finite even at two loops
\cite{Grisaru77}, its matter couplings failed to be so already at
first order \cite{vNV77}. Extended supergravities (containing $N >
1$ supercharges) were then developed, the promotion of their
$N(N-1)/2$ abelian gauge fields to $SO(N)$-gauge fields, yielding
the so-called gauged supergravities, being subsequently explored.
Among all extended supergravities, maximally extended $D=4$, $N=8$
supergravity \cite{CJ78,dWitNic82} had very attractive features:
gravity, gauge fields and matter were all included in the same
supergravity multiplet and, since no $N=8$ matter multiplets
existed, a truly unified theory (expected to be renormalizable) of
matter and all interactions could be achieved.

Developments in supergravity research also included the
construction of supergravity theories in diverse spacetime
dimensions (see \cite{SalSez89} for a collection of reprints), in
particular, in ten and eleven dimensions. Three supergravity
theories could exist in ten dimensions, a supergravity with $N=1$
supersymmetry (Type I supergravity) and two versions with $N=2$:
Type IIA (non-chiral) and Type IIB (chiral) supergravities (see
\cite{SalSez89} and references therein). On the contrary, only one
supergravity theory existed in eleven dimensions, which was proved
to be the maximal dimension in which supergravity could exist if
higher spin fields were to be excluded \cite{Nahm78}.
Eleven-dimensional supergravity was then constructed by Cremmer,
Julia and Scherk (CJS) in \cite{CJS}. Maximal lower dimensional
supergravities (those with a maximum amount of supersymmetry),
like Type IIA in $D=10$ or $N=8$ in $D=4$,  were shown to arise as
dimensional reductions ({\it i.e.}, toroidal compactifications) of
eleven-dimensional supergravity\footnote{As a matter of fact, one
of the main original motivations to build up $D=11$ supergravity
\cite{CJS} was to circumvent the technical difficulties arising in
the application of the standard Noether procedure to the
construction of the $D=4$, $N=8$ supergravity lagrangian
\cite{dWitFreed77}; the full $N=8$ lagrangian was actually
obtained \cite{CJ78} by dimensionally reducing its $D=11$
counterpart.}.

These developments encompassed a revival and update of the old
(Nordstr\"om and) Kaluza-Klein ideas, and compactifications on
non-trivial manifolds were explored (see \cite{DNP}) where the
features of the effective four-dimensional theories were dictated
by the properties of the compactifying manifold. For instance, the
gauge group and the preserved supersymmetry in four dimensions
were related to the isometry group and the holonomy group
\cite{ADP} of the compactifying manifold, respectively. Eleven was
not only the maximal dimension allowed by supersymmetry, but also
the minimal dimension that, upon compactification of the extra
seven dimensions, could accommodate the $SU(3) \times SU(2) \times
U(1)$ gauge group of the Standard Model \cite{Witten81} as a
subgroup of the isometry group. Moreover, it allowed for
`spontaneous' compactifications \cite{FR}, in which compact
seven-manifolds arose in a natural way. However, chiral fermion
families could not be obtained \cite{Witten85} from
compactification of (non-chiral) $D=11$ supergravity\footnote{This
issue was revised later on, with the discovery that
compactification spaces with singularities allowed for chiral
fermions \cite{AchWit01}.}. These facts, together with their
non-renormalizable character, damped the interest in supergravity
theories as candidates for a quantum theory of gravity.

On the other hand, new results were being obtained regarding a
formulation of the other interactions. While the electroweak
forces were successfully described by the spontaneously broken
$SU(2) \times U(1)$ Yang-Mills theory, a description for the
strong interactions was much less clear. Curiously enough, a
description was proposed in terms of a theory of strings, since
the results of the Veneziano amplitudes and Regge slopes suggested
that hadrons could be described as vibrations of a fundamental
string \cite{Veneziano68}. This setting assumed the bold proposal
of substituting point particles for one-dimensional extended
objects. However, the success of the application of Yang-Mills
theory to the description of the strong interactions in terms of
Quantum Chromodynamics (QCD), made the stringy description fall
out of favour, while satisfactorily put the formulation of strong
and electroweak forces on the same footing.

String Theory (see \cite{GSW87, Pol98}) recovered from this
setback when the realization that the spin two field contained in
its spectrum could be interpreted as the graviton \cite{SS74} (the
quantum of the gravitational field) provided that the string scale
was moved up, from the scale of the strong interactions to that of
quantum gravity. Moreover, strings provided a renormalizable
quantum theory of gravity because their interaction was smeared
over a region of spacetime, instead of taking place at a point.
The real String Theory explosion  would come in the mid eighties,
when the {\it first superstring revolution} took place.

The classical theory of superstrings (incorporating supersymmetry)
was known to be well defined in 3, 4, 6 and 10 spacetime
dimensions \cite{GrSchvocsup84}, in which cases a Wess-Zumino term
(see \cite{AI95}) existed endowing the action with
$\kappa$-symmetry\footnote{In the particle case, the existence of
a fermionic gauge symmetry was shown in \cite{AL82} in the massive
case and in \cite{Siegel83} in the massless case.} (see
\cite{GSW87}), a local fermionic symmetry that allowed for the
correct Bose-Fermi matching of degrees of freedom. At the quantum
level, the theory was shown to be consistent only in ten
dimensions, since only in that case it was anomaly free
\cite{GSchw84}. The anomaly cancelation left, moreover, five
different possible string theories in ten dimensions, namely, Type
IIA, Type IIB, Type I, SO(32) heterotic and $E_8 \times E_8$
heterotic (see \cite{GSW87} and references therein). Also,
supergravity was incorporated into String Theory: an analysis of
the massless modes (describing the low energy dynamics) in the
spectrum of the different string theories showed that they
consisted in the fields belonging to the different supergravity
multiplets in ten dimensions, possibly coupled to super Yang-Mills
multiplets. More precisely, the low energy limits of Type IIA and
IIB string theories were found to be, respectively, the
supergravities of the same name; and that of Type I and the
heterotic strings, Type I supergravity coupled to the $N=1$ vector
multiplet in ten dimensions with gauge group $SO(32)$ (for Type I
and the corresponding heterotic string) or $E_8 \times E_8$ (for
the other heterotic string).

Although a Lorentzian signature had to be imposed, the consistency
of the five perturbative string theories provided for the first
time a theoretical argument for a particular value of the
spacetime dimension. A Kaluza-Klein-type program was applied to
string compactifications, that sought for realistic models in
which ten-dimensional spacetime split into ordinary
four-dimensional spacetime, and a compact six-dimensional
Euclidean manifold. Interestingly enough, the compactifying
manifold could be chosen so that realistic models close to the
Standard Model were obtained in four dimensions. The heterotic
\cite{GHMR84} $E_8 \times E_8$ string was seen as particularly
suitable for Standard Model building, since its compactifications
managed to provide $E_6$ gauge symmetries\footnote{Notice that
ten-dimensional heterotic compactifications on six-manifolds do
not contradict the fact mentioned above that only
compactifications from eleven dimensions on seven-manifolds allow
for the Standard Model gauge group in the resulting
four-dimensional theory. The resulting gauge symmetry in heterotic
compactifications is a result of the presence of $E_8 \times E_8$
gauge fields already in the ten-dimensional theory, and not a
consequence of the isometries of the compactifying manifold.
Calabi-Yau manifolds have, indeed, no isometries at all. See
\cite{VST83} for a derivation of the Standard Model gauge group
from compactification in a IIA context.} (a candidate gauge group
in Grand Unification Theories). Moreover, choosing the
compactification manifold to be Calabi-Yau \cite{CHSW85}, the
phenomenological requirement of chiral fermion families provided
by $N=1$ supersymmetry in $D=4$ was also fulfilled. All this added
to the generalized enthusiasm that made String Theory the most
promising candidate for the unified picture of the fundamental
interactions since, proposed as a quantum theory of gravity, it
also seemed to include the Standard Model as a result of its own
selfconsistency conditions.

Despite all this headway, many issues were still unresolved. First
of all, the (five different) theories of superstrings were only
defined at the perturbative level. Moreover,  the existence of
five different perturbative theories was not too appealing if
String Theory had to unify all interactions. In fact, the usual
prescription for the computation of scattering amplitudes involves
both the sum over all possible topologies the string worldsheet
can display (the genus expansion) and, in the presence of
non-trivial backgrounds,  the loop expansion in the string
coupling constant $\alpha^\prime$ for each one of those
topologies. Although this perturbative approach can be intensively
exploited to obtain a great deal of information about the theory
it describes, it was apparent  that there was a lot of
non-perturbative structure which could not be reached by this
prescription.

That is indeed the case. There exist states that are naturally
non-perturbative but similar, instead, to the solitonic solutions
present in other field theories.  There are some special solitonic
solutions, the so-called Bo\-go\-mol'\-nyi--Prasad--Sommerfield
(BPS) states\footnote{The terminology comes from
\cite{Bog76,PradSom75}.} which, despite being classical, can be
considered as true solutions to the non-perturbative quantum
theory. BPS multiplets arise in massive representations of the $N
\geq 1$ extended supersymmetry algebras in which the Bogomol'nyi
bound, relating the eigenvalues of the momentum (mass) and of the
central charges, is saturated. Their basic feature is that they
are characterized by containing significatively fewer states than
a usual massive multiplet. Now, although both mass and charge may
undergo renormalization in perturbation theory, the BPS condition
is protected from quantum corrections. A heuristic argument to
support this claim is that, otherwise, as quantum effects are
being switched on a BPS multiplet could turn into a non-BPS one
containing many more states than the former, and such drastic
appearance of states is not expected to happen. It is in this
sense that BPS states, albeit usually described by solutions of
the classical equations of motion, are stable under quantum
corrections and can be lifted, therefore, to solutions of the
full, non-perturbative theory. This is why they are so useful to
probe the non-perturbative structure of the theory.

Thus, the notion of BPS states turned out to be crucial in order
to explore the non-perturbative regime of the five string
theories. Such research derived, in the mid nineties, in important
discoveries concerning the relations among them. Some perturbative
relations, known as T-dualities (see \cite{GiPoRa94}), among the
compactifications of the different string theories were
nevertheless already known at the time: Type IIA and Type IIB were
known to be T-dual (equivalent) when compactified on a circle of
radius $R$ and $\alpha^\prime / R$, respectively, and viceversa.
The two versions of the heterotic string were also T-dual.
S-dualities \cite{FILQ}, in contrast, were of a different nature
and provided, instead, a generalization of the conjectured
electric-magnetic duality of classical electrodynamics
\cite{MonOl77}. S-dualities related the strong coupling regime of
a string theory (where the usual perturbative prescriptions break
down) with the weak coupling regime (that could be treated in
perturbation theory) of another string theory. Type I and $SO(32)$
heterotic were shown to be S-dual, and Type IIB to be S-selfdual.
T and S-dualities were unified by U-duality \cite{HT95},  and the
resulting web of dualities managed to give a unified picture of
all string theories, which were then reinterpreted as perturbative
expansions around five different vacua of the same underlying
non-perturbative theory. The importance of this fact motivated
that its discovery were marked as the beginning of the {\it second
superstring revolution} (see \cite{SchwMTh}).

Two decisive developments came along with the discovery of the web
of dualities. The first one was the realization that $p$-branes
\cite{Achucarro:1987nc} (see also \cite{Evans88}), supersymmetric
extended objects sweeping out a worldvolume of $p$ spacelike
dimensions, were to be included in the theory \cite{Topbrane}.
These branes are solutions of the classical supergravity equations
and, since they carry topological charges \cite{JdAT} (see also
\cite{ST97}), associated to the central charges of the
supersymmetry algebra, they keep on being solutions at the quantum
level. That is why they are so valuable to probe the
non-perturbative sectors of the theory. The existence of these
extended objects in String Theory suggests a {\it $p$-brane
democracy} \cite{M-alg}, according to which they should be treated
on the same footing as the strings themselves. The second
development was the realization that Type IIA and $E_8 \times E_8$
heterotic string theories were dual to a non-perturbative theory
in eleven dimensions \cite{Witten95M,HW96}, subsequently dubbed M
Theory (see \cite{SchwMTh} for a review and \cite{Duff11dim} for a
collection of reprints). The coupling constants of those string
theories were taken as growing functions of a compactification
radius so that, in their strong coupling regime, an eleventh
dimension arose. This duality was supported by other facts; as
already mentioned, Type IIA supergravity was known to be the
dimensional reduction of eleven-dimensional supergravity, and the
IIA string could be derived \cite{DHIS87} from the
eleven-dimensional supermembrane. This added reasons to conjecture
eleven-dimensional supergravity as the low energy limit of M
Theory.

Eleven is, in fact, the maximum spacetime dimension that local
spacetime supersymmetry permits (as already mentioned) and that
can contain supersymmetric extended objects (see \cite{Duff94}).
Also, the relevant supermultiplet in eleven dimensions that
determines the field content of $D=11$ supergravity is both unique
and far simpler than its ten-dimensional counterparts. It only
contains three fields: the metric (or graviton) $g_{\mu\nu}$, its
supersymmetric partner, the gravitino $\psi^\alpha$, and a
three-form $A_3$. From this perspective, eleven dimensions are
more natural than ten, although the selfconsistency arguments
provided by String Theory were irrefutable. After the mid
nineties, however, the place of String Theory was redefined, from
a theory of vibrating one-dimensional extended objects, to a
theory of extended objects in general. And, rather than being
fundamental theories in themselves, string theories arose as five
different ten-dimensional perturbative corners of a new and truly
fundamental eleven-dimensional theory, M Theory.

These insights about the non-perturbative sectors of the theory
allowed for new activity. For instance, D-branes \cite{Pol95} (see
\cite{CJDbranes}) superseded heterotic string compactifications in
Standard Model building (see \cite{Kir03}).  On the other hand,
other developments took place, exploring the non-perturbative
sectors of the theory. That was the case of the AdS/CFT
correspondence \cite{Maldacena97} (see \cite{AGMOO00} for a
review), originally formulated in a Type IIB context. According to
it, String Theory on a ten-dimensional background containing
five-dimensional Anti-de Sitter space, $AdS_5$, is dual to a
superconformal field theory (CFT) in the conformal boundary of
$AdS_5$, namely, four-dimensional Minkowski space $M_4$. The
original formulation related IIB string theory on the maximally
supersymmetric background $AdS_5 \times S^5$, where $S^5$ is the
round five-sphere, to $N=4$ super Yang-Mills theory. Subsequent
generalizations were proposed; in particular, a dual $N=1$
superconformal field theory is obtained if $S^5$ is replaced by a
Sasaki-Einstein manifold  \cite{AFFHS99}.

The current knowledge of M Theory includes the
non-per\-tur\-ba\-tive dualities it displays and the fact that its
low energy limit is eleven-dimensional supergravity. Properties
such as the stability of BPS states allow to probe the full M
Theory from supergravity solutions: the well-known
eleven-dimensional branes can be described by their worldvolume
actions \cite{BST,blnpst}, or considered as supergravity solutions
\cite{DuffStelle91,Gu92} (see \cite{Duff94,Stelle98} for reviews),
that arguably remain solutions of the full non-perturbative M
Theory. $D=11$ supergravity thus provides a laboratory to explore
basic features of M Theory. Brane solutions preserve one-half  of
the maximum amount of supersymmetry (see \cite{JdAT}), and their
intersections preserve smaller fractions \cite{interscbranes}.
Getting to know more supersymmetric solutions of supergravity,
preserving different fractions $\nu$ of supersymmetry, would still
provide further insight into the theory.

Although supersymmetric solutions to $D=11$ supergravity involving
both bosonic and fermionic fields are known (see \cite{Hull84}),
the search for supergravity solutions has been usually restricted,
for simplicity, to purely bosonic configurations. Being bosonic,
the fermion fields can be set to zero in the equations of motion
and the requirement that the solution  be supersymmetric is
achieved provided the supersymmetry transformation of the fermions
also vanishes. Supersymmetric purely geometrical solutions, in
which the metric is the only non-vanishing field, can be
classified by Riemannian holonomy. More general supergravity
solutions can be suggestively discussed by an extension of
Riemannian holonomy in terms of generalized holonomy
\cite{Duff03,DuffStelle91}, $G$-structures \cite{GMPW04,GP02}, or
{\it spinorial geometry} \cite{Gillard:2004xq}. We shall be
discussing some related issues in this Thesis.

Another piece of valuable information about M Theory can be
obtained, also at the supergravity level, by studying the symmetry
algebra on which it is based \cite{M-alg,vHvP82}. The brane
solutions of $D=11$ supergravity are often viewed as the
fundamental objects of M Theory, in much the same way strings were
the basic objects of String Theory. In particular the topological
charges \cite{JdAT} of the M5 \cite{ST97} and M2 branes can be
naturally included in the supersymmetry algebra to give the so
called M Theory algebra \cite{M-alg}. The lack of an action
principle for M Theory can be partially overcome by group
theoretical methods and, as a matter of fact, the study of the
representations of the M Theory superalgebra suggests that states
preserving 31 supersymmetries could be treated as fundamental, the
rest being composed of them. These states, introduced in
\cite{BPS01} and called {\it preons}, could be considered  as
fundamental constituents of M Theory. As it will be shown in this
Thesis, these notions also lead naturally to consider enlarged
superspaces \cite{30/32} and supertwistors (see \cite{Bandos05}
for a review); see \cite{JdA00} for earlier ideas on superbranes
and enlarged superspaces and \cite{Azcarraga05} for a review in
this context.

The study of the symmetries of supergravity is, arguably, a useful
tool to obtain insights into the structure of M Theory. The
obvious (bosonic) local symmetry group of $D=11$ supergravity is
the Lorentz group $SO(1,10)$ and, hence, Lorentz covariant
derivatives of the supergravity fields arise naturally in the
lagrangian. However, another {\it supercovariant derivative}
taking values on the Lie algebra of $SL(32, \mathbb{R})$
\cite{Hull03} arises in order to express the variation under local
supersymmetry of its {\it gauge field}, the gravitino
$\psi^\alpha$. Indeed, the suggestion has been made \cite{Hull03}
that $D=11$ supergravity has a hidden $SL(32, \mathbb{R})$
symmetry, that might become relevant in M Theory\footnote{Several
groups may also play a role, as the rank 11 Kac-Moody group
$E_{11}$ \cite{WestE11}, $OSp(1|64)$ \cite{BDM99,West00} or
$OSp(1|32)$ (see \cite{CS}).}. However, no explicit formulation of
$D=11$ supergravity has been explicitly achieved so far exhibiting
this symmetry.

This argument applies to the original, CJS formulation of $D=11$
supergravity and, in particular, assumes a fundamental character
for the three-form $A_3$ of $D=11$ supergravity. The observation
was made in \cite{D'A+F} that the lack of a clear formulation for
the gauge group of $D=11$ supergravity could be put down,
precisely, to the presence of $A_3$. Being a three-form, it did
not admit an interpretation as a gauge potential of some symmetry
group. In consequence, $A_3$ was proposed \cite{D'A+F} to be
composed of gauge one-form potentials of suitable groups which
could play a role \cite{Lett, AnnP04} in the formulation of the
fully-fledged M Theory. This formulation leads naturally to
supersymmetry algebras {\it larger} than the standard
superPoincar\'e algebras. Other natural setting to formulate the
symmetries of supergravity theories is achieved by a Chern-Simons
(CS) formulation \cite{CS,Ha-Tr-Za-04}, in which the lagrangians
are obtained as CS forms of suitable supergroups.

In summary, a great deal of information about M Theory can be
obtained from the analysis of its low energy limit, $D=11$
supergravity, the symmetries and structure of which are hence
worth further study. This Thesis aims to make a modest progress
towards a formulation of these symmetries and the identification
of the fundamental constituents of M Theory from the analysis of
$D=11$ supergravity.


\section{The contents of this Thesis}

The plan of the Thesis is as follows.

In chapter \ref{chapter2}, the elements of $D=11$ CJS supergravity
that will be needed in the rest of the Thesis are reviewed. The
superPoincar\'e algebra is presented, and extended into the M
Theory superalgebra, containing the central charge generators that
couple to the basic M branes. The action of $D=11$ supergravity is
then introduced, in a first order formalism that treats the
vielbein, gravitino and three-form of the corresponding
supergravity multiplet, as dynamical fields. The spin connection
is composed out of them, and an additional auxiliary four-form
becomes related, on-shell, to the curvature of the three-form. The
various symmetries of the action are discussed, with particular
emphasis on supersymmetry. The variation of the gravitino under
supersymmetry allows us to introduce a {\it generalized
connection} taking values on the Lie algebra of $SL(32,
\mathbb{R})$. The local (bosonic) symmetry of supergravity (at
least when the three-form field is regarded as fundamental) is,
however, only its subgroup $SO(1,10)$, hence the name of
generalized connection. Nevertheless, the analogy can be pushed
forward, and a {\it generalized curvature} and its corresponding
holonomy, consequently called {\it generalized holonomy}, can be
introduced as a useful tool to discuss supersymmetric solutions.
Interestingly enough, the generalized curvature is shown to encode
the bosonic equations of motion of $D=11$ supergravity, not only
in the purely bosonic limit but also when the gravitino is not
vanishing \cite{BAPV05}.

Further study about generalized holonomy is carried out in chapter
\ref{chapter3}. It is usually claimed that the holonomy of a
connection on a given fiber bundle is generated by the curvature.
A more precise statement, that is usually underemphasized, is that
the Lie algebra of the holonomy group at a point $p$ is generated
by the curvature at $p$ and at any other points that can be
reached from the former by parallel transport. The effect of the
curvature at neighbouring points of $p$ can be measured by the
successive covariant derivatives of the curvature {\it at} $p$.
These covariant derivatives of the curvature are, thus, also
involved in the definition of the Lie algebra of the holonomy
group. Translated into the problem of counting the supersymmetries
of a vacuum, this means that, in general, the first order
integrability of the Killing spinor equation (which determines the
supersymmetries preserved by a bosonic supergravity solution)
might not be not enough to ensure that the equation is satisfied,
and higher order integrability conditions could be needed to solve
the equation.

Using these ideas, the generalized holonomy of several
supersymmetric solutions of supergravity is revisited in chapter
\ref{chapter3}. The generalized holonomy of the usual brane
solutions \cite{BDLW03} is reviewed showing that, in these cases,
successive covariant derivatives of the corresponding generalized
curvatures only close the algebra obtained from the curvature. In
this sense, higher order integrability for the M branes does not
add significant new information to the Lie algebra of the holonomy
group. Freund-Rubin compactifications, on the other hand, are
given as examples in which the supercovariant derivatives of the
generalized curvature are crucial to determine the Lie algebra of
the generalized holonomy group. In fact, the curvature algebra for
the Freund-Rubin compactification on the squashed $S^7$ turns out
to be the Lie algebra of $G_2$. It is argued that this cannot be
the right result, since a $G_2$ holonomy does not describe
correctly the preserved supersymmetry of the solutions. On the
contrary, the generators provided by the supercovariant derivative
of the generalized curvature enhance the holonomy algebra to
$so(7)$ \cite{Duff:2002rw,BLVW03}, which is also argued by
supersymmetry to be the right result.

Our study of generalized holonomy continues in chapter
\ref{chapter4} from a different point of view, in the context of
the {\it preon} hypothesis \cite{BPS01}. BPS states preserving $k$
supersymmetries out of 32 are argued to be composed of $\n=32-k$
of preons, a number that coincides with that of {\it broken}
supersymmetries so that $k=32$ indicates a fully supersymmetric
vacuum. Preons themselves are characterized by $k=31$: they are
$\nu=31/32$ states and preserve all supersymmetries but one. It is
shown that a set of $\n$ bosonic spinors can be introduced in
order to describe these states. For simplicity, it will be assumed
that these states are purely bosonic so that, being
$k$-supersymmetric, they are also characterized by $k$ Killing
spinors. The preonic and Killing spinors are shown to be
orthogonal and, thus, provide an alternative description of the
preserved supersymmetries. In fact, this orthogonality can be
further exploited \cite{BAIPV03}: the set of preonic spinors, on
the one hand, and the set of Killing spinors, on the other hand,
can be completed, respectively, to two bases in the space of
spinors, dual to each other.

Either one of these two bases define a {\it moving $G$-frame}
\cite{BAIPV03}, where $G$ is a group that can be chosen for
convenience. The M Theory superalgebra, the algebraic analysis of
which leads to, and can be made in terms of,  the preon
conjecture, has a maximal automorphism group of $GL(32,
\mathbb{R})$ and, thus, it is natural to choose $G=GL(32,
\mathbb{R})$. Other groups are, however, possible and the more
restrictive options $G=SL(32, \mathbb{R})$ (the relevant group in
the generalized holonomy approach) or $G=Sp(32, \mathbb{R})$ may
also be taken. The $G$-frame method is then applied to the
characterization of the generalized holonomy of preonic
supergravity solutions, namely, hypothetical supergravity
solutions preserving 31 supersymmetries, associated to the BPS
preon states. No definitive answer about their existence in
ordinary CJS supergravity can be given from this analysis.
However, preonic configurations are shown to exist \cite{BAIPV03}
in the context of Chern-Simons supergravities. Chapter
\ref{chapter4} then concludes with the introduction of a brane
action preserving 31 supersymmetries and describing, hence, a
preon state. This brane is not formulated on standard CJS
supergravity, though, but in the context of D'Auria-Fr\'e approach
to supergravity \cite{D'A+F} (that assumes a composite structure
of the three-form $A_3$ in terms of suitable one-form gauge
fields).

This result gives us reasons to further explore $D=11$
supergravity {\it \`a la} D'Auria-Fr\'e. Before doing so, however,
a break is done to introduce in chapter \ref{chapter5} the {\it
expansion method} \cite{JdAIMO} for Lie (super)algebras, since it
will be useful in this context. The mathematical ({\it
vs}.~physical) content of chapter \ref{chapter5} is
significatively higher than that of the rest of the Thesis. It is
a technical chapter giving the details of how the expansion method
works, and describing the features of the algebras obtained. First
of all, a review is done of the existing methods (contractions,
deformations and extensions) to obtain and derive new Lie algebras
(and superalgebras) from given ones. Then the expansion method for
Lie algebras ${\cal G}$ is introduced in general. Like the
contraction method, it relies on a redefinition of the group
coordinates by a parameter $\lambda$ that makes the Maurer-Cartan
(MC) one-forms of the (dual) Lie algebra expand in infinite power
series of $\lambda$ with one-form coefficients. The series can be
consistently truncated at suitable orders, provided the cutting
orders fulfil some conditions, and the retained one-form
coefficients then correspond to the MC forms of the new, {\it
expanded} algebras. The method is then applied to Lie algebras
with a particular structure of subspaces that, in the end, makes
straightforward its generalization to Lie superalgebras. As a
first application of the method, the M Theory superalgebra
(including its Lorentz automorphism part) is derived \cite{JdAIMO}
as the expansion $osp(1|32)(2,1,2)$ of $osp(1|32)$ (see chapter
\ref{chapter5} and \cite{JdAIMO} for the notation).

Chapter \ref{chapter6} returns to the main subject of this Thesis,
$D=11$ supergravity. The gauge symmetry underlying supergravity is
revised, in a formulation {\it \`a la} D'Auria and Fr\'e
\cite{D'A+F}. In general, a lagrangian theory with local symmetry
algebra ${\cal G}$ is described in terms of gauge one-form fields,
associated to the MC
 forms of ${\cal G}$, and its curvatures. However, as
opposed to its $D=4$, $N=1$ counterpart, eleven-dimensional
supergravity contains, as already mentioned,  besides the vielbein
$e^a$ and gravitino $\psi^\alpha$ one-forms, a three-form $A_3$
that, as such,  cannot be associated to a symmetry generator.
However, two new bosonic one-form fields $B^{ab}$, $B^{a_1 \ldots
a_5}$ and one fermionic $\eta^\alpha$ can be introduced to express
$A_3$, together with the former one-forms $e^a$, $\psi^\alpha$, as
a composite of these one-forms. All these one-forms can be
associated to Maurer-Cartan one-forms of a one parameter family of
Lie superalgebras. These MC forms are defined on the corresponding
group manifolds, or enlarged rigid superspaces. They are to be
interpreted as describing the underlying gauge symmetry of $D=11$
supergravity: the symmetry is hidden when $A_3$ is considered as a
fundamental field, but becomes manifest when $A_3$ is treated as a
composite field.

Free differential algebras (FDAs) are brought into the picture to
deal with this problem. FDAs \cite{Su77,D'A+F,Ni83,Cd'AF91} are a
natural generalization of (the dual point of view of) Lie
algebras, containing $p$-forms $\pi_p$ of rank $p > 1$. For a
particular case of FDAs (the minimal ones), the differentials of
the higher rank $p$-forms $\pi_p$ are nontrivial
Chevalley-Eilenberg (CE) cocycles $\omega_{p+1}$ of a certain Lie
algebra ${\cal G}$. If there exists another ({\it larger}) algebra
$\tilde{\cal G}$ in terms of the Maurer-Cartan forms of which the
non-trivial cocycles $\omega_{p+1}$ for ${\cal G}$ become trivial,
then the MC forms of $\tilde{\cal G}$ will allow us to express
$\pi_p$ as composites of them.  The problem of the composite
structure of $A_3$ and the underlying symmetry of supergravity
fits naturally in this language and, in fact, it is further
studied \cite{D'A+F,Lett,AnnP04} using these arguments.

The possible dynamical consequences of a composite $A_3$ are also
analyzed in chapter \ref{chapter6}, by substituting its composite
expression into the supergravity first order action. The equations
of motion of the new fields are shown to imply those of $A_3$, but
now considering the later as composed of them, rather than as
fundamental. Although the new fields carry more degrees of freedom
than $A_3$ does, the formulation of supergravity due to  D'Auria
and Fr\'e can be regarded as dynamically equivalent to the
standard CJS formulation, since there exist gauge symmetries that
make these extra degrees of freedom pure gauge. The chapter ends
emphasizing how enlarged superspaces can be found for some
theories such that their dimension coincides with the number of
fields present in the theory: the gauge structure of $D=11$
supergravity is an example of this extended superspace
coordinates/fields correspondence \cite{JdA00}.

In chapter \ref{chapter7}, enlarged superspaces are again used.
The relevant superspace there is actually the group manifold of
the M Theory superalgebra and its generalizations with $n$
fermionic and $\ft12 n(n+1)$ bosonic coordinates (also called
`maximal', `maximally enlarged' or `tensorial' superspace).
Extended objects in enlarged superspaces are known to provide
models for preonic objects; this is, in fact, our motivation for
the study of this system. A model for a supersymmetric tensionful
string moving in maximally enlarged superspace is proposed in
chapter \ref{chapter7}, which can be interpreted as a higher spin
generalization of the Green-Schwarz superstring. The model neither
involves Dirac matrices nor has a Wess-Zumino (WZ) term. Instead,
it is formulated in terms of two bosonic spinors, a counterpart of
the preonic spinors introduced in chapter \ref{chapter4}. The
claim \cite{30/32} that (in $D=11$) the model preserves 30
supersymmetries out of 32 (or, in general, $n-2$ out of $n$) is
supported by the fact that it possess 30 $\kappa$-symmetries, in
spite of its lack of a WZ term. The number of bosonic and
fermionic degrees of freedom of the model is worked out resorting
to a hamiltonian analysis, which can be simplified with the use of
orthosymplectic supertwistors. The chapter concludes with an
extension of these ideas to the construction of super-$p$-branes
models in maximally enlarged superspaces.

Chapter \ref{chapter8} contains our conclusions. Some technical
details are relegated to the appendices.


\chapter{Eleven-dimensional supergravity} \label{chapter2}

The general framework in which the rest of the Thesis is developed
is introduced in this chapter, devoted to the review of a number
of topics about $D=11$ Cremmer-Julia-Scherk (CJS) supergravity in
order to fix the conventions and notation used in most of the
subsequent chapters. After describing in section
\ref{sec:MTsuperalg} the supersymmetry algebras and groups
relevant for supergravity (with particular emphasis on the
eleven-dimensional case), the action principle of the theory is
introduced in section \ref{firstaction}, in a first order
formalism that turns out to be convenient for subsequent
developments. The symmetries of the action are also discussed in
that section, devoting the following section \ref{susyhol} to the
supersymmetry of the theory and to the related notions of
generalized connection, curvature and holonomy. The equations of
motion are described both in general, in section \ref{eom}, and in
the purely bosonic limit, in section \ref{eqd11s}. Finally, in
section \ref{curvandeom} the equations of motion of
eleven-dimensional supergravity are shown to be encoded in the
generalized curvature even when the gravitino is non-vanishing
\cite{BAPV05}.

\section{The M Theory superalgebra} \label{sec:MTsuperalg}

Eleven-dimensional CJS supergravity \cite{CJS} is the locally
supersymmetric field theory based on the (only) massless
supermultiplet of the superPoicar\'e group  in eleven spacetime
dimensions containing fields up to helicity two \cite{Nahm78}. Not
only the supergravity multiplet is unique in $D=11$, but the
theory does not allow modifications such as the presence of a
cosmological constant \cite{BaDeHeSe97}. It is thus worth starting
the discussion about eleven-dimensional supergravity taking a look
at the symmetry algebra on which it is based. The $D=11$
superPoincar\'e (or standard supersymmetry) algebra\footnote{The
symbol $\rtimes$ will be used throughout, either denoting
semidirect {\it sum} (when used in a Lie {\it algebra} context) or
semidirect {\it product} (when used in a Lie {\it group}
context).} $\mathfrak{E} \rtimes so(1,10)$ is made up of the usual
supertranslations algebra $\mathfrak{E}$ acted on semidirectly by
the Lorentz algebra in eleven dimensions. The supertranslations
algebra $\mathfrak{E}$ exponentiates into the supertranslations
group $\Sigma$, the group manifold of which (also denoted
$\Sigma$) corresponds to rigid superspace. In spacetime dimension
$D$, the even part of the ($N$-extended) supertranslations algebra
$\mathfrak{E}$ is generated by  $D$ bosonic (translation)
generators $P_a$, $a=1, \ldots , D$, and the odd part by $N$
fermionic (supertranslation) generators, or supercharges,
$Q^i_\alpha$ with $n$ components, $i=1, \ldots , N$,
$\alpha=1,\ldots, n$. The number $n$ of components of each
supercharge is that of the minimal spinor in spacetime dimension
$D$, and the number $N$ of supercharges has an upper bound
depending on $D$ (see below). In eleven dimensions, $D=11$, there
is only one supercharge, $N=1$, that is a Majorana spinor with
$n=32$ components. Accordingly, we shall usually assume $N=1$, in
which case $D$ and $n$ are the bosonic and fermionic dimensions of
superspace. When the dimensions are needed explicitly, to avoid
confusion it will be written $\mathfrak{E}^{(D|n)}$ (for the
superalgebra) and $\Sigma^{(D|n)}$ (for the supergroup, or
superspace); in eleven dimensions, thus, the supertranslations
algebra is $\mathfrak{E}^{(11|32)}$ and superspace is
$\Sigma^{(11|32)}$. As a supergroup, the $\Sigma^{(11|32)}$
superspace can be regarded as a central extension\footnote{See
section \ref{fourw} in chapter \ref{chapter5} for a brief review
of Lie algebra extensions.} \cite{JdA00} by the generator $P_a$ of
the abelian fermionic translations group $\Sigma^{(0|32)}$
generated by $Q_\alpha$. Further extensions and enlargements of
the algebra are possible, as we shall shortly see.

As for the structure of the superPoincar\'e algebra, the
supertranslations (anti)commutation relations defining
$\mathfrak{E}$ are
\begin{eqnarray}
\label{QQ=P} {} \{ Q_\alpha , Q_\beta \} = \Gamma^a_{\alpha\beta}
P_a \; , \quad [P_a, Q_\alpha]=0 \; , \quad [P_a, P_b]=0 \; ,
\end{eqnarray}
where $\Gamma^a{}_\alpha{}^\beta$ are $32 \times 32$
eleven-dimensional Dirac matrices defining the Clifford algebra
$Cl(1,10)$,
\begin{equation} \label{Diracclifford}
\{\Gamma_a , \Gamma_b \} = 2 \eta_{ab} \ I_{32} \; ,
\end{equation}
$\eta_{ab}$ being the Minkowski metric and $I_{32}$ the $32 \times
32$ identity matrix. The spinor indices are raised and lowered
with the $32 \times 32$ skewsymmetric charge conjugation matrix
$C_{\alpha \beta}$, which is understood in (\ref{QQ=P}):
$\Gamma^a_{\alpha\beta} \equiv \Gamma^a{}_\alpha{}^\gamma
C_{\gamma \beta}$. The Lorentz group $SO(1,10)$ has generators
$J_{ab}$, and its corresponding algebra $so(1,10)$ is defined by
the commutation relations
\begin{eqnarray} \label{Lorentz}
[J_{ab}, J^{cd}] = -4 J_{[a}{}^{[c} \delta_{b]}{}^{d]} \;  .
\end{eqnarray}
Its semidirect action on the supertranslations algebra
$\mathfrak{E}$ is given by
\begin{eqnarray} \label{Lorentztrans}
[J_{ab} , Q_\alpha]  = \ft14 \Gamma_{ab\; \alpha}{}^\beta Q_\beta
\; , \quad [J_{ab} , P_c] = 2 \eta_{c[a} P_{b]} \; ,
\end{eqnarray}
where $\Gamma^{ab}$ is the antisymmetrized product of two Dirac
matrices of the Clifford algebra $Cl(1,10)$; in general,
\begin{equation}
\Gamma^{a_1 \ldots a_k} := \Gamma^{[a_1} \cdots \Gamma^{a_k]} \; ,
\end{equation}
where the brackets denote antisymmetrization with weight one. The
set of (an\-ti)com\-mu\-ta\-tors (\ref{QQ=P}), (\ref{Lorentz}),
(\ref{Lorentztrans}) defines the superPoincar\'e algebra
$\mathfrak{E} \rtimes so(1,10)$.

The number of supersymmetries of a supergravity theory in any
dimension must be at most 32, if interacting fields of spin higher
than 2 are to be avoided\footnote{See section \ref{models} of
chapter \ref{chapter7}, and references therein, for some remarks
about higher spin theories.}. This is the requirement that places
an upper bound depending on the number of components $n$ of the
minimal spinor in dimension $D$.  The 32-component supercharge
$Q_\alpha$ of $D=11$ supergravity makes its equations display
maximal supersymmetry; the maximally supersymmetric supergravity
in four dimensions has, instead, $N=8$ 4-component supercharges
$Q_\alpha^i$, $i=1, \ldots,  8$, $\alpha=1, \ldots , 4$, so that
it also displays 32 supersymmetries. $N$-extended supergravities
in lower dimensions allow for supersymmetry algebras with a richer
structure, since new generators commuting with the rest of the
superPoincar\'e generators, and consequently called {\it central
charges}, can be introduced on the right-hand-side of the
anticommutator of two supercharges (the first equation in
(\ref{QQ=P})). Eleven-dimensional supergravity has only one
supercharge, $N=1$, but extensions nevertheless do exist
generalizing the anticommutator of two supercharges.

In fact, the anticommutator in (\ref{QQ=P}) is symmetric in the
spinor indices $(\alpha \beta)$ and takes values on the (even part
$Cl(1,10)_+$ of the) Clifford algebra $Cl(1,10)$ generated by
\begin{equation} \label{generatorscliff}
\{ I_{32}, \Gamma^{[1]}, \Gamma^{[2]}, \Gamma^{[3]}, \Gamma^{[4]},
\Gamma^{[5]} \} \; ,
\end{equation}
where the shorthand notation $\Gamma^{[k]}$ has been used to
denote generically the antisymmetrized products of Dirac matrices,
$\Gamma^{[k]} \equiv \Gamma^{a_1 \ldots a_k}$. In eleven
dimensions and Lorentzian signature, the matrices
$\Gamma^{[1]}_{\alpha \beta} \equiv (\Gamma^{[1]} C)_{\alpha
\beta}$, $\Gamma^{[2]}_{\alpha \beta} \equiv (\Gamma^{[2]}
C)_{\alpha \beta}$ and $\Gamma^{[5]}_{\alpha \beta} \equiv
(\Gamma^{[5]} C)_{\alpha \beta}$ are symmetric in $(\alpha
\beta)$, whereas the rest in (\ref{generatorscliff}) are
skewsymmetric. The standard supertranslations algebra
$\mathfrak{E}$ can accordingly be extended by adding two more
antisymmetric tensorial generators $Z_{ab} = Z_{[ab]}$, $Z_{a_1
\ldots a_5}= Z_{[a_1 \ldots a_5]}$ to the right-hand-side of the
anticommutator of two supercharges \cite{vHvP82}:
\begin{equation} \label{eq:extsuper}
\{ Q_\alpha , Q_\beta \} = \Gamma^a_{\alpha\beta} P_a +
i\Gamma^{ab}_{\alpha\beta} Z_{ab} + \Gamma^{a_1\ldots
a_5}_{\alpha\beta} Z_{a_1\ldots a_5} \; .
\end{equation}
The generators $Z_{ab}$, $Z_{a_1 \ldots a_5}$ commute among
themselves and with the rest of supertranslation generators. They
are, thus, central if the Lorentz group is ignored and, in fact,
are also called central charges.

The extension (\ref{eq:extsuper}) of the standard
supertranslations algebra $\mathfrak{E} \equiv
\mathfrak{E}^{(11|32)}$ by the bosonic generators $Z_{ab}$,
$Z_{a_1 \ldots a_5}$ gives the superalgebra
$\mathfrak{E}^{(528|32)}$, with generators
\begin{equation}
 \, P_a  \, , \, Q_\alpha  \, , \,  Z_{a_1a_2}  \, ,
\, Z_{a_1 \ldots a_5}   \; ,
\end{equation}
and bosonic dimension ${11 \choose 1} +  {11 \choose 2} + {11
\choose 5}= 11 +55 +462 = 528$. Being maximally extended (in the
bosonic sector\footnote{Further extensions are, however, possible
in the fermionic sector: see chapter \ref{chapter6}.}),
$\mathfrak{E}^{(528|32)}$ generalizes the superPoincar\'e algebra
in eleven dimensions and is usually called the {\it M Theory
superalgebra} or {\it M-algebra}\footnote{See
\cite{Se97,JdA00,D'A+F} for further generalizations of the M
Theory superalgebra and for their structure.} \cite{M-alg} (see
\cite{vHvP82,Se97,JdA00}). Its associated group manifold
$\Sigma^{(528|32)}$ corresponds to the maximally extended rigid
superspace. The bosonic generators $P_a$, $Z_{ab}$ and $Z_{a_1
\ldots a_5}$ can be collected in a {\it generalized momentum}
$P_{\alpha \beta} = P_{\beta \alpha}$ generator,
\begin{equation} \label{n32}
P_{\alpha \beta} = \Gamma^a_{\alpha\beta} P_a +
i\Gamma^{a_1a_2}_{\alpha\beta} Z_{a_1a_2} + \Gamma^{a_1\ldots
a_5}_{\alpha\beta} Z_{a_1\ldots a_5} \; ,
\end{equation}
in terms of which the (anti)commutation relations of the M Theory
superalgebra $\mathfrak{E}^{(528|32)}$ can be written succinctly
as
\begin{eqnarray} \label{QQP}
\{ Q_\alpha , Q_\beta \} = P_{\alpha \beta} \; , \quad [Q_\alpha ,
P_{\beta \gamma}] = 0 \; .
\end{eqnarray}

In terms of the generalized momentum $P_{\alpha \beta}$, these
(anti)commutation relations (\ref{QQP}) exhibit a $GL(32,
\mathbb{R})$ automorphism symmetry. When the decomposition
(\ref{n32}) is used to write $P_{\alpha \beta}$ in terms of Dirac
matrices, the $GL(32, \mathbb{R})$ automorphism symmetry is
reduced down to the Lorentz group $SO(1,10)$. In some applications
(see chapter \ref{chapter4}), it is interesting to consider the
maximal automorphism group of the M Theory superalgebra, $GL(32,
\mathbb{R})$. For other developments, however, it is convenient to
consider the reduced automorphism group $SO(1,10)$ for the M
algebra since, after all, the supergravity equations only display
a local Lorentz symmetry. The semidirect sum
$\mathfrak{E}^{(528|32)} \rtimes so(1,10)$ becomes, then, the
counterpart of the superPoincar\'e algebra $\mathfrak{E}^{(11|32)}
\rtimes so(1,10)$ in the presence of additional tensorial central
charges. In chapter \ref{chapter5}, the M Theory superalgebra with
$SO(1,10)$ automorphisms will be revisited in connection with the
orthosymplectic superalgebra $osp(1|32)$ and shown to be an {\it
expansion} of this group \cite{JdAIMO}.

The M Theory superalgebra contains complete information about the
non-perturbative BPS states of the hypothetical underlying M
Theory: the additional bosonic generators $Z_{ab}$, $Z_{a_1\ldots
a_5}$ of the M-algebra (\ref{QQP}) are related to the topological
charges \cite{JdAT} of the supermembrane  and the
super-M5-brane\footnote{This result was extended in \cite{Hull97}
by showing that these generators also contain a contribution from
the topological charges of the eleven-dimensional Kaluza-Klein
monopole ($Z_{0\mu_1\ldots \mu_4}\propto \epsilon_{0\mu_1\ldots
\mu_4 \nu_1\ldots \nu_6} \tilde{Z}^{\nu_1\ldots \nu_6}$) and of
the M9-brane ($Z_{0\mu }\propto \epsilon_{0\mu \nu_1\ldots \nu_9}
\tilde{Z}^{\nu_1\ldots \nu_9}$) which is usually identified with
the Ho\v rava-Witten hyperplane \cite{HW96} (for the Kaluza-Klein
monopole and the M9 brane only bosonic actions are known
\cite{BergshoeffKK,BergshoeffM9}).} \cite{ST97} (see also
\cite{H-JPaSm03}). These `single brane' BPS states can be
associated with $D=11$ supergravity solutions
\cite{Duff94,Stelle98} or with fundamental M Theory objects
described by their worldvolume actions \cite{BST,blnpst}. Although
the M-algebra (\ref{QQP}) leads naturally to a $D=11$
Lorentz-covariant interpretation when the splitting (\ref{n32}) is
used, it also allows both for a IIA and a IIB treatment. In the
first case, this is allowed because the (relevant) Dirac matrices
coincide in ten and eleven dimensions; in the IIB case, a
counterpart of equation (\ref{n32}) \cite{M-alg,Bars97} can be
written if the spinor indices $\alpha$ are split as $\alpha^\prime
i$, where $\alpha^\prime=1, \ldots , 16$ is a $D=10$ Majorana-Weyl
spinor index and $i=1,2$. As a result, the information about
non-perturbative BPS states of the $D=10$ superstring theories
(including D-branes) can also be
 extracted from the algebra (\ref{QQP}). This means that the
 M-algebra also encodes all the duality relations between different $D=10$
and $D=11$ superbranes. These facts add further reasons to call
(\ref{QQP}) the M Theory superalgebra \cite{M-alg}.

\vspace*{1.5em}

To conclude this section, let us write, for future reference, the
dual version of the algebras introduced above. It is usually
convenient to resort to a dual point of view to deal with Lie
algebras, especially to construct lagrangians invariant or
quasi-invariant ({\it i.e.}, invariant up to a total derivative)
under the symmetry transformations of the Lie algebra. This dual
point of view will be particularly relevant in chapters
\ref{chapter5} and \ref{chapter6}. Let $G$ be a Lie group with
parameters $g^i$, $i=1, \ldots , \textrm{dim} G$, and ${\cal G}$
its Lie algebra, generated by the (left-, say) invariant vector
fields $X_i(g)$ on the group manifold $G$, with commutation
relations $[ X_i , X_j ] =c_{ij}^k X_k$. The coalgebra ${\cal
G}^*$ is then spanned by the dual ($\omega^i (X_j) = \delta^i_j$),
left-invariant Maurer-Cartan (MC) one-forms $\omega^i(g)$ on the
group manifold $G$, subject to the Maurer-Cartan equations
\begin{equation} \label{MCchap2}
d \omega^k = -\ft12 c_{ij}^k \omega^i \wedge \omega^j \; ,
\end{equation}
which contain the same information than the commutation relations
in terms of generators $X_i$. In particular, the Jacobi identities
$c^k_{i[j} c^i_{lm]}=0$ arise from the requirement that the MC
equations (\ref{MCchap2}) be consistent with the nilpotency of the
exterior differential, $dd \equiv 0$.

Introducing the MC one-forms $\Pi^a$, $\pi^\alpha$, $\sigma^{ab}$
dual, respectively, to the generators $P_a$, $Q_\alpha$, $J_{ab}$,
the superPoincar\'e group can be described by the MC equations
{\setlength\arraycolsep{0pt}
\begin{eqnarray}\label{eq:MCsuperP}
&& d\Pi^a = \Pi^b\wedge \sigma_b{}^a  - i\pi^{\alpha} \wedge
\pi^{\beta} \Gamma^a_{\alpha\beta}  \; , \nonumber  \\
&& d\pi^\alpha = \pi^\beta \wedge \sigma_\beta{}^\alpha \quad
\left( \sigma_\alpha{}^\beta=\ft14 \sigma^{ab} \Gamma_{ab\;
\alpha}{}^\beta \right) \; , \nonumber
\\
&& d \sigma^{ab}  =  \sigma^{ac} \wedge \sigma_c{}^{b} \; ,
\end{eqnarray}
}offering a counterpart of the (anti)commutation relations
(\ref{QQ=P}), (\ref{Lorentz}), (\ref{Lorentztrans}). Finally,
introducing the MC forms $\Pi^{ab}$, $\Pi^{a_1 \ldots a_5}$ dual,
respectively, to the generators $Z_{ab}$, $Z_{a_1 \ldots a_5}$,
the whole set
\begin{equation}
 \, \Pi^a  \, , \, \pi^\alpha  \, , \,  \Pi^{a_1a_2}  \, ,
\, \Pi^{a_1 \ldots a_5}   \; ,
\end{equation}
provides, setting aside the automorphisms part, the Maurer-Cartan
one-forms of the M Theory superalgebra $\mathfrak{E}^{(528|32)}$,
left-invariant on the corresponding group manifold (maximally
extended rigid superspace) $\Sigma^{(528|32)}$. The one forms
$\Pi^a$, $\Pi^{ab}$, $\Pi^{a_1 \ldots a_5}$ can again be collected
into the symmetric spin-tensor one-form
{\setlength\arraycolsep{0pt}
\begin{eqnarray}
 \label{cEff=def}
\Pi^{\alpha\beta}&=& \ft1{32} \left(\Pi^a \Gamma_{a}^{\alpha\beta}
- \ft{i}{2} \Pi^{a_1a_2}\Gamma_{a_1a_2}{}^{\alpha\beta} +
\ft{1}{5!} \Pi^{a_1\ldots a_5} \Gamma_{a_1\ldots
a_5}{}^{\alpha\beta} \right)\; ,
\end{eqnarray}
}dual to $P_{\alpha \beta}$, in terms of which the MC equations of
the M Theory superalgebra, containing the same information as the
(anti)commutation relations (\ref{QQP}), can be written in compact
form as {\setlength\arraycolsep{0pt}
\begin{eqnarray} \label{eq:MCMTsuperalg}
 d\Pi^{\alpha\beta}=- i \pi^\alpha  \wedge \pi^\beta \; , \quad
d\pi^\alpha=0  \; .
\end{eqnarray}

}

\section{First order action of $D=11$ supergravity
}\label{firstaction}

The field content of $N$-extended supergravity in $D$ dimensions
is determined by the so-called supergravity multiplet, determined
by the massless representation of the corresponding
superPoincar\'e algebra containing fields up to helicity two. In
particular, for the construction of the eleven-dimensional
supergravity action, the central charges $Z_{ab}$, $Z_{a_1 \ldots
a_5}$ can be ignored. The fields involved in $D=11$ supergravity
\cite{CJS} are, specifically, a Lorentzian metric (corresponding
to the graviton) $g_{\mu\nu}$, a three-form $A_3$ and a Majorana
Rarita-Schwinger field $\psi^\alpha$ (the gravitino). Actually,
the presence of spinor fields makes it necessary to work in the
vielbein approach, in which the metric is replaced by a vielbein
field $e_\mu^a$ in tangent space, satisfying $g_{\mu\nu}= e_\mu^a
e_\nu^b \eta_{ab}$, where $\eta_{ab}$ is the Minkowski metric.
Except in chapter \ref{chapter3}, a {\it mostly minus} signature
for the metric will be used throughout this Thesis. In the
component approach, these fields are to be regarded as forms on
eleven-dimensional spacetime\footnote{We shall be concerned with
the spacetime component formulation of supergravity. For a review
of the superspace formulation of supergravity, see {\it e.g.}
\cite{AnnP04}.} $M^{11}$, {\setlength\arraycolsep{2pt}
\begin{eqnarray}\label{formssugra}
 e^a(x)  & = &  dx^\mu e_\mu^a(x)\; , \nonumber \\*
\psi^\alpha(x) & = & dx^\mu \psi_\mu^\alpha(x) \equiv e^a
\psi_a^\alpha(x) \; , \nonumber \\*
 A_3(x) & = & \ft1{3!} \ dx^{\mu_1} \wedge dx^{\mu_2} \wedge
dx^{\mu_3} A_{\mu_3\mu_2\mu_1}(x) \nonumber \\* & \equiv  &
\ft1{3!} \ e^{a_1} \wedge e^{a_2} \wedge e^{a_3} A_{a_3a_2a_1}(x)
\; .
\end{eqnarray}
}Notice the `superspace' reverse order convention for the
components of the $p$-forms. The differential $d$ will be taken to
act from the right,
\begin{equation}
d \alpha_p = \ft{1}{p!} d x^{\mu_p} \wedge \ldots \wedge d
x^{\mu_1} \wedge d x^{\nu} \partial_\nu \alpha_{\mu_1 \ldots
\mu_p} \; .
\end{equation}
As usual in supersymmetric theories, the number of bosonic and
fermionic degrees of freedom match. In $D=11$, $e^a$ has
$\frac{(D-2)(D-1)}{2}-1 =44$ on-shell degrees of freedom which,
together with the ${D-2 \choose 3}=84$ on-shell degrees of freedom
provided by $A_3$, makes up 128 bosonic on shell degrees of
freedom. That is the same number of on-shell degrees of freedom of
fermionic character, associated to the gravitino $\psi^\alpha$,
namely, $\frac12 2^{[D/2]}(D-3)=128$.

In addition to the forms (\ref{formssugra}), the first order
action for $D=11$ supergravity \cite{D'A+F,J+S99},
{\setlength\arraycolsep{0pt}
\begin{eqnarray}\label{S11=}
S=\int_{M^{11}}{\cal L}_{11}[e^a, \psi^\alpha, A_3, \omega^{ab},
F_4] \; ,
\end{eqnarray}
}contains the Lorentz connection one-form $\omega^{ab}$ and the
auxiliary four-form\footnote{The first order formulation of
\cite{D'A+F} involved no four-form $F_4$ but a tensor zero-form
$F_{a_1 \ldots a_4}$. The later can actually be replaced
throughout by its contraction with four vielbeins to give an $F_4$
and, hence, both formulations are equivalent.} $F_4$,
{\setlength\arraycolsep{2pt}
\begin{eqnarray}\label{formssugraaux}
 \omega^{ab}(x) & = & dx^\mu \omega_\mu^{ab}(x)\; , \nonumber \\
 F_4(x) & = & \ft1{4!} \ dx^{\mu_1} \wedge dx^{\mu_2} \wedge
dx^{\mu_3} \wedge dx^{\mu_4} F_{\mu_4\mu_3\mu_2\mu_1}(x)
\nonumber \\
&\equiv & \ft1{4!} \  e^{a_1} \wedge e^{a_2} \wedge e^{a_3} \wedge
e^{a_4} F_{a_4a_3a_2a_1}(x) \; ,
\end{eqnarray}
}that must be treated as independent fields in the variational
problem, and acquire their usual, second order formalism values
when considered on shell (see section \ref{eom}). Notice that the
on-shell counting of degrees of freedom coincides in the first and
second order formalisms, since the auxiliary fields in the former
become, on shell, functions of the fields defining the later.

The action (\ref{S11=}) is defined on eleven-dimensional spacetime
$M^{11}$ and the lagrangian that determines it reads
\cite{D'A+F,J+S99} {\setlength\arraycolsep{2pt}
\begin{eqnarray}\label{L11=}
{\cal L}_{11} &=& \ft14  R^{ab}\wedge e^{\wedge 9}_{ab} -
D\psi^\alpha \wedge \psi^\beta   \wedge
\bar{\Gamma}^{(8)}_{\alpha\beta} \nonumber \\* && + \ft14
\psi^\alpha \wedge \psi^\beta   \wedge (T^a + \ft{i}{2} \, \psi
\wedge \psi \, \Gamma^a) \wedge e_a \wedge
\bar{\Gamma}^{(6)}_{\alpha\beta}   \nonumber \\* && + (dA_3- a_4)
\wedge (\ast F_4 + b_7) +  \ft12 a_4 \wedge b_7 \nonumber \\* && -
\ft12 F_4 \wedge \ast F_4  - \ft13 A_3 \wedge dA_3\wedge dA_3 \; .
\end{eqnarray}
}Both  the Riemann tensor and the torsion,
{\setlength\arraycolsep{0pt}
\begin{eqnarray}
\label{RL=def} && R^{ab} := d\omega^{ab} - \omega^{ac}\wedge
\omega_c{}^b \; , \\
 \label{Torsion} && T^a := De^a =de^a-e^b\wedge
\omega_b{}^a \;
\end{eqnarray}
}(where, in the last equation, $D$ is the standard Lorentz
covariant derivative) enter the first order lagrangian, the
earlier in the Einstein-Hilbert term (the first one in the r.h.s.
of (\ref{L11=})) characteristic of any gravitational lagrangian.
Together with the curvature of $A_3$, these curvatures
(\ref{RL=def}), (\ref{Torsion}) are the basic ingredients of the
free differential algebra approach to $D=11$ supergravity (see
chapter \ref{chapter6}).

The derivative acting on the gravitino in its kinetic term, the
second of (\ref{L11=}), is again the Lorentz covariant derivative,
{\setlength\arraycolsep{2pt}
\begin{eqnarray}\label{hDpsiLor}
D \psi^\alpha &:=& d\psi^\alpha - \psi^\beta \wedge
\omega_\beta{}^\alpha  \; ,
\end{eqnarray}
}now defined in terms of the spin connection,
\begin{equation} \label{spincon}
\omega_\beta{}^\alpha = \ft14 \omega^{ab}
(\Gamma_{ab})_\beta{}^\alpha \; ,
\end{equation}
taking values on $so(1,10)$, the Lie algebra of the double cover
of the eleven-dimensional Lorentz group, $Spin(1,10)$, generated
by $\Gamma^{ab}$.

{}Following \cite{J+S99} (see also \cite{BAPV05,AnnP04}), we have
introduced in the lagrangian (\ref{L11=}) the notation
{\setlength\arraycolsep{2pt}
\begin{eqnarray}\label{a4}
a_4&:=& \ft12 \psi^\alpha \wedge \psi^\beta   \wedge
\bar{\Gamma}^{(2)}_{\alpha\beta}\; ,  \\
\label{b7}   b_7 &:=& \ft{i}{2} \psi^\alpha \wedge \psi^\beta
\wedge \bar{\Gamma}^{(5)}_{\alpha\beta} \; ,
\end{eqnarray}
}for the bifermionic 4- and 7-forms built up from the gravitino,
as well as the compact notation {\setlength\arraycolsep{2pt}
\begin{eqnarray}
\label{Gammak} \bar{\Gamma}^{(k)}_{\alpha\beta} &:=& \ft{1}{k!}
e^{a_k} \wedge \ldots \wedge e^{a_1} \Gamma_{a_1 \ldots a_k}
{}_{\alpha\beta} \; .
\end{eqnarray}
}Finally, $\ast F_4$ is the Hodge dual of $F_4$,
{\setlength\arraycolsep{0pt}
\begin{eqnarray}\label{F4:=}
&& \ast F_4:= - \ft{1}{4!} e^{\wedge 7}_{a_1\ldots a_4}
F^{a_1\ldots a_4} \; ,
\end{eqnarray}
}and the $(11-k)$-form {\setlength\arraycolsep{2pt}
\begin{eqnarray}
\label{E11-n} e^{\wedge (11-k)}_{a_1\ldots a_k} & := &
\ft{1}{(11-k)!} \epsilon_{a_1\ldots a_kb_1\ldots b_{11-k}}
e^{b_1}\wedge \ldots \wedge e^{b_{11-k}} \;
\end{eqnarray}
}has been introduced for convenience\footnote{See \cite{BAPV05}
for the correspondence of this notation to that of \cite{J+S99}.}.

As for the symmetries of the action, it should be noted that the
usual general covariance of any gravitational action is
implemented in this formalism by using differential forms to write
the lagrangian (\ref{L11=}). Local Lorentz symmetry is also
straightforwardly implemented in the vielbein approach. The action
is also invariant under abelian gauge symmetries of the three-form
$A_3$, and locally supersymmetric, as we now discuss.

\section{Supersymmetry and generalized holonomy} \label{susyhol}

The action (\ref{S11=}), (\ref{L11=}) is locally supersymmetric,
{\it i.e.}~it is invariant under the following local supersymmetry
transformations $\delta_{\epsilon}$ parameterized by a fermionic
$Spin(1,10)$-spinor parameter $\epsilon(x)$:
{\setlength\arraycolsep{0pt}
\begin{eqnarray}
\label{susye} && \delta_{\epsilon} e^a = - 2i  {\psi}^\alpha
\Gamma^a_{\alpha\beta}{\epsilon}^\beta \; ,
\\
\label{susyf} && \delta_{\epsilon}\psi^\alpha = {\cal
D}\epsilon^\alpha(x) \; ,
\\
\label{susyA} && \delta_{\epsilon}A_3 = \psi^\alpha \wedge
\bar{\Gamma}^{(2)}_{\alpha\beta}{\epsilon}^\beta \; ,
\end{eqnarray}
}besides more complicated expressions for
$\delta_{\epsilon}\omega^{ab}$ and $\delta_{\epsilon}F_{abcd}$,
which can be found in \cite{J+S99} and that will not be needed
below.  Let us stress that, as shown in \cite{J+S99}, the
supersymmetry transformation rules of the physical fields are the
same in the second and in the first order formalisms. The
transformations (\ref{susye})--(\ref{susyA}) have the usual form
expected from supersymmetry: the bosonic fields $e^a$ and $A_3$
transform into the (only, in this case) fermionic field
$\psi^\alpha$ which, in turn, transforms into $e^a$ and $A_3$
(included, on shell, inside the supercovariant derivative ${\cal
D}$: see below). Precisely, the transformation (\ref{susyf}) in
terms of the supersymmetric covariant derivative ${\cal D}$ is
characteristic of locally realized supersymmetry, and allows for
an interpretation of the gravitino as the {\it gauge field} of
local supersymmetry.

The introduction of the {\it generalized covariant} (or {\it
supersymmetric covariant}, or {\it supercovariant}) {\it
derivative} ${\cal D}$ allows for a simple expression for the
transformation rule (\ref{susyf}) of the gravitino under
supersymmetry. It can be written explicitly as
{\setlength\arraycolsep{2pt}
\begin{eqnarray}
\label{susypsi=} \delta_{\epsilon}\psi^\alpha = {\cal
D}\epsilon^\alpha(x)&:=& D\epsilon^\alpha(x)
-\epsilon^\beta(x) t_{\beta}{}^\alpha(x) = \nonumber \\
&=& d\epsilon^\alpha(x) - \epsilon^\beta(x)
\Omega_\beta{}^\alpha(x) \; ,
\end{eqnarray}
}in terms of the {\it generalized} (or {\it supersymmetric}) {\it
connection} one-form {\setlength\arraycolsep{0pt}
\begin{eqnarray}\label{CJSom}
&& \Omega_\beta{}^\alpha = \ft14 \omega^{ab}
\Gamma_{ab}{}_\beta{}^\alpha + \ft{i}{144} e^a \left(
\Gamma_{ab_1b_2b_3b_4}{}_\beta{}^\alpha + 8 \ \delta_{a[b_1}
\Gamma_{b_2b_3b_4]}{}_\beta{}^\alpha \right) F^{b_1b_2b_3b_4} \; ,
\nonumber \\ &&
\end{eqnarray}
}that differs from the spin connection $\omega_\alpha{}^\beta=
\ft14 \omega^{ab}\Gamma_{ab}{}_\beta{}^\alpha$ (equation
(\ref{spincon})) by the additional tensor one-form
{\setlength\arraycolsep{2pt}
\begin{eqnarray} \label{t}
t_\beta{}^\alpha &=& \ft{i}{144} e^a \left(
\Gamma_{ab_1b_2b_3b_4}{}_\beta{}^\alpha + 8 \ \delta_{a[b_1}
\Gamma_{b_2b_3b_4]}{}_\beta{}^\alpha \right) F^{b_1b_2b_3b_4},
\end{eqnarray}
}depending on the auxiliary form $F_4$ (which, on-shell, reduces
to the supercovariant field strength of $A_3$; see equation
(\ref{dA3=a+F=}) below).

A connection one-form takes values on the Lie algebra ${\cal G}$
of the {\it structure group} $G$ of a fiber bundle (see, {\it
e.g.} \cite{AI95}). The spin connection $\omega_\alpha{}^\beta$,
for instance, takes values on the Lie algebra $so(1,10)$ of the
structure group $Spin(1,10)$ of the spin bundle on
eleven-dimensional spacetime $M^{11}$. It is, therefore, natural
to ask what is the {\it generalized structure group}, on the Lie
algebra of which the generalized connection
$\Omega_\alpha{}^\beta$ takes values \cite{Duff03}. To this end,
notice that when $F_4=0$ then $t_\alpha{}^\beta=0$, and the
generalized connection (\ref{CJSom}) reduces to the
$so(1,10)$-valued spin connection (\ref{spincon}). But, in
general, $F_4 \neq 0$ and $t_\alpha{}^\beta$, as defined in
(\ref{t}), is non-vanishing. In this case, the presence of
additional Dirac matrices makes the generalized connection to take
values not on $so(1,10)$, but on the whole $Cl(1,10)_+$ generated
by the antisymmetrized products of Dirac matrices in
(\ref{generatorscliff}), namely,  $\{ I_{32}, \Gamma^{[1]},
\Gamma^{[2]}, \Gamma^{[3]}, \Gamma^{[4]}, \Gamma^{[5]} \}$. The
dimension of the relevant even part $Cl(1,10)_+$ of the Clifford
algebra is
\begin{equation}
\textrm{dim} \ Cl(10,1)_+ = {11 \choose  0} + {11 \choose  1} +
{11 \choose  2} + {11 \choose  3} + {11 \choose  4} + {11 \choose
5} =1024 \; .
\end{equation}

The problem can still be analyzed in terms of Lie algebras,
though. In fact, when $Cl(1,10)_+$ is endowed with the usual Lie
bracket provided by matrix commutation, $[ A  , B  ] =AB-BA$, it
coincides with $gl(32, \mathbb{R})$, the Lie algebra of the
general linear group $GL(32, \mathbb{R})$, of dimension
$\textrm{dim} \ gl(32, \mathbb{R}) = 32^2= 1024$. The Lie algebra
$so(1,10)$, generated by $\Gamma^{[2]}$, on which the spin
connection (\ref{spincon}) takes values, is a subalgebra of
$gl(32, \mathbb{R})$. Similarly, one may wonder what is the Lie
subalgebra of $gl(32, \mathbb{R})$ on which the generalized
connection $\Omega_\alpha{}^\beta$ takes values. An explicit
computation reveals that the generators $\{ \Gamma^{[2]},
\Gamma^{[3]}, \Gamma^{[5]} \}$ defining $\Omega_\alpha{}^\beta$ in
equation (\ref{CJSom}) do not close into a Lie algebra by
themselves, and that the presence of $\{ \Gamma^{[1]},
\Gamma^{[4]} \}$ (not that of $I_{32}$) is also required to ensure
closure under commutation \cite{Hull03}. In conclusion, the
generalized connection $\Omega_\alpha{}^\beta$ takes values on the
1023--dimensional Lie subalgebra of $gl(32, \mathbb{R})$ spanned
by its traceless generators,
\begin{equation}
\{ \Gamma^{[1]}, \Gamma^{[2]}, \Gamma^{[3]}, \Gamma^{[4]},
\Gamma^{[5]} \} \; .
\end{equation}
These are the generators of $sl(32, \mathbb{R})$, the Lie algebra
of $SL(32, \mathbb{R})$ which is, therefore, to be interpreted as
the {\it generalized structure group} of $D=11$ supergravity
\cite{Hull03}.

A remark about terminology is now in order. The local symmetry of
$D=11$ supergravity is not $SL(32, \mathbb{R})$; as mentioned in
the previous section, it is only $SO(1,10)$. In this sense, the
generalized connection $\Omega_\alpha{}^\beta$ as defined by
equation (\ref{CJSom}) may be said not to be a bona fide
connection. However, it reduces to the spin connection when
$F_4=0$ and, in a sense, generalizes it when $F_4$ is
non-vanishing. Moreover, the role played by the Riemannian
holonomy of the spin connection $\omega_\alpha{}^\beta$ in the
classification of purely geometrical supersymmetric bosonic
solutions of supergravity (for which the metric is the only
non-vanishing field) can be taken over by its generalized
counterpart $\Omega_\alpha{}^\beta$ when $F_4$ is turned on
\cite{Duff03,DuffStelle91,Hull03} (see chapter \ref{chapter3}).
This adds further reasons to call $\Omega_\alpha{}^\beta$
generalized connection.

Pushing this analogy further, the curvature two-form of the
generalized connection $\Omega_\alpha{}^\beta$,
{\setlength\arraycolsep{2pt}
\begin{eqnarray}\label{calR}
{\cal R}_\beta{}^{\alpha} &:=& d \Omega_\beta{}^{\alpha} -
\Omega_\beta{}^{\gamma}\wedge  \Omega_\gamma{}^{\alpha} \nonumber \\
&=& \ft14 R^{ab}(\Gamma_{ab})_{\alpha}{}^{\beta} +
Dt{}_{\alpha}{}^{\beta} - t_{\alpha}{}^\gamma \wedge
t_{\gamma}{}^{\beta} \; , \qquad
\end{eqnarray}
}can be introduced, and consequently referred to as the {\it
generalized curvature}\footnote{A full expression for the
generalized curvature ${\cal R}_{\alpha}{}^{\beta}$ corresponding
to purely bosonic solutions of CJS supergravity can be found  in
\cite{BEWN83,GP02}.}. In general, the curvature two-form of a
connection $w$ takes values on a subalgebra ${\cal H} \equiv
\hol(w)$ of the Lie algebra ${\cal G}$ of the structure group $G$.
The corresponding group $H \equiv \Hol (w)$ is a subgroup of $G$
and is called the {\it holonomy group} (of the connection $w$);
its corresponding Lie algebra $\hol(w)$ will sometimes be called
the holonomy algebra. Accordingly, the {\it generalized holonomy
group} $\Hol (\Omega)$ \cite{Duff03} (see also
\cite{DuffStelle91,Hull03,BDLW03,P+T03,P+T031,FFP02,BW,Duff:2002rw,BLVW03,BAIPV03}\footnote{See
\cite{LPS05} for the role of generalized holonomy when vanishing
$F_4$ is considered but higher order corrections to the
supergravity equations of motion are taken into account.}) is the
subgroup of $SL(32, \mathbb{R})$ on the Lie algebra of which the
generalized curvature ${\cal R}_\alpha{}^\beta$ takes values. In
general, however, the curvature {\it at a point} does not
determine completely the Lie algebra of the holonomy group, but
its successive covariant derivatives are needed to determine it
(see, {\it e.g.} \cite{Besse}). Generalized holonomy is no
exception \cite{BLVW03}, as shown in chapter \ref{chapter3}. See
also section \ref{Killingspin} for the role of generalized
holonomy in the determination of the number of supersymmetries
preserved by a bosonic solution of supergravity.

\section{Equations of motion}\label{eom}

\subsubsection{Algebraic equations}

Let us return to the analysis of the first order action of $D=11$
supergravity, to obtain the equations of motion. The variations of
the action (\ref{S11=}), (\ref{L11=}) with respect to the Lorentz
connection $\omega^{ab}$ and the auxiliary four-form $F_4$ give
algebraic constraints that can be used to define these auxiliary
fields in terms of those of the supergravity multiplet
(\ref{formssugra}). Indeed, from the variation of (\ref{S11=}),
(\ref{L11=}) with respect to the Lorentz connection,
{\setlength\arraycolsep{0pt}
\begin{eqnarray}\label{varS=Ta}
&& {\delta {S} \over \delta{\omega}^{ab}} = \ft14 e^{\wedge
8}_{abc} \wedge (T^c + i\psi^{\alpha}\wedge \psi^{\beta}\,
\Gamma^c_{\alpha\beta}) = 0 \; ,
\end{eqnarray}
}the torsion is seen to be given by
\begin{eqnarray} \label{STa=}
T^a = - i\psi^{\alpha}\wedge \psi^{\beta}\, \Gamma^a_{\alpha\beta}
\; ,
\end{eqnarray}
which, upon use of its definition (\ref{Torsion}), gives an
algebraic equation for the Lorentz connection $\omega^{ab}$, which
allows us to solve for it in terms of the vielbein and the
gravitino.

On the other hand, the variation of the action with respect to
$F_4$, {\setlength\arraycolsep{2pt}
\begin{eqnarray}\label{deFS}
\delta_{F} S &=& \int_{M^{11}} (dA_3- a_4-F_4) \wedge \ast \delta
F_4 = \nonumber
\\ &=& - \ft1{4!} \int_{M^{11}}   (dA_3- a_4-F_4) \wedge
e^{\wedge 7}_{a_1\ldots a_4} \, \delta  F^{a_1\ldots a_4}\; ,
\end{eqnarray}
}produces an algebraic equation of motion, $\delta S/\delta
F_4=0$, that makes of  $F_4$  the `supersymmetric' field strength
of $A_3$,
\begin{equation} \label{dA3=a+F=}
dA_3=a_4+F_4 \; .
\end{equation}
Making use of the expressions (\ref{STa=}) and (\ref{dA3=a+F=})
for the torsion $T^a$ and four-form $F_4$ into the first order
lagrangian (\ref{L11=}), the original second order CJS lagrangian
\cite{CJS} is recovered.

\subsubsection{Dynamical equations}

The variation of (\ref{S11=}) with respect to the rest of the
fields yields the dynamical equations of motion: the Einstein
equations (arising from the variation with respect to $e^a$), the
(generalization of the Maxwell) equation for $A_3$ and the
Rarita-Schwinger equation for the gravitino $\psi^\alpha$.

The explicit form of the Einstein equations,
\begin{eqnarray}
 \label{EqmE}
M_{10\, a}  :=  R^{bc} \wedge e^{\wedge 8}_{abc} + \ldots =0 \; ,
\end{eqnarray}
will not be needed in the remainder, so we will not be concerned
with it\footnote{See \cite{J+S99} for the explicit expression of
the Einstein equation in this formalism.}. The variation of the
action with respect to the three-form $A_3$,
\begin{eqnarray}\label{defcG}
\delta_{A} S= \int_{M^{11}} {\cal G}_8 \wedge \delta A_3 \; ,
\qquad {\delta S \over \delta A_3} := {\cal G}_8 \; ,
\end{eqnarray}
gives the eight-form
\begin{eqnarray}\label{defcG=}
{\cal G}_8 = d( \ast F_4 +b_7 - A_3\wedge dA_3) \; ,
\end{eqnarray}
and thus the equation of motion of $A_3$ is
\begin{eqnarray} \label{EqmA3}
{\cal G}_8 = d( \ast F_4 +b_7 - A_3\wedge dA_3) =0 \; .
\end{eqnarray}

Finally, the fermionic variation of the lagrangian (\ref{L11=})
reads ({\it cf.} \cite{J+S99}) {\setlength\arraycolsep{2pt}
\begin{eqnarray}
\label{vpsiL} \delta_{\psi} {\cal L}_{11} &=& -2 {\cal D}
\psi^\alpha \wedge \bar{\Gamma}^{(8)}_{\alpha\beta}  \wedge \delta
\psi^\beta
 + i (dA_3- a_4-F_4) \wedge
\bar{\Gamma}^{(5)}_{\alpha\beta}  \wedge \psi^\alpha \wedge \delta
\psi^\beta  \nonumber \\ && +  \left( i_a
\bar{\Gamma}^{(8)}_{\alpha\beta}  + \ft12 e_a \wedge
\bar{\Gamma}^{(6)}_{\alpha\beta} \right) \wedge (T^a +
i\psi^{\alpha}\wedge \psi^{\beta}\, \Gamma^a_{\alpha\beta}) \wedge
\psi^\alpha \wedge  \delta \psi^\beta   \nonumber \\
&& - d\; \left[ \psi^\alpha \wedge
\bar{\Gamma}^{(8)}_{\alpha\beta} \wedge \delta \psi^\beta \right]
\; ,
\end{eqnarray}
}where $i_a$ is defined by   $i_a e^b=\delta_a^b$ so that, for
$\alpha_p={1\over p!} e^{a_p} \wedge \ldots \wedge
e^{a_1}\alpha_{a_1\ldots a_p}$,
\begin{equation} \label{innerprod}
i_a\alpha_p=\ft{1}{(p-1)!} e^{a_{p}} \wedge \ldots \wedge
e^{a_2}\alpha_{aa_2\ldots a_{p}} \quad .
\end{equation}
Imposing the algebraic constraints (\ref{STa=}) and
(\ref{dA3=a+F=}) and ignoring the total derivative term, equation
(\ref{vpsiL}) gives the gravitino equation of \cite{CJS} written,
as in \cite{J+S99}, in differential form,
\begin{eqnarray}
 \label{Eqmpsi} \Psi_{10\; \beta}  := {\cal D}
\psi^\alpha \wedge \bar{\Gamma}^{(8)}_{\alpha\beta}  =0 \; ,
\end{eqnarray}
in terms of the supercovariant derivative
\begin{eqnarray}\label{hDpsi}
{\cal D} \psi^\alpha := d\psi^\alpha - \psi^\beta \wedge
\Omega_\beta{}^\alpha \equiv  D\psi^\alpha - \psi^\beta \wedge
t_\beta{}^\alpha \; ,
\end{eqnarray}
 defined for  the generalized connection (\ref{CJSom}).

\section{The purely bosonic limit}\label{eqd11s}

For many applications, it is interesting to consider the purely
bosonic limit of $D=11$ supergravity in which the gravitino
vanishes, $\psi^\alpha =0$. A torsion-free spacetime is then
recovered (see equation (\ref{STa=})), described by the Einstein
equations (\ref{EqmE}) which, in this limit, reduce to
\begin{equation}\label{EiCJS}
E_{ab}:=  \textrm{Ric}_{ab} - \ft13
F_{ac_1c_2c_3}F_{b}{}^{c_1c_2c_3}
 + \ft{1}{36} \eta_{ab}F_{c_1c_2c_3c_4}F^{c_1c_2c_3c_4}
=0\; ,
\end{equation}
where $\textrm{Ric}_{ab}$ is the Ricci tensor. The equation of
motion (\ref{EqmA3}) of $A_3$ reduces to
\begin{equation} \label{GFCJS}
{\cal G}_8 =  d*F_4 - F_4\wedge F_4 = 0 \; .
\end{equation}
The four-form $F_4$ that enters both (\ref{EiCJS}) and
(\ref{GFCJS}) reduces, by virtue of the algebraic constraint
(\ref{dA3=a+F=}), to the field strength of $A_3$, $F_4=dA_3$;
consequently, it is subject to the Bianchi identity
\begin{equation}\label{BIGFCJS}
dF_4 \equiv  0 \; .
\end{equation}

Interestingly enough, these bosonic equations are encoded in the
generalized curvature ${\cal R}_\alpha{}^\beta$ of the generalized
connection $\Omega_\alpha{}^\beta$, now still given by equations
(\ref{calR}) and (\ref{CJSom}), respectively, but setting
$\psi^\alpha=0$ (and, consequently, $F_4=dA_3$) in them. With
these restrictions, ${\cal R}_{\alpha}{}^{\beta}$ obeys
\cite{GP02,BAIPV03} {\setlength\arraycolsep{2pt}
\begin{eqnarray}\label{bEqm.}
{\cal N}_{\beta}{}^{\alpha} :=  i_a {\cal
R}_{\alpha}{}^{\gamma}\Gamma^a{}_{\gamma}{}^{\beta} &=& -
\ft{1}{4}  e^b R_{b[c_1 c_2c_3]}
\Gamma^{c_1c_2c_3}{}_\alpha{}^\beta + \ft{1}{2}
e^a  E_{ab}  \Gamma^b{}_{\alpha}{}^{\beta}  \nonumber \\
&& +  \ft{i}{36}  e^a\, [*{\cal G}_8]_{b_1b_2b_3}
(\Gamma_{a}{}^{b_1b_2b_3} + 6
\delta_{a}^{[b_1} \Gamma^{b_2b_3]})_{\alpha}{}^{\beta}   \nonumber \\
&& + \ft{i}{720} \, e^a\, [dF_4]_{b_1\ldots b_5}\,
(\Gamma_{a}{}^{b_1\ldots b_5} + 10 \delta_{a}^{[b_1}
\Gamma^{b_2\ldots b_5]})_{\alpha}{}^{\beta} \; , \nonumber \\ &&
\qquad
\end{eqnarray}
}where $E_{ab}$, ${\cal G}_8$ are the  r.h.s's of the Einstein and
the gauge field equations as defined in (\ref{EiCJS}),
(\ref{GFCJS}) and $i_a$ is defined in (\ref{innerprod}); in
particular, $i_a {\cal R}_{\alpha}{}^{\beta}=e^b {\cal R}_{ab
\alpha}{}^\beta$. The equality (\ref{bEqm.}) implies that the set
of the  free bosonic equations for CJS supergravity,
(\ref{EiCJS}), (\ref{GFCJS}), (\ref{BIGFCJS}), is equivalent to
the following simple equation for the generalized curvature
(\ref{calR}), $e^b {\cal R}_{ab\alpha}{}^\gamma
\Gamma^a{}_\gamma{}^\beta=0$, or
\begin{equation}\label{EqCJS=0}
i_a {\cal R}_{\alpha}{}^{\gamma}\; \Gamma^a{}_{\gamma}{}^{\beta}
=0\; ,
\end{equation}
since the r.h.s.~of equation (\ref{bEqm.}) is zero on account of
the equations of motion (\ref{EiCJS}), (\ref{GFCJS}) and the
Bianchi identities for $F_4$ (equation (\ref{BIGFCJS})) and for
the Riemann tensor, $R_{b[c_1 c_2c_3]} =0$.

Especially relevant are those solutions of the purely bosonic
equations (\ref{EiCJS}), (\ref{GFCJS}), (\ref{BIGFCJS}) of $D=11$
supergravity preserving some supersymmetry (see section
\ref{Killingspin} of chapter \ref{chapter3} for the conditions
that a purely bosonic supergravity solution must meet in order to
preserve supersymmetry). In eleven dimensions, supergravity
displays the maximum amount, 32, of supersymmetries permitted (see
section \ref{sec:MTsuperalg}) and hence, a supersymmetric bosonic
solution of $D=11$ supergravity preserves a number $k$ of
supersymmetries between 1 and 32. A supersymmetric solution can be
referred to by the fraction of preserved supersymmetry as a
$\nu=k/32$ solution.

The $\nu= 1/2$-su\-per\-sym\-me\-tric solutions are usually
regarded as the basic solutions of $D=11$ supergravity\footnote{On
the other hand, states with $\nu=31/32$ ({\it BPS preons}
\cite{BPS01}) can be argued, on purely algebraic grounds, to be
fundamental in M Theory: see sec.~\ref{sec:BPSpreons} of chapter
\ref{chapter4}.}. These are the M-wave \cite{Hull84}, the
Kaluza-Klein monopole \cite{Gross:1983hb,Sorkin:1983ns,Hull97} and
the elementary brane solutions, namely, the M2-brane
\cite{DuffStelle91} and the M5-brane \cite{Gu92} (the existence of
an M9-brane has been conjectured in \cite{BergshoeffM9,Hull97}).
See also \cite{Duff94,Stelle98} and references therein.  The
M2-brane solution\footnote{See equations (\ref{M2eqs}) and
(\ref{M5eqs}) of chapter \ref{chapter3} for the expressions of the
metric and four-form corresponding, respectively, to the M2- and
M5-brane solutions of $D=11$ supergravity; that section,
\ref{sec:Mbranes}, also discusses the generalized holonomy of
these brane solutions.} solves the Einstein equation
${E}_{ab}={\cal T}_{ab} - {1\over 9} \eta_{ab} \, {\cal
T}_{c}{}^c$ with a singular energy-momentum tensor density source
${\cal T}_{ab} \propto \delta^3(x -\hat{x}(\xi))$ ($\hat{x}(\xi)$
being pulled-backed on the M2-brane worldvolume, parameterized by
coordinates $\xi$). The gauge field equation also receives a
singular source contribution $J_8$ in the r.h.s., ${\cal G}_8 =
J_8$, similar to that of the electric current to the r.h.s.~of
Maxwell equations. In this sense, the M2-brane carries a
supergravity counterpart of the electric charge in Maxwell
electrodynamics (see \cite{Duff91} for a discussion). The other
basic $\nu=1/2$ brane solution of $D=11$ supergravity, the
M5-brane, is a counterpart of the Dirac monopole, {\it i.e.} of
the magnetically charged particle. It is characterized by a
modification of the Bianchi identities (equation (\ref{BIGFCJS}))
with the analogue of a magnetic current in the r.h.s., $dF_4 =
{\cal J}_5$.

Intersecting branes preserve less than one-half of the maximum
supersymmetry, {\it i.e.}, they correspond to $\nu < 1/2$
supergravity solutions \cite{interscbranes}. On the other hand,
there also exist maximally supersymmetric solutions ($\nu =1$)
preserving, thus, all 32 supersymmetries. Four solutions exhaust
the complete list of $\nu=1$ solutions of $D=11$ supergravity
\cite{FFP02}: eleven-dimensional Minkowski space, the
compactifications $AdS_4 \times S^7$, $AdS_7 \times S^4$
\cite{FR,Pilch:1984xy} on round-spheres and the pp-wave of
\cite{KowGlikwave}. In spite of the fact that the supersymmetry
algebra allowed, in principle, for all the fractions $\nu =k /32$,
$k=1, \ldots , 32$, to be preserved \cite{GauntHull00}, no
explicit solutions (other than those maximally supersymmetric)
preserving more that $\nu=1/2$ were known for some time. Solutions
with extra supersymmetry were found indeed as pp-waves
\cite{CLP02,GH02,BJ02,Michelson} or G\"odel universes
\cite{Gauntlett:2002nw,Harmark:2003ud}, preserving $k=18, 20, 22,
24, 26$ and (in IIB supergravity) $28$ supersymmetries.

Since the supergravity multiplet is the only one without higher
spin fields in $D=11$, no usual field-theoretical matter
contribution to the r.h.s.'s of the equations of motion
(\ref{EiCJS}), (\ref{GFCJS}), (\ref{BIGFCJS}) may appear.
Modifications to the equations might arise, however, not only due
to the presence of the branes just mentioned, but also if higher
order corrections to the curvature \cite{W00,Howe03,LPST,LPS05} (a
counterpart of the string $\alpha^\prime$ corrections \cite{Zw85}
in $D=10$) are taken into account. These corrections should have
an M Theoretical interpretation.

\section{Equations of motion and generalized curvature} \label{curvandeom}

Let us now return to the general case of non-vanishing gravitino,
$\psi^\alpha \neq 0$, and  show that there exists a counterpart of
equation (\ref{EqCJS=0}) collecting the equations of motion of the
bosonic fields in terms of the generalized curvature
\cite{BAPV05}. The gravitino equation of motion (\ref{Eqmpsi}),
$\Psi_{10\; \beta}=0$, is expressed in terms of the supercovariant
derivative ${\cal D}$ of $\psi^\alpha$ (equation (\ref{hDpsi})),
defined in terms of the generalized connection
$\Omega_\alpha{}^\beta$ (equations (\ref{CJSom}), (\ref{t})). As a
result, the integrability/selfconsistency condition for equation
(\ref{Eqmpsi}) may be written in terms of the generalized
curvature ${\cal R}_\alpha{}^\beta$ of equation (\ref{calR}).
Using ${\cal D} {\cal D} \psi^\alpha = - \psi^\beta \wedge {{\cal
R}}_\beta{}^\alpha$ and\footnote{This follows  from direct
calculation: $t^{\quad\gamma}_{1\, \alpha} \wedge
\bar{\Gamma}^{(8)}_{\gamma\beta} = -{i\over 2} F_4 \wedge
\bar{\Gamma}^{(5)}_{\alpha\beta}+ {1\over 2} \ast F_4 \wedge
\bar{\Gamma}^{(2)}_{\alpha\beta}$.} $t^{\quad\gamma}_{1[\beta}
\wedge \bar{\Gamma}^{(8)}{}_{\alpha]\gamma} = 0$ which implies
${\cal D} \bar{\Gamma}^{(8)}_{\beta\alpha} = {D}
\bar{\Gamma}^{(8)}_{\beta\alpha} = T^a \wedge i_a
\bar{\Gamma}^{(8)}_{\beta\alpha}$, we obtain
{\setlength\arraycolsep{0pt}
\begin{eqnarray}
\label{DEqmpsi} && {\cal D} \Psi_{10\; \alpha} = {\cal D}
\psi^\beta \wedge (T^a + i \psi \wedge \psi \Gamma^a) \wedge i_a
\bar{\Gamma}^{(8)}_{\beta\alpha} - \nonumber
\\ && \qquad  - \ft{i}{6} \psi^\beta \wedge \left[ {\cal R}_\beta{}^\gamma
\wedge e^{\wedge 8}_{abc} \Gamma^{abc}_{\gamma\alpha}
 + i {\cal D} \psi^\delta \wedge \psi^\gamma \wedge
e^{\wedge 7}_{a_1\ldots a_4} \Gamma^{[a_1a_2a_3}_{\delta\alpha}
\Gamma^{a_4]}_{\beta\gamma}\right] = 0\; . \nonumber \\ &&
\end{eqnarray}
}The first term in the second part of equation (\ref{DEqmpsi})
vanishes due to the algebraic constraint (\ref{STa=}). Hence on
the surface of constraints, the selfconsistency of the gravitino
equation is guaranteed when \cite{BAPV05}
\begin{eqnarray}
\label{SfEq} {\cal M}_{10\; \alpha\beta} := {\cal
R}_\beta{}^\gamma \wedge e^{\wedge 8}_{abc}
\Gamma^{abc}_{\gamma\alpha} + i {\cal D} \psi^\delta \wedge
\psi^\gamma \wedge  e^{\wedge 7}_{a_1\ldots a_4}
 \Gamma^{[a_1a_2a_3}_{\delta\alpha} \Gamma^{a_4]}_{\beta\gamma} = 0 \; . \qquad
\end{eqnarray}
As it will now be shown, {\it equation (\ref{SfEq}) collects all
the equations of motion of the bosonic fields, (\ref{EqmE}),
(\ref{EqmA3}), and the corresponding Bianchi identities for the
$A_3$ gauge field and for the Riemann curvature tensor}
\cite{BAPV05}. Equation (\ref{SfEq}) is, thus, the counterpart of
equation (\ref{EqCJS=0}) when the gravitino is non-vanishing. Let
us stress that we distinguish between the algebraic equations or
constraints (equations (\ref{STa=}) and (\ref{dA3=a+F=})) from the
true dynamical equations ((\ref{EqmE}), (\ref{EqmA3})) and that
our statement above refers to the dynamical equations; thus it is
also true for the second order formalism.

To show this it is sufficient to use the second Noether theorem
and/or the fact that the purely bosonic limit of (\ref{SfEq})
implies equation (\ref{EqCJS=0}), which is equivalent to the set
of all bosonic equations and Bianchi identities when
$\psi^\alpha=0$. According to the second Noether theorem, the
local supersymmetry under (\ref{susye})--(\ref{susyA}) reflects
(and is reflected by) the existence of an interdependence among
the bosonic and fermionic equations of motion; such a relation is
called a Noether identity. Furthermore, since the local
supersymmetry variation of the gravitino (\ref{susyf}) is given by
the supercovariant derivative ${\cal D} \epsilon^\alpha$, the
gravitino equation $\Psi^\alpha$ should enter the corresponding
Noether identity through ${\cal D} \Psi^\alpha$. Thus, ${\cal
D}\Psi^\alpha$ should be expressed in terms of the equations of
motion for the bosonic fields, in our case including the algebraic
equations for the auxiliary fields. Hence, due to the equations
(\ref{DEqmpsi}), (\ref{STa=}), the l.h.s.~of equation (\ref{SfEq})
vanishes when {\it all} the bosonic equations are taken into
account.

Indeed, schematically, ignoring for simplicity the purely
algebraic equations and neglecting the boundary contributions, the
variation of the action (\ref{S11=}), (\ref{L11=}) (considered now
in the second order formalism) reads
\begin{eqnarray}
\label{vS11=} \delta S = \int_{M^{11}} \left(-2\Psi_{10\, \alpha}
\wedge \delta  \psi^\alpha + {\cal G}_8 \wedge \delta A_3 + M_{10
\, a} \wedge \delta e^a \right) \; . \qquad
\end{eqnarray}
For the local supersymmetry transformations $\delta_{\epsilon}$,
equations (\ref{susye})--(\ref{susyA}), one finds,
 integrating by parts
{\setlength\arraycolsep{2pt}
\begin{eqnarray}
\label{vsusyS11=0} \delta_{\epsilon} S &=& \! \int_{M^{11}}
\!\left(-2\Psi_{10\, \alpha} \wedge {\cal D} {\epsilon}^\alpha +
{\cal G}_8 \wedge \delta_{\epsilon}A_3 + M_{10 \, a} \wedge
\delta_{\epsilon} e^a \right) =  \nonumber \\ &=&  - \int_{M^{11}}
(-2{\cal D}\Psi_{10\, \alpha} -  {\cal G}_8 \wedge
 \psi^\beta \wedge
\bar{\Gamma}^{(2)}_{\beta\alpha}
  +2i  M_{10 \, a} \wedge
 \psi^\beta  \Gamma^{a}_{\beta\alpha}
) \, \epsilon^\alpha  =0 \, . \nonumber \\
\end{eqnarray}
}Since  $\delta_{\epsilon} S=0$ is satisfied for an arbitrary
fermionic function $\epsilon^\alpha(x)$, it follows that
\begin{eqnarray}
\label{DPsi=1} {\cal D}\Psi_{10\, \alpha} = -\ft{1}{2} \psi^\beta
\wedge \left( - 2i \Gamma^{a}_{\beta\alpha} M_{10 \, a} +  {\cal
G}_8 \wedge \bar{\Gamma}^{(2)}_{\beta\alpha} \right) \; . \quad
\end{eqnarray}

By virtue of equations (\ref{DEqmpsi}) and (\ref{DPsi=1}), and
after the algebraic equations (\ref{STa=}), (\ref{dA3=a+F=})  are
taken into account, {\setlength\arraycolsep{2pt}
\begin{eqnarray}
\label{SfEq=} {\cal M}_{10\; \alpha\beta} &:=&   {\cal
R}_\beta{}^\gamma \wedge e^{\wedge 8}_{abc}
\Gamma^{abc}_{\gamma\alpha} + i {\cal D} \psi^\delta \wedge
\psi^\gamma \wedge  e^{\wedge 7}_{a_1\ldots a_4}
 \Gamma^{[a_1a_2a_3}_{\delta\alpha} \Gamma^{a_4]}_{\beta\gamma} =   \nonumber \\
 & & =-3i \left( - 2i \Gamma^{a}_{\beta\alpha}
M_{10 \, a} +  {\cal G}_8 \wedge \bar{\Gamma}^{(2)}_{\beta\alpha}
\right) \; . \qquad
\end{eqnarray}
}It then follows that the equation of motion for the bosonic
fields (\ref{SfEq}), ${\cal M}_{10\; \alpha\beta}=0$, is
satisfied,
\begin{equation}
\label{SfEq0.} {\cal R}_\beta{}^\gamma \wedge e^{\wedge 8}_{abc}
\Gamma^{abc}_{\gamma\alpha} = -i {\cal D} \psi^\delta \wedge
\psi^\gamma \wedge  e^{\wedge 7}_{a_1\ldots a_4}
\Gamma^{[a_1a_2a_3}_{\delta\alpha} \Gamma^{a_4]}_{\beta\gamma} \;
,
\end{equation}
after the dynamical equations (\ref{EqmE}), (\ref{EqmA3}) are
used. Getting rid of the vielbein forms, the equation (\ref{SfEq})
(or (\ref{SfEq0.})) can be written in terms of the components
${\cal R}_{ab}{}_{\alpha}{}^{\beta}$, $({\cal
D}\psi)_{ab}{}^\alpha$ of the two-forms ${\cal
R}_{\alpha}{}^{\beta}$, ${\cal D} \psi^\alpha$,
{\setlength\arraycolsep{0pt}
\begin{eqnarray} \label{compcalR}
&& {\cal R}_{\alpha}{}^{\beta} = \ft12 e^b \wedge e^a {\cal
R}_{ab}{}_{\alpha}{}^{\beta} \; , \\
&& {\cal D}\psi^\alpha = \ft12 e^b \wedge e^a ({\cal
D}\psi)_{ab}{}^\alpha \; ,
\end{eqnarray}
}as
\begin{equation} \label{SfEq1.}
{\cal R}_{bc}{}_{\alpha}{}^{\gamma} \Gamma^{abc}{}_{\gamma\beta} =
4i (({\cal D}\psi)_{bc}\Gamma^{[abc})_\beta \,
(\psi_d\Gamma^{d]})_\alpha \; .
\end{equation}

Equation (\ref{SfEq=}) also shows what Lorentz-irreducible parts
of the concise bosonic equations ${\cal M}_{10\; \alpha\beta}=0$
coincide with the Einstein and with the 3-form gauge field
equations. These are given, respectively,  by
{\setlength\arraycolsep{0pt}
\begin{eqnarray}
\label{SfEq=Ei}
 && M_{10 \, a}= -\ft{1}{192} \mathrm{tr} (\Gamma_a{\cal M}_{10}) \; ,
 \qquad \\ \label{SfEq=GF} && {\cal G}_8 \wedge e^a \wedge e^b
 = \ft{i}{96} \mathrm{tr} (\Gamma^{ab} {\cal M}_{10}) \;  . \qquad
\end{eqnarray}
}All other Lorentz-irreducible parts in equation (\ref{SfEq}),
${\cal M}_{10\; \alpha\beta}=0$, are satisfied either identically
or due to the Bianchi identities that are the integrability
conditions for the algebraic equations (\ref{STa=}),
(\ref{dA3=a+F=}) used in the derivation of (\ref{SfEq=}).

In conclusion, we have proven  that  equation (\ref{SfEq})
collects all the dynamical bosonic equations of motion in the
second order approach to supergravity. To see that it collects all
the Bianchi identities as well, one may either perform a direct
calculation or study its purely bosonic limit. The latter way is
simpler and it also provides an alternative proof of the above
statement as we now show.

For bosonic configurations, $\psi^\alpha=0$, equation (\ref{SfEq})
reduces to
\begin{eqnarray}
\label{SfEq0}  {\cal R}_\beta{}^\gamma \wedge e^{\wedge 8}_{abc}
\Gamma^{abc}_{\gamma\alpha} =  0 \; .
\end{eqnarray}
Decomposing ${\cal R}_\alpha{}^\beta$ on the vielbein basis as in
(\ref{compcalR}), equation (\ref{SfEq0}) implies
\begin{eqnarray}
\label{EqcRb1} {\cal R}_{ab\; \beta}{}^\gamma
\Gamma^{abc}_{\gamma\alpha} =0 \; .
\end{eqnarray}
Contracting (\ref{EqcRb1}) with $\Gamma_{c}^{\alpha\delta}$ one
finds
\begin{eqnarray}
\label{EqcRb2} {\cal R}_{ab\; \beta}{}^\gamma
\Gamma^{ab}{}_{\gamma}{}^{\delta} =0 \; .
\end{eqnarray}
Then, contracting again with the Dirac matrix
$\Gamma_{d}^{\alpha\delta}$ and using $\Gamma^{ab}\Gamma_{d}=
\Gamma^{ab}{}_{d} + 2 \Gamma^{[a}\delta_d{}^{b]}$  as well as
equation (\ref{EqcRb1}), one recovers equation (\ref{bEqm.}),
${\cal N}_{a\beta}{}^\alpha=0$, namely,
\begin{eqnarray}
\label{SfEq01}   i_a {\cal R}_\beta{}^\gamma
\Gamma^{a}_{\gamma}{}^{\alpha} \equiv e^b {\cal
R}_{ab\beta}{}^\gamma \Gamma^{a}_{\gamma}{}^{\alpha} =  0 \; .
\end{eqnarray}
Since (\ref{SfEq01}) collects all the bosonic equations of $D=11$
CJS supergravity as well as all the Bianchi identities in the
purely bosonic limit \cite{GP02,BAIPV03}, $\psi^\alpha =0$, the
equivalence of equations (\ref{SfEq01}) and (\ref{SfEq0})
 will imply that
${\cal M}_{10\; \alpha\beta}=0$, equation (\ref{SfEq}), does the
same for the case of non-vanishing fermions, $\psi^\alpha\not=0$
\cite{BAPV05}.

The Bianchi identities $R_{a[bcd]}\equiv 0$ and $dF_4\equiv 0$
appear as the irreducible parts
$\textrm{tr}(\Gamma_{c_1c_2c_3}{\cal N}_a)$ and
$\textrm{tr}(\Gamma_{c_1\ldots c_5}{\cal N}_a)$ of equation
(\ref{bEqm.}) \cite{BAPV05}; more precisely, in the later case the
relevant part in  ${\cal N}_a$ is proportional to
$[dF_4]_{b_1\ldots b_5} (\Gamma_a{}^{b_1\ldots b_5} + 10
\delta_a{}^{[b_1} \Gamma^{b_2\ldots b_5]})$, but the two terms in
the brackets are independent. Knowing this, one may also reproduce
the  terms that include the Bianchi identities in the concise
equation (\ref{SfEq0.}) (equivalent to (\ref{SfEq1.}) or
(\ref{SfEq=})) with a non-vanishing gravitino.


\chapter{Subtleties about generalized holonomy}
\label{chapter3}

The generalized holonomy of some solutions of eleven-dimensional
supergravity is reviewed in this chapter. It is done by paying
particular attention to a feature of holonomy already mentioned in
section \ref{susyhol}, namely, that covariant derivatives of the
curvature might be needed to define the Lie algebra of the
holonomy group. In section \ref{Killingspin}, the supersymmetry
transformations discussed in general in chapter \ref{chapter2} are
particularized for purely bosonic solutions of supergravity. The
Killing spinor equation that results as a consistency condition
from the vanishing of the gravitino variation is presented and the
usefulness of the integrability conditions of the equation
exhibited. These (first order) integrability conditions are
related to the generalized curvature. In section \ref{higherinteg}
it is argued that ordinary, first order integrability is in
general not enough to characterize the holonomy, and that iterated
commutators of the supercovariant derivatives  may be needed to
properly define the holonomy algebra.

To check for possible consequences of the higher order
integrability conditions, the generalized holonomy of the usual
M-branes is reviewed in section \ref{sec:Mbranes}. It is found
that, in these cases, successive commutators of the supercovariant
derivatives only help to close the algebra obtained at first order
(the curvature algebra) and that successive commutators do not add
significant information. The situation is, however, different for
other supergravity solutions: as section \ref{squashed} shows,
second order integrability conditions are necessary to compute the
generalized holonomy of Freund-Rubin compactifications. Knowledge
of the embedding of the generalized holonomy group in the
generalized structure group is, moreover, needed to determine
correctly the number of preserved supersymmetries. Some details
are relegated to Appendix \ref{holS7}.

This chapter follows closely reference \cite{BLVW03}, and uses the
conventions therein. In particular, we temporarily resort to a
mostly plus metric $g_{MN}$, $M, N, \ldots$, denoting
eleven-dimensional spacetime indices ($\mu , \nu , \ldots ,$ and
$a, b, \ldots ,$ will be reserved for lower dimensions) . With
these conventions, the generalized connection (\ref{CJSom}) will
be denoted $\Omega_M$ and its associated supercovariant derivative
will act from the left and will be defined as
\begin{equation}
\D_M\equiv\partial_M +\ft14\Omega_M=
D_M-\ft1{288}(\Gamma_M{}^{NPQR}-8\delta_M^N\Gamma^{PQR}) F_{NPQR},
\label{eq:gencd}
\end{equation}
where $D_M$ denotes the L\'evi-Civita covariant derivative
associated to the spin connection $\omega_M$. The purely
bo\-so\-nic equations of motion (\ref{EiCJS}) and (\ref{GFCJS})
will read in this chapter: {\setlength\arraycolsep{0pt}
\begin{eqnarray} \label{Einstein} &&
\textrm{Ric}_{MN}=\ft1{12}\left(F_{MPQR}F_N{}^{PQR}-\ft1{12}g_{MN}
F^{PQRS}F_{PQRS}\right),\\
&& d*F_4+\ft12 F_4 \wedge F_4=0. \label{4form}
\end{eqnarray}
}Spinor indices will be omitted and derivatives will act from the
left.

\section{Killing spinors, holonomy and supersymmetry}
\label{Killingspin}

For purely bosonic supergravity solutions, $\psi_M=0$, the
supersymmetry transformations simplify considerably. The bosonic
fields, $e^a$ and $A_3$, of such a solution are clearly invariant
under supersymmetry, {\setlength\arraycolsep{0pt}
\begin{eqnarray}
&& \delta_{\epsilon} e^a = 0 \; ,
\\
&& \delta_{\epsilon}A_3 = 0 \; ,
\end{eqnarray}
}since their transformation rules, (\ref{susye}) and
(\ref{susyA}), respectively,  are proportional to a vanishing
gravitino. On the other hand, the invariance of the bosonic
solution under supersymmetry implies, in particular, that the
solution cannot change its bosonic character after the
transformation, {\it i.e.}, that no gravitino is generated by the
transformation. This amounts to requiring that the variation
(\ref{susyf}) of the gravitino under supersymmetry also vanishes.
Namely, with the convention of (\ref{eq:gencd}),
\begin{equation} \label{eq:kse}
\delta_{\epsilon}\psi_M \equiv  {\cal D}_M \epsilon =0 \; .
\end{equation}
%

It should be remarked that the expression (\ref{eq:kse}) is not an
identity, since the non-trivial character of the transformation of
the gravitino, equation (\ref{susyf}), will not allow for it to be
identically satisfied for any spinor field $\epsilon$. Equation
(\ref{eq:kse}) is, instead, a consistency requirement and only the
spinors $\epsilon$ solving the equation will parameterize unbroken
supersymmetries. The equation (\ref{eq:kse}) is usually called
{\it Killing spinor equation}, and its solutions, {\it Killing
spinors}. The number $k$ of preserved supersymmetries of a bosonic
supergravity solution is, thus, given by the number of Killing
spinors\footnote{In lower dimensional supergravities, or in
compactifications of $D=11$ supergravity, further spin $1/2$
fermions might exist, their supersymmetry transformations being
algebraic, instead of differential, in $\epsilon$. In these cases,
the invariance of purely bosonic solutions under supersymmetry
requires that the variation of these fermions also vanishes,
setting further algebraic constraints on the parameters
$\epsilon_J$ if they are to parameterize preserved
supersymmetries. We shall not encounter this situation in our
discussion.} $\epsilon_J$, $J=1, \ldots, k$.

In a fiber bundle, the notions of constancy with respect to the
covariant derivative, invariance under parallel transport and
invariance under the holonomy group come down to the same thing
(see, {\it e.g.} \cite{Besse}): in fact, the holonomy group is a
measure of how vectors and tensors on the fiber transform under
parallel transport around a closed loop at a point. Let us
momentarily set $F_4 = 0$, so that, since we are dealing with
bosonic supergravity solutions ($\psi_M=0$), the only
non-vanishing field is the metric; these configurations therefore
correspond to purely geometrical solutions, to which the results
of Riemannian geometry can be applied. In this case, the
supercovariant derivative ${\cal D}_M$ (equation (\ref{eq:gencd}))
acting on spinors reduces to the covariant derivative associated
to the L\'evi-Civita-induced spin connection $D_M$ (see
(\ref{spincon})) taking values on the Lie algebra $so(1,10)$ of
the tangent space group $SO(1,10)$ (the structure group). The
Killing spinor equation (\ref{eq:kse}) accordingly reduces to
\begin{equation} \label{paralellsp}
D_M \epsilon =0 \; .
\end{equation}
Unbroken supersymmetries of purely geometrical supergravity
solutions are, thus, parameterized by spinors parallel with
respect to the spin connection (that is, satisfying
(\ref{paralellsp})). Riemannian holonomy controls in this case the
number of solutions to the equation (\ref{paralellsp}) and,
consequently, the number of preserved supersymmetries: solutions
to (\ref{paralellsp}) exist if, and only if,  the spinor
representation {\bf 32} of the structure group $SO(1,10)$, to
which the spinor $\epsilon$ belongs, is not only reducible under
the Riemannian holonomy group $\textrm{Hol}(\omega)$, but also the
identity representation arises in the decomposition of the {\bf
32} of $SO(1,10)$ under $\textrm{Hol}(\omega)$. The number $k$ of
times that the identity shows up in this decomposition ({\it
i.e.}, the number of {\it singlets} in this decomposition)
corresponds to the number of invariant spinors $\epsilon_J$, $J=1,
\ldots , k$, under the action of $\textrm{Hol}(\omega)$. These are
the spinors invariant under parallel transport and, thus,
satisfying equation (\ref{paralellsp}).

A heuristic argument can be given to support this result. A
simpler equation for the parallel spinors is obtained if
(\ref{paralellsp}) is further differentiated,
\begin{equation} \label{intRiemann}
[D_M , D_N ] \epsilon =0 \; .
\end{equation}
{}From the computational point of view, this {\it (first order)
integrability condition} of the spinor equation (\ref{paralellsp})
is more convenient, because it is only algebraic, whereas
(\ref{paralellsp}) is differential in $\epsilon$. The commutator
$[D_M , D_N ]$ of two L\'evi-Civita covariant derivatives is
proportional to the Riemann tensor which, according to the
Ambrose-Singer theorem \cite{ambrose} (see also \cite{KobNom63}),
determines the Lie algebra of the holonomy group. Obviously,
equation (\ref{intRiemann}) is only necessary for equation
(\ref{paralellsp}); however, for the relevant cases usually
encountered in supergravity (including vanishing-flux
compactifications), it is also sufficient\footnote{Were
(\ref{intRiemann}) not sufficient for (\ref{paralellsp}), the
holonomy argument would keep on being correct: further
integrability conditions would then be needed to determine the
holonomy (see next section).}. The spinors solving equation
(\ref{intRiemann}) and, hence, invariant under the holonomy group,
solve the equation (\ref{paralellsp}) and correspond to preserved
supersymmetries.

Notice that the existence of parallel spinors imply a holonomy
reduction: the generic holonomy of a Riemannian manifold coincides
with the structure group $SO(1,10)$. If parallel spinors exist,
only when $\textrm{Hol}(\omega) \subset SO(1,10)$ with strict
inclusion, the spinor representation can be reducible under
$\textrm{Hol}(\omega)$. Riemannian holonomy groups have been
classified by Berger \cite{Berger} in the Euclidean case, such
classification having been partially extended to the Lorentzian
case by Bryant \cite{Bryant}.

Let us now return to the case of non-vanishing four-form, $F_4
\neq 0$. This is the generic case in supergravity and, in fact,
the presence of $F_4$ allows for supergravity solutions preserving
exotic fractions of supersymmetry. As already discussed, the
preserved supersymmetries of a bosonic solution when $F_4 \neq 0$
are now parameterized by the Killing spinors solving the Killing
spinor equation (\ref{eq:kse}). The relevant covariant derivative
is not any longer the L\'evi-Civita covariant derivative, but the
supercovariant derivative (\ref{eq:gencd}) associated to the
generalized connection $\Omega_M$ taking values on the Lie algebra
of the generalized structure group $SL(32, \mathbb{R})$
\cite{Hull03} (see section \ref{susyhol}). The presence of $F_4$
terms in the supercovariant derivative does not hamper, however,
an analysis of the Killing spinor equation similar to that of its
Riemannian counterpart. Again, the (first order) integrability
condition of (\ref{eq:kse}),
\begin{equation}
M_{MN}\epsilon\equiv[\D_M,\D_N]\epsilon=0 \; , \label{eq:intc}
\end{equation}
is an algebraic, rather than a differential, equation for the
Killing spinors. The commutator $M_{MN} = [\D_M,\D_N]$ of
supercovariant derivatives now defines the generalized curvature
${\cal R}$ (in fact, $M_{MN}$ contains the same information than
equation (\ref{calR})) taking values, again by the Ambrose-Singer
theorem \cite{ambrose}, in the Lie algebra $\textrm{hol}(\Omega)$
of the generalized holonomy group. The proposal was then put
forward in \cite{Duff03} (see also \cite{DuffStelle91}) that the
role of Riemannian holonomy in the determination of unbroken
supersymmetries of a supergravity solution with non-vanishing
$F_4$ was taken over by generalized holonomy. In particular, in
analogy with the purely geometrical, Riemannian case,  the number
of Killing spinors and, thus, the number of preserved
supersymmetries of a purely bosonic solution of eleven-dimensional
supergravity ought to be given by the number of singlets in the
decomposition of the {\bf 32} representation of the generalized
structure group ($SL(32, \mathbb{R})$ \cite{Hull03}) under the
generalized holonomy group $\textrm{Hol}(\Omega)$
\cite{Duff03,DuffStelle91}. Notice that this argument does not
apply to hypothetical preonic ($31/32$-supersymmetric) solutions
\cite{BPS01}, for which both the 31 unbroken supersymmetries and
the only broken one are singlets. See
\cite{BDLW03,P+T03,P+T031,FFP02,BW,BLVW03,Duff:2002rw,BAIPV03,LPS05}
for further discussion about generalized holonomy.

Two remarks are in order. Firstly, both in the Riemannian and the
generalized cases, the relevant structure and holonomy groups can
be smaller: this is the case, {\it e.g.}, in compactification. In
this case, the relevant representations of these groups are
involved in the supersymmetry counting (see section \ref{squashed}
for an example). Secondly, spinors are assumed to be globally
defined on the manifolds we are dealing with; namely, the
manifolds $M$ fulfilling the Einstein equations (\ref{Einstein})
(or (\ref{EiCJS}) with the notation of chapter \ref{chapter2}) are
endowed with a spin structure\footnote{This could actually be a
subtle issue: different spin structures on a manifold could yield
different number of preserved supersymmetries \cite{FoFG05}.} and,
consequently, fulfil the topological restriction of having
vanishing Stiefel-Whitney class (see \cite{DNP} and references
therein). The promotion of spinors from the (spinor
representation) {\bf 32} of $SO(1,10)$ to the (fundamental
representation) {\bf 32} of $SL(32, \mathbb{R})$ may encompass the
loss of the information contained in the spin structure
\cite{GMSW04}. A different approach to deal with supersymmetric
supergravity solutions, in which the spin structure is naturally
incorporated, is that of $G$-structures \cite{GMPW04,GP02} (see
also
\cite{GaGuPa03,GaMaSpWa04,GauntMaSpWa04,MacConamhna:2004fb,GoNeWa03}).
The later approach has proved to be very useful to build up
explicit supergravity solutions (see \cite{GMSW04,Gauntlett05} for
reviews, and \cite{Grana06,BeCvL05} for $G$-structures in the
context of flux compactifications). See \cite{Gillard:2004xq} for
another recent approach to deal with the Killing spinor equation.

As in the Riemannian case, the presence of Killing spinors entails
a generalized holonomy reduction:  as shown in \cite{P+T03,
Hull03}, for a $D=11$ supergravity solution to preserve $k$
supersymmetries, the generalized holonomy group must be such
that\footnote{That is also the case in Type II $D=10$
supergravities \cite{P+T031}.} $\Hol (\Omega)  \subseteq
SL(32-k,\R)\ltimes ( \R^{32-k} \otimes \stackrel{k}{\ldots}
\otimes \ \R^{32-k} )
  \equiv SL(32-k,\R)\ltimes  \R^{k(32-k)}$ or, from the Lie
algebra point of view, {\setlength\arraycolsep{2pt}
\begin{eqnarray}
\hol (\Omega )  \subseteq   sl(32-k,\R)\ltimes ( \R^{32-k} \oplus
\stackrel{k}{\ldots} \oplus \ \R^{32-k} ) \; , \label{reducgenhol}
\end{eqnarray}
}where $sl(32-k,\R)$ acts on each of the $k$ copies of $\R^{32-k}$
through the same, fundamental representation. The issue of
classifying supersymmetric vacua may thus be mapped into one of
classifying the generalized holonomy groups as subgroups of
$SL(32,\R)$. An investigation of basic supersymmetric
configurations of M Theory was performed in \cite{BDLW03} (see
also \cite{Duff03,Hull03}), where a large variety of generalized
holonomy groups were obtained. However, one of the striking
results of the analysis of \cite{BDLW03} was the fact that
identical generalized holonomies may yield different amounts of
supersymmetries. This shows that knowledge of the holonomy group
is insufficient to fully classify the supergravity solution, and
that knowledge of its embedding into the generalized structure
group is also needed; in other words, knowledge of the
decomposition of the $32$-component spinor under $\Hol(\Omega)$ is
also needed.

\section{Higher order integrability} \label{higherinteg}

Being only algebraic in $\epsilon$, the (first order)
integrability condition (\ref{eq:intc}) is more convenient than
the Killing spinor equation itself, (\ref{eq:kse}), in order to
determine Killing spinors for a particular supergravity solution.
It might happen, however, that the integrability condition
(\ref{eq:intc}) were only necessary, and not sufficient, for the
Killing spinor equation (\ref{eq:kse}). That is indeed the case
for Freund-Rubin compactifications \cite{FR} of $D=11$
supergravity, for which the preserved supersymmetry depends, in
general, on the orientation chosen for the compactifying manifold
(see \cite{DNP}). Freund-Rubin compactification on the squashed
seven-sphere (the coset space $SO(5) \times SU(2) / SU(2) \times
SU(2)$) \cite{ADP,DNP83}, for instance, preserves $N=1$
supersymmetry for one orientation (that can be referred to as
left-squashing) while breaks it all for the other orientation
(right-squashing). Accordingly, the Killing spinor equation
(\ref{eq:kse}) has solutions in the first case, but no solutions
in the second one. And yet, both orientations share the same
(first order) integrability condition (\ref{eq:intc}) which is,
therefore, not sufficient for (\ref{eq:kse}). This issue can be
resolved by going beyond first order integrability: successive
covariant derivatives of equation (\ref{eq:kse}) ({\it i.e.},
higher order integrability conditions) can give a set of
additional algebraic equations for $\epsilon$, sufficient for
(\ref{eq:kse}) \cite{vNW}.

This discussion can be put in a (generalized) holonomy context, by
asking whether the Lie algebra generated by the curvature
(expressed in the first order integrability condition
(\ref{eq:intc})) agrees with the Lie algebra of the holonomy
group. Actually, as shown in \cite{BDLW03}, in many cases the
complete Lie algebra of $\Hol(\Omega)$ was not obtained from first
order integrability (\ref{eq:intc}), so that in particular the
algebra had to be closed by hand.  This issue is rather suggestive
that the generalized curvature at a local point carries incomplete
information of the generalized holonomy group, in apparent
violation of the Ambrose-Singer theorem (but in agreement with the
issue of left- versus right-squashing of $S^7$ mentioned above).
However, the Ambrose-Singer theorem really indicates that
$\Hol_p(\Omega)$ at a point $p$ is spanned by elements of the
generalized curvature (\ref{eq:intc}) not just at point $p$, but
at all points $q$ connected to $p$ by parallel transport (see {\it
e.g.}  \cite{KobNom63,Besse,joyce}).  Thus there is in fact no
contradiction. Furthermore, this is rather suggestive that
satisfying higher order integrability (representing motion from
$p$ to $q$, an information encoded in the successive covariant
derivatives of the curvature) is in fact a necessary condition for
identifying the proper generalized holonomy group \cite{BLVW03}.

In the remainder of this chapter, the interplay of higher order
integrability and generalized holonomy will be explored, resorting
to specific examples. We begin by revisiting the generalized
holonomy of the M5 and M2-brane solutions of supergravity, and
show that higher order integrability yields precisely the
`missing' generators that were needed to close the algebra.  Other
than this, however, the generalized holonomy groups for the
M-branes identified in \cite{BDLW03} are unchanged.  Following
this, we turn to the squashed $S^7$ \cite{ADP,DNP83}, where the
situation is considerably different.

The importance of higher order integrability was of course
previously recognized in \cite{vNW} for the case of the squashed
$S^7$.  Here, we reinterpret the result of \cite{vNW} in the
language of generalized holonomy, and confirm the statement of
\cite{Duff:2002rw} that while first order integrability yields the
incorrect result $\hol^{(1)}(\Omega)=G_2\subset so(7) \subset
so(8)$, higher order integrability corrects this to $\hol(\Omega)=
so_\pm(7) \subset so(8)$, where the two distinct possibilities
$so(7)_-$ and $so(7)_+$ arise from left- and right-squashing,
respectively, and correspond to the two different embeddings of
$so(7)$ into $so(8)$. Since the spinor decomposes as either
$\mathbf 8_s \to \mathbf 7 + \mathbf 1$ or $\mathbf 8_s \to
\mathbf 8$ in the two cases, this explains the resulting $N=1$ or
$N=0$ supersymmetry in four dimensions \cite{Duff:2002rw,BLVW03}
(see section \ref{squashed}).

Let us now introduce a convenient notation for the Lie algebra
generators associated to the $n$-th order integrability
conditions. For the supercovariant derivative (\ref{eq:gencd})
associated to the generalized connection $\Omega$, first order
integrability (\ref{eq:intc}) of the Killing spinor equation
(\ref{eq:kse}) yields the generators
\begin{equation}
M_{MN}\equiv [\D_M(\Omega),\D_N(\Omega)]
=\ft14(\partial_M\Omega_N-\partial_N\Omega_M
+\ft14[\Omega_M,\Omega_N])\equiv\ft14\Rm_{MN}(\Omega),
\label{eq:1int}
\end{equation}
where $\Rm_{MN}(\Omega)$ is the generalized curvature, {\it i.e.},
the curvature of $\Omega$ (see equation (\ref{calR})). Higher
order integrability expressions may be obtained by taking
generalized covariant derivatives of (\ref{eq:1int}). The
corresponding generators will be taken to be
{\setlength\arraycolsep{0pt}
\begin{eqnarray}
&& M_{MN_1N_2} \equiv [\D_M,M_{N_1N_2}] \; ,
\label{eq:2int}\\
&& M_{MN_1N_2N_3} \equiv [\D_N,M_{N_1N_2N_3}] \; ,
\label{eq:3int}\\
&& M_{MN_1N_2N_3N_4} \equiv [\D_M,M_{N_1N_2N_3N_4}] \; ,
\label{eq:4int}\\* &&  \vdots \nonumber
\end{eqnarray}

}

Higher order integrability conditions correspond to measuring the
generalized curvature $\Rm_{MN}(\Omega)$ parallel transported away
from the original base point $p$.  In this sense, the information
obtained from higher order integrability is precisely that
required by the Ambrose-Singer theorem in making the connection
between $\Hol_p(\Omega)$ and the curvature of the generalized
connection.


\section{Generalized holonomy of the M-branes} \label{sec:Mbranes}

As examples of how higher order integrability may affect
determination of the generalized holonomy group, we first revisit
the case of the M5- and M2-brane solutions of supergravity.  The
generalized holonomy of these solutions, as well as several
others, was originally investigated in \cite{BDLW03}.  For vacua
with non-vanishing flux, including the brane solutions, it was
seen that the Lie algebra generators obtained from first order
integrability, (\ref{eq:1int}), are insufficient for the closure
of the algebra.  In particular, additional generators must be
obtained by further commutators.  In \cite{BDLW03}, this was done
by closing the algebra by hand.  In the present context, however,
additional commutators are readily available from the higher order
integrability expressions, (\ref{eq:2int})--(\ref{eq:4int})
\cite{BLVW03}.

\subsection{Generalized holonomy of the M5-brane}

The metric and four-form corresponding to the M5-brane solution of
$D=11$ supergravity are given by \cite{Gu92}
{\setlength\arraycolsep{0pt}
\begin{eqnarray} \label{M5eqs}
&&ds^2_{11}=H_5^{-1/3} (dx^{\mu})^2+H_5^{2/3} (d y^i)^2 \; ,\nonumber\\
&&F_{ijkl}=\epsilon_{ijklm}\partial^m H_5 \; ,
\end{eqnarray}
}where $x^\mu$, $\mu=0,1, \ldots, 5$, are coordinates
corresponding to the worldvolume directions, $y^i$, $i=1,\ldots ,
5$, are transverse space coordinates and $\epsilon_{ijklm}=\pm1$
is the L\'evi-Civita symbol, and $H_5(y^i)$  a function, in
transverse space.  Preservation of supersymmetry requires both the
metric and four-form to be determined by the same function $H_5$
which is, in turn, demanded to be harmonic by the equations of
motion (\ref{Einstein}), (\ref{4form}).

When acting on spinors, the generalized connection $\Omega_M$
defining the supercovariant derivative (\ref{eq:gencd}) for the
solution (\ref{M5eqs}) reads \cite{BDLW03}
\begin{equation} \label{M5conn}
\Omega_\mu=\Omega_\mu^{\nu i}K_{\nu i} \; ,\quad
\Omega_i=-\ft13\partial_i\ln H_5\Gamma^{(M5)}+\ft12\Omega_i^{jk}
T_{jk} \; ,
\end{equation}
where
\begin{equation}
\Omega_\mu^{\nu i}= -\ft23H_5^{-1/2} \delta_\mu^\nu \partial^i \ln
H_5 \; ,\quad \Omega_i^{jk}=\ft83
\delta_i^{[j}\partial^{k]}\ln{H_5} \; ,
\end{equation}
and $T_{ij}$, $K_{\mu i}$ belong to the set
\begin{equation}
T_{ij}=\Gamma_{ij}P_5^+,\quad K_\mu=\Gamma_\mu P_5^+,\quad K_{\mu
i}=\Gamma_{\mu i}P_5^+,\quad K_{\mu ij}=\Gamma_{\mu i j}P_5^+ \; ,
\label{eq:m5gen}
\end{equation}
of generators of a Lie algebra to be specified below (see equation
(\ref{HolM5})). In (\ref{M5conn}), $\Gamma^{(M5)}\equiv
\frac{1}{5!}\epsilon_{ijklm}\Gamma^{ijklm}$ and, in
(\ref{eq:m5gen}), $P_5^+ \equiv \frac{1}{2}(1+\Gamma^{(M5)})$ is
the M5-brane $1/2$-supersymmetry projector. The generalized
connection $\Omega_M$ of (\ref{M5conn}) includes the generator
$\Gamma^{(M5)}$ in addition to $T_{ij}$ and $K_{\mu i}$. However,
the connection itself is not physical and, in fact, the terms
containing $\Gamma^{(M5)}$ drop out from the expression of the
generalized curvature (see below) and hence do not contribute to
generalized hol\-o\-no\-my.

The integrability conditions of the Killing spinor equation
(\ref{eq:kse}), posed with the supercovariant derivative
associated to the generalized connection (\ref{M5conn}) of the
M5-brane, can now be discussed. The first order integrability of
the Killing spinor equation provides the set of generators
(\ref{eq:1int}) corresponding to the Lie algebra of the
generalized curvature. For the M5-brane solution, these generators
read \cite{BDLW03}
{\setlength\arraycolsep{2pt}
\begin{eqnarray}
M_{\mu\nu} & \equiv & \ft14\Rm_{\mu\nu}= 0,\nonumber\\
M_{\mu i} & \equiv & \ft14\Rm_{\mu i} \nonumber
\\ &=& H_5^{-1/2}\left[\ft16
(\partial_i\partial^j\ln{H_5}-\ft23\partial_i\ln{H_5}\partial^j\ln{H_5})
+\ft1{18}\delta_i^j(\partial\ln{H_5})^2\right]K_{\mu j}, \nonumber\\
M_{ij} & \equiv & \ft14\Rm_{ij} \nonumber
\\ &=& \left[\ft23(\partial_l\partial_{[i\vphantom{]}} \ln{H_5}
-\ft23 \partial^l\ln{H_5}\partial_{[i}\ln{H_5})\delta_{j]}^k
-\ft29(\partial\ln H_5)^2\delta_{[i}^k\delta_{j]}^l\right] T_{kl}
\; . \nonumber
\\ && \label{eq:m51order}
\end{eqnarray}
}Only the generators $T_{ij}$ and $K_{\mu i}$ show up in the
expression for the Lie algebra (\ref{eq:m51order}) corresponding
to the generalized curvature. As noticed in \cite{BDLW03}, the
remaining generators $K_\mu$ and $K_{\mu ij}$ of (\ref{eq:m5gen})
have to be obtained by closing the algebra defined by
(\ref{eq:m51order}) `by hand'. Alternatively, higher order
integrability conditions, expressed as
(\ref{eq:2int})--(\ref{eq:4int}), can be used to obtain the
remaining generators that ensure closure of the algebra
\cite{BLVW03}.

In fact, the generators defining the second order integrability
conditions, obtained for the M5-brane upon insertion of the
corresponding generalized connection (\ref{M5conn}) into
(\ref{eq:2int}), take on the form \cite{BLVW03}
{\setlength\arraycolsep{0pt}
\begin{eqnarray}
&&M_{\mu\nu\lambda}=M_{\mu\nu\lambda}^{\rho i}K_{\rho i},\quad
M_{\mu\nu i}=\ft12M_{\mu\nu i}^{jk} T_{jk},\quad M_{\mu ij}=M_{\mu
ij}^{\nu k}K_{\nu k}+\ft12M_{\mu ij}^{\nu kl}K_{\nu kl},
\nonumber\\
&&M_{i\mu\nu}=0,\quad M_{i\mu j}=M_{i\mu j}^{\nu k}K_{\nu
k}+\ft12M_{i\mu j}^{\nu kl}K_{\nu kl}, \quad
M_{ijk}=\ft12M_{ijk}^{lm} T_{lm}, \label{eq:m5ms}
\end{eqnarray}
}where the component factors $M_{AMN}^{\cdots}$ are functions of
$H_5$ and its derivatives.  For example,
{\setlength\arraycolsep{0pt}
\begin{eqnarray}
&& \kern-1em M_{\mu\nu\lambda}^{\rho
i}=\ft1{36}H^{-3/2}[\partial^j\ln H_5
\partial_j\partial^i\ln H_5-\ft13\partial^i\ln H_5(\partial H_5)^2
]\eta_{\mu[\nu}\delta_{\lambda]}^\rho,\nonumber\\
&& \kern-1em M^{jk}_{\mu\nu i} = \ft49H^{-1}[\partial^{[j}\ln H_5
\partial_i\partial^{k]}\ln H_5
-\delta_i^{[j|}\partial^l\ln H_5\partial_l\partial^{|k]}\ln
H_5]\eta_{\mu\nu} \; . 
\end{eqnarray}
}The other factors arising in (\ref{eq:m5ms}) are similar and
their explicit forms will not be needed. An additional generator
$K_{\mu ij}$ arises at second order through the expressions
$M_{\mu ij} \equiv [{\cal D}_\mu , {\cal R}_{ij}]$ and $M_{i \mu
j} \equiv [{\cal D}_i , {\cal R}_{\mu j}]$ in (\ref{eq:m5ms}).
However, this does not still suffice to close the algebra. Pushing
this procedure one step further into third order integrability
(\ref{eq:3int}), it is found that  the generator $K_\mu$ arises
through $M_{ki \mu j} \equiv [{\cal D}_k, [{\cal D}_i , {\cal
R}_{\mu j}]]$. The complete set of generators (\ref{eq:m5gen}) is
then obtained and, actually,  no new generator is found beyond
third order \cite{BLVW03}.

The generators (\ref{eq:m5gen}) thus generate the Lie algebra
$\hol_{M5}$ of the generalized holonomy group of the M5-brane
\cite{BDLW03}. The ${5 \choose 2} = 10$ generators $T_{ij}$
correspond to $so(5)$, whereas the remaining $6+ 6 \cdot 5 + 6
\cdot {5 \choose 2} = 96$ generators $K_\mu$, $K_{\mu i}$, $K_{\mu
ij}$ in (\ref{eq:m5gen}) span the abelian Lie algebra $\R^{96}$,
on which $so(5)$ acts semidirectly, {\it i.e.}, through a
96-dimensional representation. According to the general rule
(\ref{reducgenhol}), as a supergravity solution preserving $k=16$
supersymmetries, the M5-brane (\ref{M5eqs}) must have its
generalized holonomy in $sl(32-k,\R)\ltimes ( \R^{32-k} \oplus
\stackrel{k}{\ldots} \oplus \ \R^{32-k})$, namely,
\begin{eqnarray}
\hol_{M5} \ \subseteq \ sl(16,\R)\ltimes ( \R^{16} \oplus
\stackrel{16}{\ldots} \oplus \ \R^{16}) \; . \label{M5holcont}
\end{eqnarray}
The 96-dimensional representation of $so(5) \subset sl(16,\R)$ on
$\R^{96}$ must be, therefore, reducible into at most $k=16$ copies
of the same (reducible or irreducible) representation of dimension
$32-k=16$; thus, in this case, $\R^{96} = (\R^{16} \oplus
\stackrel{6}{\ldots} \oplus \ \R^{16} ) \subset ( \R^{16} \oplus
\stackrel{16}{\ldots} \oplus \ \R^{16})$, where each of the six
copies of $\R^{16}$ carries the same 16-dimensional representation
of $so(5)$. This representation turns out to be further reducible
into four 4-dimensional (spinor) representations $\mathbf{4}$ of
$so(5)$. Introducing the convenient notation $\R^{4(\mathbf{4})}$
to denote this splitting of $\R^{16}$, the generalized holonomy
algebra of the M5-brane solution of $D=11$ supergravity
(\ref{M5eqs}) is then \cite{BDLW03}
\begin{equation} \label{HolM5}
\hol_{M5}=so(5) \ltimes (\R^{4(\mathbf{4})} \oplus
\stackrel{6}{\ldots} \oplus \ \R^{4(\mathbf{4})}) \; .
\end{equation}

For the M5-brane case, higher order integrability conditions just
provide the generators missing at first order, that can
nevertheless be obtained by closing the generalized curvature
algebra (defined by first order integrability) `by hand'. In
particular, higher order integrability conditions do not change
the generalized holonomy (\ref{HolM5}) of the M5-brane, which
remains the same as in \cite{BDLW03}.

\subsection{Generalized holonomy of the M2-brane}

The analysis of the M2-brane is similar to that of the M5-brane.
The supergravity solution corresponding to the M2-brane is given
by \cite{DuffStelle91}
{\setlength\arraycolsep{0pt}
\begin{eqnarray} \label{M2eqs}
&&ds_{11}^2=H_2^{-2/3}(dx^\mu)^2+H_2^{1/3} (d y^i)^2, \nonumber\\
&&F_{\mu\nu\rho i}=\epsilon_{\mu\nu\rho}\partial_i H_2^{-1},
\end{eqnarray}
}where $x^\mu$, $\mu=0,1,2$, are coordinates corresponding to the
worldvolume directions, $y^i$, $i=1,\ldots , 8$, are transverse
space coordinates and  $\epsilon_{\mu\nu\rho}=\pm1$. $H_2 (y^i)$
is a harmonic function in transverse space.

Denoting by $P_2^+=\fft12(1+\Gamma^{(M2)})$ the
$1/2$-supersymmetry projector of the M2-brane, where
$\Gamma^{(M2)} \equiv
\fft1{3!}\epsilon_{\mu\nu\rho}\Gamma^{\mu\nu\rho}$, the following
generators
\begin{equation}
T_{ij}=\Gamma_{ij}P_2^+ \; , \quad K_{\mu i}=\Gamma_{\mu
i}P_2^+,\quad K_{\mu ijk}=\Gamma_{\mu ijk}P_2^+, \label{eq:m2gen}
\end{equation}
can be introduced in order to express the generalized connection
of the M2-brane solution (\ref{M2eqs}) \cite{BDLW03}:
\begin{equation} \label{M2con}
\Omega_\mu=\Omega_\mu^{\nu i}K_{\nu i},\quad
\Omega_i=\ft23\partial_i\ln H_2\Gamma^{(M2)}+\ft12\Omega_i^{jk}
T_{jk} \; .
\end{equation}
Here, the components of $\Omega_M$ are
\begin{equation}
\Omega_\mu^{\nu i}=-\ft43 H_2^{-1/2} \delta_\mu^\nu \partial^i \ln
H_2 \; ,\quad \Omega_i^{jk}=\ft43\delta_i^{[j}\partial^{k]}\ln
H_2.
\end{equation}
The generators of the generalized curvature algebra corresponding
to the M2-brane solution are again obtained through first order
integrability of the Killing spinor equation (\ref{eq:kse}),
written for the supercovariant derivative associated to the
generalized connection (\ref{M2con}). These generators are
\cite{BDLW03}
{\setlength\arraycolsep{2pt}
\begin{eqnarray} \label{genholalM2}
M_{\mu\nu} & \equiv & \ft14\Rm_{\mu\nu}=0,\nonumber\\
M_{\mu i} & \equiv & \ft14\Rm_{\mu i} \nonumber \\
&=& \ft1{18}H_2^{-1/2}\left[ 6(\partial_i\partial^j\ln
H_2+2\partial_i \ln H_2\partial^j \ln H_2)
-(\partial\ln H_2)^2\delta_i^j\right]K_{\mu j},\nonumber\\
M_{ij} & \equiv & \ft14\Rm_{ij} \nonumber \\
&=& \left[ -\ft13(\partial_l\partial_{[i\vphantom{]}}\ln H_2
-\ft13\partial^l\ln H_2\partial_{[i}\ln H_2)\delta_{j]}^k
-\ft1{18}(\partial\ln H_2)^2\delta_{[i}^k\delta_{j]}^l \right]
T_{kl}, \nonumber \\
&&
\end{eqnarray}
}and include terms proportional only to the generators $T_{ij}$
and $K_{\mu i}$ of (\ref{eq:m2gen}). As in the M5-brane case, the
closure of the algebra spanned by the generators
(\ref{genholalM2}) can be achieved either `by hand' \cite{BDLW03}
or by working out higher order integrability conditions
\cite{BLVW03}. The generators (\ref{eq:2int}), corresponding to
the second order integrability conditions, when considered for the
M2-brane solution have the general form \cite{BLVW03}
{\setlength\arraycolsep{2pt}
\begin{eqnarray} \label{secondM2}
&& M_{\mu\nu\lambda}=M_{\mu\nu\lambda}^{\rho i}K_{\rho i} \;
,\quad M_{\mu\nu i}=\ft12M_{\mu\nu i}^{jk} T_{jk}+M_{\mu\nu
i}^{\nu k}K_{\nu k} \; , \nonumber \\
&& M_{\mu ij}=M_{\mu ij}^{\nu k}K_{\nu k}+\ft16M_{\mu ij}^{\nu
klm}K_{\nu klm} \; ,  \nonumber \\
&& M_{i \mu j}=M_{i\mu j}^{\nu k}K_{\nu k}+\ft16M_{i\mu j}^{\nu
klm}K_{\nu klm} +\ft12M_{i\mu j}^{kl} T_{kl} \; , \nonumber \\
&&  M_{i\mu\nu}=0 \; , \quad M_{ijk}=\ft12M_{ijk}^{lm} T_{lm}.
\end{eqnarray}
}where the explicit form of their components along the generators
(\ref{eq:m2gen}) will not be needed. The generators $M_{\mu ij}
\equiv [{\cal D}_\mu , {\cal R}_{ij}]$ and $M_{i \mu j} \equiv
[{\cal D}_i , {\cal R}_{\mu j}]$ in (\ref{secondM2}) give rise to
the additional generator $K_{\mu ijk}$ of (\ref{eq:m2gen}) which,
together with $T_{ij}$ and $K_{\mu i}$ generate the Lie algebra
$\hol_{M2}$ of the generalized holonomy group of the M2-brane
 solution of $D=11$ supergravity \cite{BDLW03}.

Since the M2-brane preserves $k=16$ supersymmetries, $\hol_{M2}$
must be contained, by virtue of equation (\ref{reducgenhol}), in
$sl(16,\R)\ltimes ( \R^{16} \oplus \stackrel{16}{\ldots} \oplus \
\R^{16})$. In fact, $T_{ij}$ in (\ref{eq:m2gen}) generate $so(8)
\subset sl(16, \R )$ while $K_{\mu i}$, $K_{\mu ijk}$ are the
generators of the abelian Lie algebra $\R^{192} = (\R^{16} \oplus
\stackrel{12}{\ldots} \oplus \ \R^{16} ) \subset ( \R^{16} \oplus
\stackrel{16}{\ldots} \oplus \ \R^{16})$. The representation of
$so(8)$ on each $\R^{16}$ is reducible into two 8-dimensional
(spinor) representations $\mathbf{8}_s$, making $\R^{16}$ split as
$\R^{2(\mathbf{8}_s)}$ and yielding a generalized holonomy for the
M2-brane \cite{BDLW03}
\begin{equation}
\hol_{M2}=so(8) \ltimes (\R^{2(\mathbf{8}_s)} \oplus
\stackrel{12}{\ldots} \oplus \ \R^{2(\mathbf{8}_s)}) \; .
\end{equation}
Second order integrability is, thus, sufficient to guarantee the
closure of the Lie algebra of the generalized holonomy group of
the M2-brane.

Note that the generalized connection $\Omega_M$ contains complete
information about the generalized holonomy of the spacetime, as
the complete set of integrability conditions
(\ref{eq:1int})--(\ref{eq:4int}) may be obtained through
commutators and derivatives of $\Omega_M$.  In this sense, the
algebra of the holonomy group can never be larger than the algebra
obtained through the generators in $\Omega_M$ itself.  However it
can certainly be smaller.  This is apparent for the M5-brane,
where the $\Gamma^{(M5)}$ generator is absent in the generalized
curvature $\Rm_{MN}(\Omega)$ and its derivatives and also for the
M2-brane, where $\Gamma^{(M2)}$ is absent. For these examples, and
in fact for all vacua considered in \cite{BDLW03,BW}, the
generators appearing in $\Omega_M$ and those appearing in
$\Rm_{MN}(\Omega)$ are nearly identical.  As a result, the
generalized holonomy group may be correctly identified at first
order in integrability, and the higher order conditions only serve
to complete the set of generators needed for closure of the
algebra.

A different situation may arise, however, if for some reason (such
as accidental symmetries) a greatly reduced set of generators
appear in $\Rm_{MN}(\Omega)$.  In such cases, examination of first
order integrability may result in the misidentification of the
actual generalized holonomy group.  What happens here is that the
algebra of the curvature $\Rm_{MN}(\Omega)$ at a single point $p$
forms a subalgebra of the Lie algebra of the holonomy group.  It
is then necessary to explore the curvature at all points $q$
connected by parallel transport to $p$ in order to determine the
actual holonomy.  We demonstrate below that this incompleteness of
first order integrability does arise in the case of generalized
holonomy.

\section{Higher order integrability and the squashed $S^7$} \label{squashed}

For an example of the need to resort to higher order integrability
to characterize the generalized holonomy group $\Hol(\Omega)$, we
turn to Freund-Rubin compactifications of eleven-dimensional
supergravity. With vanishing gravitino, the Freund-Rubin ansatz
\cite{FR} for the $4$-form field strength $F_4$,
\begin{equation}
\label{Fleft} F_{\mu \nu \rho \sigma} = 3m \epsilon_{\mu \nu \rho
\sigma} , \quad \mu=0,1,2,3,
\end{equation}
with $m$ constant and all other components vanishing, leads to
spontaneous compactifications of the product form $AdS_4 \times
X^7$.  Here $X^7$ is a compact, Einstein, Euclidean 7-manifold.
Decomposing the eleven-dimensional Dirac matrices $\Gamma_M$ as
\begin{equation}
\Gamma_M = (\gamma_\mu \otimes 1 , \gamma_5 \otimes
\Gamma_m),\qquad \mu = 0,1,2,3,\quad m=1,\ldots, 7,
\end{equation}
where $\gamma_\mu$ and $\Gamma_m$ are four- and seven-dimensional
Dirac matrices, respectively, and assuming the usual
direct-product ansatz $\epsilon(x^\mu) \otimes \eta(y^m)$ for
eleven-dimensional spinors, the Killing spinor equation
(\ref{eq:kse}) splits as
{\setlength\arraycolsep{0pt}
\begin{eqnarray}
\label{kads} && \D_\mu \epsilon = \left( \partial_\mu \ +\ft14
\omega_\mu{}^{\alpha \beta} \gamma_{\alpha \beta}
+ m \gamma_\mu \gamma_5 \right) \epsilon =0 ,  \\
\label{kleft} && \D_m \eta = \left( \partial_m \ +\ft14
\omega_m{}^{ab} \Gamma_{ab} - \ft{i}2 m \Gamma_m \right) \eta =0 .
\end{eqnarray}
}Since $AdS_4$ admits the maximum number of Killing spinors (four
in this case), the number $N$ of supersymmetries preserved in the
compactification coincides with the number of Killing spinors of
the internal manifold $X^7$, that is, with the number of solutions
to the Killing spinor equation (\ref{kleft}).  Therefore we only
need to concern ourselves with the Killing spinors on $X^7$.

An orientation reversal of $X^7$ or, alternatively, a sign
reversal of $F_4$, provides another solution to the equations of
motion (\ref{Einstein}), (\ref{4form}) and, hence, another
acceptable Freund-Rubin vacuum \cite{DNP83,DNP}.  For
definiteness, we shall call {\it left}-orientation the solution
corresponding to the choice of sign of $F_4$ in (\ref{Fleft}),
that leads to the Killing spinor equation (\ref{kleft}), and {\it
right}-orientation the solution corresponding to the opposite
choice of sign of $F_4$:
\begin{equation} \label{Fright}
(\hbox{right})\qquad F_{\mu \nu \rho \sigma} = -3m \epsilon_{\mu
\nu \rho \sigma}, \quad \mu=0,1,2,3,
\end{equation}
leading to the Killing spinor equation
\begin{equation} \label{kright}
(\hbox{right})\qquad \D_m \eta = \left( \partial_m \ +\ft14
\omega_m{}^{ab} \Gamma_{ab} + \ft{i}2 m \Gamma_m \right) \eta =0 .
\end{equation}

{}From either (\ref{kleft}) or (\ref{kright}), we see that the
generalized connection defining $\D_m$ takes values in the algebra
spanned by $\{ \Gamma_{ab}, \Gamma_a \}$ and therefore the
generalized structure group is $SO(8)$. Notice, however, that both
Killing spinor equations (\ref{kleft}) and (\ref{kright}) share
the same first order integrability condition \cite{ADP,DNP}
\begin{equation} \label{int1}
M_{mn}\eta \equiv [\D_m,\D_n] \eta=\ft14\Rm_{mn}\eta \equiv \ft14
\Cm_{mn}\eta=\ft14 C_{mn}{}^{ab} \Gamma_{ab} \eta= 0,
\end{equation}
where $ C_{mn}{}^{ab}$ is the Weyl tensor of $X^7$ (thus
demonstrating that, in this case the generalized curvature tensor
is simply the Weyl tensor).  Thus first order integrability is
unable to distinguish between left and right orientations on the
sphere.  Then it might be possible that spinors $\eta$ solving the
integrability condition (\ref{int1}) will only satisfy the Killing
spinor equation for one orientation, that is, satisfy
(\ref{kleft}) but not (\ref{kright}) (or the other way around). In
fact, the skew-whiffing theorem \cite{DNP83,DNP} for Freund-Rubin
compactifications  proves that this will, in general, be the case:
it states that at most one orientation can give $N > 0$, with the
exception of the round $S^7$, for which both orientations give
maximal supersymmetry, $N=8$. Since the preserved supersymmetry
$N$ is given by the number of singlets in  the decomposition of
the $\mathbf 8_s$ of $SO(8)$ (the generalized structure group)
under the generalized holonomy group $\Hol(\Omega)$, then, in
general, each orientation must have either a different generalized
holonomy, or the same generalized holonomy but a different
decomposition of the $\mathbf 8_s$.

To illustrate this feature, consider compactifications on the
squashed $S^7$ \cite{DNP83,ADP}. This choice for $X^7$ has the
topology of the sphere, but the metric is distorted away from that
of the round $S^7$; it is instead the coset space $SO(5) \times
SU(2) / SU(2) \times SU(2)$ endowed with its Einstein metric
\cite{DNP83,ADP}.  The compactification on the left-squashed $S^7$
preserves $N=1$ supersymmetry whereas that on the right-squashed
$S^7$ has $N=0$; put another way, the integrability condition
(\ref{int1}) has one non-trivial solution, corresponding in turn
to a solution to the Killing spinor equation (\ref{kleft}) (making
the left-squashed $S^7$ preserve $N=1$), but not to a solution to
(\ref{kright}), which in fact has no solutions (yielding $N=0$ for
the right-squashed $S^7$).  On the other hand, an analysis of the
Weyl tensor of the squashed $S^7$ shows that there are only 14
linear combinations $\Cm_{mn}$ of gamma matrices in (\ref{int1}),
corresponding to the generators of $G_2$ \cite{ADP,DNP}. Though
appealing, $G_2$ cannot be, however, the generalized holonomy
since the $\mathbf 8_s$ of $SO(8)$ would decompose as $\mathbf 8_s
\rightarrow \mathbf8\rightarrow \mathbf 7+\mathbf 1$ under
$SO(8)\supset SO(7)\supset G_2$ regardless of the orientation,
giving $N=1$ for both left- and right-squashed solutions. We thus
conclude that in this case  the first order integrability
condition (\ref{int1}) is insufficient to determine the
generalized holonomy.

The resolution to this puzzle is naturally given by higher order
integrability.  In the case of the squashed $S^7$, it turns out
that the second order integrability condition (\ref{eq:2int}) is
sufficient.  For a general Freund-Rubin internal space $X^7$ this
condition reads\footnote{In (\ref{int1}), $\D_l$ (the generalized
covariant derivative in equation (\ref{kleft})) should not be
confused with $D_l$ (the L\'evi-Civita covariant derivative).}
\cite{vNW}
\begin{equation} \label{int2}
M_{lmn}\eta\equiv\ft14 [\D_l,\Cm_{mn}]\eta=
\ft14\left(D_lC_{mn}{}^{ab}\Gamma_{ab}\mp2imC_{mnl}{}^{a}\Gamma_a\right)
\eta = 0 \; ,
\end{equation}
the $-$ sign corresponding to the left solution, and the $+$ to
the right.  For the squashed $S^7$, we find that only 21 of the
$M_{lmn}$ are linearly independent combinations of the Dirac
matrices \cite{BLVW03}.  The details are provided in Appendix
\ref{holS7}. Following the notation of \cite{ADP,DNP}, we split
the index $m$ as $m=(0,i,\hat{i})$, with $i=1,2,3$,
$\hat{i}=4,5,6=\hat{1},\hat{2},\hat{3}$; then, with a suitable
normalization,  the linearly independent generators in
(\ref{int2}) may be chosen to be \cite{BLVW03}
{\setlength\arraycolsep{0pt}
\begin{eqnarray} \label{G2}
&& \Cm_{0i}=\Gamma_{0i}+\ft12\epsilon_{ikl}
\Gamma^{\hat{k}\hat{l}} \; , \quad \Cm_{ij}=
\Gamma_{ij}+\Gamma_{\hat{i}\hat{j}} \; , \nonumber \\
&& \Cm_{i\hat{j}}=-\Gamma_{i\hat{j}}-\ft12\Gamma_{j\hat{i}}
+\ft12\delta_{ij}\delta^{kl}\Gamma_{k\hat{l}}
-\ft12\epsilon_{ijk}\Gamma^{0\hat{k}} \; , \\
\label{SO7} && M_{ij}=\Gamma_{\hat{i}\hat{j}} \mp \ft23\sqrt{5} im
\epsilon_{ijk} \Gamma^{\hat{k}} \; , \quad M_i=
\Gamma_{0\hat{i}} \mp \ft23\sqrt{5} im \Gamma_i \; , \nonumber \\
&& M= \delta^{kl}\Gamma_{k\hat{l}} \pm 2\sqrt{5}im \Gamma_0 ,
\end{eqnarray}
}the $-$ sign in front of $m$ corresponding to the left solution
and the $+$ to the right. Notice that there are 8 linearly
independent generators in $\Cm_{i\hat j}$ of (\ref{G2}), since
$\delta^{kl} \Cm_{k \hat{l}} \equiv \Cm_{1\hat{1}} +\Cm_{2\hat{2}}
+ \Cm_{3\hat{3}} =0$. The $3+3+8=14$ generators $\Cm_{0i}$,
$\Cm_{ij}$, $\Cm_{i\hat{j}}$ span $G_2$ \cite{ADP,DNP}, and are
the same as those obtained from the first integrability condition
(\ref{int1}), while the $3+3+1=7$ additional generators $M_{ij}$,
$M_i$, $M$ of (\ref{SO7}) were not contained in (\ref{int1}).
Taken together, they generate the 21 dimensional algebra $so(7)$,
regardless of the orientation, provided \cite{BLVW03}
\begin{equation}
m^2=\ft{9}{20} ,
\end{equation}
in agreement with the Einstein equation for the squashed $S^7$
\cite{DNP}.

The embedding of $so(7)$ into $so(8)$ is, however, different for
each orientation. We use $so(7)_-$ to denote the embedding
corresponding to the left solution and $so(7)_+$ the right. While
the spinor $\eta$ transforms as an $\mathbf 8_s$ of the
generalized structure group $SO(8)$, the decomposition of the
$\mathbf 8_s$ is different under left- and right-squashing.  With
our Dirac conventions, it turns out that $\mathbf8_s \rightarrow
\mathbf7+\mathbf1$ under $so(8) \supset so(7)_-$, giving $N=1$ for
the left-squashed $S_7$, while $\mathbf8_s \rightarrow \mathbf8$
under $so(8)\supset so(7)_+$, giving $N=0$ for the right-squashed
$S^7$.

Since $so(7)$ is the subalgebra of $so(8)$ that yields the correct
branching rules of the $\mathbf8_s$ of $SO(8)$, we conclude that
second order integrability is sufficient in this case to identify
all generators of the Lie algebra of the holonomy $\Hol(\Omega_m)$
of the connection $\Omega_m$ defining the supercovariant
derivative $\D_m$ in (\ref{kleft}) . Hence the generalized
holonomy algebra of the Freund-Rubin compactification on the
squashed $S^7$ is given precisely by $\hol(\Omega_m)=so(7)$
\cite{Duff:2002rw, BLVW03}\footnote{For a $d$-dimensional manifold
$X_d$, the {\it cone $C(X_d)$ over $X_d$} is the
$(d+1)$-dimensional manifold defined to have topology $\R^+ \times
X_d$ and metric $g(C(X_d)) = dr^2 + r^2 g(X_d)$, where $g(X_d)$ is
the metric on $X_d$ and $r$ parameterizes $\R^+$. In a
supergravity context, {\it Killing spinors} on the Freund-Rubin
compactifying manifold $X_d$ correspond to {\it parallel spinors
with respect to the L\'evi-Civita connection} on $C(X_d)$
\cite{Bar93,AFFHS99}. Thus, in this case, the {\it generalized
holonomy} of $X_d$ corresponds to the {\it Riemannian holonomy} of
$C(X_d)$. A compactifying 7-manifold $X_7$ preserves $N=1$
supersymmetry for one orientation if its corresponding
8-dimensional cone $C(X_7)$ has Spin(7) holonomy \cite{AFFHS99},
in agreement with this result for the generalized holonomy of the
squashed $S^7$ \cite{Duff:2002rw, BLVW03}.}. In this case, it is
the embedding of $so(7)$ in $so(8)$ (with corresponding spinor
decomposition $\mathbf8_s\to\mathbf7+\mathbf1$ or
$\mathbf8_s\to\mathbf8$) that determines the number of preserved
supersymmetries.  This indicates that, for generalized holonomy,
knowledge of the holonomy group {\it and} the embedding are both
necessary in order to understand the number of preserved
supersymmetries.  While this was already observed in
\cite{BDLW03,Hull03} for non-compact groups, here we see that this
is also true when the generalized holonomy group is compact.

The analysis of the squashed $S^7$, along with that of the brane
solutions of the previous section, highlights several features of
generalized holonomy.  For the squashed $S^7$, the Lie algebra of
the generalized holonomy group is in fact larger than that
generated locally by the Weyl curvature at a point $p$.  In this
case, the algebra arising from lowest order integrability is
already closed, but is only a subalgebra of the correct holonomy
algebra.  It is then mandatory to examine the second order
integrability expression (\ref{int2}) in order to identify the
generalized holonomy group. On the other hand, for the M2 and
M5-branes, lowest order integrability, while lacking a complete
set of generators, nevertheless closes on the correct holonomy
algebra, and no really new information is gained at higher order.
Of course, in all cases, complete information is contained in the
generalized connection $\Omega_M$ itself.  However, examination of
$\Omega_M$ directly can be misleading, as it may contain gauge
degrees of freedom, which are unphysical.  This is most clearly
seen in the case of the round $S^7$, where $\Omega_m=\omega_m^{ab}
\Gamma_{ab}-2im\Gamma_m$ is certainly non-vanishing, while the
generalized curvature $\Rm_{mn}$, given by the Weyl tensor, is
trivial, $\Rm_{mn}=0$.

For generalized holonomy to be truly useful, it ought to go beyond
simply a classification scheme, and must yield methods for
constructing new supersymmetric solutions. In much the same way
that the rich structure of Riemannian holonomy teaches us a great
deal about the geometry of Killing spinors on Riemannian
manifolds, the formal analysis of generalized holonomy via
connections on Clifford bundles may lead to a similar expansion of
knowledge of supergravity structures and manifolds with fluxes.
Such an analysis is well beyond the scope of this Thesis and,
instead, we now continue  with the study of generalized holonomy
to characterize supersymmetric solutions of supergravity, from a
different point of view.


\chapter{Generalized holonomy for BPS preons} \label{chapter4}

The observation \cite{BPS01} that  BPS states that break $\n=32-k$
supersymmetries can be treated as composites of those preserving
all but one supersymmetries, suggests that the
$k=31$-supersymmetric states might be considered as fundamental
constituents of M Theory. These $\nu=31/32$ BPS states were
accordingly named {\it BPS preons} in \cite{BPS01}. In this
chapter we apply the ideas previously developed about generalized
holonomy to the study of hypothetical preonic solutions of
eleven-dimensional supergravity. In section \ref{sec:BPSpreons},
the notion of preonic states is reviewed. States composed of $\n$
preons are shown to be characterized by $\n$ bosonic spinors that
parameterize the broken supersymmetries. In section
\ref{sec:frame}, these spinors are shown to be orthogonal to the
Killing spinors characterizing the unbroken supersymmetries. A
moving $G$-frame (where the group $G$ can chosen to be
$G=GL(32,\mathbb{R})$, $SL(32,\mathbb{R})$ or $Sp(32,\mathbb{R})$)
defined by both preonic and Killing spinors can be consequently
used to describe the corresponding states. We then apply, in
section \ref{secIII}, this moving $G$-frame method to the study of
the generalized holonomies of hypothetical preonic solutions of
supergravity. Although no definite answer to the question of the
existence of preonic solutions for the standard $D=11$
supergravity is given here, we do show, in section
\ref{sec:preonsugra}, that $\nu=31/32$ supersymmetric preonic
configurations exist in Chern-Simons (CS) supergravity {\it i.e.},
that CS supergravity does have preonic solutions. To conclude this
chapter, we propose in section \ref{sec:preonbranes} a
worldvolume action for BPS preons in the background of the
D'Auria-Fr\'e formulation of $D=11$ supergravity \cite{D'A+F}. The
notation and conventions are restored to those of chapter
\ref{chapter2}. This chapter follows closely reference
\cite{BAIPV03}.

\section{BPS preons} \label{sec:BPSpreons}

Group theoretical methods usually help with the lack of a
dynamical description of M Theory; in particular,  the
representation theory of the M-algebra $\mathfrak{E}^{(528|32)}$
(see section \ref{sec:MTsuperalg} of chapter \ref{chapter2})  can
shed some light into the structure of M Theory.
Bogomoln'yi-Prasad-Sommerfield (BPS) states saturate the
Bogomoln'yi bound associated to the M Theory superalgebra
(\ref{QQP}) and are, therefore, protected from corrections as
argued in the Introduction (chapter \ref{chapter1}). They are,
thus, intrinsically non-perturbative and are expected to be
fundamental states of the fully-fledged M Theory.

A BPS state $\vert BPS\, , \, k\rangle$  described by a
supergravity solution preserving $k$ supersymmetries is
characterized by $k$ spinors $\epsilon_J^\alpha$, $J=1, \ldots , k
\leq 32$ parameterizing the supersymmetry transformations
(\ref{susye})--(\ref{susyA}) of the spacetime fields. In
particular, it will be assumed that the state $\vert BPS\, , \,
k\rangle$ corresponds to a purely bosonic supergravity solution,
so that the spinors $\epsilon_J^\alpha$ are Killing and satisfy
the Killing spinor equation (\ref{eq:kse}). These spinors
parameterize the unbroken supersymmetries that leave invariant the
supersymmetric state; that is, at the level of generators acting
on $\vert BPS\, , \, k\rangle$,
\begin{eqnarray}\label{kSUSY}
\epsilon_J{}^\alpha  Q_\alpha \vert BPS\, , \, k\rangle = 0 \, ,
\quad J=1,\ldots, k\; ,  \quad k \leq 32 \; .
\end{eqnarray}
Here, $Q_\alpha$ are the supersymmetry generators, that we shall
take to be in the maximally extended supersymmetry algebra,
namely, the M Theory superalgebra $\mathfrak{E}^{(528|32)}$, whose
(anti)commutation relations are given in (\ref{QQP}): $ \{
Q_\alpha , Q_\beta\} = P_{\alpha \beta}$, $[Q_\alpha ,
P_{\beta\gamma}]=0$, $\alpha , \beta , \gamma = 1,2, \ldots, 32$,
so that $P_{\alpha \beta}= P_{\beta\alpha}$. The generalized
momentum $P_{\alpha \beta}$ can be decomposed in the basis of
$D=11$ $Spin(1,10)$ ($32 \times 32$)
 Dirac matrices as in (\ref{n32}), namely,
 $P_{\alpha\beta} = P_a \Gamma^a_{\alpha\beta} + i Z_{ab}
\Gamma^{ab}_{\alpha\beta} +  Z_{a_1\ldots a_5} \Gamma^{a_1\ldots
a_5}_{\alpha\beta}$, containing the standard $D=11$ momentum $P_a$
and the tensorial `central' charge  generators $Z_{ab}$, $Z_{a_1
\ldots a_5}$. As discussed in section \ref{sec:MTsuperalg}, these
central charges are associated to the basic M Theory branes.

In a formal, quantum-mechanical discussion, a $\nu =
k/32$-su\-per\-sym\-me\-tric BPS state $|BPS, k \rangle$ can also
be defined as an eigenstate of the generalized momentum operator
$P_{\alpha \beta}$,
\begin{equation}
P_{\alpha \beta}| BPS, k \rangle= p^{(k)}_{\alpha \beta}|BPS, k
\rangle
\end{equation}
with eigenvalue $p^{(k)}_{\alpha \beta}$ such that $\mathrm{det} \
p^{(k)}_{\alpha \beta}=0$, as justified below. The va\-ni\-shing
determinant condition implies that the matrix $p^{(k)}_{\alpha
\beta}$ has rank less than the maximal possible rank 32. More
precisely, a $\nu=k/32$-BPS state $| BPS, k \rangle$ is such that
\begin{equation} \label{rank}
\mathrm{rank} \ p^{(k)}_{\alpha \beta} \equiv \n =  32-k \, ,
\quad 1 \leq k <32 \; .
\end{equation}
Recall from the discussion of section \ref{sec:MTsuperalg} that
the maximal automorphism group of the M-algebra
$\mathfrak{E}^{(528|32)}$ is $GL(32, \mathbb{R})$. Then the matrix
$p^{(k)}_{\alpha \beta}$ can be diagonalized by a
$GL(32,\mathbb{R})$ transformation $g_{\alpha}{}^{(\gamma)}$,
\begin{equation}\label{p=GpG}
p_{\alpha\beta}^{(k)}= g_{\alpha}{}^{(\gamma)}p_{(\gamma)(\delta)}
g_{\beta}{}^{(\delta)} \; .
\end{equation}
In (\ref{p=GpG}), $p_{(\gamma)(\delta)}$ is a diagonal matrix that
can be put in the canonical form
\begin{equation}\label{pdiag0}
p_{(\gamma)(\delta)}= \textrm{diag}
(\underbrace{1,\ldots,1,-1,\ldots, -1}_{\n=32-k},
\underbrace{0,\ldots,0}_{k})\; ,
\end{equation}
where the number of non-vanishing elements, all $+1$ or $-1$, is
equal to $\n= \textrm{rank} (p_{\alpha\beta}^{(k)})$. However, the
usual assumptions for the supersymmetric quantum mechanics
describing BPS states do not allow for negative eigenvalues of
$P_{\alpha\beta}=\{Q_\alpha , Q_\beta\}$ ($p_{11}=-1$, {\it e.g.},
would imply $(Q_1)^2 \vert BPS, k \rangle = - \vert BPS, k
\rangle$, contradicting unitarity). Thus, only positive
eigenvalues are allowed and
\begin{equation}\label{pdiag}
p_{(\gamma)(\delta)}= \textrm{diag}
(\underbrace{1,\ldots,1}_{\n=32-k}, \underbrace{0,\ldots,0}_{k})\;
.
\end{equation}

Substituting (\ref{pdiag}) into (\ref{p=GpG}), one arrives at
\begin{equation}\label{p=G1G}
p_{\alpha\beta}^{(k)}= g_{\alpha}{}^{(\gamma)}\; \textrm{diag}
(\underbrace{1,\ldots,1}_{\n=32-k},
\underbrace{0,\ldots,0}_{k})_{(\gamma)(\delta) } \;\;
g_{\beta}{}^{(\delta )} \; ,
\end{equation}
or, equivalently, introducing the $\n$ vectors $\lambda_\alpha^1,
\ldots, \lambda_\alpha^{\n}$ of $GL(32, \mathbb{R})$, defined by
$g_{\alpha}{}^{1}=
 \lambda_{\alpha}{}^{1}$, $\ldots$, $ g_{\alpha}{}^{\n}=
 \lambda_{\alpha}{}^{\n}$,
{\setlength\arraycolsep{0pt}\begin{eqnarray}\label{npreon}
P_{\alpha\beta}\vert BPS, k \rangle & = &
\sum\limits_{r=1}^{\n=32-k}
\lambda_{\alpha}{}^r \lambda_{\beta}{}^r \vert BPS, k \rangle \;  \nonumber \\
&\equiv& \left(\lambda_{\alpha}{}^1\lambda_{\beta}{}^1 + \ldots +
\lambda_{\alpha}{}^{\n} \lambda_{\beta}{}^{\n} \right) \vert BPS,
k \rangle \; . \qquad
\end{eqnarray}}

Taking suitable linear combinations of the supertranslations,
namely, $Q^{(0)}_{\alpha} = (g^{-1})^\beta{}_\alpha Q_{\beta}$,
the algebra diagonalizes on BPS states,
{\setlength\arraycolsep{0pt}
\begin{eqnarray}
& & \{Q^{(0)}_r ,  Q^{(0)}_s \} |BPS, k\rangle = \delta_{rs} |BPS,
k\rangle \; , \nonumber \\* && \{Q^{(0)}_r ,  Q^{(0)}_J \} |BPS,
k\rangle = \{Q^{(0)}_J , Q^{(0)}_K \} |BPS, k\rangle =0 \; ,
\end{eqnarray}
}where $r, s= 1, \ldots, \n$, $J,K=1 , \ldots , k$, so that the
set of 32 supercharges $Q^{(0)}_\alpha = (Q^{(0)}_r ,Q^{(0)}_J)$
acting on the BPS state $|BPS, k\rangle$ splits into $k$
generators $Q^{(0)}_J$ of supersymmetry that preserve the BPS
state (and correspond to the generators of (\ref{kSUSY}),
$Q^{(0)}_J | k \rangle = 0$),  and $\n=32-k$ generators
$Q^{(0)}_r$ corresponding the set of broken supersymmetries.

Equation (\ref{npreon}) suggests that all BPS states can be
considered as composites of states with $\mathrm{rank} \ p_{\alpha
\beta}^{(k)}=1$ \cite{BPS01}, that is, preserving $k=31$
supersymmetries. The hypothetical objects carrying these
``elementary values" of $p_{\alpha \beta}^{(31)}$ are called {\it
BPS preons} \cite{BPS01}. For a BPS preon state, the index $r$ in
equation ({\ref{npreon}) assumes only one value and can therefore
be suppressed. In summary, a BPS preon \cite{BPS01} state $\vert
BPS\, ,\, 31 \rangle \equiv \vert \lambda \rangle$ preserves 31
supersymmetries (hence the notation $\vert BPS\, ,\, 31 \rangle$)
and is characterized by the following choice of central charges
matrix
\begin{equation} \label{BPSdef}
p_{\alpha \beta}= g^{\gamma}_{\, \, \, \alpha} \ p^{(0)}_{\gamma
\delta} \ g^{\delta}_{\, \, \, \beta} = \lambda_{\alpha}
\lambda_{\beta} \; ,
\end{equation}
in terms of a single bosonic spinor\footnote{\label{fnote:spinor}
By construction, $\lambda_\alpha$ is a $GL(32 , \mathbb{R})$
vector. However, we keep the `spinor' name for it bearing in mind
the possibility of a spacetime treatment, although this is not
straightforward and would require additional study. } (hence the
notation $\vert \lambda \rangle$) such that
\begin{eqnarray}\label{preon}
P_{\alpha \beta} \vert \lambda \rangle = \lambda_\alpha
\lambda_\beta \vert \lambda \rangle \; .
\end{eqnarray}
Equation (\ref{npreon}) may be looked at as a manifestation of the
{\it composite structure} of the $\nu = k/32$ BPS state $\vert
BPS, k \rangle$,
\begin{eqnarray}\label{k=npreon}
\vert BPS, k \rangle = \vert \lambda^1 \rangle \otimes \ldots
\otimes  \vert \lambda^{\n} \rangle \; ,
\end{eqnarray}
where $ \vert \lambda^1 \rangle$, $\ldots$, $ \vert \lambda^{\n}
\rangle$, $\n=32 - k$, are BPS elementary, preonic states
characterized by the spinors $\lambda_\alpha{}^1$ , $\ldots$,
$\lambda_\alpha{}^{\n}$, respectively.

From this point of view, all the single-brane solutions of
11-di\-men\-sio\-nal supergravity, which preserve $16$ out of $32$
supersymmetries (see section \ref{eqd11s} of chapter
\ref{chapter2}), correspond to composites of $16$ BPS preons. By
the same token, intersecting branes, preserving {\it less} than
$16$ supersymmetries ($\nu<1/2$) correspond to composites of {\it
more} than $16$ preons, and  solutions with extra supersymmetry
($\nu > 1/2$) can be considered as composites of less than 16 BPS
preons. Initially, it seemed that solutions preserving all
supersymmetries but one, {\it i.e.} describing the excitations of
a BPS preon, could not exist, and indeed they were not found by
means of the standard brane ansatzes used to solve the usual
11-dimensional supergravity \cite{CJS} equations. A more general
study in the context of standard $D=11$ supergravity has shown
that the existence of such solutions is not ruled out
\cite{Hull03,Duff03}.

The possible existence of brane solutions with extra
supersymmetries  should not be excluded, although these solutions
would describe quite unusual branes. The reason why the `standard'
brane solutions (like M-waves, M2 and M5-branes in $D=11$) always
break $1/2$ of the supersymmetry is that their $\kappa$-symmetry
projector (the bosonic part of which is identical to the projector
defining the preserved supersymmetries \cite{BKOP97,ST97}) has the
form $(1-\bar{\Gamma})$ with $tr \bar{\Gamma}=0$,
$\bar{\Gamma}^2=I$. However, worldvolume actions for branes with a
different form for the $\kappa$-symmetry projector are known
\cite{BL98,B02,ZU,30/32,Bandos05} although in an enlarged
superspace (see \cite{JdA00}): see chapter \ref{chapter7} for an
explicit example. A question arises, whether such actions may be
written in usual spacetime or superspace.

However, and independently of whether BPS preons can be associated
with solutions of standard supergravity or there is, instead, a
{\it BPS preon conspiracy} preventing their existence in {\it
standard} $D$=11 spacetime or superspace, preons do provide an
algebraic classification of the M Theory BPS states \cite{BPS01}.
In this perspective such a BPS preon conspiracy, if it exists,
would perhaps indicate the necessity of a wider geometric
framework for a suitable description of M Theory, such as extended
superspaces and supertwistors. If, on the contrary, solitonic
solutions with the properties of BPS preons were actually found,
extended superspaces would still provide a useful tool for a
description of M Theory\footnote{\label{footn} There are also
related reasons to consider more general superspaces, as the
ensuing {\it fields/extended superspace coordinates
correspondence} \cite{JdA00,Azcarraga05} associated with extended
superspaces: see section \ref{adboscoor} of chapter \ref{chapter6}
and further references therein.}. One is led to expect that the
additional tensorial coordinates of these superspaces carry a
counterpart of the information which, in the framework of standard
$D=10,11$ supergravity, is encoded in the antisymmetric tensor
gauge fields entering the supergravity multiplets ({\it
cf.}~\cite{JdA00}). This point of view may be also supported by
the observation that in the standard topological charge treatment
of the tensorial generators of the M--algebra \cite{JdAT}, these
topological charges are associated just with these gauge fields.

\section{Moving $G$-frame} \label{sec:frame}

When a BPS state $\vert k \rangle$ is realized as a solitonic
solution of supergravity, it is characterized by $k$ Killing
spinors $\epsilon_J{}^\beta (x)$ or by the $\n=32-k$ bosonic
spinors $\lambda_\alpha{}^r (x)$ associated with the $\n$ BPS
preonic components of the state $\vert BPS, k \rangle$. The
Killing spinors and the preonic spinors are orthogonal. Indeed,
using the (anti)commutation relations (\ref{QQP}) of the
M-algebra, if the preserved supersymmetries correspond to the
generators $\epsilon_J{}^\alpha Q_\alpha$, $J=1, \ldots , k$,
equation (\ref{kSUSY}), then
\begin{equation} \sum\limits_{r=1}^{\n=32-k} \epsilon_{(J}{}^\alpha
\lambda_{\alpha}{}^r\; \epsilon_{K)}{}^\beta \lambda_{\beta}{}^r =
0\; ,
\end{equation}
which implies the orthogonality of Killing and preonic spinors
\cite{BAIPV03}, {\setlength\arraycolsep{0pt}
\begin{eqnarray} \label{eKl=0}
 \epsilon_{J}{}^\alpha \lambda_{\alpha}{}^r=0 \; , \quad J =
1,...,k\; , \quad r=1,..., \n \; ,
\end{eqnarray}
}explaining the relation $\n=32-k$ between the number of preons
$\n= \textrm{rank}(p_{\alpha\beta}^{(k)})$ and the number of
preserved supersymmetries $k$.

Then, BPS preonic ($\lambda_\alpha{}^r$) and Killing
($\epsilon_J{}^\alpha$) spinors provide an alternative (dual)
characterization of a $\nu$-supersymmetric solution; either one
can be used and, for solutions with extra supersymmetries ($\nu >
1/2$)
\cite{CLP02,GH02,BJ02,Michelson,Gauntlett:2002nw,Harmark:2003ud},
the characterization provided by BPS preons is a more economic
one. Moreover, the use of both BPS preonic spinors and Killing
 spinors allows us to develop a {\it moving
G-frame} method \cite{BAIPV03}, which we now introduce, and that
may be useful in the search for new supersymmetric solutions of
supergravity.

The set of Killing and preonic spinors can be completed to obtain
bases in the spaces of spinors with upper and with lower indices
by introducing $\n=32-k$ spinors $w_r{}^\alpha$  and $k$ spinors
$u_{\alpha}{}^L$ satisfying
\begin{eqnarray}\label{wl=}
 w_s{}^\alpha \lambda_\alpha{}^r= \delta_s^r \; , \quad
 w_s{}^\alpha  u_\alpha{}^J =0 \; , \quad
 \epsilon_J{}^\alpha  u_\alpha{}^K =  \delta_J{}^K \; .
\end{eqnarray}
Either of these two dual bases defines a {\it generalized moving
$G$-frame} described by the nondegenerate matrices
{\setlength\arraycolsep{0pt}
\begin{eqnarray}\label{g}
g_{\alpha}{}^{(\beta)} = \left( \lambda_\alpha{}^s \, ,
u_\alpha{}^J \right)\; , \qquad
g^{-1}{}_{(\beta)}{}^\alpha = \left( \begin{array}{c} w_s{}^\alpha \\
\epsilon_J{}^\alpha \end{array} \right)\; , \quad
\end{eqnarray}
}where $(\alpha)=(s,J)=(1,\ldots,32-k;J=1,\ldots,k)$. Indeed,
$g^{-1}{}_{(\beta)}{}^\gamma g_{\gamma}{}^{(\alpha)} =
\delta_{(\beta)}{}^{(\alpha)}$ is equivalent to Eqs. (\ref{wl=})
and (\ref{eKl=0}), while
\begin{eqnarray}\label{g-1g}
\delta_{\alpha}{}^{\beta} = g_{\alpha}{}^{(\gamma)}
g^{-1}{}_{(\gamma)}{}^\beta \equiv  \lambda_\alpha{}^r w_r{}^\beta
+ u_\alpha{}^J \epsilon_J{}^\beta \;
\end{eqnarray}
provides the unity $I_{32}$ decomposition or completeness relation
in terms of these dual bases.

One may consider the dual basis $g^{-1}{}_{(\beta)}{}^\alpha$ to
be constructed from the bosonic spinors in
$g_{\alpha}{}^{(\beta)}$ by solving equation~(\ref{g-1g}) or
$g^{-1}g= I_{32}$ (Eqs. (\ref{wl=}) and (\ref{eKl=0})).
Alternatively, one may think of $w_r{}^\alpha$ and $u_\alpha{}^J$
as being constructed from $\epsilon_J{}^\alpha$ and
$\lambda_\alpha{}^r$ through a solution of the same constraints.
In this sense \cite{BAIPV03} {\it the generalized moving $G$-frame
(\ref{g}) is constructed from $k$ Killing spinors
$\epsilon_J{}^\alpha$ characterizing the supersymmetries preserved
by a BPS state (realized as a solution of the supergravity
equations) and from the $\n=32-k$ bosonic spinors
$\lambda_\alpha{}^r$ characterizing the BPS preons from which the
BPS state is composed}. Although many of the considerations below
are general, we shall be mainly interested here in the cases
$G=SL(32,\mathbb{R})$ and $G=Sp(32,\mathbb{R})$.

In $D=11$, the charge conjugation matrix $C^{\alpha\beta}=-
C^{\beta\alpha}$ allows us to express explicitly the dual basis
$g^{-1}$ in terms of the original one $g$ or {\it vice versa}. In
particular, in the preonic $k=31$ case one finds that, since
$\lambda_\alpha C^{\alpha\beta}\lambda_\beta \equiv 0$, then
$\lambda^\alpha= C^{\alpha\beta}\lambda_\beta$ has to be expressed
as $\lambda^\alpha= \lambda^I \epsilon_I{}^\alpha$, for some
coefficients $\lambda^I$, $I=1,\ldots, 31$. In general (as {\it
e.g.}, in CJS supergravity with nonvanishing $F_4$), the charge
conjugation matrix is not `covariantly constant', ${\cal
D}C^{\alpha\beta}= -2 \Omega^{[\alpha\beta]} \not= 0$, where
$\Omega_\alpha{}^\beta$ is the $D=11$ supergravity generalized
connection (\ref{CJSom}) (see section \ref{susyhol} of chapter
\ref{chapter2}). This relates the coefficients $\lambda^I=
\lambda^\alpha u_\alpha{}^I$ to the antisymmetric (non-symplectic)
part of the generalized connection,
$\Omega^{[\alpha\beta]}=C^{[\alpha\gamma}\Omega_\gamma{}^{\beta]}$
by\footnote{To see this, one calculates $d\lambda^I={\cal
D}\lambda^I= ({\cal D}C^{\alpha\beta})\lambda_\beta
u_\alpha{}^I+C^{\alpha\beta}({\cal D}\lambda_\beta)u_\alpha{}^I
+C^{\alpha\beta}\lambda_\beta {\cal D}u_\alpha{}^I$ and use
equation~(\ref{Dl31}), (\ref{Du31}) to find
$d\lambda^I=A\lambda^I+2\lambda_\alpha
\Omega^{[\alpha\beta]}u_\beta{}^I$.} $d\lambda^I - A \lambda^I= 2
\lambda_\alpha \Omega^{[\alpha\beta]}u_\beta{}^I$. In
$OSp(1|32)$-related models, $\Omega^{[\alpha\beta]}=0$ and $A=0$,
hence $\lambda^I$ is constant and we may set
$\lambda^I=\delta^I_{31}$ using the {\it global} transformations
of $GL(31,\mathbb{R})$, which is a rigid symmetry of the system of
Killing spinors. This allows us to identify $\lambda^\alpha$
itself with one of the Killing spinors
{\setlength\arraycolsep{0pt}
\begin{eqnarray}\label{OSpe}
&& G= Sp(32,\mathbb{R})  :   \nonumber \\
&& \qquad  \epsilon_I{}^\alpha = (\epsilon_i{}^\alpha,
\lambda^\alpha) \; , \quad \lambda^\alpha:=
C^{\alpha\beta}\lambda_\beta \; \quad  i=1,\ldots , 30 \; .
\end{eqnarray}

}

Without specifying a solution of the constraints (\ref{g-1g}) (or
$g^{-1}g= I_{32}$), the moving frame possesses a
$G=GL(32,\mathbb{R})$ symmetry. One may impose as  additional
constraints $\mathrm{det} (g)=1$ or $\mathrm{det}(g^{-1})=1$
reducing $G$ to $SL(32,\mathbb{R})$,
\begin{eqnarray}\label{SLG}
G= SL(32,\mathbb{R})\; :  \quad \mathrm{det}(g_\beta^{\,
(\alpha)})=1= \mathrm{det} (g^{-1}_{\; (\alpha)}{}^\beta)\; .
\end{eqnarray}
 For instance, in the
preonic case $k=31$ this would imply
\begin{equation}
w^\alpha =\ft{1}{(31)!} \varepsilon^{\alpha\beta_1\ldots
\beta_{31}} u_{\beta_1}{}^1 \ldots u_{\beta_{31}}{}^{31} \; .
\end{equation}
Such a frame is most convenient to study the   bosonic solutions
of CJS supergravity, since the corresponding generalized holonomy
must be a subgroup of $SL(32, \mathbb{R})$ (see section
\ref{susyhol} of chapter \ref{chapter2} and references therein).

\section{Generalized holonomy of preonic solutions} \label{secIII}

The Killing equation (\ref{eq:kse}) for a $\nu=k/32$
supersymmetric solution,
\begin{eqnarray}\label{Killingk}
{\cal D} \epsilon_J{}^\alpha =d  \epsilon_J{}^\alpha -
\epsilon_J{}^\beta \Omega_\beta{}^\alpha =0 \; , \qquad
J=1,\ldots, k\; ,
\end{eqnarray}
implies the following equations for the other components of the
moving $G$-frame {\setlength\arraycolsep{0pt}
\begin{eqnarray}\label{Dlk}
&& {\cal D} \lambda_{\alpha}{}^r := d\lambda_{\alpha}{}^r +
\Omega_{\alpha}{}^{\beta} \, \lambda_{\beta}{}^r =
\lambda_{\alpha}{}^s \, A_s{}^r \; ,
\\ \label{Duk}
&& {\cal D} u_\alpha{}^J  :=
 du_\alpha{}^J + \Omega_{\alpha}{}^\beta
u_\beta{}^J = \lambda_{\alpha}{}^r \, B_r^J \; ,
\\ \label{Dwk}
&& {\cal D} w_r{}^\alpha  := dw_r{}^\alpha - w_r{}^\beta
\Omega_{\beta}{}^{\alpha} = - A_r{}^s w_s{}^\alpha - B_r^J
\epsilon_J{}^\alpha \; ,
\end{eqnarray}
}where $\alpha , \beta = 1, \ldots, 32$, $J = 1,\ldots, k$,
$r,s=1,\ldots, (32-k)$, and  $A_s{}^r$ and $B_r{}^I$ are $(32-k)
\times (32-k)$ and $(32-k) \times k$ arbitrary one-form matrices.
To obtain the equations (\ref{Dlk}), (\ref{Duk}), (\ref{Dwk}) one
can take firstly the  derivative ${\cal D}$ of  the orthogonality
relations (\ref{eKl=0}), (\ref{wl=}).  After using
equation~(\ref{Killingk}), this results in
{\setlength\arraycolsep{0pt}
\begin{eqnarray}\label{Dort1}
&& \epsilon_I{}^\alpha {\cal D}\lambda_\alpha{}^r=0  \; , \quad
\epsilon_I{}^\alpha {\cal D}u_\alpha{}^J=0\; , \\ \label{Dort2} &&
w_s{}^\alpha {\cal D}\lambda_\alpha{}^r = - {\cal D} w_s{}^\alpha
\; \lambda_\alpha{}^r  \; , \quad   w_s{}^\alpha {\cal
D}u_\alpha{}^J = - {\cal D} w_s{}^\alpha \, u_\alpha{}^J\, .
\end{eqnarray}
}Then, for instance, to derive  (\ref{Dlk}), one uses the unity
decomposition (\ref{g-1g}) to express ${\cal D}\lambda_\alpha{}^r$
through the contractions $w_s{}^\alpha {\cal D}\lambda_\alpha{}^r$
and $\epsilon_I{}^\alpha {\cal D}\lambda_\alpha{}^r$: $\;  {\cal
D}\lambda_\alpha{}^r \equiv \lambda_\alpha{}^s\, w_s{}^\beta {\cal
D}\lambda_\beta{}^r + u_\alpha{}^I \, \epsilon_I{}^\beta {\cal
D}\lambda_\beta{}^r$. The second term vanishes due to
(\ref{Dort1}), while the first one is not restricted by the
consequences of the Killing spinor equations and may be written as
in equation~(\ref{Dlk}) in terms of an arbitrary form $A_s{}^r
\equiv w_s{}^\alpha {\cal D}\lambda_\alpha{}^r$.

Notice that, using the unity decomposition (\ref{g-1g}),  one may
also solve formally  equations (\ref{Killingk}), (\ref{Dlk}),
(\ref{Duk}), (\ref{Dwk}) with respect to the generalized
connection $\Omega_\alpha{}^\beta$ of equation (\ref{CJSom}),
\begin{eqnarray}\label{om=A+B}
\Omega_\alpha{}^\beta = A_r{}^s \, \lambda_{\alpha}{}^r
w_s{}^\beta + B_r{}^J \lambda_{\alpha}{}^r  \epsilon_J{}^\beta
-(dg g^{-1})_\alpha{}^\beta \; ,
\end{eqnarray}
where $g_{\alpha}^{\, (\beta)}$ and $g^{-1}_{\;
(\beta)}{}^{\alpha}$ are defined in equation (\ref{g}) and, hence,
\begin{equation}\label{dgg-1G}
(dg g^{-1})_\alpha{}^\beta = d\lambda_{\alpha}{}^r \, w_r{}^\beta
+ du_\alpha{}^I \, \epsilon_I{}^\beta  \; .
\end{equation}

For a BPS $\nu= 31/32$, preonic configuration, equations
(\ref{Dlk}), (\ref{Duk}), (\ref{Dwk}) read
{\setlength\arraycolsep{0pt}
\begin{eqnarray}\label{Dl31}
&& {\cal D} \lambda_{\alpha} := d\lambda_{\alpha} +
\Omega_{\alpha}{}^\beta \lambda_{\beta} = A \lambda_{\alpha} \; ,
\\ \label{Du31}
&& {\cal D} u_\alpha{}^I := du_\alpha{}^I +
\Omega_{\alpha}{}^\beta u_\beta{}^I = B^I \lambda_{\alpha} \; ,
\\ \label{Dw31}
&& {\cal D} w^\alpha  := dw^\alpha - w^\beta
\Omega_{\beta}{}^{\alpha} = - A\,  w^\alpha - B^I
\epsilon_I{}^\alpha \;
\end{eqnarray}
}and contain $1+31=32$ arbitrary one-forms  $A$ and $B^I$.

For $G=SL(32,\mathbb{R})$ one may choose $\textrm{det}(g)=1$,
equation~(\ref{SLG}), which implies $\textrm{tr}(dg g^{-1}):= (dg
g^{-1})_\alpha{}^\alpha =0$. Then the $sl(32,\mathbb{R})$-valued
generalized connection $\Omega_\alpha{}^\beta$
($\Omega_\alpha{}^\alpha =0$) allowing for a $\nu=k/32$
supersymmetric configuration is determined by
equation~(\ref{om=A+B}) with $A_r{}^r=0$,
\begin{eqnarray}\label{trA=0}
G= SL(32,\mathbb{R})\; :  \qquad A_r{}^r=0\; .
\end{eqnarray}
In particular, the $sl(32,\mathbb{R})$-valued generalized
connection allowing for a BPS preonic, $\nu=31/32$, configuration,
should have the form \cite{BAIPV03}
\begin{eqnarray}\label{om31=B}
G= SL(32,\mathbb{R}) \; ,  \nu=31/32 \; : \; \Omega_\alpha{}^\beta
=  B^I\; \lambda_{\alpha} \epsilon_I{}^\beta - (dg
g^{-1})_\alpha{}^\beta \;
\end{eqnarray}
in terms of   $31$ arbitrary one-forms $B^I$, $I=1, \ldots , 31$.

Assuming a definite form for the generalized connection
$\Omega_\alpha{}^\beta$, one finds that Eqs.~(\ref{om=A+B}) become
differential equations for $k$ Killing spinors
$\epsilon_{J}{}^\alpha$ {\it and $n=32-k$ BPS preonic spinors
$\lambda_\alpha{}^r$} once $(dg g^{-1})=d\lambda_{\alpha}{}^r
w_r{}^\beta - u_\alpha{}^I d\epsilon_I{}^\beta$
(equation~(\ref{dgg-1G})) is taken into account. On the other
hand, one might reverse the argument and ask for the structure of
a theory allowing for $\nu=k/32$ supersymmetric solutions. This
question is especially interesting for the case of BPS preonic and
$\nu=30/32$ solutions as, for the moment, such solutions are
unknown in the standard $D=11$ CJS and $D=10$ Type II
supergravities.

The simplest application of the moving $G$-frame construction is
to find an explicit form for the general solution of the
integrability conditions,
\begin{equation} \label{eIR=0}
\epsilon_J{}^\beta {\cal R}_\beta{}^\alpha = 0\; ,
\end{equation}
which are necessary for the Killing spinor equation
(\ref{Killingk}). In (\ref{eIR=0}), ${\cal R}_\beta{}^\alpha$ is
the generalized curvature (\ref{calR}) corresponding to the $D=11$
supergravity generalized connection $\Omega_\alpha{}^\beta$ of
(\ref{CJSom}). To make things simpler, we shall consider that the
solutions we are dealing with are such that their generalized
holonomy is fully determined by ${\cal R}_\beta{}^\alpha$ and,
like in the M2 and M5-brane cases (see section \ref{sec:Mbranes}
of chapter \ref{chapter3}), further supercovariant derivatives of
${\cal R}_\beta{}^\alpha$ do not provide additional essential
information.

Since the Killing spinor equation (\ref{Killingk}) implies Eqs.
(\ref{Dlk}), (\ref{Duk}), one may solve instead the
selfconsistency conditions for these equations,
{\setlength\arraycolsep{0pt}
\begin{eqnarray}\label{DDlk}
&& {\cal D}{\cal D}\lambda_\alpha{}^r = {\cal R}_\alpha{}^{\beta}
\lambda_\beta{}^r =  \lambda_\alpha{}^s (dA-A\wedge A)_s{}^r \\
 \label{DDuk}
&& {\cal D}{\cal D}u_\alpha{}^I = {\cal R}_\alpha{}^{\beta}
u_\beta{}^I =  \lambda_\alpha{}^r (dB_r^I + B_s^I \wedge
A_r{}^s)\; .
\end{eqnarray}
}Using the unity decomposition (\ref{g-1g}), which implies ${\cal
R}_\alpha{}^{\beta}= {\cal R}_\alpha{}^{\gamma} \lambda_\gamma{}^r
\, w_r{}^\beta + {\cal R}_\alpha{}^{\gamma} u_\gamma{}^I \,
\epsilon_I{}^\beta$,
 one finds the following expression for the generalized curvature
\begin{equation}\label{kcalR=gl}
 {\cal R}_\alpha{}^\beta = G_r{}^s \, \lambda_{\alpha}{}^r w_s{}^\beta
+ \nabla B_r^I \lambda_{\alpha}{}^r  \epsilon_I{}^\beta \; ,
\end{equation}
where {\setlength\arraycolsep{0pt}
\begin{eqnarray} \label{Grs=}
 && G_r{}^s :=  (dA - A\wedge A)_r{}^s \; , \qquad \\
\label{nbB=} && \nabla B_r^I:= dB_r^I - A_r{}^s \wedge B_s^I \; ,
\end{eqnarray}
}For $k=31$, corresponding to the case of a BPS preon,
equation~(\ref{kcalR=gl}) simplifies to \cite{BAIPV03}
\begin{equation}\label{calR=gl}
 {\cal R}_\alpha{}^\beta = dA \, \lambda_{\alpha} w^\beta
+(dB^I + B^I \wedge A) \,  \lambda_{\alpha}  \epsilon_I{}^\beta \;
.
\end{equation}
Equations (\ref{kcalR=gl}) and (\ref{calR=gl}) imply ${\cal
R}_\alpha{}^\beta = \lambda_\alpha{}^r (\cdots)_r{}^\beta$ and,
thus, due to the orthogonality condition (\ref{eKl=0}) they solve
equation~(\ref{eIR=0}), $ \epsilon_I{}^\beta {\cal
R}_\beta{}^{\alpha}  = 0$.

The conditions $G\subset SL(32,\mathbb{R})$ and hence, also for
the generalized holonomy group, $\Hol(\Omega) \subset
SL(32,\mathbb{R})$, ${\cal R}_\alpha{}^\alpha=0$ (which is always
the case for bosonic solutions of `free' CJS \cite{Hull03,P+T03}
and Type II supergravities \cite{P+T031}), imply  $A_r{}^r=0$ in
equation~(\ref{kcalR=gl}) [see equation~(\ref{trA=0})], while for
$k=31$ equation~(\ref{calR=gl}) simplifies to \cite{BAIPV03}
\begin{eqnarray}\label{calR=sl}
 \Hol(\Omega) \subset SL(32,\mathbb{R})\; , \; k=31 \; :
 \quad {\cal R}_\alpha{}^\beta = dB^I  \lambda_{\alpha}
\epsilon_I{}^\beta \; .
\end{eqnarray}
Finally, for $G\subset Sp(32,\mathbb{R})$ $\Omega^{[\alpha\beta
]}=0$, then $ \Hol(\Omega) \subset Sp(32,\mathbb{R})$,  $ {\cal
R}^{\alpha\beta} := C^{\alpha\gamma}{\cal R}_\gamma{}^\beta =
{\cal R}^{(\alpha\beta )}$, and equation~(\ref{calR=sl}) reduces
to \cite{BAIPV03}
 \begin{eqnarray}\label{calR=sp}
  \Hol(\Omega) \subset Sp(32,\mathbb{R})\; , \;  k=31 \;  : \quad
  {\cal R}_\alpha{}^\beta = dB  \, \lambda_{\alpha}
\lambda^\beta \; ,
\end{eqnarray}
where only one arbitrary one-form $B$ appears [to obtain
(\ref{calR=sp}) one has to keep in mind that $\epsilon_I{}^\alpha
= (\epsilon_i{}^\alpha, C^{\alpha\beta}\lambda_\beta)$,
$I=(i,31)$, equation~(\ref{OSpe})]. {\it Eqs.~(\ref{calR=sl}),
(\ref{calR=sp}) solve equation~(\ref{eIR=0}) for preons when
$G=SL(32,\mathbb{R})$ and $G=Sp(32,\mathbb{R})$, respectively}.

Equation (\ref{kcalR=gl}) with $A_r{}^r=0$ (Eq.~(\ref{trA=0}),
and, hence,  $(dA - A\wedge A)_r{}^r=0$) provides an explicit
expression for the result of equation (\ref{M5holcont}), namely,
for the fact that a $k$-supersymmetric solution of either  $D=11$
or $D=10$ Type II supergravities must have its generalized
holonomy group contained in $\Hol(\Omega) \subset \;
SL(32-k,\mathbb{R}) \ltimes \mathbb{R}^{k(32-k)}$. For a BPS preon
$k=31$, and $\; \Hol(\Omega) \subset {\mathbb{R}}^{31}$ as
expressed by equation (\ref{calR=sl}). However, our explicit
expressions for the $[sl(32-k,\mathbb{R}) \ltimes
(\mathbb{R}^{(32-k)} \oplus \stackrel{k}{\ldots} \oplus
\mathbb{R}^{(32-k)}]$-valued generalized curvatures ${\cal
R}_\alpha{}^\beta$, Eqs.~(\ref{kcalR=gl}), (\ref{calR=sl}), given
in terms of the Killing spinors $ \epsilon_I{}^\beta$ and bosonic
spinors $\lambda_{\alpha}{}^r$ characterizing the BPS preon
contents of a $\nu=k/32$ BPS state, may be useful in searching for
new supersymmetric solutions, including preonic $\nu=31/32$ ones.
Some steps in this direction are taken in the next section.

\section{BPS preons in supergravity} \label{sec:preonsugra}

\subsection{BPS preons in Chern-Simons supergravity}
\label{SubSecBPSCS}

The first observation is that the generalized curvature allowing
for a BPS preonic ($k=31$ supersymmetric) configuration for the
case of
 $\Hol(\Omega) \subset SL(32,\mathbb{R})$ holonomy, equation (\ref{calR=sl}),
is nilpotent
\begin{eqnarray}\label{RR=0}
{\cal R}_\alpha{}^\gamma \wedge {\cal R}_\gamma{}^\beta =0
\;,\quad \mathrm{for} \quad \Hol(\Omega) \subset SL(32,\mathbb{R})
\;\; , \; k=31 \;.
\end{eqnarray}
As a result it solves \cite{BAIPV03} the purely bosonic equations
of a Chern-Simons supergravity (see \cite{CS}),
\begin{eqnarray}\label{RRRRR=0}
{\cal R}_\alpha{}^{\gamma_1} \wedge {\cal
R}_{\gamma_1}{}^{\gamma_2} \wedge {\cal R}_{\gamma_2}{}^{\gamma_3}
\wedge {\cal R}_{\gamma_3}{}^{\gamma_4} \wedge {\cal
R}_{\gamma_4}{}^\beta =0\; .
\end{eqnarray}
The same is true for $\Hol(\Omega) \subset Sp(32,\mathbb{R})
\subset SL(32,\mathbb{R})$, where ${\cal R}$ is given by equation
(\ref{calR=sp}). Thus, there exist BPS preonic solutions in CS
supergravity theories, including $OSp(1|32)$-type ones.

Note that equation (\ref{RR=0}) follows in general for a preonic
configuration only. In fact, it implies that the generalized
holonomy algebra is abelian, in agreement with the fact noted
above that 31-supersymmetric solutions have their generalized
holonomy groups $\Hol(\Omega)$ in $\mathbb{R}^{31}$. For
configurations preserving $k\leq 30$ of the $32$ supersymmetries,
the bosonic equations of a CS supergravity, Eqs. (\ref{RRRRR=0})
reduce to (see (\ref{Grs=}), (\ref{nbB=}))
\begin{eqnarray}\label{GGGGG=0}
G_{s}{}^{s_2} \wedge G_{s_2}{}^{s_3} \wedge G_{s_3}{}^{s_4} \wedge
G_{s_4}{}^{s_5} \wedge G_{s_5}{}^{r} =0 \;\; , \nonumber \\
G_{s}{}^{s_2} \wedge G_{s_2}{}^{s_3} \wedge G_{s_3}{}^{s_4} \wedge
G_{s_4}{}^{r} \wedge \nabla B_{r}{}^{I} =0 \;\; ,
\end{eqnarray}
which are not satisfied identically for $G_r{}^r=0$.
Eqs.~(\ref{GGGGG=0}) are satisfied {\it e.g.}, by configurations
with $G_s{}^r=0$, for which the generalized holonomy group is
reduced down to $\Hol(\Omega) \subset \mathbb{R}^{k(32-k)}$,
${\cal R}_\beta{}^\alpha= \nabla B^I_r \lambda_\beta{}^r
\epsilon_{I}{}^{\alpha}$. Thus, {\it only} the preonic,
$\nu=31/32$, configurations {\it always} solve the Chern-Simons
supergravity equations (\ref{RRRRR=0}).

\subsection{Searching for preonic solutions of the free bosonic
CJS equations}\label{secIVB}

We now go back to the question of whether BPS $\nu=31/32$
(preonic) solutions exist for the standard CJS supergravity
\cite{CJS}. This problem can be addressed step by step, beginning
by studying the existence of preonic solutions of the `free'
bosonic CJS equations. To this aim it is useful to observe
\cite{GP02,BAIPV03,BAPV05} that these equations may be collected
in a compact expression for the generalized curvature, $i_a {\cal
R}_\alpha{}^\gamma \Gamma^a{}_\gamma{}^\beta = 0$ (equation
(\ref{EqCJS=0}) of chapter \ref{chapter2}). The generalized
curvature of a BPS preonic configuration satisfies equation
(\ref{calR=sl}), and thus it solves the `free' bosonic CJS
supergravity equations (\ref{EqCJS=0}) if \cite{BAIPV03}
 \begin{eqnarray}\label{GldB=0}
i_adB^I \, \epsilon_{_I}{}^\alpha \Gamma^a{}_{\alpha}{}^{\beta}
=0\; .
\end{eqnarray}
Actually, equation~(\ref{calR=sl}) substituted in (\ref{EqCJS=0})
gives
\begin{equation}
\lambda_\alpha \,
i_a dB^I \, \epsilon_I{}^\gamma
 \Gamma^a{}_{\gamma}{}^{\beta}  =0 \; .
\end{equation}
However, since $\lambda_\alpha\not=0$, this is equivalent to
(\ref{GldB=0}).

Equation (\ref{GldB=0}) contains a summed $I=1, \ldots , 31$ index
and, as a result, it is not easy to handle. It would be much
easier to deal with the expression
$\Gamma^a{}_{\alpha}{}^{\gamma}i_{a} {\cal R}_{\gamma}{}^{\beta}$
which, with equation~(\ref{calR=sl}) is equal to
$\Gamma^a{}_{\alpha}{}^{\gamma} \lambda_{\gamma}i_a dB^J
 \epsilon_{J}{}^{\beta}$. Indeed, $(\Gamma^a \lambda)_{\alpha}i_a dB^J
 \epsilon_{J}{}^{\beta}=0$, for instance, would imply
 $(\Gamma^a \lambda)_{\alpha}i_a dB^J =0$ which may be shown to have
 only trivial solutions. However,
$\Gamma^a{}_{\alpha}{}^{\gamma} i_a {\cal R}_\gamma{}^{\beta} \neq
 0$ in general {\it for a solution of the `free'
bosonic CJS equations} (equation~(\ref{EqCJS=0})),
\begin{equation}\label{GaiacalR=}
\Gamma^a{}_{\alpha}{}^{\gamma}i_a{\cal R}_{\gamma}{}^{\beta} = -
\ft{i}{12} \left( D\hat{F}_{\alpha}{}^{\beta}  + \, {\cal O} (F\,
F) \right) \; , \quad
\end{equation}
where $D=e^aD_a$ is the Lorentz covariant derivative (not to be
confused with ${\cal D}$ defined  in Eqs.~(\ref{Killingk}),
(\ref{CJSom})),
\begin{eqnarray}\label{hatF=}
\hat{F}_{\alpha}{}^{\beta} = F_{a_1a_2a_3a_4}
(\Gamma^{a_1a_2a_3a_4})_{\alpha}{}^{\beta} \; , \qquad
\end{eqnarray}
and $ {\cal O} (F\, F)$ denotes the terms of second order in
$F_{c_1c_2c_3c_4}$, {\setlength\arraycolsep{0pt}
\begin{eqnarray}\label{OFF}
 {\cal O} (F\, F)&=& \ft{1}{(3!)^2 \, 4!}\, e^a \left(\Gamma_a{}^{b_1b_2b_3} +
2 \delta_a^{[b_1}\Gamma^{b_2b_3]} \right)
\epsilon_{b_1b_2b_3c_1 \ldots c_4 d_1 \ldots d_4} F^{c_1 \ldots c_4}F^{d_1 \ldots d_4} \nonumber  \\
&& +  \ft{2i}{3} e^a \left(\Gamma_a{}^{b_1b_2b_3b_4} + 3
\delta_a^{[b_1}\Gamma^{b_2b_3b_4]} \right)
F_{cdb_1b_2}F^{cd}{}_{b_3b_4} \nonumber \\
&& + \ft{8i}{9} e^a \Gamma^{b_1b_2b_3b_4b_5}
F_{acb_1b_2}F^{c}{}_{b_3b_4b_5}
 \; .
\end{eqnarray}}

Equation (\ref{calR=sl}) then implies that for a hypothetical
preonic solution of the `free' bosonic CJS equations, the gauge
field strength $F_{abcd}$ should be nonvanishing (otherwise
$dB^J=0$ and ${\cal R}_\alpha{}^\beta=0$, see above) and satisfy
\begin{eqnarray}\label{GdB31=}
\Gamma^a{}_{\alpha}{}^{\gamma} \lambda_{\gamma}\; i_adB^J \;
\epsilon_{J}{}^{\beta}= - \ft{i}{12}
\left(D\hat{F}_{\alpha}{}^{\beta} + \, {\cal O} (F\, F)\right) \;
. \qquad
\end{eqnarray}
Using (\ref{wl=}), Eqs. (\ref{GdB31=}) split into a set of
restrictions for $F_{abcd}$,
\begin{eqnarray}\label{DFl=0}
\left( D\hat{F} + \, {\cal O} (F\, F)\right){}_{\alpha}{}^{\beta}
\lambda_{\beta} =0 \; ,
\end{eqnarray}
and equations for $dB^I$,
\begin{eqnarray}\label{GdB31=u}
\Gamma^a{}_{\alpha}{}^{\gamma} \lambda_{\gamma}\; i_adB^I \; = -
\ft{i}{12} \left( D\hat{F}  + \, {\cal O} (F\, F)
\right){}_{\alpha}{}^{\beta}  \; u_\beta{}^I . \qquad
\end{eqnarray}
{\it Eq.~(\ref{GdB31=}) or, equivalently,
Eqs.~(\ref{DFl=0}),(\ref{GdB31=u}) are the equations to be
satisfied by a CJS preonic configuration} \cite{BAIPV03}. Note
that if a non-trivial solution of the above equations with some
$F_{abcd}\not=0$ and some $dB^I\not=0$ is found, one would have
then to check in particular that such a solution satisfies $ddB^I=
0$ and $D_{[e}F_{abcd]}=0$.

On the other hand, if the general solution of the above equation
turned out to be trivial, $dB^I = 0$, this would imply ${\cal
R}_\alpha{}^\beta =0$ and, thus, a trivial generalized holonomy
group, $\Hol(\Omega)=1$. However, this is the necessary condition
for fully supersymmetric, $k=32$, solutions \cite{FFP02}. Hence a
general trivial solution for Eqs. (\ref{DFl=0}), (\ref{GdB31=u})
would indicate that a solution preserving $31$ supersymmetries
possesses all $32$ ones (thus corresponding to a fully
supersymmetric vacuum) and, hence, that there are no preonic,
$\nu=31/32$ solutions of the {\it free bosonic} CJS supergravity
equations (\ref{EiCJS}), (\ref{BIGFCJS}), (\ref{GFCJS}) and
(\ref{STa=}) and (\ref{dA3=a+F=}). If this happened to be the
case, one would have to study the existence of preonic solutions
for the CJS supergravity equations with non-trivial right hand
sides. These could be produced by corrections of higher-order in
curvature \cite{W00,Howe03,HW96} and by the presence of sources
(from some possibly exotic $p$-branes).

\section{On possible preonic branes} \label{sec:preonbranes}

\subsection{Brane solutions and worldvolume
actions}\label{secIVC}

As far as supersymmetric $p$-brane solutions of supergravity
equations are concerned, the usual situation is that to $\nu=1/2$
supersymmetric solutions ($\nu =16/32$ in the $D=11$ and $D=10$
Type II cases) there also exist worldvolume actions in the
corresponding ($D=11$ or $D=10$ Type II) superspaces possessing
$16$ $\kappa$-symmetries, exactly the number of supersymmetries
preserved by the supergravity solitonic solutions. The {\it
$\kappa$-symmetry--preserved supersymmetry} correspondence was
further discussed and extended for the case of $\nu < 1/2$
multi-brane solutions in \cite{BKOP97,ST97}.

In this perspective one may expect that if  preonic $\nu=31/32$
supersymmetric solutions of the CJS equations with a source do
exist,  a worldvolume action possessing $31$ $\kappa$-symmetries
should also exist in a curved $D=11$ superspace. For the time
being, no such actions are known in the {\it standard} $D=11$
superspace, but they do exist in a superspace enlarged with
additional tensorial `central' charge coordinates (see chapter
\ref{chapter7} and \cite{30/32,BL98,B02,ZU}). One might expect
that the role of these additional tensorial coordinates could be
taken over by the tensorial fields of supergravity. But this would
imply that the corresponding action does not exist in the flat
standard $D=11$ superspace as it would require a contribution from
the above additional field degrees of freedom (replacing the
tensorial coordinate ones as in \cite{JdA00}). This lack of a
clear flat {\it standard} superspace limit hampers the way towards
a hypothetical worldvolume action for a BPS preon in the usual
curved $D=11$ superspace.

Nevertheless, a shortcut in the search for such an action may be
provided by the observation \cite{BdAIL} that the superfield
description of the dynamical supergravity--superbrane interacting
system, described by the sum of the {\it superfield} action for
supergravity (still unknown for $D=10,11$) and the super-$p$-brane
action, is gauge equivalent to the much simpler dynamical system
described by the sum of the spacetime, component action for
supergravity and the action {\it for the purely bosonic limit} of
the super-$p$-brane. This bosonic $p$-brane action carries the
memory of being the bosonic limit of a super-$p$-brane by still
possessing $1/2$ of the spacetime local supersymmetries
\cite{BdAI}; this preservation of local supersymmetry reflects the
$\kappa$-symmetry of the original super-$p$-brane action.

Thus the $\kappa$-symmetric worldvolume actions for
super-$p$-branes have a clear spacetime counterpart: the purely
bosonic actions in spacetime possessing a part of local spacetime
supersymmetry of a `free' supergravity theory. This fact, although
explicitly discussed for the standard, $\nu=1/2$ superbranes in
\cite{BdAIL}, is general since it follows from symmetry
considerations only and thus it applies to any superbrane,
including a hypothetical preonic one. The number of
supersymmetries possessed by this bosonic brane action coincides
with the number of $\kappa$-symmetries of the parent
super-$p$-brane action. Moreover, these supersymmetries are
extracted by a projector which may be identified with the bosonic
limit of the $\kappa$-symmetry projector for the superbrane. With
this guideline in mind one may simplify, in a first stage, the
search for a worldvolume action for a BPS preon in standard
supergravity (or in a model minimally extending the standard
supergravity) by discussing the bosonic limit that such a
hypothetical action should have.

\subsection{BPS preons in D'Auria-Fr\'e
supergravity}\label{secIVD}

Consider \cite{BAIPV03} a symmetric spin-tensor one-form
$e^{\alpha\beta} = e^{\beta\alpha} =dx^\mu e_\mu^{\alpha\beta}
(x)$,  transforming under local supersymmetry as
\begin{eqnarray}\label{eabsusy}
\delta_{\varepsilon} e^{\alpha\beta} = - 2 i \psi^{(\alpha} \,
\varepsilon^{\beta)} \; ,
\end{eqnarray}
where $\psi^{\alpha}$ is a fermionic one-form,
\begin{eqnarray}\label{psiF}
 \psi^{\alpha} =  dx^\mu \psi_\mu^{\alpha}(x)\; ,
\end{eqnarray}
which we may identify with the gravitino. Let us consider for
simplicity the worldline action ({\it cf. } \cite{BL98})
\begin{eqnarray}\label{BPSac}
S &=& \int_{W^1} \lambda_\alpha (\tau)\, \lambda_\beta
(\tau) \hat{e}^{\alpha\beta} \nonumber \\
&=&  \int_{W^1} d\tau \lambda_\alpha (\tau)\, \lambda_\beta (\tau)
\, e_\mu^{\alpha\beta} (\hat{x}(\tau)) \,
\partial_\tau \hat{x}^\mu (\tau)
\; ,
\end{eqnarray}
where $\tau$ parameterizes the worldline $W^1$ in $D=11$
spacetime, $\hat{e}^{\alpha\beta}:= d\tau \partial_\tau
\hat{x}^\mu (\tau) \, e_\mu^{\alpha\beta} (\hat{x}(\tau))$ and
$\lambda_\alpha (\tau)$ is an auxiliary spinor field on the
worldline $W^1$. The extended ($p\ge 1$) object counterpart of
this worldline action is the following action for tensionless
$p$-branes (cf.~\cite{B02,ZU})
\begin{eqnarray}
S_{p+1}&=& \int_{W^{p+1}} \lambda_\alpha \lambda_\beta \hat{\rho}
\wedge
\hat{e}^{\alpha\beta} \nonumber\\
&=& \int_{W^{p+1}} d^{p+1} \xi \, \rho^k  \lambda_\alpha
\lambda_\beta \hat{e}^{\alpha\beta}_\mu \partial_k \hat{x}^\mu \;,
\label{Sp+1}
\end{eqnarray}
where $\hat{\rho}(\xi)$ is a $p$-form auxiliary field, and $\rho^k
(\xi)$ is the worldvolume vector density (see
\cite{Bansuper90,BanZhel93}) related to $\hat{\rho}(\xi)$ by
$\hat{\rho}(\xi)= (1/p!) d\xi^{j_p} \wedge \ldots \wedge
d\xi^{j_1} \rho_{j_1\ldots j_p}(\xi) =(1/p!) d\xi^{j_p} \wedge
\ldots \wedge d\xi^{j_1} \epsilon_{j_1\ldots j_pk} \rho^k(\xi)$.

The action (\ref{BPSac}) possesses all but one of the local
spacetime supersymmetries\footnote{Notice that when a brane action
is considered in a supergravity {\it background}, the local
spacetime supersymmetry is not a gauge symmetry of that action but
rather a transformation of the background; it becomes a gauge
symmetry only when a supergravity action is added to the brane one
so that supergravity is dynamical.}, equation~(\ref{eabsusy}),
$31$ for $\alpha, \beta=1, \ldots, 32$ corresponding to $D=11$.
Indeed, performing a supersymmetric variation $\delta_\varepsilon$
of (\ref{BPSac}) assuming $\delta_{\varepsilon} \lambda_\alpha
(\tau)=0$, one finds
\begin{eqnarray}\label{susyBPSac}
\delta_{\varepsilon} S &=& - 2i \int_{W^1} \hat{\psi}^\alpha
\lambda_\alpha (\tau)\; \hat{\varepsilon}^\beta \lambda_\beta
(\tau)  \; .
\end{eqnarray}
Thus, one sees that $\delta_{\varepsilon} S=0$ for the
supersymmetry parameters on $W^1$ that obey ({\it cf. }
(\ref{eKl=0}))
\begin{eqnarray}\label{el=0W}
\hat{\varepsilon}^\beta \lambda_\beta (\tau) =0  \quad
(\hat{\varepsilon}^\beta:= {\varepsilon}^\beta (\hat{x}(\tau))) \;
.
\end{eqnarray}
Equation (\ref{el=0W}) possesses $31$ solutions, which may be
expressed through worldvolume spinors
$\hat{\epsilon}_I{}^\alpha(\tau)$ (the worldline counterparts of
the Killing spinors) orthogonal to $\lambda_\alpha (\tau)$,
$\hat{\epsilon}_I{}^\alpha(\tau)\lambda_\alpha (\tau)=0$, as
\begin{eqnarray}\label{ve=eI}
\hat{\varepsilon}^\beta = {\varepsilon}^I(\tau)
\hat{\epsilon}_I{}^\beta \quad , \quad I=1,\ldots,31 \quad ,
\end{eqnarray}
for some arbitrary ${\varepsilon}^I(\tau)$. The same is true for
the tensionless $p$-branes described by the action (\ref{Sp+1}).

Thus, the actions (\ref{BPSac}), (\ref{Sp+1}) possess $31$ of the
$32$ local spacetime supersymmetries (\ref{eabsusy}) and, in the
light of the discussion  of the previous subsection, can be
considered as the spacetime counterparts of a superspace
BPS-preonic action (hypothetical in the standard superspace but
known \cite{BL98,BLS99,B02,ZU} in flat maximally enlarged or
tensorial superspaces).

The question that remains to be settled is the meaning of the
symmetric spin-tensor one-form $e^{\alpha\beta}$ with the local
supersymmetry transformation rule (\ref{eabsusy}) in $D=11$
supergravity. The contraction of $e^{\alpha\beta}$ with the Dirac
matrix $\Gamma^a$,
\begin{eqnarray}\label{ea=}
e^a = e^{\alpha\beta} \Gamma^a_{\alpha\beta} \quad ,
\end{eqnarray}
may be identified with the $D=11$ vielbein. Decomposing
$e^{\alpha\beta}$ in the basis of the $D=11$ $Spin (1,10)$
gamma-matrices,
\begin{eqnarray}\label{eab=}
e^{\alpha\beta} &=& e^{\beta\alpha}= \ft{1}{32} \left( e^a
\Gamma_a^{\alpha\beta} - \ft{i}{2!} B^{ab}
\Gamma_{ab}^{\alpha\beta} + \ft{1}{5!} B^{a_1\ldots a_5}
\Gamma_{a_1\ldots a_5}^{\alpha\beta} \right) \; ,
\end{eqnarray}
one finds \cite{BAIPV03} that $e^{\alpha\beta}$ also contains the
antisymmetric tensor one-forms $B^{ab}(x)=dx^\mu B_\mu^{ab}(x)$
and $B^{a_1\ldots a_5}(x)= dx^\mu B_\mu^{a_1\ldots a_5}(x)$. Such
fields, whose supersymmetry transformation properties follow from
(\ref{eab=}) and (\ref{eabsusy}), also appear among the additional
fields introduced in \cite{D'A+F} in order to investigate the
hidden gauge symmetry of $D=11$ supergravity, which  will be
discussed in chapter \ref{chapter6}. In this case, however, the
degrees of freedom of the $B$ fields in (\ref{BPSac}) will not be
reduced by the gauge symmetry to be discussed in chapter
\ref{chapter6}. Thus, the action (\ref{BPSac}), preserving 31 out
of 32 supersymmetries, could be treated as a worldline action for
a BPS preon in the presence of supergravity with additional fields
{\it \`a la} D'Auria and Fr\'e \cite{D'A+F}.

The formulation of $D=11$ supergravity due to D'Auria and Fr\'e
\cite{D'A+F} is, actually, closely related to enlarged superspaces
so, in this sense, it is not surprising that preonic branes would
exist in such a context (given that preonic actions are known in
enlarged superspaces, see \cite{Bandos05} for a review). It is,
then, worthwhile both to take a closer look at the D'Auria-Fr\'e
approach to supergravity and to further study actions for
supersymmetric extended objects in enlarged superspaces. We shall,
thus, turn our attention to these issues in chapters
\ref{chapter6} and \ref{chapter7} respectively. In particular, the
symmetry algebras underlying the construction of supergravity {\it
\`a la} D'Auria and Fr\'e will be reviewed in chapter
\ref{chapter6}. These algebras were known to be fermionic central
extensions of the M Theory superalgebra, but their expected
relation to the orthosymplectic superalgebra $osp(1|32)$ was quite
unclear. In chapter \ref{chapter6} this relation will be discussed
in terms of {\it Lie algebra expansions}, a new method of building
up new algebras from given ones. It seems appropriate, then, to
stop our physical discussion momentarily and open up a purely
mathematical parenthesis to introduce the expansion method in the
next chapter.


\chapter{Interlude: Lie algebra expansions} \label{chapter5}

Setting aside the problem of finding whether an algebra is a
subalgebra of another one there are, essentially, three different
ways of relating and/or obtaining new algebras from given ones:
contractions, deformations and extensions. In this chapter we
explore a fourth way to obtain new algebras of increasingly higher
dimensions from a given one $\mathcal G$. The idea, originally
considered in \cite{HaSa01} in a less general context and
developed in general in \cite{JdAIMO} (see also \cite{JdA02})
consists in looking at the algebra ${\cal G}$ as described by the
Maurer-Cartan (MC) forms\footnote{ See section
\ref{sec:MTsuperalg} of chapter \ref{chapter2} for the dual
formulation of Lie algebras in terms of MC one-forms.} on the
manifold of its associated group $G$ and, after rescaling some of
the group parameters by a factor $\lambda$, in expanding the MC
forms as a series in $\lambda$. The resulting {\it expansion}
method is different from the three above albeit, when the algebra
dimension does no change in the process, it may lead to a simple
In\"on\"u-Wigner (IW) or IW-generalized contraction (see section
\ref{fourw}), but not always. Furthermore, the algebras to which
it leads have in general a higher dimension than the original one
(hence the {\it expansion} name), in which case they cannot be
related to it by any contraction or deformation process.

A description of the expansion method is given in this rather
technical chapter. Our main concern will be its application to Lie
superalgebras, so we proceed step by step towards that goal. First
of all, the brief review in section \ref{fourw} of the three
already known methods to obtain new Lie algebras from given ones
will be useful in order to discuss the properties and structure of
the algebras encountered in the rest of the chapter (and other
parts of this Thesis). Section \ref{dos} introduces the expansion
method for Lie algebras ${\cal G}$. When further assumptions are
made about the structure of the original Lie algebras, the results
provided by the expansion method are more interesting. That is why
the existence of a subalgebra in ${\cal G}$ is assumed in section
\ref{tres}, and the further existence of a symmetric coset is
assumed in subsection \ref{scoset}. Section \ref{general}
generalizes in a convenient way the case in which ${\cal G}$
contains a subalgebra, by assuming that there is a certain
subspace splitting of ${\cal G}$. All these cases are actually
combined in section \ref{sec:susy} to discuss the expansions of
Lie superalgebras. The chapter concludes with an explicit example:
the derivation of the M Theory algebra from an expansion of
$osp(1|32)$. Appendix \ref{ap:expansion} contains some technical
details.  This chapter follows closely references \cite{JdAIMO}
and \cite{JdA02}.

\section{Three well-known ways to relate Lie (super)algebras}
\label{fourw}

\subsubsection{Contractions}

         The {\it first} one is the {\it contraction} procedure
\cite{Seg51,IW53,Sal61}. In its \.In\"on\"u and Wigner (IW) simple
form \cite{IW53}, the contraction  $\mathcal{G}_c$ of a Lie
algebra $\mathcal G$ is performed with respect to a subalgebra
$\mathcal{L}_0$ by rescaling the basis generators of the coset
$\mathcal{G} / \mathcal{L}_0$ by means of a parameter, and then by
taking a singular limit for this parameter. The generators in
$\mathcal{G} / \mathcal{L}_0$ become abelian in the contracted
algebra $\mathcal{G}_c$, and the subalgebra $\mathcal{L}_0 \subset
\mathcal{G}_c$ acts on them. As a result, $\mathcal{G}_c$ has a
semidirect structure, and the abelian generators determine an
ideal of $\mathcal{G}_c$; obviously, $\mathcal{G}_c$ has the same
dimension as $\mathcal{G}$. The contraction process has well known
physical applications as {\it e.g.}, in understanding the
non-relativistic limit from a group theoretical point of view, or
to explain the appearance of dimensionful generators when the
original algebra $\mathcal{G}$ is semisimple (and hence with
dimensionless generators). This is achieved by using a
dimensionful contraction parameter, as in the derivation of the
Poincar\'e group from the de Sitter groups (there, the parameter
is the radius $R$ of the universe, and the limit is $R \rightarrow
\infty$). There have been many discussions and variations of the
IW contraction procedure (see \cite{AC79, CelTar, Lord85,MonPat91,
HMOS94, Wei:00} to name a few), but all of them have in common
that $\mathcal{G}$ and $\mathcal{G}_c$ have, necessarily, the same
dimension as vector spaces.

This procedure can be extended to generalized IW contractions in
the sense of Weimar-Woods (W-W) \cite{Wei:00}. These are defined
when $\mathcal{G}$ can be split in a sum of vector subspaces
\begin{equation}
      \mathcal{G}= V_0\oplus V_1\oplus \dots\oplus V_n = \bigoplus_{s=0}^n V_s ,
\label{gc1}
\end{equation}
($V_0$ being the vector space of the subalgebra $\mathcal{L}_0$),
such that the following conditions are satisfied:
\begin{equation}
c^{k_s}_{i_pj_q}= 0 \;\; \text{if} \ s>p+q \qquad \textrm{\it
i.e.}\qquad  [V_p,V_q]\subset \bigoplus_s V_s,\ s\leq p+q \ ,
\label{gc2}
\end{equation}
where $i_p$ labels the generators of $\mathcal{G}$ in $V_p$, and
$c^k_{ij}$ are structure constants of $\mathcal{G}$. Then the W-W
\cite{Wei:00} contracted algebra is obtained by rescaling the
group parameters as $g^{i_p}\mapsto \lambda^p g^{i_p}$,
$p=0,\dots,n$, and then by taking a singular limit for $\lambda$.
The contracted Lie algebra obtained this way, $\mathcal{G}_c$, has
the same dimension as $\mathcal{G}$. The case $n=1$ corresponds to
the simple IW contraction.

\subsubsection{Deformations}

The {\it deformation} of algebras, and Lie algebras in particular
\cite{Gerst64,NijRich66,NijRich67b,Rich67} (see also
\cite{Le67,Her70}), allows us to obtain algebras {\it close}, but
not isomorphic, to a given one.  This leads to the important
notion of rigidity \cite{Gerst64,NijRich66, Rich67} (or physical
stability): an algebra is called {\it rigid} when any attempt to
deform it leads to an equivalent (isomorphic) one. From a physical
point of view, the deformation process is essentially the inverse
to the contraction one (see \cite{Le67} and the second ref. in
\cite{Wei:00}), and the dimensions of the original and deformed
Lie algebras are again the same. For instance, the Poincar\'e
algebra is not rigid, but the de Sitter algebras, being
semisimple, have trivial second cohomology group by the Whitehead
lemma and, as a result, they are rigid. One may also consider the
Poincar\'e algebra as a deformation of the Galilei algebra, so
that this deformation may be read as a group theoretical
prediction of relativity. Thus, the mathematical deformation may
be physically considered as a tool for developing a physical
theory from another pre-existing one.

Deformations are performed by modifying the r.h.s.~of the original
commutators by adding new terms that depend on a parameter $t$ in
the form
\begin{equation}
\hspace{-1cm}  [X,Y]_t =[X,Y]_0+\sum^\infty_{i=1} \omega_i(X,Y)
t^i \ , \quad X,Y \in {\cal G}\;,\quad \omega_i(X,Y) \in {\cal G}
\; . \label{def1}
\end{equation}
Checking the Jacobi identities up to $O(t^2)$, it is seen that the
expression satisfied by $\omega_1$ characterizes it as a
two-cocycle so that the second Lie algebra cohomology group
$H^2(\mathcal{G}, \mathcal{G})$ of $\mathcal{G}$ with coefficients
in the Lie algebra $\mathcal{G}$ itself is the group of
infinitesimal deformations of $\mathcal{G}$. Thus
$H^2(\mathcal{G}, \mathcal{G})=0$ is a sufficient condition for
rigidity \cite{Gerst64,NijRich66,NijRich67b,Rich67}.

\subsubsection{Extensions}

 In contrast with the previous procedures, the initial
data of the extension problem include {\it two} algebras ${\cal
G}$ and ${\cal A}$. A Lie algebra $\tilde{\mathcal{G}}$ is an
extension of the Lie algebra $\mathcal{G}$ by the Lie algebra
$\mathcal{A}$ if $\mathcal{A}$ is an ideal of
$\tilde{\mathcal{G}}$ and
$\tilde{\mathcal{G}}/\mathcal{A}=\mathcal{G}$. As a result,
$\text{dim}\, \tilde{\mathcal{G}}= \text{dim}\, \mathcal{G} +
\text{dim}\, \mathcal{A}$, so this process is also `dimension
preserving'.

Given $\mathcal{G}$ and $\mathcal{A}$, in order to obtain an
extension $\tilde{\mathcal{G}}$ of $\mathcal{G}$ by $\mathcal{A}$
it is necessary to specify first an action $\rho$ of $\mathcal{G}$
on $\mathcal{A}$ {\it i.e.}, a Lie algebra homomorphism $\rho: \,
\mathcal{G} \longrightarrow \text{End}\, \mathcal{A}$. The
different possible extensions $\tilde{\mathcal{G}}$ for
$(\mathcal{G},\mathcal{A} ,\rho)$ and the possible obstructions to
the extension process are, once again, governed by cohomology (see
\cite{AI95} and references therein). To be more explicit, let
$\mathcal{A}$ be abelian. The extensions are governed by
$H^2_\rho(\mathcal{G},\mathcal{A})$. Some special cases are: 1)
trivial action $\rho=0$, $H^2_0(\mathcal{G},\mathcal{A})\neq 0$.
These are central extensions, in which $\mathcal{A}$ belongs to
the centre of $\tilde{\mathcal{G}}$; they are determined by
non-trivial ${\cal A}$-valued two-cocycles on ${\cal G}$, and
non-equivalent extensions correspond to non-equivalent cocycles;
2) non-trivial action $\rho\neq 0$,
$H^2_\rho(\mathcal{G},\mathcal{A})= 0$ (semidirect extension of
$\mathcal{G}$ by $\mathcal{A}$); and 3) $\rho=0$,
$H^2(\mathcal{G},\mathcal{A})=0$ (direct sum of $\mathcal{G}$ and
$\mathcal{A}$, $\tilde{\mathcal{G}} = \mathcal{G} \oplus
\mathcal{A}$, or trivial extension).

Well-known examples of extensions in Physics include the centrally
extended Galilei algebra, which is relevant in quantum mechanics,
or the M Theory superalgebra that, without the Lorentz
automorphisms part, is the maximal central extension of the
abelian $D=11$ supertranslations algebra (see section
\ref{sec:MTsuperalg} of chapter \ref{chapter2} and
\cite{vHvP82,M-alg,JdA00}).

\section{The expansion method} \label{dos}

Let $G$ be a Lie group, of local coordinates $g^i$, $i=1, \ldots
,r=\textrm{dim}\,G$. Let $\mathcal{G}$ be its Lie
algebra\footnote{Calligraphic ${\mathcal G}$, ${\mathcal L}$,
${\mathcal W}$ will denote both the Lie algebras and their
underlying vector spaces; $V$, $W$ etc. will be used for vector
spaces that are not necessarily Lie algebras.} of basis $\{X_i
\}$, which may be realized by left-invariant generators $X_i(g)$
on the group manifold. Let  ${\mathcal G}^*$ be the coalgebra, and
let  $\{ \omega ^i (g) \}$, $i=1,\dots,r= \mathrm{dim} \, G$ be
the basis determined by the (dual, left-invariant) Maurer-Cartan
(MC) one-forms on $G$. Then, when $[X_i,X_j]=c_{ij}^k X_k$, the MC
equations read
\begin{equation} \label{eq:mc}
d\omega^k(g)=-\frac{1}{2}c_{ij}^k \omega^i(g) \wedge \omega^j(g)
\; , \quad i,j,k=1,\ldots,r \quad .
\end{equation}
\noindent

We wish to show in this section how we may obtain new algebras by
means of a redefinition $g^{l}\rightarrow \lambda g^l$ of some of
the group parameters and by looking at the power series expansion
in $\lambda$ of the resulting one-forms $\omega^{i}(g,\lambda)$.
Let $\theta$ be the left-invariant canonical form on $G$,
\begin{equation}
       \theta(g)=g^{-1}dg=
e^{-g^i X_i}\, d e^{g^i X_i} \equiv \omega^i X_i \; .
\end{equation}
\noindent Since {\setlength\arraycolsep{2pt}
\begin{eqnarray}
e^{-A}\, d e^{A} & = & dA + \frac12 [dA,A]+\frac{1}{3!}[[dA,A],A]+
\frac{1}{4!}[[[dA,A],A],A]+\ldots \nonumber \\ & = & dA +
\sum_{n=1}^{\infty} \frac{1}{(n+1)!}[\stackrel{n}{\ldots}
[dA,A],\ldots,A], A] \; ,
\end{eqnarray}
}one obtains, for $A \equiv g^k X_k$, $dA =(dg^j)X_j$, the
expansion of $\theta(g)$ and of the MC forms $\omega^i(g)$ as
polynomials in the group coordinates $g^i$ :
{\setlength\arraycolsep{0pt}
\begin{eqnarray} \label{eq:theta}
&&\theta(g) = \left[ \delta_j^i +\frac{1}{2!} c_{j k}^i g^k
\right.
\nonumber \\
&& \qquad \qquad + \left. \frac{1}{3!} c_{j k_1}^{h_1}c_{h_1
k_2}^i g^{k_1} g^{k_2} +
       \frac{1}{4!} c_{j k_1}^{h_1}c_{h_1 k_2}^{h_2}c_{h_2 k_3}^{i}
g^{k_1} g^{k_2} g^{k_3}+ \ldots \right] dg^j X_i \; , \nonumber \\
&& \\
\label{eq:serie2} && \omega^i(g) =  \left[ \delta_j^i
+\frac{1}{2!} c_{j k}^i g^k
\right. \nonumber \\
&& \left. \qquad  + \sum_{n=2}^{\infty} \frac{1}{(n+1)!} c_{j
k_1}^{h_1}c_{h_1 k_2}^{h_2} \ldots
       c_{h_{n-1} k_{n-1}}^{h_{n-1}}c_{h_{n-1} k_n}^{i}
g^{k_1} g^{k_2} \ldots g^{k_{n-1}} g^{k_n} \right] dg^j \ .
\nonumber \\ &&
\end{eqnarray}
}Looking at (\ref{eq:serie2}), it is evident that the redefinition
\begin{equation}
g^l \rightarrow \lambda g^l
\end{equation}
of {\it some} coordinates $g^l$ will produce an expansion of the
MC one-forms $\omega^i(g, \lambda)$ as a sum of one-forms
$\omega^{i,\alpha}(g)$ on $G$ multiplied by the corresponding
powers $\lambda^\alpha$ of $\lambda$.

\subsection{The Lie algebras $\mathcal{G}(N)$ expanded from $\mathcal{G}$}

Consider, as a first example, the splitting of ${\mathcal G}^*$
into the sum of two (arbitrary) vector subspaces,
\begin{equation} \label{split}
{\mathcal G}^*  = V_0^* \oplus V_1^* \; ,
\end{equation}
$V^*_0$, $V^*_1$ being generated by the MC forms
$\omega^{i_0}(g)$, $\omega^{i_1}(g)$ of ${\mathcal G}^*$ with
indices corresponding, respectively, to the unmodified and
modified parameters,
\begin{equation} \label{eq:redefinition}
g^{i_{0}} \rightarrow  g^{i_{0}} \; , \; g^{i_{1}} \rightarrow
\lambda g^{i_{1}}  \quad,\quad i_0 \; (i_1) = 1, \ldots,
\textrm{dim} \, V_0 \; (\textrm{dim} \, V_1) \,.
\end{equation}
In general, the series of $\omega^{i_0}(g,\lambda) \in V_0^*$,
$\omega^{i_1}(g,\lambda) \in V_1^*$, will involve all powers of
$\lambda$,
\begin{equation} \label{eq:series}
\omega^{i_{ p}}(g,\lambda)  =  \sum_{\alpha=0}^{\infty}
\lambda^\alpha \omega^{i_{p},\alpha}(g) = \omega^{i_{ p},0}(g) +
\lambda \omega^{i_{ p},1}(g) + \lambda^2 \omega^{i_{ p},2}(g) +
\ldots \; ,
\end{equation}
for $p=0,1$ and $\omega^{i_p}(g,1)=\omega^{i_p}(g)$. We will see
in the following sections what restrictions on $\mathcal G$ make
zero certain coefficient one-forms $\omega^{i_{p},\alpha}$.

With the above notation, the MC equations (\ref{eq:mc}) for
$\mathcal G$
      can be rewritten as
\begin{equation} \label{eq:MC}
d\omega^{k_{ s}}= -\frac{1}{2} c_{i_{ p}j_{ q}}^{k_{ s}} \,
\omega^{i_{ p}} \wedge \omega^{j_{ q}} \quad (p,q,s=0,1)
\end{equation}
or, explicitly {\setlength\arraycolsep{2pt}
\begin{eqnarray}
\label{eq:MCzero} d\omega^{k_{ 0}} & = & -\frac{1}{2} c_{i_{ 0}j_{
0}}^{k_{ 0}} \omega^{i_{ 0}} \wedge \omega^{j_{ 0}} - c_{i_{ 0}j_{
1}}^{k_{ 0}} \omega^{i_{ 0}} \wedge \omega^{j_{ 1}} -\frac{1}{2}
c_{i_{ 1}j_{ 1}}^{k_{ 0}}
\omega^{i_{ 1}} \wedge \omega^{j_{ 1}} \; , \\
\label{eq:MCone} d\omega^{k_{ 1}} & = & -\frac{1}{2} c_{i_{ 0}j_{
0}}^{k_{ 1}} \omega^{i_{ 0}} \wedge \omega^{j_{ 0}} - c_{i_{ 0}j_{
1}}^{k_{ 1}} \omega^{i_{ 0}} \wedge \omega^{j_{ 1}} -\frac{1}{2}
c_{i_{ 1}j_{ 1}}^{k_{ 1}} \omega^{i_{ 1}} \wedge \omega^{j_{ 1}}
\; .
\end{eqnarray}
}

\noindent Inserting now the expansions (\ref{eq:series}) into the
MC equations (\ref{eq:MC}) and using (\ref{eq:sumatorio}) in
appendix \ref{ap:expansion}, the MC equations are expanded in
powers of $\lambda$:
\begin{equation} \label{eq:MCexpanded}
\sum_{\alpha=0}^{\infty} \lambda^\alpha d\omega^{k_{ s}, \alpha}=
\sum_{\alpha=0}^{\infty} \lambda^\alpha \left[ -\frac{1}{2} c_{i_{
p}j_{ q}}^{k_{ s}} \sum_{\beta=0}^{\alpha} \omega^{i_{ p}, \beta}
\wedge \omega^{j_{ q}, \alpha-\beta} \right] \; .
\end{equation}
The equality of the two $\lambda$-polynomials in
(\ref{eq:MCexpanded}) requires the equality of the coefficients of
equal power $\lambda^{\alpha}$. This implies that the coefficient
one-forms $\omega^{i_{ p}, \alpha}$ in the expansions
(\ref{eq:series}) satisfy the identities:
\begin{equation} \label{eq:MCG}
d\omega^{k_{ s}, \alpha}= -\frac{1}{2} c_{i_{ p}j_{ q}}^{k_{ s}}
\sum_{\beta=0}^{\alpha} \omega^{i_{ p}, \beta} \wedge \omega^{j_{
q}, \alpha-\beta} \quad (p,q,s=0,1) \quad .
\end{equation}
We can rewrite (\ref{eq:MCG}) in the form
\begin{equation} \label{eq:cnts}
d\omega^{k_{ s}, \alpha}= -\frac{1}{2} C_{i_{ p},\beta \; j_{
q},\gamma}^{k_{ s},\alpha}\;
      \omega^{i_{ p}, \beta} \wedge
\omega^{j_{ q}, \gamma} \;, \;\;\; C_{i_{ p},\beta \; j_{
q},\gamma}^{k_{ s},\alpha}= \left\{
\begin{array}{lll} 0\,, &
\mathrm{if} \ \beta + \gamma \neq \alpha  \\
c_{i_{ p}j_{ q}}^{k_{s}}\,, & \mathrm{if} \ \beta + \gamma =
\alpha
\end{array} \right. .
\end{equation}

We now ask ourselves whether we can use the expansion coefficients
$\omega^{k_{ 0}, \alpha}$, $\omega^{k_{ 1}, \beta}$ up to given
orders $N_{ 0} \geq 0$, $N_{1} \geq 0$, $\alpha=0,1,\ldots, N_0$,
$\beta=0,1,\ldots, N_1$, so that equation~(\ref{eq:cnts}) (or
(\ref{eq:MCG})) determines the MC equations of a new Lie algebra.
The answer is affirmative. More precisely, {\it the vector space
generated by}
\begin{equation} \label{eq:largerbasis1}
\{ \omega^{i_{ 0}, 0}, \omega^{i_{ 0}, 1}, \ldots, \omega^{i_{ 0},
N}, \omega^{i_{ 1}, 0}, \omega^{i_{ 1}, 1}, \ldots, \omega^{i_{
1}, N} \} \; ,
\end{equation}
{\it together with the MC equations~(\ref{eq:cnts}) for the
structure constants}
\begin{equation} \label{eq:cnt2}
C_{i_{ p},\beta \; j_{ q},\gamma}^{k_{ s},\alpha}= \left\{
\begin{array}{lll}
0, & \mathrm{if} \ \beta + \gamma \neq \alpha  \\
c_{i_{ p}j_{ q}}^{k_{s}}, & \mathrm{if} \ \beta + \gamma = \alpha
\end{array} \right.
      \quad ( \alpha, \beta, \gamma = 1, \ldots ,N\;;\;p,q,s=0,1) \; ,
\end{equation}
{\it determines a Lie algebra $\mathcal{G}(N)$ for each expansion
order $N \geq 0$  of dimension} $\textrm{dim} \, \mathcal{G}(N) =
(N+1) \,  \textrm{dim} \, \mathcal{G}$ \cite{JdAIMO}.

\noindent To see why, consider the one-forms
\begin{equation} \label{eq:lb1}
\{\omega^{i_0,\alpha_0}\,;\,\omega^{i_1,\alpha_1}\}= \{
\omega^{i_{ 0}, 0}, \omega^{i_{ 0}, 1}, \ldots, \omega^{i_{ 0},
N_{ 0}}; \omega^{i_{ 1}, 0}, \omega^{i_{ 1}, 1}, \ldots,
\omega^{i_{ 1}, N_{ 1}} \}
\end{equation}
where we have not assumed {\it a priori} the same range for the
expansions of the one-forms of $V_0^*$ and $V_1^*$. To see whether
the vector space  $V^*(N_{ 0}, N_{ 1})$ of basis (\ref{eq:lb1})
determines a Lie algebra $\mathcal{G}(N_0,N_1)$, it is sufficient
to check that a) the exterior algebra generated by (\ref{eq:lb1})
is closed\footnote{An algebra of forms closed under $d$ defines in
general a free differential algebra (FDA): see chapter
\ref{chapter6} and references therein.} under the exterior
derivative $d$ and that b) the Jacobi identities for ${\mathcal
G}$ are satisfied.

To have closure under $d$ we need that the r.h.s.~of equations
(\ref{eq:cnts}) does not contain one-forms that are not already
present in (\ref{eq:lb1}). Consider the forms
$\omega^{i_s,\beta_s}$, $s=0,1$, that contribute to $d \omega^{k_{
s},\alpha_s}$ up to order $\alpha=N_s$. Looking at
equations~(\ref{eq:MCG}) it follows trivially that
\begin{equation} \label{eq:N}
N_0=N_1 \qquad (= N) \; .
\end{equation}
\noindent To check the Jacobi identities for $\mathcal{G}(N)$, it
is sufficient to see that $dd\omega^{k_s,\alpha}\equiv 0$ in
(\ref{eq:cnts}) is consistent with the definition of $C_{i_p,\beta
\; j_q,\gamma}^{k_s,\alpha}\,$. Equation (\ref{eq:cnts}) gives
\begin{equation} \label{eq:jacobi}
0 = C_{i_p,\beta\;j_q,\gamma}^{k_s,\alpha} C_{l_t,\rho\;
m_u,\sigma}^{i_p,\beta}\omega^{j_q,\gamma}\wedge
\omega^{l_t,\rho}\wedge\omega^{m_u,\sigma} \quad
(\alpha,\beta,\gamma,\rho,\sigma=1,\ldots,N)\;,
\end{equation}
\noindent which implies
\begin{equation}\label{eq:bjacobi}
C_{i_p,\beta\;[j_q,\gamma}^{k_s,\alpha} C_{l_t,\rho\;
m_u,\sigma]}^{i_p,\beta}=0\; .
\end{equation}
\noindent Now, on account of definition (\ref{eq:cnt2}), the terms
in the l.h.s.~above are either zero (when
$\alpha\not=\gamma+\rho+\sigma$) or give zero due to the Jacobi
identities for ${\mathcal G}$, $c_{i_{ p} [ j_{ q}}^{k_s} c_{l_t
m_u ]}^{i_p} =0 $. Thus, the $C_{i_p,\beta
\;j_q,\gamma}^{k_s,\alpha}\,$ satisfy the Jacobi identities
(\ref{eq:bjacobi}) and define the Lie algebra $\mathcal{G}(N,N)
\equiv \mathcal{G}(N)$ \cite{JdAIMO}.

Explicitly, the resulting algebras for the first few orders are \cite{JdAIMO}:\\[6pt]
\noindent $N=0\;,\;{\cal G}$(0):
\begin{equation} \label{eq:Nzero}
d\omega^{k_{ s}, 0}= -\frac{1}{2} c_{i_{ p}j_{ q}}^{k_{ s}}
\omega^{i_{ p}, 0} \wedge \omega^{j_{ q}, 0} \quad (p,q,s=0,1) \;
,
\end{equation}
\noindent{\it i.e.}, $\mathcal{G}(0)$ reproduces the original
algebra $\mathcal{G}$.
\\[6pt]
\noindent $N=1\;,\;{\cal G}$(1): {\setlength\arraycolsep{2pt}
\begin{eqnarray} \label{eq:N10}
d\omega^{k_{ s}, 0}&=& -\frac{1}{2}  c_{i_{ p}j_{ q}}^{k_{
s}} \omega^{i_{ p}, 0} \wedge \omega^{j_{ q}, 0}  \; ,  \\
\label{eq:N11} d\omega^{k_{ s}, 1}&=& -c_{i_{ p}j_{ q}}^{k_{ s}}
\omega^{i_{ p}, 0} \wedge \omega^{j_{ q}, 1} \quad (p,q,s=0,1) \;
.
\end{eqnarray}} \\[-10pt]
\noindent $N=2\;,\;{\cal G}$(2): {\setlength\arraycolsep{0.5pt}
\begin{eqnarray} \label{eq:N20}
\!\!\!\!\!\!\!\!\!\!\!\!\! d\omega^{k_{ s}, 0}&=& -\frac{1}{2}
c_{i_{p}j_{ q}}^{k_{
s}} \omega^{i_{ p}, 0} \wedge \omega^{j_{ q}, 0} \; , \\
\label{eq:N21} \!\!\!\!\!\!\!\!\!\!\!\!\! d\omega^{k_{ s}, 1}&=& -
c_{i_{ p}j_{ q}}^{k_{s}} \omega^{i_{ p}, 0} \wedge \omega^{j_{ q},
1}
\; ,\\
\label{eq:N22} \!\!\!\!\!\!\!\!\!\!\!\!\! d\omega^{k_{ s}, 2}&=&
-c_{i_{ p}j_{ q}}^{k_{ s}} \omega^{i_{ p}, 0} \wedge \omega^{j_{
q}, 2} -\frac{1}{2} c_{i_{ p}j_{q}}^{k_{ s}} \omega^{i_{ p}, 1}
\wedge \omega^{j_{ q}, 1} \;\; (p,q,s=0,1) \quad .
\end{eqnarray}}

In sight of the above results, the following remark is in order.
Since $\omega^{i_{ p}, 0}(g) \neq \omega^{i_{ p}}(g)$, one might
wonder how the MC equations for $\mathcal{G}(0)=\mathcal{G}$ can
be satisfied by $\omega^{i_{ p},0}(g)$. The
$\mathrm{dim}\,\mathcal{G}$ MC forms $\omega^{i_{p}}(g)$ are
left-invariant forms on the group manifold $G$ of $\mathcal{G}$.
The $(N+1)\mathrm{dim}\,\mathcal{G}$ $\omega^{i_{p}, \alpha}(g)$
($\alpha=0,1,\ldots,N$) determined by the expansions
(\ref{eq:series})  are also one-forms on $G$, but they are no
longer left-invariant under $G$-translations. They cannot be,
since there are only $\mathrm{dim} \, G=r$ linearly independent MC
forms on $G$. Nevertheless, equations (\ref{eq:cnts}) determine
the MC relations that will be satisfied by the MC forms on the
manifold of the {\it higher dimensional} group $G(N)$ associated
with $\mathcal{G}(N)$. These MC forms on $G(N)$ will depend on the
$(N+1)\mathrm{dim} \, \mathcal{G}(N)$ coordinates of $G(N)$
associated with the generators (forms) $X_{i_p, \alpha}$
($\omega^{i_p,\alpha}$) that determine $\mathcal{G}(N)$
(${\mathcal G}^*(N)$).

\subsection{Structure of the expanded algebras $\mathcal{G}(N)$}
\label{s:structure}

Let $V_{p,\alpha}$ be, at each order $\alpha=0,1,\ldots, N$, the
vector space spanned by the generators $X_{{i_p},\alpha}$,
$p=0,1$; clearly, $V_{p,\alpha} \approx V_{p}$. Let
\begin{equation} \label{def:W}
W_{\alpha}=V_{0,\alpha} \oplus V_{1,\alpha} \quad , \quad
{\mathcal G}(N) = \bigoplus_{\alpha=0}^N W_{\alpha} \quad .
\end {equation}
We first notice that ${\mathcal G}(N-1)$ is a vector subspace of
${\mathcal G}(N)$, but not a subalgebra for $N \geq 2$. Indeed,
for $N \geq 2$ there always exist $\alpha,\beta \leq N-1$ such
that $\alpha+\beta=N$. Denoting by $C_{i_p,\alpha
\;j_q,\beta}^{(N)\,k_s,\gamma}$ and $C_{i_p,\alpha
\;j_q,\beta}^{(N-1)\,k_s,\gamma}$ the structure constants of
$\mathcal{G}(N)$ and $\mathcal{G}(N-1)$ respectively, one sees
that, for $\alpha+\beta=N$, $C_{i_p,\alpha \;
j_q,\beta}^{(N-1)\,k_s,\alpha+\beta}=0$ in $\mathcal{G}(N-1)$
(since $\alpha+\beta > N-1$) while, in general, $C_{i_p,\alpha \;
j_q,\beta}^{(N)\,k_s,\alpha+\beta} \neq 0$ in $\mathcal{G}(N)$. In
other words, $\mathcal{G}(N-1)$ is not a subalgebra of
$\mathcal{G}(N)$ because the structure constants for the elements
of the various subspaces $V_{p,\alpha}$ depend on $N$ and they are
different, in general, for $\mathcal{G}(N-1)$ and
$\mathcal{G}(N)$. Likewise, $\mathcal{G}(M)$ for $1 \leq M < N$ is
not a subalgebra of $\mathcal{G}(N)$.

We now show that the Lie algebras $\mathcal{G}(N)$ have a Lie
algebra extension structure for $N \geq 1$. More precisely, {\it
the Lie algebra $\mathcal{G}(0)$ is a subalgebra of
$\mathcal{G}(N)$, for all $N \geq 0$. For $N \geq 1$, $W_N$ is an
abelian ideal $\mathcal{W}_N \subset \mathcal{G}(N)$ and
$\mathcal{G}(N) / \mathcal{W}_N = \mathcal{G}(N-1)$ i.e.,
$\mathcal{G}(N)$ is an extension of $\mathcal{G}(N-1)$ by
$\mathcal{W}_N$ which is not semidirect for $N \ge 2$}
\cite{JdAIMO}. To prove this result, notice that $\mathcal{G}(0)
\subset \mathcal{G}(N)$ is a subalgebra by construction, since
$C_{i_p,0 \; j_q,0}^{(N)\,k_s,\alpha}=0$, $\alpha=1,\ldots,N$, by
equation~(\ref{eq:cnts}). For the second part, notice that, since
$\alpha+N > N$ for $\alpha \neq 0$, $[W_{\alpha}, \,
\mathcal{W}_N] =0 $; in particular, $\mathcal{W}_N$ is an abelian
subalgebra. Furthermore $[\mathcal{W}_0, \, \mathcal{W}_N] \subset
\mathcal{W}_N$, so that $\mathcal{W}_N$ is an ideal of
$\mathcal{G}(N)$. Now, the vector space $\mathcal{G}(N) / W_N$ is
isomorphic to $\mathcal{G}(N-1)$. $\mathcal{G}(N-1)$ is a Lie
algebra the MC equations of which are (\ref{eq:cnts}), and
$\mathcal{G}(N) / \mathcal{W}_N \approx \mathcal{G}(N-1)$. Since
$\mathcal{G}(N-1)$ is not a subalgebra of $\mathcal{G}(N)$ for
$N\geq 2$, the extension is not semidirect.

\subsection{Limiting cases}

Let us discuss the limiting cases $V_0=0,V_1=V$ and $V_0=V,V_1=0$
When $V_1=V$, all the group parameters are modified by
(\ref{eq:redefinition}). In this case $\mathcal{G}(0)$ is the
trivial $\mathcal{G}(0)=0$ subalgebra of $\mathcal{G}(N)$. The
first order $N=1$, $\omega^{i_1,1}=dg^{i_1}$, corresponds to an
abelian algebra with the same dimension as $\mathcal{G}$ (in fact,
$\mathcal{G}(1)$ is the IW contraction of $\mathcal{G}$ with
respect to the trivial $V_0=0$ subalgebra). For $N \ge 2 $ we will
have extensions with the structure in section \ref{s:structure}.

For the other limiting case, $V_1=0$, there is obviously no
expansion and we have $\mathcal{G}(0)=\mathcal{G}$.

\section{The case in which $\mathcal{G}$ contains a subalgebra} \label{tres}

Let ${\mathcal G} = V_{ 0} \oplus V_{ 1}$ as before, where now
$V_0$ is a subalgebra $\mathcal{L}_0$ of $\mathcal{G}$. Then,
\begin{equation} \label{eq:subalgebra}
c_{i_0 j_0}^{k_1} = 0 \quad \quad (i_p = 1, \ldots, \textrm{dim}
\, V_p \, , \; p=0,1) \; ,
\end{equation}
and the  basis one-forms $\omega^{i_0}$ are associated with the
(sub)group parameters $g^{i_0}$ unmodified under the rescaling
(\ref{eq:redefinition}). The MC equations for $\mathcal{G}$ become
{\setlength\arraycolsep{2pt}
\begin{eqnarray}
\label{eq:MCzero1} d\omega^{k_{0}} & = & -\frac{1}{2}
c_{i_{0}j_{0}}^{k_{0}} \omega^{i_{0}} \wedge \omega^{j_{0}} -
c_{i_{0}j_{1}}^{k_{0}} \omega^{i_{0}} \wedge \omega^{j_{1}}
-\frac{1}{2} c_{i_{1}j_{1}}^{k_{0}}
\omega^{i_{1}} \wedge \omega^{j_{1}} \; , \\
\label{eq:MCone1} d\omega^{k_{1}} & = & - c_{i_{0}j_{1}}^{k_{1}}
\omega^{i_{0}} \wedge \omega^{j_{1}} -\frac{1}{2}
c_{i_{1}j_{1}}^{k_{1}} \omega^{i_{1}} \wedge \omega^{j_{1}} \; .
\end{eqnarray}}

Using (\ref{eq:subalgebra}) in equation (\ref{eq:serie2}), one
finds that the expansions of $\omega^{i_0}(g,\lambda)$
($\omega^{i_1}(g, \lambda)$)  start with the power  $\lambda^0$
($\lambda^1$): {\setlength\arraycolsep{2pt}
\begin{eqnarray} \label{eq:serieszero}
\omega^{i_{0}}(g,\lambda) & = & \sum_{\alpha=0}^{\infty}
\lambda^\alpha \omega^{i_{0},\alpha}(g) = \omega^{i_{0},0}(g) +
\lambda \omega^{i_{0},1}(g) + \lambda^2 \omega^{i_{0},2}(g) +
\ldots \nonumber  \\ && \\
\label{eq:seriesone} \omega^{i_{1}}(g,\lambda) & = &
\sum_{\alpha=1}^{\infty} \lambda^\alpha \omega^{i_{1},\alpha}(g) =
\lambda \omega^{i_{1},1}(g) + \lambda^2 \omega^{i_{1},2}(g) +
\lambda^3 \omega^{i_{1},3}(g) + \ldots  \; . \nonumber \\ &&
\end{eqnarray}}

\noindent Inserting them into the MC equations (\ref{eq:MCzero1})
and (\ref{eq:MCone1}) and using equation~(\ref{eq:sumatorio}) of
appendix \ref{ap:expansion} when the double sums begin with
$(0,0)$, $(0,1)$ and $(1,1)$, we get {\setlength\arraycolsep{0pt}
\begin{eqnarray}
\label{eq:MCins0} && \sum_{\alpha=0}^{\infty} \lambda^\alpha
d\omega^{k_{0}, \alpha}  =  -\frac{1}{2} c_{i_{0}j_{0}}^{k_{0}}
\omega^{i_{0}, 0} \wedge \omega^{j_{0}, 0} \nonumber \\
&& \qquad  + \lambda \left[ -c_{i_{0}j_{0}}^{k_{0}}
      \omega^{i_{0}, 0} \wedge \omega^{j_{0}, 1}
-c_{i_{0}j_{1}}^{k_{0}}
      \omega^{i_{0}, 0} \wedge \omega^{j_{1}, 1} \right]
\nonumber \\
&  & \qquad  + \sum_{\alpha=2}^{\infty} \lambda^\alpha \left[
-\frac{1}{2} c_{i_{0}j_{0}}^{k_{0}} \sum_{\beta=0}^{\alpha}
\omega^{i_{0}, \beta} \wedge \omega^{j_{0}, \alpha-\beta} \right.
\nonumber \\
&& \qquad \qquad \qquad  \left.   -c_{i_{0}j_{1}}^{k_{0}}
\sum_{\beta=0}^{\alpha-1} \omega^{i_{0}, \beta} \wedge
\omega^{j_{1}, \alpha-\beta} - \frac{1}{2} c_{i_{1}j_{1}}^{k_{0}}
      \sum_{\beta=1}^{\alpha-1} \omega^{i_{1}, \beta} \wedge
\omega^{j_{1}, \alpha-\beta} \right] \; , \nonumber \\ && \\
\label{eq:MCins1} && \sum_{\alpha=1}^{\infty} \lambda^\alpha
d\omega^{k_{1}, \alpha}  =  -\lambda c_{i_{0}j_{1}}^{k_{1}}
      \omega^{i_{0}, 0} \wedge \omega^{j_{1}, 1}
\nonumber  \\
&& \qquad + \sum_{\alpha=2}^{\infty} \lambda^\alpha \left[
-c_{i_{0}j_{1}}^{k_{1}} \sum_{\beta=0}^{\alpha-1} \omega^{i_{0},
\beta} \wedge \omega^{j_{1}, \alpha-\beta} -\frac{1}{2}
c_{i_{1}j_{1}}^{k_{1}}
      \sum_{\beta=1}^{\alpha-1} \omega^{i_{1}, \beta} \wedge
\omega^{j_{1}, \alpha-\beta} \right] \, . \nonumber \\ &&
\end{eqnarray}}

\noindent Again, the equality of the coefficients of equal power
$\lambda^\alpha$ in
(\ref{eq:MCins0}), (\ref{eq:MCins1}) leads to the equalities: \\[5pt]
$\alpha=0$:
\begin{equation} \label{eq:00}
d\omega^{k_{0}, 0} = -\frac{1}{2} c_{i_{0}j_{0}}^{k_{0}}
\omega^{i_{0}, 0} \wedge \omega^{j_{0}, 0} \quad ;
\end{equation}
$\alpha=1$: {\setlength\arraycolsep{2pt}
\begin{eqnarray}
\label{eq:01} d\omega^{k_{0}, 1} & = & -c_{i_{0}j_{0}}^{k_{0}}
      \omega^{i_{0}, 0} \wedge \omega^{j_{0}, 1}
-c_{i_{0}j_{1}}^{k_{0}}
      \omega^{i_{0}, 0} \wedge \omega^{j_{1}, 1} \quad , \\
\label{eq:11} d\omega^{k_{1}, 1} & = & -c_{i_{0}j_{1}}^{k_{1}}
      \omega^{i_{0}, 0} \wedge \omega^{j_{1}, 1} \quad ;
\end{eqnarray}}

\noindent $\alpha \geq 2$: {\setlength\arraycolsep{0pt}
\begin{eqnarray}
\label{eq:0r} && \kern-2em d\omega^{k_{0}, \alpha}  = -\frac{1}{2}
c_{i_{0}j_{0}}^{k_{0}} \sum_{\beta=0}^{\alpha} \omega^{i_{0},
\beta} \wedge \omega^{j_{0}, \alpha-\beta} -c_{i_{0}j_{1}}^{k_{0}}
\sum_{\beta=0}^{\alpha-1} \omega^{i_{0}, \beta} \wedge
\omega^{j_{1}, \alpha-\beta} \nonumber \\ && \kern-2em \qquad
\qquad -\frac{1}{2} c_{i_{1}j_{1}}^{k_{0}}
      \sum_{\beta=1}^{\alpha-1} \omega^{i_{1}, \beta} \wedge
\omega^{j_{1}, \alpha-\beta} \, , \\
 \label{eq:1r} && \kern-2em d\omega^{k_{1}, \alpha} =
-c_{i_{0}j_{1}}^{k_{1}} \sum_{\beta=0}^{\alpha-1} \omega^{i_{0},
\beta} \wedge \omega^{j_{1}, \alpha-\beta} -\frac{1}{2}
c_{i_{1}j_{1}}^{k_{1}}
      \sum_{\beta=1}^{\alpha-1} \omega^{i_{1}, \beta} \wedge
\omega^{j_{1}, \alpha-\beta} \, .
\end{eqnarray}}

To allow for a different range in the orders $\alpha$ of each
$\omega^{i_p, \alpha}$, we now denote the coefficient one-forms in
(\ref{eq:serieszero}) ((\ref{eq:seriesone})) $\omega^{i_0 ,
\alpha_0}$ ($\omega^{i_1 , \alpha_1}$), $\alpha_0=0,1,\ldots,N_0$
($\alpha_1=1,2,\ldots,N_1$). With this notation, the above
relations take the generic form
\begin{equation} \label{eq:MCsub}
d\omega^{k_{ s}, \alpha_s}= -\frac{1}{2} C_{i_{ p},\beta_p \;
j_{q},\gamma_q}^{k_s,\alpha_s}\; \omega^{i_{ p}, \beta_p} \wedge
\omega^{j_{ q}, \gamma_q} \quad ,
\end{equation}
where
\begin{equation} \label{eq:Csub}
C_{i_{ p},\alpha_p \; j_{ q},\alpha_q}^{k_{ s},\alpha_s}= \left\{
\begin{array}{lll} 0, &
\mathrm{if} \ \beta_p + \gamma_q \neq \alpha_s  \\
c_{i_{ p}j_{ q}}^{k_{s}}, & \mathrm{if} \ \beta_p + \gamma_q =
\alpha_s  \end{array} \right. \quad \begin{array}{l}
p,q,s=0,1 \\
i_{p,q,s}=1,2,\ldots, \textrm{dim} \, V_{p,q,s} \\
\alpha_0,\beta_0,\gamma_0=0,1, \ldots, N_0 \\
\alpha_1,\beta_1,\gamma_1=1,2, \ldots, N_1 \;  .
\end{array}
\end{equation}

As in the preceding case, we now ask ourselves whether the
expansion coefficients $\omega^{k_{0}, \alpha_0}$, $\omega^{k_{1},
\alpha_1}$ up to a given order $N_0,N_1$ determine the MC
equations (\ref{eq:MCsub}) of a new Lie algebra
$\mathcal{G}(N_0,N_1)$. It is obvious from (\ref{eq:00}) that the
zeroth order of the expansion in $\lambda$ corresponds to
$N_0=0=N_1$ (omitting all $\omega^{i_1,\alpha_1}$ and thus
allowing $N_1$ to be zero), and that ${\mathcal G}(0,0)={\mathcal
L}_0$. It is seen directly that the terms up to first order give
two possibilities: ${\mathcal G}(0,1)$ (equations (\ref{eq:00}),
(\ref{eq:11}) for $\omega^{k_0,0}\,, \omega^{k_1,1}$) and
${\mathcal G}(1,1)$ (equations (\ref{eq:00}), (\ref{eq:01}),
(\ref{eq:11}) for
$\omega^{k_0,0}\,,\omega^{k_1,1}\,,\omega^{k_0,1}$). Thus, we see
that now (and due to (\ref{eq:subalgebra})) one does not need to
retain {\it all} $\omega^{i_p,\alpha_p}$ up to a given order to
obtain a Lie algebra. To look at the general $N_0\geq 0, N_1\geq
1$ case, consider the vector space $V^*(N_{0}, N_{1})$, generated
by
\begin{equation} \label{eq:largerbasis}
\{ \omega^{i_0, \alpha_0}\, ; \, \omega^{i_1, \alpha_1} \} = \{
\omega^{i_{0}, 0}, \omega^{i_{0}, 1}, \omega^{i_{0}, 2}, \ldots,
\omega^{i_{0}, N_{0}}; \omega^{i_{1}, 1}, \omega^{i_{1}, 2},
\ldots, \omega^{i_{1}, N_{1}} \} \; .
\end{equation}
To see that it determines a Lie algebra ${\mathcal G}(N_0,N_1)$ of
dimension
\begin{equation} \label{eq:dimsub}
\textrm{dim} \, \mathcal{G}(N_0,N_1)= (N_0+1) \, \textrm{dim} \,
V_0 + N_1 \textrm{dim} \, V_1 \; ,
\end{equation}
we first notice that the Jacobi identities in $\mathcal{G}(N_{0},
N_{1})$ will follow from those in $\mathcal{G}$. To find the
conditions that $N_{0}$ and $N_{1}$ must satisfy to have closure
under $d$, we look at the orders $\beta_p$ of the forms
$\omega^{i_{p},\beta_p}$ that appear in the expression
(\ref{eq:MCsub}) of $d \omega^{k_{s},\alpha_s}$ up to a given
order $\alpha_s \geq s$. Looking at equations~(\ref{eq:00}) to
(\ref{eq:1r}) we find the following table:

\begin{center}
\begin{tabular}{|c|cc|}
\hline
$\alpha_s \geq s$ & $\omega^{i_{0}, \beta_0}$ & $\omega^{i_{1}, \beta_1}$ \\
\hline $d \omega^{k_{0}, \alpha_0}$ & $\beta_0 \leq \alpha_0$ &
$\beta_1 \leq \alpha_0$ \\
$d \omega^{k_{1}, \alpha_1}$ & $\beta_0 \leq \alpha_1-1$ &
$\beta_1 \leq \alpha_1$ \\
\hline
\end{tabular}
\\[6pt]
{\footnotesize Table 5.1. Orders $\beta_p$ of the forms
$\omega^{i_{p}, \beta_p}$ that contribute to $d\omega^{k_{s},
\alpha_s}$}
\end{center}

\noindent Since there must be enough one-forms in
(\ref{eq:largerbasis}) for the MC equations (\ref{eq:MCsub}) to be
satisfied, the $N_{0}+1$ and $N_{1}$ one-forms $\omega^{i_{0},
\alpha_0}$  ($\alpha_0=0, 1,\ldots,N_0$) and $\omega^{i_{1},
\alpha_1}$ ($\alpha_1=1,2,\ldots,N_1$) in (\ref{eq:largerbasis})
should include, at least, those appearing in their differentials.
Thus, the previous table 5.1 implies the reverse inequalities
\begin{center}
\begin{tabular}{|c|cc|}
\hline
$\alpha_s \geq s$ & $\omega^{i_{0}, \beta_0}$ & $\omega^{i_{1}, \beta_1}$ \\
\hline
$d \omega^{k_{0}, \alpha_0}$ & $N_{0} \geq N_{0}$ & $N_{1} \geq N_{0}$ \\
$d \omega^{k_{1}, \alpha_1}$ & $N_{0}  \geq N_{1}-1$ & $N_{1} \geq N_{1}$ \\
\hline
\end{tabular}
\\[6pt]
{\footnotesize Table 5.2. Conditions on the number $N_0$ $(N_1)$
of one-forms $\omega^{i_{0}, \alpha_0} (\omega^{i_{1},
\alpha_1})$}
\end{center}
Hence, in this case there are two ways of cutting the expansions
(\ref{eq:serieszero}), (\ref{eq:seriesone}), namely for
{\setlength\arraycolsep{2pt}
\begin{eqnarray}
\label{eq:order1}
N_1 & = & N_0 \;, \\
\label{eq:order2} \textrm{or}\qquad N_1 & = & N_0 + 1 \; .
\end{eqnarray}}

Besides (\ref{eq:N}) there is now an additional type of solutions,
equation~(\ref{eq:order2}). For the $N_0=0, N_1=1$ values
equation~(\ref{eq:dimsub}) yields $\textrm{dim} \,
\mathcal{G}(0,1)= \textrm{dim} \, \mathcal{G}$. Then, $\alpha_0=0$
and $\alpha_1=1$ only, the label $\alpha_p$ may be dropped and the
structure constants (\ref{eq:Csub}) for $\mathcal{G}(0,1)$ read
\begin{equation} \label{eq:CIW}
C_{i_{ p}\,j_{ q}}^{k_{ s}}= \left\{ \begin{array}{lll} 0, &
\mathrm{if} \ p + q \neq s  \\
c_{i_{ p}j_{ q}}^{k_{s}}, & \mathrm{if} \ p + q = s  \end{array}
\right. \quad \begin{array}{l}
p=0,1 \\
i_{p,q,s}=1,2,\ldots, \textrm{dim} \, V_{p,q,s} \quad ,
\end{array}
\end{equation}
which shows that $V_1$ is an abelian ideal of $\mathcal{G}(0,1)$.
Hence, $\mathcal{G}(0,1)$ is just the (simple) IW contraction of
$\mathcal{G}$ with respect to the subalgebra $\mathcal{L}_0$, as
it may be seen by taking the $\lambda\rightarrow 0$ limit in
(\ref{eq:MCins0})-(\ref{eq:MCins1}), which reduce to equations
(\ref{eq:00}) and (\ref{eq:11}).

To summarize, {\it let $\mathcal{G}=V_{0} \oplus V_{1}$, where
$V_0$ is a subalgebra $\mathcal{L}_{0}$ and let the coordinates
$g^{i_p}$ of $G$ be rescaled by $g^{i_0} \rightarrow
g^{i_{0}},g^{i_1} \rightarrow \lambda g^{i_{1}}$
(equation~(\ref{eq:redefinition})). Then, the coefficient
one-forms $\{\omega^{i_0,\alpha_0}$, $\omega^{i_1, \alpha_1}\}$ of
the expansions (\ref{eq:serieszero}), (\ref{eq:seriesone}) of the
Maurer-Cartan forms of $\mathcal{G}^*$ determine Lie algebras
$\mathcal{G}(N_{0}, N_{1})$ when $N_1=N_0$ {\it or} $N_1=N_0+1$ of
dimension} $\textrm{dim} \, \mathcal{G}(N_0,N_1)= (N_0+1) \,
\textrm{dim} \, V_0 + N_1 \textrm{dim} \, V_1$ {\it and with
structure constants (\ref{eq:Csub}),}
\begin{displaymath}
C_{i_{ p},\beta_p \; j_{ q},\gamma_q}^{k_{ s},\alpha_s}= \left\{
\begin{array}{lll} 0, &
\mathrm{if} \ \beta_p + \gamma_q \neq \alpha_s  \\
c_{i_{ p}j_{ q}}^{k_{s}}, & \mathrm{if} \ \beta_p + \gamma_q =
\alpha_s  \end{array} \right. \quad \begin{array}{l}
p,q,s=0,1 \\
i_{p,q,s}=1,2,\ldots, \textrm{dim} \, V_{p,q,s} \\
\alpha_0,\beta_0,\gamma_0=0,1, \ldots, N_0 \\
\alpha_1,\beta_1,\gamma_1=1,2, \ldots, N_1 \;  .
\end{array}
\end{displaymath}
      {\it In particular, $\mathcal{G}(0,0)=\mathcal{L}_0$ and
$\mathcal{G}(0,1)$ (equation~(\ref{eq:order2}) for $N_{0} =0$) is
the simple IW contraction of $\mathcal{G}$ with respect to the
subalgebra $\mathcal{L}_{0}$} \cite{JdAIMO}.

\subsection{The case in which $\mathcal{G}$ contains a symmetric coset} \label{scoset}
Let us now  particularize to the case in which $\mathcal{G}/
\mathcal{L}_0=V_1$ is a symmetric coset {\it i.e.},
\begin{equation} \label{eq:z2grad2}
[V_{ 0}, \, V_{ 0}] \subset V_{ 0} \; , \quad [V_{ 0}, \, V_{ 1}]
\subset V_{ 1} \; , \quad [V_{ 1}, \, V_{ 1}] \subset V_{ 0} \; ,
      \end{equation}
$\left( [V_{ p}, \, V_{ q}] \subset V_{ p +  q} \; , (p+q)
\textrm{mod}\, 2 \right)$. This applies, for instance, to all
superalgebras where $V_0$ is the bosonic subspace and $V_1$ the
fermionic one. Then, if $c_{i_{ p}j_{ q}}^{k_{ s}}$ ($ p,q,s
=0,1$; $i_p=1,\ldots\textrm{dim} \, V_p$) are the structure
constants of $\mathcal{G}$, $c_{i_{ p}j_{q}}^{k_{ s}}=0$ if $ s
\neq (p+q) \textrm{mod} \, 2$, the MC equations reduce to
{\setlength\arraycolsep{2pt}
\begin{eqnarray}
\label{eq:Z2MCzero} d\omega^{k_{0}} & = & -\frac{1}{2}
c_{i_{0}j_{0}}^{k_{0}} \omega^{i_{0}} \wedge \omega^{j_{0}}
-\frac{1}{2} c_{i_{1}j_{1}}^{k_{0}}
\omega^{k_{1}} \wedge \omega^{j_{1}} \; , \\
\label{eq:Z2MCone} d\omega^{k_{1}} & = & - c_{i_{0}j_{1}}^{k_{1}}
\omega^{i_{0}} \wedge \omega^{j_{1}} \; .
\end{eqnarray}}
\\[-8pt]
In this case, the rescaling (\ref{eq:redefinition}) leads to an
even (odd) power series in $\lambda$ for the MC forms
$\omega^{i_0}(g,\lambda)$ ($\omega^{i_1}(g,\lambda)$):
\begin{eqnarray} \label{eq:z2splitseries1}
\omega^{i_{ 0}}(g,\lambda) & = & \omega^{i_{ 0},0}(g) + \lambda^2
\omega^{i_{ 0},2}(g) + \lambda^4 \omega^{i_{ 0},4}(g)
+ \ldots \nonumber \\
\omega^{i_{ 1}}(g,\lambda) & = & \lambda \omega^{i_{ 1},1}(g) +
\lambda^3 \omega^{i_{ 1},3}(g) + \lambda^5 \omega^{i_{ 1},5}(g) +
\ldots \; ,
\end{eqnarray}

\noindent namely, $\omega^{i_{ \overline{\alpha}}}(g, \lambda) =
\sum_{\alpha=0}^{\infty} \lambda^{\alpha}
\omega^{i_{\overline{\alpha}},\alpha}(g)\,;\, \overline{ \alpha}=
\alpha \, (\mathrm{mod} \, 2)$.

Indeed, under (\ref{eq:redefinition}) $ dg^{i_{ 0}} \rightarrow
dg^{i_{ 0}} \, , \; dg^{i_{ 1}} \rightarrow \lambda \, dg^{i_{
1}}$, which contributes with $\lambda^0$ ($\lambda$) to
$\omega^{i_{ 0}}(g,\lambda)$ ($\omega^{i_{ 1}}(g,\lambda)$);
$c_{j_{ q} k_{ s}}^{i_{ p}}$ vanish trivially unless $p=(q+s)\,
\textrm{mod} \, 2$ . Then, under (\ref{eq:redefinition}), the
$g^{k_{ s}} dg^{j_{ q}}$ terms in (\ref{eq:serie2}) with one
$g^{k_s}$ rescale as
\begin{eqnarray}
      p = 0 \; & : & \;
c_{j_{ 0} k_{{0}}}^{i_{ 0}} \, g^{k_{ 0}} dg^{j_{ 0}} \rightarrow
c_{j_{ 0} k_{{0}}}^{i_{ 0}} \, g^{k_{ 0}} dg^{j_{ 0}} \nonumber \;
, \; c_{j_{ 1} k_{{1}}}^{i_{ 0}} \, g^{k_{ 1}} dg^{j_{ 1}}
\rightarrow \lambda^2 c_{j_{ 1} k_{{1}}}^{i_{ 0}} \,
g^{k_{{1}}} dg^{j_{ 1}} \; ;\nonumber \\
      p =  1 \; & : & \;
c_{j_{ 0} k_{{1}}}^{i_{ 1}} \, g^{k_{ 1}} dg^{j_{ 0}}  \rightarrow
\lambda \,  c_{j_{ 0} k_{{1}}}^{i_{ 1}} \, g^{k_{{1}}} dg^{j_{ 0}}
\; ,
\end{eqnarray}

\noindent so that the  powers $\lambda^0$ and $\lambda^2$
($\lambda$) contribute to $\omega^{i_{ 0}}(g,\lambda)$
($\omega^{i_{ 1}}(g,\lambda)$). For the terms in (\ref{eq:serie2})
involving the products of $n$ $g^{k_s}$'s,

\begin{equation}
c_{j_{ q} k_{{s}_1}}^{h_{{t}_1}} c_{h_{{t}_1}
k_{{s}_2}}^{h_{{t}_2}} \ldots
       c_{h_{{t}_{n-2}} k_{{t}_{n-1}}}^{h_{{t}_{n-1}}}
c_{h_{{t}_{n-1}} k_{{s}_n}}^{i_{ p}} g^{k_{{s}_1}} g^{k_{{s}_2}}
\ldots g^{k_{{s}_{n-1}}}
       g^{k_{{s}_n}} dg^{j_{ q}} \; ,
\end{equation}
\noindent the fact that $V_1=\mathcal{G}/\mathcal{L}_0$ is a
symmetric space requires that $p=q + {s}_1 + {s}_2 \ldots + {s}_n
\, \mathrm{(mod 2)}$. Thus, after the rescaling
(\ref{eq:redefinition}), only even (odd) powers of $\lambda$, from
$\lambda^0$ ($\lambda$) up to the closest (lower or equal to)
$n+1$ even (odd) power $\lambda^{n+1}$, contribute to $\omega^{i_{
0}}(g,\lambda)$ ($\omega^{i_{ 1}}(g,\lambda)$).

\subsubsection{Structure of ${\mathcal G}(N_0,N_1)$ in the symmetric
coset case}

Inserting the power series above into the MC equations
(\ref{eq:Z2MCzero}) and (\ref{eq:Z2MCone}), we arrive at the
equalities: {\setlength\arraycolsep{0pt}
\begin{eqnarray}
\label{eq:z2even} && d \omega^{k_{0}, 2\sigma} = -\frac{1}{2}
c_{i_{0}j_{0}}^{k_{0}} \sum_{\rho=0}^{\sigma} \omega^{i_{0},2
\rho} \wedge \omega^{j_{0},2(\sigma-\rho)} \nonumber \\
&& \qquad \qquad \; \; -\frac{1}{2} c_{i_{1}j_{1}}^{k_{0}}
      \sum_{\rho=1}^{\sigma} \omega^{i_{1},2\rho-1} \wedge
\omega^{j_{1}, 2(\sigma-\rho)+1} \; ,  \\
\label{eq:z2odd} && d \omega^{k_{1}, 2\sigma+1} =
-c_{i_{0}j_{1}}^{k_{1}} \sum_{\rho=0}^{\sigma} \omega^{i_{0},
2\rho} \wedge \omega^{j_{1},2(\sigma-\rho)+1} \; ,
\end{eqnarray}}

\noindent where the expansion orders $\alpha$ are either
$\alpha=2\sigma$ or $\alpha=2\sigma+1$. From them it follows that
the vector spaces generated by
\begin{equation} \label{eq:z2largerbasis}
\{ \omega^{i_{0}, 0}, \omega^{i_{0}, 2}, \omega^{i_{0}, 4},
\ldots, \omega^{i_{0}, N_{0}}; \omega^{i_{1}, 1}, \omega^{i_{1},
3}, \ldots, \omega^{i_{1}, N_{1}} \} \; ,
\end{equation}
\noindent where $N_0 \geq 0$ (and even) and $N_1 \geq 1$ (and
odd), will determine a Lie algebra when
{\setlength\arraycolsep{2pt}
\begin{eqnarray}
\label{eq:Nzeroz2}
N_1&=&N_0-1 \quad ,\\
\label{eq:Nonez2} \textrm{or} \quad N_1&=&N_0+1 \quad .
\end{eqnarray}
}Notice that we have a new type of  solutions (\ref{eq:Nzeroz2})
with respect to the preceding case (equations~(\ref{eq:order1}),
(\ref{eq:order2})), and that the previous solution $N_0=N_1$ is
not allowed now since $N_0$ ($N_1$) is necessarily even (odd).
Then, for the symmetric case, the algebras ${\mathcal G}(N_0,N_1)$
may also be denoted $\mathcal{G}(N)$, where $N=\textrm{max}
\{N_0,N_1 \}$, and are obtained at each order by adding
alternatively copies of $V_0$ and $V_1$. Its structure constants
are given by
\begin{equation}
C_{i_{\overline{\beta}},\beta \;
j_{\overline{\gamma}},\gamma}^{k_{\overline{\alpha}},\alpha}=
\left\{
\begin{array}{lll}
0,                        & \mathrm{if} \ \beta + \gamma \neq \alpha  \\
c_{i_{\overline{\beta}}j_{\overline{\gamma}}}^{k_{\overline{\alpha}}},
& \mathrm{if} \ \beta + \gamma = \alpha \; ; \; \overline{\alpha}
, \overline{\beta} , \overline{\gamma}= \alpha , \beta , \gamma \,
(\textrm{mod}2) \; .
\end{array} \right.
\end{equation}

Let us write explicitly the MC equations~for the first algebras
obtained. If we allow for $N_1=0$, we get the trivial case

$\mathcal{G}(0,0)=\mathcal{G}(0)$:
\begin{equation} \label{eq:z2Nzero}
d\omega^{k_{0},0}= -\frac{1}{2} c_{i_{ 0}j_{ 0}}^{k_{ 0}}
\omega^{i_{ 0}, 0} \wedge \omega^{j_{ 0}, 0}
\end{equation}
{\it i.e.}, $\mathcal{G}(0,0)$ is the subalgebra $\mathcal{L}_0$
of the original algebra $\mathcal{G}$.

$\mathcal{G}(0,1)=\mathcal{G}(1)$: {\setlength\arraycolsep{2pt}
\begin{eqnarray} \label{eq:z2N10}
d\omega^{k_{0},0}&=& -\frac{1}{2} c_{i_{ 0}j_{ 0}}^{k_{ 0}}
\omega^{i_{ 0}, 0} \wedge \omega^{j_{
0}, 0} \; , \\
\label{eq:z2N11} d\omega^{k_{ 1}, 1}&=& -c_{i_{ 0}j_{ 1}}^{k_{ 1}}
\omega^{i_{ 0}, 0} \wedge \omega^{j_{ 1}, 1}\; ,
\end{eqnarray}}
\\[-8pt]
\noindent so that $\mathcal{G}(0,1)$ is again the IW contraction
of $\mathcal{G}$ with respect to $\mathcal{L}_{0}$.

$\mathcal{G}(2,1)=\mathcal{G}(2)$: {\setlength\arraycolsep{0.5pt}
\begin{eqnarray} \label{eq:z2N20}
d\omega^{k_{ 0}, 0}&=& -\frac{1}{2}  c_{i_{ 0}j_{ 0}}^{k_{
0}} \omega^{i_{ 0}, 0} \wedge \omega^{j_{ 0}, 0} \; , \\
\label{eq:z2N21} d\omega^{k_{ 1}, 1}&=& -c_{i_{ 0}j_{ 1}}^{k_{1}}
\omega^{i_{ 0}, 0} \wedge \omega^{j_{ 1}, 1}\; ,\\
\label{eq:z2N22} d\omega^{k_{ 0}, 2}&=& -c_{i_{ 0}j_{ 0}}^{k_{ 0}}
\omega^{i_{ 0}, 0} \wedge \omega^{j_{ 0}, 2} -\frac{1}{2} c_{i_{
1}j_{1}}^{k_{ 0}} \omega^{i_{ 1}, 1} \wedge \omega^{j_{ 1}, 1}\; .
\end{eqnarray}}

The structure of the Lie algebras $\mathcal{G}(N)$ can be
summarized as follows. {\it The Lie algebra
$\mathcal{G}(0)=\mathcal{L}_0$ is a subalgebra of $\mathcal{G}(N)$
for all $N \ge 0$. $W_{\alpha}$ in (\ref{def:W}) reduces here to}
\begin{equation}
W_{\alpha}=\left\{ \begin{array}{lll} V_{0,\alpha} \; , & \quad
\mathrm{if} \ \alpha \, \mathrm{even} \\
V_{1,\alpha} \; , & \quad \mathrm{if} \ \alpha \; \mathrm{odd}
\quad .
\end{array} \right.
\end{equation}
{\it For $N \ge 1$, $W_N$  is an abelian ideal $\mathcal{W}_N$ of
$\mathcal{G}(N)$ and $\mathcal{G}(N) / \mathcal{W}_N =
\mathcal{G}(N-1)$, {\it i.e.}, $\mathcal{G}(N)$ is an extension of
$\mathcal{G}(N-1)$ by $\mathcal{W}_N$.} {\it Further, for $N$ even
and $\mathcal{L}_{ 0}$ abelian, the extension $\mathcal{G}(N)$ of
$\mathcal{G}(N-1)$ by $\mathcal{W}_N$ is central} \cite{JdAIMO}.

\noindent The proof of the first part of the claim proceeds as in
section \ref{s:structure}. For the second part, notice that, for
$N \geq 1$, the only thing that prevents the abelian ideal
$\mathcal{W}_N$ from being central is its failure to commute with
$\mathcal{W}_0 \approx \mathcal{L}_{ 0}$, since $[W_{\alpha}, \,
\mathcal{W}_N] =0$ for $\alpha=1,2,\ldots, N$. But for $N$ even,
$C_{i_0,0\;j_0,N}^{k_0,N}= c_{i_{0} j_{ 0}}^{k_{ 0}}$ , which
vanish for $\mathcal{L}_0$ abelian. Thus $\mathcal{W}_N$ becomes a
central ideal, and $\mathcal{G}(N)$ a central extension of
$\mathcal{G}(N-1)$ by $\mathcal{W}_N$.

\section{Rescaling with several different powers} \label{general}

Let us extend now the above results to the case where the group
parameters are multiplied by arbitrary integer powers of
$\lambda$. Let $\mathcal{G}$ be split into a sum of $n+1$ vector
subspaces,
\begin{equation} \label{eq:splitn}
{\mathcal G}=V_{0} \oplus V_{1} \oplus  \cdots \oplus V_{n}
=\bigoplus_0^n V_p \; ,
\end{equation}
and let the rescaling {\setlength\arraycolsep{0pt}
\begin{eqnarray} \label{eq:nredef}
&& g^{i_{ 0}} \rightarrow  g^{i_{ 0}} \; , \quad  g^{i_{ 1}}
\rightarrow \lambda g^{i_{ 1}} \; , \; \ldots \; , \quad  g^{i_{
n}} \rightarrow \lambda^n g^{i_{ n}} \nonumber \\ &&  (g^{i_{ p}}
\rightarrow \lambda^p g^{i_{ p}}, \; p=0,\ldots,n)
\end{eqnarray}
}of the group coordinates $g^{i_{ p}}$ be subordinated to the
splitting (\ref{eq:splitn}) in an obvious way. We found in the
previous section ($p=0,1$) that, when the rescaling
(\ref{eq:redefinition}) was performed,  having $V_0$ as a
subalgebra $\mathcal{L}_0$ proved to be convenient (though not
necessary) since it led to more types of solutions
((\ref{eq:order1})-(\ref{eq:order2}),{\it cf.}~(\ref{eq:N})).
Furthermore, the first order algebra $\mathcal{G}(0,1)$ for that
case was found to be the simple IW contraction of $\mathcal{G}$
with respect to $\mathcal{L}_0$. By the same reason, we will
consider here conditions on $\mathcal{G}$ that will lead to a
richer new algebras structure, including the generalized IW
contraction of $\mathcal{G}$ in the sense \cite{Wei:00} of
Weimar-Woods (W-W). In terms of the structure constants of
$\mathcal G$ we will then require that they fulfil the condition
(\ref{gc2}), namely,
\begin{equation} \label{eq:cont}
c_{i_p j_q}^{k_s}=0 \quad \textrm{if $s > p+q$} \;
\end{equation}
{\it i.e.}, that the Lie bracket of elements in $V_p$, $V_q$ is in
$\oplus_s V_s$ for $s \leq p+q$. This condition leads, through
(\ref{eq:serie2}), to a power series expansion of the one-forms
$\omega^{i_{ p}}$ in $V^*_{ p}$ that, for each $p=0,1,\ldots,n$,
starts precisely with the power $\lambda^p$,
{\setlength\arraycolsep{2pt}
\begin{eqnarray} \label{eq:szero}
\omega^{i_{ 0}}(g,\lambda) & = & \sum_{\alpha=0}^{\infty}
\lambda^\alpha \omega^{i_{ 0},\alpha}(g) = \omega^{i_{ 0},0}(g) +
\lambda \omega^{i_{ 0},1}(g) + \lambda^2 \omega^{i_{ 0},2}(g) +
\ldots \; , \nonumber \\* && \\ \label{eq:sone} \omega^{i_{
1}}(g,\lambda) & = & \sum_{\alpha=1}^{\infty} \lambda^\alpha
\omega^{i_{ 1},\alpha}(g) = \lambda \omega^{i_{ 1},1}(g) +
\lambda^2 \omega^{i_{ 1},2}(g) +
\lambda^3 \omega^{i_{ 1},3}(g) + \ldots \; , \nonumber \\* && \\
\vdots & & \nonumber \\
\label{eq:sn} \omega^{i_{ n}}(g,\lambda) & = &
\sum_{\alpha=n}^{\infty} \lambda^\alpha \omega^{i_{ n},\alpha}(g)
= \lambda^n \omega^{i_{ n},n}(g) + \lambda^{n+1} \omega^{i_{
n},n+1}(g) + \ldots \; . \nonumber \\ &&
\end{eqnarray}}

We may extend all the sums so that they begin at $\alpha=0$ by
setting $ \omega^{i_p,\alpha}\equiv 0$ when $\alpha<p$. Then,
inserting the expansions of $\omega^{i_p,\alpha}$ in the MC
equations~and using (\ref{eq:sumatorio}) we get (\ref{eq:MCG}) for
$p,q,s=0,1,\ldots,n$. If we now introduce the notation
$\omega^{i_p,\alpha_p}$ with different ranges for the expansion
orders, $\alpha_p=p,p+1,\ldots N_p$ for each $p$, we see that the
MC equations take the form
\begin{equation} \label{eq:MCn}
d\omega^{k_{ s}, \alpha_s}= -\frac{1}{2} C_{i_{ p},\beta_p \;
j_{q},\gamma_q}^{k_{ s},\alpha_s}\; \omega^{i_{ p}, \beta_p}
\wedge \omega^{j_{ q}, \gamma_q} \quad ,
\end{equation}
where
\begin{equation} \label{eq:Cn}
C_{i_{ p},\beta_p \; j_{ q},\gamma_q}^{k_{ s},\alpha_s}= \left\{
\begin{array}{lll} 0, &
\mathrm{if} \ \beta_p + \gamma_q \neq \alpha_s  \\
c_{i_{ p}j_{ q}}^{k_{s}}, & \mathrm{if} \ \beta_p + \gamma_q =
\alpha_s  \end{array} \right. \quad \begin{array}{l}
p,q,s=0,1,\ldots, n \\
i_{p,q,s}=1,2,\ldots, \textrm{dim} \, V_{p,q,s} \\
\alpha_p,\beta_p,\gamma_p=p,p+1, \ldots, N_p
\end{array}
\end{equation}
and the $c_{i_p j_q}^{k_s}$ satisfy (\ref{eq:cont}). To find now
the $\omega^{i_p ,\beta_p}$'s that enter $d
\omega^{k_s,\alpha_s}$, $s=0,1,\ldots,n$, we need an explicit
expression for it. This is found  in appendix \ref{ap:expansion},
equations (\ref{facil})-(\ref{eq:MCexp2}).
   From them we read that $d \omega^{k_{s}, \alpha_s}$,
$s=0,1,\ldots,n$, is expressed in terms of products of the forms
$\omega^{i_{p}, \beta_p}$ in the following table:

\begin{center}
{\footnotesize
\begin{tabular}{|c|ccccc|}
\hline
      $\alpha_s \geq s$ & $\omega^{i_{0}, \beta_0}$ & $\omega^{i_{1},
\beta_1}$
& $\omega^{i_{2}, \beta_2}$ & $\cdots$ & $\omega^{i_{n}, \beta_n}$  \\
\hline $d \omega^{k_{0}, \alpha_0}$ & $\beta_0 \leq \alpha_0$ &
$\beta_1 \leq \alpha_0$ &
$\beta_2 \leq \alpha_0$ & $\cdots$ & $\beta_n \leq \alpha_0$   \\
$d \omega^{k_{1}, \alpha_1}$ & $\beta_0 \leq \alpha_1-1$ &
$\beta_1 \leq \alpha_1$ & $\beta_2 \leq \alpha_1$ & $\cdots$
& $\beta_n \leq \alpha_1$ \\
$d \omega^{k_{2}, \alpha_2}$ & $\beta_0 \leq \alpha_2-2$ &
$\beta_1 \leq \alpha_2-1$ & $\beta_2 \leq \alpha_2$ & $\cdots$
& $\beta_n \leq \alpha_2$ \\
$\vdots$ & $\vdots$ & $\vdots$ & $\vdots$ & & $\vdots$  \\
$d \omega^{k_{n}, \alpha_n}$ & $\beta_0 \leq \alpha_n-n$ &
$\beta_1 \leq \alpha_n-n+1$ & $\beta_2
      \leq \alpha_n-n+2$ & $\cdots$ & $\beta_n \leq \alpha_n$ \\ \hline
\end{tabular}}
\\[6pt]
{\footnotesize Table 5.3. Types and orders of the forms
$\omega^{i_{p}, \beta_p}$ needed to express $d\omega^{k_{s},
\alpha_s}$}
\end{center}

\noindent Now let $V^*(N_0, \ldots, N_n)$ be the vector space
generated by {\setlength\arraycolsep{0pt}
\begin{eqnarray} \label{eq:larger}
&& \{ \omega^{i_0 , \alpha_0} \  ; \omega^{i_1 , \alpha_1} \ ;
\ldots
; \omega^{i_n , \alpha_n} \}  =  \nonumber \\
& & = \{ \omega^{i_{ 0}, 0}, \omega^{i_{ 0}, 1}, \stackrel{N_0
+1}{\ldots}, \omega^{i_{ 0}, N_{ 0}}; \, \omega^{i_{ 1}, 1},
\stackrel{N_1}{\ldots}, \omega^{i_{ 1}, N_{ 1}}; \, \ldots; \,
\omega^{i_{n}, n}, \stackrel{N_n-n+1}{\ldots}, \omega^{i_{n},
N_{n}} \} \; . \nonumber \\ &&
\end{eqnarray}
}These one-forms determine a Lie algebra
$\mathcal{G}(N_0,N_1,\ldots,N_n)$, of dimension
\begin{equation} \label{eq:dim}
\textrm{dim} \, \mathcal{G}(N_0, \ldots, N_n) = \sum_{p=0}^{n}
(N_p -p+1) \, \textrm{dim} \, V_p \; .
\end{equation}
More precisely, {\it let $\mathcal{G}=V_{0} \oplus V_{1} \oplus
\cdots \oplus V_n$ be a splitting of $\mathcal{G}$ into $n+1$
subspaces and let $\mathcal{G}$ fulfil the Weimar-Woods
contraction condition (\ref{eq:cont}) subordinated to this
splitting, $c_{i_p j_q}^{k_s}=0$ if $s>p+q$. The one-form
coefficients $\omega^{i_p , \alpha_p}$ of (\ref{eq:larger})
resulting from the expansion of the Maurer-Cartan forms
$\omega^{i_p}$ in which $g^{i_{ p}} \rightarrow \lambda^p g^{i_{
p}}, \; p=0,\ldots,n$ (equation~(\ref{eq:nredef})), determine Lie
algebras $\mathcal{G}(N_0,N_1,\ldots,N_n)$ of dimension
(\ref{eq:dim}) and structure constants}
\begin{displaymath}
C_{i_{ p},\beta_p \; j_{ q},\gamma_q}^{k_{ s},\alpha_s}= \left\{
\begin{array}{lll} 0, &
\mathrm{if} \ \beta_p + \gamma_q \neq \alpha_s  \\
c_{i_{ p}j_{ q}}^{k_{s}}, & \mathrm{if} \ \beta_p + \gamma_q =
\alpha_s  \end{array} \right. \quad \begin{array}{l}
p,q,s=0,1,\ldots, n \\
i_{p,q,s}=1,2,\ldots, \textrm{dim} \, V_{p,q,s} \\
\alpha_p,\beta_p,\gamma_p=p,p+1, \ldots, N_p \;  ,
\end{array}
\end{displaymath}
{\it (equation(\ref{eq:Cn})) if $N_q=N_{q+1}$ {\it or}
$N_q=N_{q+1}-1$ ($q=0,1,\ldots,n-1$) in $(N_0,N_1,\ldots,N_n)$. In
particular, the $N_p=p$ solution determines the algebra
$\mathcal{G}(0,1,\ldots, n)$, which is the generalized
\.In\"on\"u-Wigner contraction of $\mathcal{G}$} \cite{JdAIMO}.

Let us prove this statement. To enforce the closure under $d$ of
the exterior algebra generated by the one-forms in
(\ref{eq:larger}) and to find the conditions that the various
$N_p$ must meet, we require, as in section \ref{tres}, that all
the forms $\omega^{i_p,\beta_p}$ present in $d\omega^{k_s,
\alpha_s}$ are already in (\ref{eq:larger}). Looking at
equations~(\ref{facil})-(\ref{eq:MCexp2}) and at table 5.3 above,
we find the restrictions

\begin{center} \label{table4}
{\footnotesize
\begin{tabular}{|c|ccccc|}
\hline
      $\alpha_s  \geq s$ & $\omega^{i_{0}, \beta_0}$ & $\omega^{i_{1},
\beta_1}$
& $\omega^{i_{2}, \beta_2}$ & $\cdots$ & $\omega^{i_{n}, \beta_n}$  \\
\hline $d \omega^{k_{0}, \alpha_0}$ & $N_0 \geq N_0$ & $N_1 \geq
N_0$ &
$N_2 \geq N_0$ & $\cdots$ & $N_n \geq N_0$   \\
$d \omega^{k_{1}, \alpha_1}$ & $N_0 \geq N_1 -1$ & $N_1 \geq N_1$
& $N_2 \geq N_1$
& $\cdots$ & $N_n \geq N_1$ \\
$d \omega^{k_{2}, \alpha_2}$ & $N_0 \geq N_2 -2$ & $N_1 \geq N_2
-1$ & $N_2 \geq N_2$
      & $\cdots$ & $N_n \geq N_2$ \\
$\vdots$ & $\vdots$ & $\vdots$ & $\vdots$ & & $\vdots$  \\
$d \omega^{k_{n}, \alpha_n}$ & $N_0 \geq N_n -n$ & $N_1 \geq N_n
-n+1$ & $N_2 \geq N_n -n+2$  & $\cdots$ & $N_n \geq N_n$ \\ \hline
\end{tabular}}
\\[6pt]
{\footnotesize Table 5.4. Closure conditions on the number $N_p$
of one-forms $\omega^{i_{p}, \alpha_p}$}
\end{center}

\noindent It then follows that there are $2^n$ types of
solutions\footnote{\label{note} This number may be found, {\it
e.g.} for $n=3$, by writing symbolically the solution types in
(\ref{eq:gralcond}) as [0,0,0,0] for $N_0=N_1=N_2=N_3$; [0,0,0,1]
for $N_0=N_1=N_2, N_3=N_2+1$; [0,0,1,0] for $N_0=N_1,
N_2=N_1+1=N_3$; [0,0,1,1] for $N_0=N_1, N_2=N_1+1,N_3=N_2+1$;
[0,1,0,0] for $N_0,N_1=N_0+1=N_2=N_3$; [0,1,0,1] for
$N_0,N_1=N_0+1=N_2,N_3=N_2+1$; [0,1,1,0] for
$N_0,N_1=N_0+1,N_2=N_1+1=N_3$ and [0,1,1,1] for
$N_0,N_1=N_0+1,N_2=N_1+1,N_3=N_2+1$. This notation numbers the
solutions in base 2; since [0,1,1,1] corresponds to $2^3-1$ we
see, adding [0,0,0,0], that there are $2^3$ ways of cutting the
expansions that determine Lie algebras ${\mathcal
G}(N_0,N_1,N_2,N_3)$, and $2^n$ in the general ${\mathcal
G}(N_0,N_1,\ldots,N_n)$ case.} characterized by
$(N_0,N_1,\ldots,N_n)$, $N_p \geq p$, $p=0,1,\ldots,n,\,$ where

\begin{equation}
\label{eq:gralcond} N_{q+1} = N_q \quad \textrm{or} \quad N_{q+1}
= N_q +1 \quad (q=0,1,\ldots, n-1) \; .
\end{equation}
The Jacobi identities for $\mathcal{G}(N_{0},\ldots, N_n)$,
{\setlength\arraycolsep{2pt}
\begin{eqnarray}\label{eq:granjacobi}
&& C_{i_p,\beta_p\;[j_q,\gamma_q}^{k_s,\alpha_s} C_{l_t,\rho_t\;
m_u,\sigma_u]}^{i_p,\beta_p} =0 = \nonumber \\
&&  C_{i_p,\beta_p\;j_q,\gamma_q}^{k_s,\alpha_s} C_{l_t,\rho_t\;
m_u,\sigma_u}^{i_p,\beta_p}  +
C_{i_p,\beta_p\;m_u,\sigma_u}^{k_s,\alpha_s}
C_{j_q,\gamma_q\;l_t,\rho_t}^{i_p,\beta_p}+
C_{i_p,\beta_p\;l_t,\rho_t}^{k_s,\alpha_s}
C_{m_u,\sigma_u\;j_q,\gamma_q}^{i_p,\beta_p} \; ,\nonumber \\
\end{eqnarray}
}are again satisfied through the ones for $\mathcal{G}$. This is a
consequence of the fact that, for ${\mathcal G}$, the exterior
derivative of the $\lambda$-expansion of the MC equations is the
$\lambda$-expansion of their exterior derivative, but it may also
be seen directly.

Indeed, we only need to check that (\ref{eq:granjacobi}) reduces
to the Jacobi identities for $\mathcal{G}$ when the order in the
upper index is the sum of those in the lower ones since the $C$'s
are zero otherwise. First we see that, when
$\alpha_s=\gamma_q+\rho_t+\sigma_u$, all three terms in the
r.h.s.~of (\ref{eq:granjacobi}) give non-zero contributions. This
is so because the range of $\beta_p$ is only limited by
$\beta_p\leq\alpha_s$, which holds when $\beta_p=\rho_t+\sigma_u$,
$\beta_p=\gamma_q+\rho_t$ and $\beta_p=\sigma_u+\gamma_q$.
Secondly, and since $\beta_p\geq p$, we also need that the terms
in the $i_p$ sum that are suppressed in (\ref{eq:granjacobi}) when
$p>\beta_p$ be also absent in the Jacobi identities for
$\mathcal{G}$ so that (\ref{eq:granjacobi}) does reduce to the
Jacobi identities for $\mathcal{G}$. Consider {\it e.g.}, the
first term in the r.h.s.~of (\ref{eq:granjacobi}). If $p>\beta_p$,
then $p>\rho_t+\sigma_u$ and hence $p>t+u$. Thus, by the W-W
condition (\ref{eq:cont}), this term will not contribute to the
Jacobi identities for $\mathcal{G}$ and no sum over the subspace
$V_p$ index $i_p$ will be lost as a result. The argument also
applies to the other two terms for their corresponding
$\beta_p$'s.

 A particular solution to (\ref{eq:gralcond}) is
obtained by setting $N_p = p$, $p=0,1,\ldots,n$, which defines
$\mathcal{G}(0,1,\ldots, n)$, with $\textrm{dim}
\,\mathcal{G}(0,1,\ldots, n)=\textrm{dim}\,\mathcal{G}=r$ (from
(\ref{eq:dim})). Since in this case $\alpha_p$ takes only one
value ($\alpha_p=N_p=p$) for each $p=0,1,\ldots,n$, we may drop
this label. Then, the structure constants (\ref{eq:Cn}) for
$\mathcal{G}(0,1, \ldots,n)$ read
\begin{equation} \label{eq:CIWn}
C_{i_{ p}\, j_{ q}}^{k_{ s}}= \left\{ \begin{array}{lll} 0, &
\mathrm{if} \ p + q \neq s  \\
c_{i_{ p}j_{ q}}^{k_{s}}, & \mathrm{if} \ p + q = s  \end{array}
\right. \quad \begin{array}{l}
p=0,1,\ldots,n \\
i_{p,q,s}=1,2,\ldots, \textrm{dim} \, V_{p,q,s} \; ,
\end{array}
\end{equation}
which shows that $\mathcal{G}(0,1,\ldots,n)$ is the generalized IW
contraction of $\mathcal{G}$, in the sense of \cite{Wei:00},
subordinated to the splitting (\ref{eq:splitn}). Of course, when
$n=1$ ($p=0,1$), $ V=V_0 \oplus V_1$,  $\mathcal{L}_0$ is a
subalgebra and equations (\ref{eq:gralcond}) ((\ref{eq:CIWn}))
reduce to (\ref{eq:order1}) or (\ref{eq:order2}) ((\ref{eq:CIW})),
what concludes the proof of the statement.

    For instance, for the case ${\mathcal G}=V_0 \oplus V_1 \oplus V_2$
there are four types of algebras\footnote{With the notation of
footnote \ref{note}, these correspond, respectively, to [0,0,0],
[0,0,1], [0,1,0] and [0,1,1].} ${\mathcal G}(N_0,N_1,N_2)$
{\setlength\arraycolsep{2pt}
\begin{eqnarray}
\label{eq:tres4}
N_0 & = & N_1 = N_2  \; \\
\label{eq:tres3}
N_0 & = & N_1 = N_2-1 \; ,  \\
\label{eq:tres2}
N_0 & = & N_1-1=N_2-1 \; ,  \\
\label{eq:tres1} N_0 & = & N_1-1= N_2-2 \; .
\end{eqnarray}

}

Since in the above theorem $\alpha_p\geq p$ for all $p=0,\ldots,n$
was assumed, all types of one-forms $\omega^{i_p,\alpha_p}$ with
indices $i_p$ in all subspaces $V_p$ were present in the basis of
$\mathcal{G}(N_0,N_1,\ldots,N_n)$. However, one may consider
keeping terms in the expansion up to a certain order $l,\; l<n$ in
which case due to (\ref{eq:sn}), the forms $\omega^{i_p,\alpha_p}$
with $p > l$ will not appear. Those with $p\leq l$ will determine
the vector space $V^*(N_0,N_1,\ldots,N_l)$ where $N_l$ is the
highest order $l$ and hence $\alpha_l$ takes only the value
$N_l=l=\alpha_l$. This vector space, of dimension
\begin{equation}\label{dimcajitas}
\textrm{dim}\,V^*(N_0, \ldots, N_l) = \sum_{p=0}^{l} (N_p -p+1) \,
\textrm{dim} \, V_p \quad,
\end{equation}
determines a Lie algebra $\mathcal{G}(N_0,N_1,\ldots,N_l)$, as
claims the following statement.

{\it Let $\mathcal{G}=\oplus_0^n V_p$, satisfy the Weimar-Woods
conditions (\ref{eq:cont}). Then, up to a certain order $N_l=l<n$,
the one-forms} {\setlength\arraycolsep{0pt}
\begin{eqnarray} \label{lbasis4}
&& \{ \omega^{i_0 , \alpha_0} \ ; \omega^{i_1 , \alpha_1} \ ;
\ldots ; \omega^{i_l , \alpha_l} \} = \nonumber \\
&& \quad = \{ \omega^{i_{ 0}, 0}, \omega^{i_{ 0}, 1},
\stackrel{N_0 +1}{\ldots}, \omega^{i_{ 0}, N_{ 0}}\, ; \,
\omega^{i_{ 1}, 1}, \stackrel{N_1}{\ldots}, \omega^{i_{ 1}, N_{
1}}\, ; \, \ldots; \omega^{i_{l}, N_{l}} \} \; ,
\end{eqnarray}
}{\it where $N_l=l=\alpha_l$, determine a Lie algebra
$\mathcal{G}(N_0,N_1,\ldots N_l)$ of dimension (\ref{dimcajitas})
and structure constants given by} {\setlength\arraycolsep{0pt}
\begin{equation} \label{eq:cnts4}
C_{i_{ p},\beta_p \; j_{ q},\gamma_q}^{k_{ s},\alpha_s}= \left\{
\begin{array}{lll} 0, &
\mathrm{if} \ \beta_p + \gamma_q \neq \alpha_s  \\
c_{i_{ p}j_{ q}}^{k_{s}}, & \mathrm{if} \ \beta_p + \gamma_q =
\alpha_s  \end{array} \right. \quad  \begin{array}{l}
p,q,s=0,1,\ldots, l \\
i_{p,q,s}=1,2,\ldots, \textrm{dim} \, V_{p,q,s} \\
\alpha_p,\beta_p,\gamma_p=p,p+1, \ldots, N_p \ \le l,
\end{array}
\end{equation}
}{\it if $N_q=N_{q+1}$ or $N_q=N_{q+1}-1$, ($q=0,1,\ldots ,l-1$)}
\cite{JdAIMO}.

To see that this is indeed the case, notice that the restriction
$\alpha_p \le N_l=l < n$ on the order $\alpha_p$ of the one-forms
$\omega^{i_p,\alpha_p}$ implies, due to (\ref{eq:sn}), that $V_l$
is monodimensional and that $\omega^{i_l,l}$ is the last form
entering (\ref{lbasis4}). Then, looking at the closure conditions
in table 5.4, we can restrict ourselves to the box delimited by
$\omega^{i_p,\beta_p}$, $d\omega^{k_s,\alpha_s}$ with $p,s \le l$.
This box will give spaces $V^*(N_0,N_1,\ldots, N_l)$, where
$N_q=N_{q+1}$ or $N_q=N_{q+1}-1$ ($q=0,1,\ldots,l-1$), and these
spaces will determine Lie algebras if the Jacobi identities for
(\ref{eq:cnts4})
\begin{equation} \label{JIcaja}
C_{i_p,\beta_p\; [ j_q,\gamma_q}^{k_s,\alpha_s} C_{l_t,\rho_t\;
m_u,\sigma_u ]}^{i_p,\beta_p} =0 \; ,\;\; i_{p,q,s}=1,2,\ldots,
\mathrm{dim} V_{p,q,s}
\end{equation}
{\it i.e.}, if $c_{i_p [ j_q}^{k_s} c_{l_t m_u]}^{i_p}=0\,, \;
s,q,t,u \le l$, is satisfied when
$\alpha_s=\gamma_q+\rho_t+\sigma_u$ above. Note that this is not
the Jacobi identities for $\mathcal{G}$ since $i_p$ now runs over
the basis of $\oplus_0^l V_p \subset \mathcal{G}$ only since $p
\le l$, and we are thus removing the values corresponding to the
basis of $\oplus_{l+1}^n V_p$. However, if $p>l$ it is also {\it
e.g.} $p>\beta_p=\rho_t+\sigma_u \ge t+u$ in which case $c_{l_t\;
m_u}^{i_p}=0$ by (\ref{eq:cont}), what concludes the proof.

Notice that, since the structure constants (\ref{eq:cnts4}) are
obtained from those of $\mathcal{G}$ by restricting the $i_p$
indices to be in the subspaces $V_p,\, p\leq l$,
$\mathcal{G}(N_0,N_1,\ldots,N_l)$ is {\it not} a subalgebra of
$\mathcal{G}(N_0,N_1,\ldots,N_n)$.

\section{Superalgebra expansions}
\label{sec:susy}

The above general procedure of generating Lie algebras from a
given one does not rely on the antisymmetry of the structure
constants of the original Lie algebra. Hence, with the appropriate
changes to account for the grading, the method is applicable when
$\mathcal{G}$ is a Lie superalgebra, a case which we consider
explicitly in this section.

Let $G$ be a supergroup and $\mathcal G$ its superalgebra. It is
natural to consider a splitting of $\mathcal G$ into the sum of
three subspaces $\mathcal G=V_0 \oplus V_1 \oplus V_2$, $V_1$
being the fermionic part of $\mathcal{G}$ and $V_0 \oplus V_2$ the
bosonic part, so that the notation reflects the
$\mathbb{Z}_2$-grading of $\mathcal{G}$. The even space is always
a subalgebra of $\mathcal{G}$ but it may be convenient to consider
it further split into the sum $V_0 \oplus V_2$ to allow for the
case in which a subspace ($V_0$) of the bosonic space is itself a
subalgebra $\mathcal{L}_0$.

Notice that, since $V_0$ is a Lie algebra $\mathcal{L}_0$, the
$\mathbb{Z}_2$-graduation of $\mathcal{G}$ implies that the
splitting $\mathcal G=V_0 \oplus V_1 \oplus V_2$ satisfies the W-W
contraction conditions (\ref{eq:cont}). Indeed, let $c_{i_p
j_q}^{k_s}$ ($i_{p,q,s}= 1, \ldots, \textrm{dim} \, V_{p,q,s}$,
$p,q,s=0,1,2$) be the structure constants of $\mathcal{G}$. The
$\mathbb{Z}_2$-graduation of $\mathcal G$ obviously  implies
{\setlength\arraycolsep{2pt}
\begin{eqnarray}
c_{i_0 j_0}^{k_1} &=& c_{i_0 j_1}^{k_2} =0 \; ,\label{eq:c1}\\
c_{i_0 j_1}^{k_0} &=&c_{i_1 j_1}^{k_1}= c_{i_0 j_2}^{k_1} =c_{i_2
j_1}^{k_0}= c_{i_2 j_1}^{k_2}= c_{i_2 j_2}^{k_1} =0
   \; .\label{eq:c2}
\end{eqnarray}
}The first set of restrictions (\ref{eq:c1}), together with the
assumed subalgebra condition for $V_0$ (which, in addition,
requires $c_{i_0 j_0}^{k_2} =0$), are indeed the W-W conditions
(\ref{eq:cont}) for $\mathcal{G}$; note that these conditions
alone allow for $c_{i_1 j_1}^{k_0} \neq 0$, and $c_{i_1j_1}^{k_2}
\neq 0$ (and for $c_{i_1 j_1}^{k_1} \neq 0$, although here $c_{i_1
j_1}^{k_1} = 0$ due to the $\mathbb{Z}_2$-grading).

To apply now the above general procedure one must rescale the
group parameters. The rescaling (\ref{eq:nredef}) for $V=V_0
\oplus V_1 \oplus V_2$ takes the form
\begin{equation} \label{eq:superredef1}
g^{i_{ 0}} \rightarrow  g^{i_{ 0}} \; , \; g^{i_{ 1}} \rightarrow
\lambda g^{i_{ 1}} \; , \; g^{i_{ 2}} \rightarrow \lambda^2 g^{i_{
2}} \; .
\end{equation}
The present $\mathbb{Z}_2$-graded case fits into the preceding
general discussion for $n=2$, but with additional restrictions
besides the W-W ones that follow from the $\mathbb{Z}_2$-grading.
This situation is described by the following statement.

{\it Let $\mathcal{G}=V_{0}\oplus V_{1} \oplus V_2$ be a Lie
superalgebra, $V_{1}$ its odd part, and $V_{0} \oplus V_{2}$ the
even one. Let further $V_0$ be a subalgebra $\mathcal{L}_0$. As a
result, $\mathcal{G}$ satisfies the W-W conditions (\ref{eq:cont})
and, further, $V_1$ is a symmetric coset. Then, the coefficients
of the expansion of the Maurer-Cartan forms of $\mathcal{G}$
rescaled by (\ref{eq:superredef1}) determine Lie superalgebras
$\mathcal{G}(N_{0},N_{1},N_2)$, $N_p\geq p,\, p=0,1,2$, of
dimension}
\begin{equation}
\label{eq:dimsuper} \textrm{dim}\,\mathcal{G}(N_0,N_1,N_2)=
\left[\frac{N_0+2}{2}\right]\textrm{dim}V_0 +
\left[\frac{N_1+1}{2}\right]\textrm{dim}V_1 +
\left[\frac{N_2}{2}\right]\textrm{dim}V_2 \; ,
\end{equation}
\noindent {\it and structure constants}
\begin{equation} \label{eq:Csuper}
C_{i_{ p},\beta_p \; j_{ q},\gamma_q}^{k_{ s},\alpha_s}= \left\{
\begin{array}{lll} 0, &
\mathrm{if} \ \beta_p + \gamma_q \neq \alpha_s  \\
c_{i_{ p}j_{ q}}^{k_{s}}, & \mathrm{if} \ \beta_p + \gamma_q =
\alpha_s  \end{array} \right. \quad \begin{array}{l}
p,q,s=0,1,2 \\
i_{p,q,s}=1,2,\ldots, \textrm{dim} \, V_{p,q,s} \; ,
\end{array}
\end{equation}
{\it and $\alpha_p,\beta_p,\gamma_p=p,p+2, \ldots, N_p-2, N_p$,
where} [$\quad$] {\it denotes integer part and the $N_0,N_2$
(even) and $N_1$ (odd) integers satisfy one of the three
conditions below} {\setlength\arraycolsep{2pt}
\begin{eqnarray}
\label{eq:tress1}
N_0 & = & N_1+1 = N_2  \; \\
\label{eq:tress2}
N_0 & = & N_1-1 = N_2 \; ,  \\
\label{eq:tress3} N_0 & = & N_1-1= N_2-2 \; .
\end{eqnarray}}
\\[-14pt]
\noindent {\it In particular, the superalgebra
$\mathcal{G}(0,1,2)$ (equation~(\ref{eq:tress3}) for $N_{0}=0$) is
the generalized \.In\"on\"u-Wigner contraction of $\mathcal{G}$}
\cite{JdAIMO}.

Indeed, since $V_1$ is a symmetric coset the rescaling
(\ref{eq:superredef1}) leads to an even (odd) power series in
$\lambda$ for the one-forms $\omega^{i_0}(g,\lambda)$ and
$\omega^{i_2}(g,\lambda)$ ($\omega^{i_1}(g,\lambda)$), as in
Sec.~\ref{scoset} (equations~(\ref{eq:z2splitseries1})). Thus, the
conditions $N_0, N_2$ even, $N_1$ odd, have to be added to those
that follow from the closure inequalities in table 5.4. This gives
the conditions
\begin{eqnarray}
N_0+1 &\ge& N_1 \ge N_0-1 \\
N_1+1 &\ge& N_2 \ge N_1-1 \\
N_0+2 &\ge& N_2 \ge N_0 \; ,
\end{eqnarray}
\noindent from which equations
(\ref{eq:tress1})--(\ref{eq:tress3}) follow.

\section{The M Theory superalgebra as an expansion of $osp(1|32)$}
\label{seis}

Let us work out an explicit example to illustrate the expansion
method. The M Theory superalgebra (see section
\ref{sec:MTsuperalg} of chapter \ref{chapter2} and references
therein) is sometimes regarded (see {\it e.g.} \cite{M-alg}) as an
IW contraction of the superalgebra $osp(1|32)$. That is indeed the
case if the 55 Lorentz generators are excluded, otherwise there
are not enough generators in $osp(1|32)$ to give the M-algebra by
the dimension-preserving method of contraction (see section
\ref{fourw}). In other words, the M Theory superalgebra, when its
Lorentz automorphism generators are included,
$\mathfrak{E}^{(528|32)} \rtimes so(1,10)$, cannot be obtained as
a contraction of $osp(1|32)$. Let us show that, in contrast, the
former is an expansion of the later \cite{JdAIMO}.

The orthosymplectic superalgebra is defined by the 528 bosonic MC
forms $\rho^{\alpha\beta}=\rho^{\beta\alpha}$ of the symplectic
algebra $sp(32)$ and by the 32 fermionic MC forms $\nu^\alpha$
satisfying the MC equations {\setlength\arraycolsep{0pt}
\begin{eqnarray}
 &&        d\rho^{\alpha\beta} =
-i {\rho^\alpha}_\gamma \wedge \rho^{\gamma\beta}- i \nu^\alpha
\wedge
\nu^\beta \nonumber\\
 &&   d\nu^\alpha = -i {\rho^\alpha}_\beta\wedge \nu^\beta
    \qquad (\alpha,\beta=1,\ldots,32)\; .
\label{ospmaurer}
\end{eqnarray}
}The spinor indices are raised and lowered by the $32 \times 32$
symplectic form $C_{\alpha \beta}$, which can be interpreted, as
in section \ref{sec:MTsuperalg}, as the $D=11$ charge conjugation
matrix. It is useful to use Dirac matrices to decompose
$\rho^{\alpha\beta}$ as
\begin{equation}
\rho^{\alpha\beta}=\ft{1}{32} \left(  \rho^a \Gamma_a -\ft{i}{2}
\rho^{ab} \Gamma_{ab}+ \ft{1}{5!} \rho^{a_1\dots a_5}
\Gamma_{a_1\dots a_5}\right)^{\alpha\beta} \; .\label{generalrho}
\end{equation}
In terms of the one-forms $\rho^a$, $\rho^{ab}$, $\rho^{a_1 \ldots
a_5}$ entering the decomposition (\ref{generalrho}) of
$\rho^{\alpha\beta}$ the MC equations (\ref{ospmaurer}) of
$osp(1|32)$ can be rewritten as {\setlength\arraycolsep{0pt}
\begin{eqnarray}   \label{ospmaurerd}
&&      d\rho_{a} = -\ft{1}{16} \rho_n \wedge {\rho^b}_a
+\ft{1}{32(5!)^2} {\epsilon^{b_1\dots b_{10}}}_a \rho_{b_1\dots
b_5} \wedge \rho_{b_6\dots b_{10}} -\nu^\alpha
(\Gamma_a)_{\alpha\beta}\wedge \nu^\beta \; , \nonumber\\
&& d\rho_{ab} = -\ft{1}{16} e_a \wedge e_b - \ft{1}{16} \rho_{a
c}\wedge {\rho^c}_{b} - \ft{1}{16(4!)}\rho_{a c_1\dots c_4}\wedge
{\rho^{c_4\dots c_1}}_b  \nonumber \\* && \qquad \qquad -
\nu^\alpha
(\Gamma_{ab})_{\alpha\beta}\wedge \nu^\beta \; , \nonumber\\
&&   d\rho_{a_1\dots a_5} = \ft{1}{16(5!)}{\epsilon^{b c_1\dots
c_5}}_{a_1 \dots a_5} \rho_b \wedge \rho_{c_1\dots c_5} +
\ft{5}{16}{\rho^b}_{[a_1\dots a_4} \wedge
\rho_{a_5] b}\nonumber \\
  & & \qquad \qquad +\ft{1}{4(4!)^2}
{\epsilon^{b_1\dots b_6}}_{a_1\dots a_5} \rho_{b_1 b_2 b_3 c_1
c_2}\wedge {\rho^{c_2 c_1}}_{b_4 b_5 b_6}- \nu^\alpha
(\Gamma_{a_1\dots a_5})_{\alpha\beta}\wedge \nu^\beta \; ,
\nonumber\\
&& d\nu^\alpha = \ft{1}{32} \left(  \rho^a \Gamma_a -\ft{i}{2}
\rho^{ab} \Gamma_{ab}+ \ft{1}{5!} \rho^{a_1\dots a_5}
\Gamma_{a_1\dots a_5}\right)^\alpha{}_\beta \wedge \nu^\beta \ .
\end{eqnarray}
}This form of the MC equations of $osp(1|32)$ suggests a splitting
of the underlying vector space into three subspaces
$osp(1|32)=V_0\oplus V_1\oplus V_2$, where $V_0$ is the space
generated by the 55 MC forms $\rho^{ab}=\rho^{\alpha\beta}
(\gamma^{ab})_{\alpha\beta}$ of the Lorentz subalgebra of
$osp(1|32)$, $V_1$ the fermionic subspace generated by
$\nu^\alpha$,  and $V_2$ the space generated by the remaining
11+462 bosonic generators
$\rho^a=\rho^{\alpha\beta}(\Gamma^a)_{\alpha\beta}$,
$\rho^{a_1\dots a_5}=\rho^{\alpha\beta} (\Gamma^{a_1\dots
a_5})_{\alpha\beta}$. Moreover, this splitting fulfils the general
conditions discussed for superalgebras in section \ref{sec:susy}.
It then follows that, after the redefinition
(\ref{eq:superredef1}) of the group parameters of $osp(1|32)$, the
expansions of the forms in $V_0$ contain even powers of $\lambda$
starting from $\lambda^0$, that those of the forms in $V_1$
include only odd powers in $\lambda$ starting from $\lambda^1$,
and that those of $V_2$ contain even orders starting with
$\lambda^2$, {\it i.e.},
\begin{eqnarray}
\label{eq:Vcero}
   V_0 & : & \quad
   \rho^{ab} = \sum^\infty_{n=0} \lambda^{2n} \rho^{ab,2n} = \rho^{ab,0}+\lambda^2 \rho^{ab,2}+\cdots \;   ;  \\
\label{Vuno} V_1 & : &  \quad \nu^\alpha = \sum^\infty_{n=0}
\lambda^{2n+1}\nu^{\alpha,2n+1} = \lambda\nu^{\alpha,1} +
\lambda^3\nu^{\alpha,3}+\cdots
   \; ; \\
\label{eq:Vdos} V_2 & : &  \quad \left\{ \begin{array}{l}
     \rho^a = \sum^\infty_{n=1}\lambda^{2n}\rho^{a,2n} = \lambda^2 \rho^{a,2}+\cdots \;
     , \\ \\
\rho^{a_1\dots a_5} = \sum^\infty_{n=1}\lambda^{2n}\rho^{a_1 \dots
a_5,2n} = \lambda^2 \rho^{a_1\dots a_5,2}+ \cdots \; . \end{array}
\right.
\end{eqnarray}

The restriction (\ref{eq:tress1}) allow to cut the series
(\ref{eq:Vcero})--(\ref{eq:Vdos}) at orders $N_0= 2$, $N_1=1$,
$N_2=2$, respectively, to obtain the MC equations of the expansion
$osp(1|32) (2,1,2)$: {\setlength\arraycolsep{0pt}
\begin{eqnarray}   \label{expanMTheory}
 &&     d\rho^{ab,0}=-\ft{1}{16}
 \rho^{ac,0}\wedge {\rho_c}^{b\,,0} \; , \nonumber \\
&&      d\rho^{a\,,2}=-\ft{1}{16} \rho^{b,2}\wedge {\rho_b}^{a,0}
-i \nu^{\alpha,1}
\wedge \nu^{\beta,1} \Gamma^{a}_{\alpha\beta} \quad, \nonumber\\
&&    d\rho^{ab,2}=-\ft{1}{16} \left( \rho^{ac,0}\wedge
{\rho_c}^{b,2} + \rho^{ac,2}\wedge {\rho_c}^{b,0} \right) -
\nu^{\alpha,1}
\wedge \nu^{\beta,1} \Gamma^{ab}_{\alpha\beta} \quad, \nonumber\\
&&    d\rho^{a_1\dots a_5}{}^{,2}=\ft{5}{16} \rho^{b[a_1\dots
a_4|\,,2} \wedge \rho_b{}^{|a_5],0}\nonumber - i \nu^{\alpha,1}
\wedge \nu^{\beta,1} \Gamma^{a_1\dots
a_5}_{\alpha\beta} \quad, \nonumber\\
&&      d\nu^{\alpha,1}=-\ft{1}{64} {\nu}^{\beta,1} \wedge
\rho^{ab,0}{\Gamma_{ab}}_\beta{}^\alpha \quad,
\end{eqnarray}
}Now, setting $\rho^{ab,0} \equiv -16 \sigma^{ab}$ and identifying
$\rho^{a,2} \equiv \Pi^a$, $\rho^{ab,2} \equiv \Pi^{ab}$,
$\rho^{a_1 \cdots a_5,2} \equiv \Pi^{a_1\cdots a_5}$ and
$\nu^{\alpha,1} \equiv \pi^\alpha$, the set of equations
(\ref{expanMTheory}) coincides with the MC equations of the M
Theory superalgebra containing the Lorentz group
$\mathfrak{E}^{(528|32)} \rtimes so(1,10)$ (equations
(\ref{eq:MCMTsuperalg}) when the Lorentz part is restored)
\cite{JdAIMO}. As a check, notice that the dimensional counting is
correct since, by equation (\ref{eq:dimsuper}),
{\setlength\arraycolsep{0pt}
\begin{eqnarray}
 \textrm{dim}\, osp(1|32)(2,1,2) &=& 2\cdot55+ 32+473=583+32 =
\nonumber \\ &=& \textrm{dim}\, \left( \mathfrak{E}^{(528|32)}
\rtimes so(1,10) \right) \; .
\end{eqnarray}
}In conclusion, from the supergroup point of view \cite{JdAIMO},
\begin{equation}
\Sigma^{(528|32)} \rtimes  SO(1,10)\approx OSp(1|32)(2,1,2) \; .
\end{equation}

This concludes this mathematical parenthesis, and we now return to
$D=11$ supergravity. In chapter \ref{chapter4} we were able to
write down a worldvolume action for a {\it preonic} brane in a
D'Auria and Fr\'e supergravity background. This formulation of
supergravity is closely related to the notions of enlarged
supersymmetry algebras and superspaces. In chapter \ref{chapter7},
a worldvolume action for a string describing the excitations of
two preons will be formulated, in fact, in an enlarged superspace.
In the next chapter, D'Auria-Fr\'e supergravity will be revisited,
and the expansion method for Lie algebras will be find useful to
describe the origin of the underlying symmetry algebras.


\chapter[The underlying symmetry of $D=11$ supergravity]
{The underlying symmetry \\ of $D=11$ supergravity}
\label{chapter6}

The problem of the hidden or underlying geometry of $D=11$
supergravity was raised already in the pioneering paper by
Cremmer-Julia-Scherk (CJS) \cite{CJS} (see also \cite{Kallosh84}),
where the possible relevance of $OSp(1|32)$ was suggested. It was
specially considered by D'Auria and Fr\'e \cite{D'A+F}, where the
search for the local supergroup of $D=11$ supergravity was
formulated as a search for a composite structure of its three-form
$A_3$. Indeed, while the graviton and gravitino are given by
one-form fields $e^a=dx^\mu e_\mu^a(x)$, $\psi^\alpha= dx^\mu
\psi^\alpha_\mu(x)$ and can be considered, together with the spin
connection $\omega^{ab}=dx^\mu\omega_\mu^{ab}(x)$,  the gauge
fields for the standard superPoincar\'e group \cite{DoMa77}, the
$A_{\mu_1\mu_2\mu_3}(x)$ abelian gauge field is not associated
with a symmetry generator and it rather corresponds to a
three-form $A_3$. However, one may ask whether it is possible to
introduce a set of additional one-form fields such that they,
together with $e^a$ and $\psi^\alpha$, can be used to express
$A_3$ in terms of products of one-forms. If so, the `old' and
`new' one-form fields may be considered as gauge fields of a
larger supergroup, and all the CJS supergravity fields can then be
treated as gauge fields, with $A_3$ expressed in terms of them.
This is what is meant here by the underlying gauge group structure
of $D=11$ supergravity: it is hidden when the standard $D=11$
supergravity multiplet is considered, and manifest when $A_3$
becomes a composite of the one-form gauge fields associated with
the extended group. The solution to this problem is equivalent to
the trivialization of a standard $D=11$ supersymmetry algebra
four-cocycle (related to $dA_3$) on an enlarged superalgebra.

The notion of free differential algebras (FDAs) is a natural
extension of that of Lie algebras, particularly suitable to
account for the $p$-form fields present in supergravity theories.
The notion of FDA and their construction as a process governed by
cohomology will be reviewed in section \ref{FDAandLie}. All this
is put in its due context in section \ref{D=11FDA}, where the FDA
of $D=11$ supergravity is presented, and $dA_3$ seen to be related
to a non-trivial supersymmetry algebra cocycle. We then apply
these ideas to discuss the trivialization of FDAs (the process of
obtaining {\it Lie} algebras from FDAs) and its implications for
the physics they may describe. To that end, a family of extensions
$\tilde{\mathfrak{E}}(s)$ of the supersymmetry superalgebra is
described in section \ref{Sectriv} and its relation to $osp(1|32)$
discussed in section \ref{osprel}. In section \ref{nature}, the
cocycle associated to $dA_3$ is shown to be trivialized by any
member of the family, except for $s \neq 0$ \cite{Lett,AnnP04},
extending previous results \cite{D'A+F}. Section \ref{dyncomp}
analyzes the possible dynamical consequences of a composite
structure of $A_3$ and shows the presence of additional gauge
symmetries in the action for a composite $A_3$. Section
\ref{adboscoor} concludes this chapter with some remarks about a
conjectured fields/enlarged superspace coordinates correspondence.
The main results of this chapter can be found in references
\cite{Lett,AnnP04}.

\section{Free differential algebras, Lie algebras and cohomology}
\label{FDAandLie}

The presence of forms of orders higher than one in the
supergravity lagrangians makes it especially convenient to resort
to free differential algebras in order to discuss the geometry
associated to those theories. In fact, the discussion of this
section about the relation of free differential algebras and Lie
algebras can be straightforwardly extended to account for their
superalgebra counterparts, the structures of interest in
supergravity theories.

A free differential algebra (FDA) \cite{Su77,D'A+F,Cd'AF91,Ni83}
(termed Cartan integrable system in \cite{D'A+F}) is an exterior
algebra with constant coefficients, generated by a set of forms
(not necessarily of the same rank) closed under the action of the
exterior differential $d$. The dual formulation of a Lie algebra
${\cal G}$, in terms of Maurer-Cartan (MC) one-forms\footnote{In
chapter \ref{chapter5}, the MC one-forms of the Lie algebra ${\cal
G}$ were denoted as $\omega^i$. Here, the notation $\pi^i$ is
preferred to reserve $\omega$ for non-trivial Chevalley-Eilenberg
cocycles.} $\pi^i$ left-invariant on the corresponding group
manifold, provides the simplest example of an FDA. As an FDA,
${\cal G}$ is to be regarded as generated by a collection of
one-forms $\pi^i$, $i=1, \ldots , \textrm{dim} {\cal G}$, and
two-forms $d \pi^i$, related through the MC equations of ${\cal
G}$ (equation (\ref{eq:mc})) and closed under $d$ due to the
Jacobi identity.

A more interesting application of FDAs is the description of the
local symmetry of a theory through the {\it gauging} of Lie
algebras. The gauge FDA associated to the Lie algebra ${\cal G}$
is obtained by replacing the MC one-forms $\pi^i$ of ${\cal G}$ by
their {\it gauge field} or {\it soft} (see \cite{Cd'AF91}) {\it
one-form} counterparts $A^i$, and by introducing {\it two-form
curvatures} satisfying a generalization of the MC equations of
${\cal G}$ (see equation (\ref{eq:mc})),
\begin{equation} \label{eq:FDACartan}
F^k = d A^k + \ft12 c^k_{ij} A^i \wedge A^j  .
\end{equation}
The curvatures then satisfy the consistency conditions expressed
by the Bianchi identities
\begin{equation} \label{eq:FDABianchi}
dF^k=c_{ij}^k F^i \wedge A^j \; .
\end{equation}
The structure equations (\ref{eq:FDACartan}) and the Bianchi
identities (\ref{eq:FDABianchi}) then define the gauge FDA
associated with the Lie algebra ${\cal G}$. Dynamically, the
relevance of the FDA constructed this way from the Lie algebra
${\cal G}$ is reflected by the fact that the lagrangian of a
theory with local symmetry ${\cal G}$  is built up from the gauge
potentials $A^i$ and their curvatures $F^i$.

An FDA is called {\it minimal} \cite{Su77} when the differential
of any $p$-form in the FDA is expressible {\it only} in terms of
sums of wedge products of $q$-forms in the FDA, with $q \leq p$.
The FDA is {\it contractible} if it is generated by pairs of forms
$\pi_p$, $\pi_{p+1}$ such that $d \pi_p = \pi_{p+1}$, $d \pi_{p+1}
= 0$. According to {\it Sullivan's first theorem} \cite{Su77},
which is the counterpart for FDAs of the L\'evi-Mal'\v cev theorem
(see \cite{AI95}) for Lie algebras, the most general FDA is the
semidirect sum of a contractible with a minimal one. For instance,
regarded as an FDA, a Lie algebra ${\cal G}$ is minimal, whereas
the FDA (\ref{eq:FDACartan}), (\ref{eq:FDABianchi}) is
contractible. If the `flat limit' of the contractible algebra
(\ref{eq:FDACartan}), (\ref{eq:FDABianchi}) is considered, in
which all the curvatures are set to zero, $F^i =0$, the gauge
potentials $A_i$ turn out to satisfy the same equations than
$\pi^i$ ({\it i.e.}, (\ref{eq:FDACartan}) reduces to
(\ref{eq:mc})), and the minimal algebra, which in this case
coincides with the Lie algebra ${\cal G}$, is recovered.

The minimal FDA does not need to be a Lie algebra, though. Indeed,
it is the typical case in supergravity theories that their
lagrangian contains not only one-forms and their curvature
two-forms, but also\footnote{Here, the superindex $i$ is again
used to label the forms, and a subindex showing their rank is
added.} $p$-forms $A^i_p$, $p> 1$, and their curvature
$(p+1)$-forms $F^i_{p+1}$. This is precisely the case of $D=11$
supergravity, the lagrangian (\ref{L11=}) of which involves not
only the one-forms $e^a$ and $\psi^\alpha$ and their curvatures,
but also the three-form $A_3$ and, in the first order approach, an
auxiliary four-form $F_4$ which is related, on-shell, to its
curvature. For notational analogy with the gauging of Lie
algebras, it is convenient to introduce the {\it rigid} $p$-form
counterparts $\pi^i_p$ of the {\it soft} \cite{Cd'AF91} $p$-forms
$A^i_p$, such that, in the `flat limit' in which all the
curvatures are set to zero, $F^i_{p+1}=0$, the forms $A^i_p$
satisfy the same structure equations than $\pi^i_p$. The FDA thus
reduces to the minimal FDA, which is nevertheless not a Lie
algebra since it contains forms $\pi^i_p$ of rank higher than one.
And yet, from a physical point of view, despite the unclear
relation in this case of the minimal FDA with a Lie algebra, a
lagrangian built up from $A^i_p$ and $F^i_{p+1}$ possesses a local
symmetry: that is the case in supergravity theories, which are the
field theories of local supersymmetry.

Consider a Lie algebra ${\cal G}$ defined through the MC, left
invariant, one-forms $\pi^i_1$ on the group manifold $G$ of ${\cal
G}$ satisfying the MC equations (\ref{eq:mc}). {\it Sullivan's
second theorem} \cite{Su77} determines the structure of minimal
FDAs, built by using the MC forms $\pi^i_1$, through an iterative
process \cite{Su77,Cd'AF91} (see also \cite{Fre05}). First, the MC
one-forms $\pi_1^i$ of any minimal FDA close into the original Lie
algebra ${\cal G}$. The structure equations for additional
$p$-forms $\pi_p^i$ of the minimal FDA can be written as
\begin{equation} \label{eq:cocgen}
d \pi^i_p = \omega^i_{p+1}(\pi^j_1) \; ,
\end{equation}
where $\omega^i_{p+1}$ are nontrivial Chevalley-Eilenberg (CE)
\cite{CE48,AI95} $(p+1)$-cocycles on the Lie algebra ${\cal G}$.
In other words, for each $i$, $\omega^i_{p+1}$ is a  closed
$(p+1)$-form built up as a sum of exterior products ({\it i.e.},
as an {\it exterior polynomial}) of the MC one-forms $\pi_1^i$
(and, thus, invariant under ${\cal G}$) which is not the
differential of a $p$-form invariant under ${\cal G}$, namely,
$\pi^i_p$ is not an (exterior) polynomial in $\pi_1^i$. The
process can be iterated by adding new $q$-forms $\pi^{\prime i}_q$
such that their differentials are non-trivial $(q+1)$-cocycles
depending on $(\pi^i_1, \pi^j_p)$, then on $(\pi^i_1, \pi^j_p,
\pi^{\prime i}_q)$, and so on.

In general, the  non-trivial character of the cocycles
$\omega^i_{p+1}$ defining a minimal FDA can be interpreted by
saying that there are not enough MC one-forms $\pi^i_1$ in the Lie
algebra ${\cal G}$ to write the $p$-forms  $\pi^i_p$ in equation
(\ref{eq:cocgen}) in terms of them. But it may happen that the
introduction of an algebra ${\cal \tilde{G}}$ {\it larger} than
${\cal G}$ allows for the ${\cal G}$-cocycles to be written in
terms of the new MC one-forms of ${\cal \tilde{G}}$, so that they
are (left-)invariant under the corresponding group $\tilde{G}$.
Bearing this in mind, the question of whether there exists a Lie
algebra describing the same local symmetry than a given FDA can be
put in precise mathematical terms, at least for minimal FDAs: if
there exists an extension\footnote{See section \ref{fourw} of the
previous chapter and references therein.} ${\cal \tilde{G}}$ of
the Lie algebra ${\cal G}$ for which the cocycles $\omega^i_{p+1}$
become trivial, {\it i.e.}, such that for each $i$, the $p$-forms
$\pi^i_ p$ (related to $\omega^i_{p+1}$ through equation
(\ref{eq:cocgen})) can be expressed as (exterior) polynomials in
the MC one-forms of ${\cal \tilde{G}}$, then the FDA can be
`trivialized', by writing its forms in terms of the MC forms of
the Lie algebra ${\cal \tilde{G}}$. In a supergravity context, it
is in this sense that ${\cal \tilde{G}}$ can be said to be the
{\it underlying gauge symmetry} of the theory under consideration.

Notice that the trivialization problem might have either no
solution at all (see \cite{Ni83} for an example) or more than one
solution: there might exist more than one enlarged algebra ${\cal
\tilde{G}}$ that trivializes the cocycles defining the minimal
FDA. The later is the case for $D=11$ supergravity, for which two
superalgebras were already found in \cite{D'A+F} to account for
its underlying symmetry. It will be shown in section \ref{nature}
that, in fact, not only the two algebras of \cite{D'A+F} solve the
problem, but that there exists a whole one-parameter family of Lie
superalgebras describing the underlying symmetry of $D=11$
supergravity. First, we shall introduce the FDA corresponding to
$D=11$ supergravity.

\section{The $D=11$ supergravity FDA} \label{D=11FDA}

Let us first consider, momentarily,  the case of four-dimensional
simple supergravity, where the only fields involved are the
graviton and gravitino (in the $N=1$, $D=4$ supergravity
multiplet) and the Lorentz connection. These can actually be
considered as the gauge fields of simple $D=4$ supergravity
\cite{DoMa77} and can be described by a gauge (super)FDA
constructed as discussed in section \ref{FDAandLie}. Indeed,
replacing the MC one-forms $\Pi^a$, $\pi^\alpha$, $\sigma^{ab}$ of
the superPoincar\'e algebra by the gauge field one-forms $e^a$,
$\psi^\alpha$, $\omega^{ab}$, respectively, and introducing their
corresponding curvatures, ${\mathbf R}^a$, ${\mathbf R}^\alpha$,
$\mathbf{R}^{ab}$, the superPoincar\'e MC equations
(\ref{eq:MCsuperP}) can be promoted to the structure equations
(see equation (\ref{eq:FDACartan})) {\setlength\arraycolsep{0pt}
\begin{eqnarray}\label{CartanFDAsuperP}
&& {\mathbf R}^a :=  de^a -e^b\wedge \omega_b{}^a  +
i\psi^{\alpha} \wedge \psi^{\beta} \Gamma^a_{\alpha\beta} = T^a  +
i\psi^{\alpha} \wedge \psi^{\beta}
\Gamma^a_{\alpha\beta} \; , \nonumber  \\
&& {\mathbf R}^\alpha  :=   d\psi^\alpha - \psi^\beta \wedge
\omega_\beta{}^\alpha  \quad \left( \omega_\alpha{}^\beta=\ft14
\omega^{ab} \Gamma_{ab\; \alpha}{}^\beta \right) \; , \nonumber
\\
&& \mathbf{R}^{ab} :=   d \omega^{ab}  - \omega^{ac} \wedge
\omega_c{}^{b}  \; ,
\end{eqnarray}
}where $T^a:=De^a=de^a -e^b\wedge \omega_b{}^a$ is the torsion
(see equation (\ref{Torsion})). The equations
(\ref{CartanFDAsuperP}), together with their selfconsistency or
integrability conditions (see (\ref{eq:FDABianchi}))
{\setlength\arraycolsep{0pt}
\begin{eqnarray}\label{BianchiFDAsuperP}
&& D \mathbf{R}^a = - e^b \wedge  \mathbf{R}_b{}^a + 2i
\psi^\alpha \wedge  \mathbf{R}^\beta \Gamma^a_{\alpha\beta} \; ,
\nonumber
\\
&& D \mathbf{R}^\alpha  = - \ft14 \psi^\beta \wedge
\mathbf{R}^{ab} \Gamma_{ab}{}_\beta{}^\alpha  \; , \nonumber
\\
&& D \mathbf{R}^{ab} = 0 \; ,
\end{eqnarray}
}where $D$ is the Lorentz covariant derivative, form the gauge FDA
of the superPoincar\'e group. When all the curvatures are set to
zero, ${\mathbf R}^a=0$, ${\mathbf R}^\alpha =0$, ${\mathbf
R}^{ab} =0$, the Bianchi identities (\ref{BianchiFDAsuperP}) are
trivially satisfied and, as discussed in the previous section, the
structure equations (\ref{CartanFDAsuperP}) of the FDA reduce to
the MC equations (\ref{eq:MCsuperP}) of the superPoincar\'e
algebra. The minimal FDA is, in this case, a Lie superalgebra
(superPoincar\'e), the local symmetry of simple supergravity.

This FDA description is, however, incomplete for $D=11$
supergravity, due to the presence of the three-form field $A_3$.
When $A_3$ is taken into account, the FDA defined by equations
(\ref{CartanFDAsuperP}) must be completed by the definition of the
four-form field strength \cite{D'A+F}
\begin{eqnarray}\label{R4=} \mathbf{R}_4 &:=&
dA_3 + \ft14 \psi^\alpha \wedge \psi^\beta \wedge e^a \wedge e^b
\Gamma_{ab}{}_{\alpha\beta} \; ,
\end{eqnarray}
supplemented by its corresponding Bianchi identity\footnote{See
section \ref{sec:MTsuperalg} for the notation.},
\begin{eqnarray} \label{BI:F4}
d\mathbf{R}_4 &=& - \psi^\alpha \wedge \mathbf{R}^\beta \wedge
\bar{\Gamma}^{(2)}_{\alpha\beta} - \ft12 \psi^\alpha \wedge
\psi^\beta \wedge e^b \wedge \mathbf{R}^a
{\Gamma}_{ab}{}_{\alpha\beta} \; .
\end{eqnarray}
Equations (\ref{CartanFDAsuperP}), (\ref{R4=}) are the structure
equations for the  {\it $D=11$ supergravity FDA}, their
corresponding Bianchi identities being (\ref{BianchiFDAsuperP}),
(\ref{BI:F4}). The definition (\ref{R4=}) of the curvature
$\mathbf{R}_4$ of $A_3$ is obviously inspired in the algebraic
constraint (\ref{dA3=a+F=}) that relates $F_4$ to $A_3$. Indeed,
resorting to the superspace formulation of supergravity and
setting $\mathbf{R}^a=0$ and $\mathbf{R}_4=F_4:= \ft{1}{4!}
e^{a_4}\wedge \ldots \wedge e^{a_1} F_{a_1\ldots a_4}$, the
on-shell $D=11$ superspace supergravity constraints
\cite{CremmerFerrara80,BrinkHowe80} are recovered (see also
\cite{AnnP04}).

In contrast with the $D=4$ case, the above FDA for vanishing
curvatures cannot be associated with the MC equations of a {\it
Lie} superalgebra due to the presence of the {\it three}-form
$A_3$. In fact, according to the general discussion in section
\ref{FDAandLie}, for vanishing curvatures the bi-fermionic
four-form
\begin{eqnarray}\label{a4=}
a_4= - \ft14 \psi^\alpha \wedge \psi^\beta \wedge e^a \wedge e^b
\Gamma_{ab}{}_{\alpha\beta} \;
\end{eqnarray}
entering the definition (\ref{R4=}) of the curvature
$\mathbf{R}_4$ of $A_3$ (see also equation (\ref{a4})) becomes a
CE four-cocycle on the supertranslations algebra $\mathfrak{E}
\equiv \mathfrak{E}^{(11|32)}$ given by
\begin{eqnarray}\label{a4=0}
\omega_4 (x^a, \theta^\alpha) = - \ft14 \pi^\alpha \wedge
\pi^\beta \wedge \Pi^a \wedge \Pi^b
\Gamma_{ab}{}_{\alpha\beta}=d\omega_3 (x^a, \theta^\alpha) \; ,
\end{eqnarray}
where $\Pi^a = dx^a -id\theta^{\alpha } \Gamma^a_{\alpha\beta}
\theta^{\beta }$ and $\pi^\alpha = d\theta^{\alpha }$. In equation
(\ref{a4=0}), the dependence of the forms $\omega_3$ and
$\omega_4$ on the coordinates $Z=(x^a, \theta^\alpha)$ of rigid
superspace $\Sigma \equiv \Sigma^{(11|32)}$, the group manifold of
the $D=11$ supertranslations group, has been written explicitly.
The Lorentz group, being simple and not adding anything to the
cohomology, can be neglected in this discussion.

As discussed in general, the $\mathfrak{E}$-invariant and closed
four-cocycle $\omega_4$ is, furthermore, non-trivial in the CE
cohomology, since the three-form
$\omega_3=\omega_3(x^a,\theta^\alpha)$ in (\ref{a4=0})  cannot be
expressed in terms of the invariant MC forms $\Pi^a$, $\pi^\alpha$
of $\mathfrak{E}$. Now, we may ask whether there exists an
extension $\tilde{\mathfrak{E}}$ of the standard $D=11$
supersymmetry algebra $\mathfrak{E}$, with MC forms defined on its
associated enlarged  superspace $\tilde{\Sigma}$, on which the CE
four-cocycle $\omega_4$ becomes trivial. In this way, the problem
of writing the original $A_3$ field in terms of one-form fields
becomes purely geometrical: it is equivalent to looking, in the
spirit of the fields/extended superspace variables correspondence
of \cite{JdA00} (see section \ref{adboscoor}), for an {\it
enlarged} supergroup manifold $\tilde{\Sigma}$ on which one can
find a new three-form $\tilde{\omega}_3$ (corresponding to $A_3$)
that can be expressed in terms of sums of exterior products of
$\tilde{\mathfrak{E}}$ MC forms on $\tilde{\Sigma}$ (that will
correspond to one-form gauge fields), hence depending on the
coordinates $\tilde{Z}$ of $\tilde{\Sigma}$. That such a
$\tilde{\Sigma}$-invariant  form $\tilde{\omega}_3(\tilde{Z})$
should exist is also not surprising if we recall that the CE
$(p+2)$-cocycles on $\mathfrak{E}$ that characterize \cite{AT89}
the Wess-Zumino terms of the super-$p$-brane actions and their
associated FDAs, can also be trivialized on larger superalgebras
$\tilde{\mathfrak{E}}$ \cite{BESE,JdA00} (see also
\cite{Anguelova:2003sn}) associated to extended superspaces
$\tilde{\Sigma}$, and that the pull-back of
$\tilde{\omega}_3(\tilde{Z})$ to the supermembrane worldvolume
defines an invariant Wess-Zumino term.

To summarize, the minimal FDA of $D=11$ supergravity is obtained
by enlarging the supertranslations algebra $\mathfrak{E}$
(containing the MC one-forms $\Pi^a$, $\pi^\alpha$, corresponding
to the gauge fields $e^a$, $\psi^\alpha$, respectively) with the
three-form $\omega_3$ (corresponding to $A_3$) such that its
differential is the CE four-cocycle $\omega_4$ on $\mathfrak{E}$
(equation (\ref{a4=0})). Notice, however, that further
enlargements are possible, within the FDA construction scheme of
section \ref{FDAandLie}. Actually, the closed seven-form
\begin{equation}  \label{R7=0}
\omega_7 =-  \omega_3 \wedge \omega_4 + \ft{i}{ 2 \cdot 5!}
\pi^\alpha \wedge \pi^\beta \wedge \Pi^{a_5} \wedge \ldots \wedge
\Pi^{a_1} \, {\Gamma}_{a_1\ldots a_5\; \alpha\beta}
 \;
\end{equation}
is a non-trivial cocycle\footnote{Notice that $\omega_7$ is not a
CE seven-cocycle {\it on the standard, $D=11$ supertranslations
algebra $\mathfrak{E}\equiv \mathfrak{E}^{(11|32)}$}, since it
explicitly involves $\omega_3$.} {\it on the FDA generated by
$(\Pi^a, \pi^\alpha, \omega_3)$} (\cite{Cd'AF91}, vol.~II,
p.~866). The seven-cocycle $\omega_7$ is nothing but the `flat
limit', $\mathbf{R}_7 =0$, of $dA_6$, where the six-form $A_6$ is
the dual six-form  of $A_3$, defined by $F_7 =
* F_4$, where $F_4 = dA_3$ and $F_7 =dA_6 + A_3 \wedge dA_3$. The
structure equation of $A_6$,
\begin{equation}
\label{CJS:R7=} \mathbf{R}_7 := dA_6 + A_3\wedge dA_3 -
 \ft{i}{2} \psi^\alpha \wedge \psi^\beta
\wedge \bar{\Gamma}^{(5)}_{\alpha\beta}
\end{equation}
(see (\ref{Gammak}) for the notation), together  with its
corresponding Bianchi identity, {\setlength\arraycolsep{2pt}
\begin{eqnarray}
\label{BI:F7} d\mathbf{R}_7 &= & \left(\mathbf{R}_4 + \ft12
\psi\wedge \psi\wedge\bar{\Gamma}^{(2)}\right) \wedge
\left(\mathbf{R}_4 + \ft12 \psi\wedge
\psi\wedge\bar{\Gamma}^{(2)}\right) \nonumber \\ && + i
\psi^\alpha \wedge \mathbf{R}^\beta \wedge
\bar{\Gamma}^{(5)}_{\alpha\beta} - \ft{i}{2 \cdot 4!} \psi^\alpha
\wedge  \psi^\beta \wedge e^{c_4} \wedge
 \ldots \wedge  e^{c_1} \wedge  \mathbf{R}^a
{\Gamma}_{ac_1\ldots c_4}{}_{\alpha\beta}  \nonumber \\
&& - \ft14 \psi^\alpha \wedge  \psi^\beta \wedge \psi^\gamma
\wedge  \psi^\delta \wedge \bar{\Gamma}^{(2)}_{\alpha\beta} \wedge
\bar{\Gamma}^{(2)}_{\gamma\delta} \equiv 0 \; ,
\end{eqnarray}
}may thus be added to the FDA (\ref{CartanFDAsuperP}),
(\ref{R4=}), (\ref{BianchiFDAsuperP}), (\ref{BI:F4}). We shall,
however, ignore this enlargement of the algebra and work with the
FDA generated by $(\Pi^a, \pi^\alpha, \omega_3)$ whose gauging,
described (neglecting the Lorentz part) by the structure equations
(\ref{CartanFDAsuperP}) and (\ref{R4=}) and their Bianchi
identities (\ref{BianchiFDAsuperP}), (\ref{BI:F4}), involves only
$e^a$, $\psi^\alpha$, $A_3$ and their curvatures $\mathbf{R}^a$,
$\mathbf{R}^\alpha$, $\mathbf{R}_4$. Nevertheless, it would be an
interesting question for further study to determine whether the
Lie algebras $\tilde{\mathfrak{E}}(s)$ introduced below to
trivialize the four-cocycle $\omega_4$ (of equation (\ref{a4=0}))
also allow for the trivialization of $\omega_7$ (of equation
(\ref{R7=0})). Namely, whether there exists a six-form
$\tilde{\omega}_6$ (corresponding to the `flat limit' of $A_6$)
constructed as an exterior polynomial of the MC one-forms of
$\tilde{\mathfrak{E}}(s)$ and such that $\omega_7= d
\tilde{\omega}_6$. This would correspond to the problem of finding
the {\it underlying gauge symmetry} of the duality-symmetric
formulation of $D=11$ supergravity (see \cite{BNS98} for the
action).

\setcounter{equation}0
\section{A family of extended superalgebras}
\label{Sectriv}

As stated in  \cite{D'A+F}, the problem is whether the $D=11$
supergravity FDA (\ref{CartanFDAsuperP}), (\ref{R4=}), may be
completed with a number of additional {\it one}-forms and their
curvatures in such a way that the three-form $A_3$ obeying
(\ref{R4=}) is constructed from one-forms, becoming composite
rather than fundamental or `elementary'. This problem, when
attacked in the flat limit achieved by setting  all the curvatures
to zero, is equivalent to trivializing the $\Sigma$ four-cocycle
$\omega_4$ (equation (\ref{a4=0})) on the algebra
$\tilde{\mathfrak{E}}$ of an {\it enlarged} superspace group
$\tilde{\Sigma}$. A one-parameter family of Lie superalgebra
extensions $\tilde{\mathfrak{E}}(s)$ (with the notation of
\cite{Lett}) of the M Theory superalgebra was first proposed by
D'Auria and Fr\'e in \cite{D'A+F} as an ansatz to solve the
problem. All the superalgebras in the family contain a set of 528
bosonic and $32+32=64$ fermionic generators,
\begin{equation} \label{generators}
 \, P_a  \, , \, Q_\alpha  \, , \,  Z_{a_1a_2}  \, ,
\, Z_{a_1 \ldots a_5}  \, , \, Q^\prime_\alpha \; ,
\end{equation}
including the M Theory superalgebra ones (see section
\ref{sec:MTsuperalg} of chapter \ref{chapter2}) plus a central
fermionic generator $Q^\prime_\alpha$, and are defined through the
(anti)commutation relations {\setlength\arraycolsep{0pt}
\begin{eqnarray} \label{susyalg}
&& \{Q_\alpha,Q_\beta\}=\Gamma^a_{\alpha\beta} P_a +
i\Gamma^{a_1a_2}_{\alpha\beta} Z_{a_1a_2} + \Gamma^{a_1\ldots
a_5}_{\alpha\beta} Z_{a_1\ldots a_5} \; , \nonumber \\
&& [ P_a , Q_\alpha ] = \delta \;  \Gamma_{a \; \alpha}{}^\beta
Q^\prime_\beta \; , \quad \nonumber
\\ && [ Z_{a_1a_2} , Q_\alpha ]=i\gamma_1
\Gamma_{a_1a_2 \; \alpha}{}^\beta Q^\prime_\beta \; ,  \quad
\nonumber \\
&& [ Z_{a_1 \ldots a_5} , Q_\alpha ]=\gamma_2 \Gamma_{a_1 \ldots
a_5 \; \alpha}{}^\beta Q^\prime_\beta \;   , \nonumber \\
&& [Q^\prime_\alpha , \textrm{all} \} =0 \; ,
\end{eqnarray}
}that display their structure as central extensions of the M
Theory superalgebra by the fermionic generator $Q^\prime_\alpha$.
With the notation of section \ref{sec:MTsuperalg} of chapter
\ref{chapter2}, the family of enlarged superalgebras
$\tilde{\mathfrak{E}}(s)$ could be denoted as
$\mathfrak{E}^{(528|64)}(s)$ or, in order to emphasize the
presence of two independent fermionic generators, as
$\mathfrak{E}^{(528|32+32)}(s)$. The corresponding group manifolds
can accordingly be denoted $\tilde{\Sigma}(s)$ or
$\Sigma^{(528|32+32)}(s)$. The notation $\tilde{\mathfrak{E}}(s)$,
$\tilde{\Sigma}(s)$ will be preferred, although
$\mathfrak{E}^{(528|32+32)}(s)$, $\Sigma^{(528|32+32)}(s)$ will
sometimes be used to avoid confusion.

In (\ref {susyalg}), $\delta$, $\gamma_1$, $\gamma_2$ are real
parameters only restricted by the requirement that (\ref{susyalg})
be indeed a superalgebra, {\it i.e.}, that the Jacobi identities
are satisfied. This translates into a relation for the parameters
\cite{D'A+F}:
\begin{equation}
\label{idg} \delta + 10 \gamma_1- 6! \gamma_2=0 \; .
\end{equation}
One parameter ($\gamma_1$ if nonvanishing, $\delta$ otherwise) can
be removed by rescaling the new fermionic generator
$Q^\prime_\alpha$ and it is thus inessential. Hence equations
(\ref{susyalg}) describe, effectively, a one-parameter family of
Lie superalgebras that may be denoted $\tilde {\mathfrak E}(s)$ by
using a parameter $s$ given by {\setlength\arraycolsep{0pt}
\begin{eqnarray}
\label{s-def} s:= {\delta \over 2\gamma_1} - 1 \;,\; \gamma_1\neq0
 \qquad & \Rightarrow & \qquad \left\{
{\setlength\arraycolsep{2pt}\begin{array}{lll} \delta
&=& 2\gamma_1(s+1) \, , \\
\gamma_2 &=& 2\gamma_1({s \over 6!} + {1 \over 5!}) \; .
\end{array}} \right.    \;
\end{eqnarray}
}This notation also accounts for the case $\gamma_1 = 0$, by
considering the limit $\gamma_1 \rightarrow 0$,  $s \rightarrow
\infty$ and $\gamma_1 s \rightarrow \delta/2 \neq 0$, and the
corresponding algebra can be denoted
$\tilde{\mathfrak{E}}(\infty)$. In terms of $s$, the algebra
(\ref{susyalg}) reads: {\setlength\arraycolsep{0pt}
\begin{eqnarray}\label{Sigma(s)}
&& \{Q_\alpha,Q_\beta\}=\Gamma^a_{\alpha\beta} P_a +
i\Gamma^{a_1a_2}_{\alpha\beta} Z_{a_1a_2} + \Gamma^{a_1\ldots
a_5}_{\alpha\beta} Z_{a_1\ldots a_5} \; , \nonumber \\
&& [ P_a , Q_\alpha ] = 2\gamma_1(s+1) \;  \Gamma_{a \;
\alpha}{}^\beta Q^\prime_\beta \; , \quad \nonumber
\\ && [ Z_{a_1a_2} , Q_\alpha ]=i\gamma_1
\Gamma_{a_1a_2 \; \alpha}{}^\beta Q^\prime_\beta \; , \nonumber \\
&&  [ Z_{a_1 \ldots a_5} , Q_\alpha ]=2\gamma_1({s \over 6!} + {1
\over 5!}) \Gamma_{a_1 \ldots a_5 \; \alpha}{}^\beta
Q^\prime_\beta \; , \quad \nonumber \\
&& [Q^\prime_\alpha  , \textrm{all} \} =0 \; .
\end{eqnarray}}

Introducing the MC one-forms
\begin{equation} \label{formsdual}
 \, \Pi^a  \, , \, \pi^\alpha  \, , \,  \Pi^{a_1a_2}  \, ,
\, \Pi^{a_1 \ldots a_5}  \, , \, \pi^{\prime \alpha} \; ,
\end{equation}
dual to the generators (\ref{generators}), and left-invariant on
the corresponding group manifolds $\tilde{\Sigma}(s) \equiv
\Sigma^{(528|32+32)}(s)$, the family of superalgebras $\tilde
{\mathfrak E}(s)$ can be equivalently described by the MC
equations
\begin{eqnarray} \label{Sigma(s)dual}
 && d\Pi^a =  - i\pi^{\alpha} \wedge
\pi^{\beta} \Gamma^a_{\alpha\beta}  \; , \nonumber \\
&& d\pi^\alpha = 0 \; , \nonumber
\\
&& d\Pi^{a_1a_2} = - \pi^\alpha \wedge \pi^\beta \,
\Gamma^{a_1a_2}_{\alpha\beta} \; ,  \nonumber \\
&& d\Pi^{a_1\ldots a_5} = - i \pi^\alpha \wedge
\pi^\beta \, \Gamma^{a_1\ldots a_5}_{\alpha\beta} \; , \nonumber  \\
&& d\pi^{\prime \alpha} = \pi^\beta \wedge \left(- i \, \delta \,
\Pi^a \Gamma_{a\, \beta}{}^\alpha - \gamma_1 \, \Pi^{ab}
\Gamma_{ab\, \beta}{}^\alpha - i \, \gamma_2 \, \Pi^{a_1\ldots
a_5} \Gamma_{a_1\ldots a_5 \beta}{}^\alpha \right) \; . \nonumber
\\ &&
\end{eqnarray}
In the dual, MC formulation of the family of superalgebras $\tilde
{\mathfrak E}(s)$, the parameters $\delta$, $\gamma_1$, $\gamma_2$
are only involved in the last equation of (\ref{Sigma(s)dual}),
the MC equation for the extra fermionic MC one-form $\pi^{\prime
\alpha}$, and the relation (\ref{idg}) among the parameters is
obtained from the integrability condition $dd\pi^{\prime
\alpha}=0$. In terms of the parameter $s$ defined in
(\ref{s-def}), the last equation in (\ref{Sigma(s)dual}) reads
{\setlength\arraycolsep{0pt}
\begin{eqnarray}
d\pi^{\prime \alpha}  &=&  -2\gamma_1 \pi^\beta \wedge \left(i
(s+1)  \Pi^a \Gamma_{a} + \ft12 \Pi^{ab} \Gamma_{ab} +   i \left(
\ft{s}{6!} + \ft{s}{5!} \right)  \Pi^{a_1\ldots a_5}
\Gamma_{a_1\ldots a_5} \right)_\beta{}^\alpha . \nonumber \\
&& \label{deta=}
\end{eqnarray}

}

Finally, introducing the `soft' one-form fields,
\begin{equation} \label{formsdualsoft}
 \, e^a  \, , \, \psi^\alpha  \, , \,  B^{a_1a_2}  \, ,
\, B^{a_1 \ldots a_5}  \, , \, \eta^\alpha \; ,
\end{equation}
corresponding to the MC one-forms (\ref{formsdual}), and their
corresponding curvatures,
\begin{equation} \label{formsdualcurv}
 \, {\mathbf R}^a  \, , \, {\mathbf R}^\alpha  \, , \,  {\cal B}^{a_1a_2}  \, ,
\, {\cal B}^{a_1 \ldots a_5}  \, , \, {\cal B}^\alpha \; ,
\end{equation}
the family of gauge FDAs corresponding to $\tilde {\mathfrak
E}(s)$ is described by the equations (\ref{CartanFDAsuperP}),
(\ref{R4=}) together with the corresponding equations for the new
one-forms and their curvatures, namely, by
\begin{eqnarray}
&& \kern-2em {\mathbf R}^a :=  de^a -e^b\wedge \omega_b{}^a  +
i\psi^{\alpha} \wedge \psi^{\beta} \Gamma^a_{\alpha\beta} = T^a  +
i\psi^{\alpha} \wedge \psi^{\beta}
\Gamma^a_{\alpha\beta} \; , \nonumber  \\
&& \kern-2em {\mathbf R}^\alpha  :=   d\psi^\alpha - \psi^\beta
\wedge \omega_\beta{}^\alpha  \quad \left( \omega_\alpha{}^\beta=
\ft14 \omega^{ab} \Gamma_{ab\; \alpha}{}^\beta \right) \; ,
\nonumber
\\
&& \kern-2em \mathbf{R}^{ab} :=  d \omega^{ab}  - \omega^{ac}
\wedge
\omega_c{}^{b}  \; , \nonumber \\
&& \kern-2em {\cal B}_2^{ab} = DB^{ab} + \psi^\alpha  \wedge
\psi^\beta \, \Gamma^{ab}_{\alpha\beta} \; , \nonumber
\\
&& \kern-2em {\cal B}_2^{a_1\ldots a_5} = DB^{a_1\ldots a_5} + i
\psi^\alpha  \wedge \psi^\beta \,
\Gamma^{a_1\ldots a_5}_{\alpha\beta} \; , \nonumber  \\
&& \kern-2em {\cal B}_2^\alpha  = D\eta^\alpha + \psi^\beta \wedge
\left( i \, \delta \, e^a \Gamma_{a\, \beta}{}^\alpha + \gamma_1
\, B^{ab} \Gamma_{ab\, \beta}{}^\alpha + i \, \gamma_2 \,
B^{a_1\ldots a_5} \Gamma_{a_1\ldots a_5 \ \beta}{}^\alpha \right)
\; , \nonumber \\* && \label{trivializingFDA}
\end{eqnarray}
and their corresponding Bianchi identities.

\section{The relation of $\tilde{\mathfrak E}(s)$ with
$osp(1|32)$} \label{osprel}

For $s\not=0$, the superalgebras $\tilde {\mathfrak E}(s)$ are
non-trivial deformations (see section \ref{fourw}) of $\tilde
{\mathfrak E}(0)$. Actually, the superalgebra $\tilde {\mathfrak
E}(0)$ is singled out within the family $\tilde {\mathfrak E}(s)$
for having an enhanced automorphism group. Introducing, as in
(\ref{n32}), the generalized momentum $P_{\alpha \beta} =
\Gamma^a_{\alpha\beta} P_a + i\Gamma^{a_1a_2}_{\alpha\beta}
Z_{a_1a_2} + \Gamma^{a_1\ldots a_5}_{\alpha\beta} Z_{a_1\ldots
a_5}$, the $D=11$ decomposition {\setlength\arraycolsep{0pt}
\begin{eqnarray}
 \label{II=GG}
&& \delta_{(\alpha}{}^{\gamma} \delta_{\beta)}{}^{\delta} =
\ft{1}{32} \left( \Gamma^a_{\alpha\beta} \Gamma_a^{\gamma\delta} -
\ft12
\Gamma^{a_1a_2}{}_{\alpha\beta}\Gamma_{a_1a_2}{}^{\gamma\delta} +
\ft{1}{ 5!} \Gamma^{a_1\ldots a_5}{}_{\alpha\beta}
\Gamma_{a_1\ldots a_5}{}^{\gamma\delta} \right) \nonumber \\* &&
\end{eqnarray}
}allows us to write the superalgebra (\ref{Sigma(s)}) for $s=0$ as
{\setlength\arraycolsep{0pt}
\begin{eqnarray}
\label{osp-exp} && \{Q_\alpha,Q_\beta\}= P_{\alpha\beta} \; ,
\nonumber \\
&& [P_{\alpha\beta} , Q_{\gamma}] = 64 \; \gamma_1 \; C_{\gamma
(\alpha } Q^\prime_{\beta )} \; , \quad \nonumber \\
&& [Q^\prime_\alpha  , \textrm{all} \} =0 \; .
\end{eqnarray}
}Similarly, it is possible to collect the MC one-forms $\Pi^a$,
$\Pi^{a_1a_2}$, $\Pi^{a_1\cdots a_5}$ in a symmetric spin-tensor
one-form (\ref{cEff=def}), $\Pi^{\alpha\beta} = \ft{1}{32} (\Pi^a
\Gamma_{a} - \ft{i}{2} \Pi^{a_1a_2}\Gamma_{a_1a_2}
\\  + \ft{1}{5!} \Pi^{a_1\ldots a_5} \Gamma_{a_1\ldots
a_5})^{\alpha\beta}$ that allows us to write, for $s=0$, the MC
equations (\ref{Sigma(s)dual}) of $\tilde{\mathfrak{E}}(0)$ in
compact form as {\setlength\arraycolsep{0pt}
\begin{eqnarray}\label{compacts0}
&& d\Pi^{\alpha\beta}=- i \pi^\alpha  \wedge \pi^\beta \; ,
\nonumber
\\
&& d\pi^\alpha=0  \; , \nonumber \\
&& d\pi^{\prime \alpha}=-64i \gamma_1 \, \pi^\beta \wedge
\Pi_\beta{}^\alpha \; .
\end{eqnarray}
}The explicit appearance in equation (\ref{osp-exp}) of the
$Sp(32)$-invariant eleven-dimensional $32 \times 32$ charge
conjugation matrix $C_{\alpha \beta}$ or, alternatively, its
concealed appearance in the contraction of spinor indices in
(\ref{compacts0}), exhibits $Sp(32)$ as the automorphism symmetry
of $\tilde{\mathfrak E}(0)$. In contrast, the rest of
superalgebras $\tilde {\mathfrak E}(s)$, $s \neq 0$, have a
reduced automorphism symmetry $SO(1,10)$, since they involve
explicitly the $SO(1,10)$ Dirac matrices.

Hence, the generalizations of the superPoincar\'e group $\Sigma
\rtimes SO(1,10)$ for the $s\neq 0$ and $s=0$ cases are,
respectively, the semidirect products ${\tilde{\Sigma}(s) \rtimes
SO(1,10)}$ and ${\tilde{\Sigma}(0) \rtimes Sp(32)}$. Precisely for
$s=0$, both ${\tilde{\Sigma}(0) \rtimes SO(1,10)}$ and
${\tilde{\Sigma}(0) \rtimes Sp(32)}$ can be obtained from
$OSp(1|32)$ by the expansion method of chapter \ref{chapter5};
they are given, respectively, by the expansions $Osp(1|32)(2,3,2)$
and $Osp(1|32)(2,3)$ \cite{Lett} as it will now be shown.

The derivation of ${\tilde{\mathfrak{E}}(0) \rtimes so(1,10)}$ as
an expansion of $osp(1|32)$ fits into the general discussion of
section \ref{sec:susy} of the expansion method for superalgebras.
In fact, it  follows the same steps that led to the M Theory
superalgebra in section \ref{seis}, the only difference being the
way the cutting orders of the series expansions of the MC
one-forms of $osp(1|32)$ are chosen. Consider the 528 $sp(32)$
bosonic $\rho^{\alpha \beta}$ and 32 fermionic $\nu^\alpha$ MC
forms of $osp(1|32)$, satisfying the MC equations
(\ref{ospmaurer}). Again, it is useful to decompose
$\rho^{\alpha\beta}$ in terms of Dirac matrices as in
(\ref{generalrho}), $\rho^{\alpha\beta}=\frac1{32} \left(  \rho^a
\Gamma_a -\frac{i}{2} \rho^{ab} \Gamma_{ab}+ \frac{1}{5!}
\rho^{a_1\dots a_5} \Gamma_{a_1\dots a_5}\right)^{\alpha\beta}$;
this decomposition is adapted to the splitting
$osp(1|32)=V_0\oplus V_1\oplus V_2$, where $V_0$ is generated by
$\rho^{ab}$, $V_1$ by $\nu^\alpha$ and $V_2$ by $\rho^a$ and
$\rho^{a_1\dots a_5}$. In terms of $\rho^{ab}$, $\nu^\alpha$,
$\rho^a$ and $\rho^{a_1\dots a_5}$, the superalgebra $osp(1|32)$
takes the form (\ref{ospmaurerd}).

After the redefinition (\ref{eq:superredef1}) of the group
parameters of $osp(1|32)$, the MC forms expand as in
(\ref{eq:Vcero})--(\ref{eq:Vdos}), namely,
{\setlength\arraycolsep{0pt}
\begin{eqnarray} \label{fullMexp}
&& \rho^{ab}=\rho^{ab,0}+\lambda^2 \rho^{ab,2} +\lambda^4
\rho^{ab,4} +\cdots \;  , \nonumber
\\
&& \rho^a = \lambda^2 \rho^{a,2}+ \lambda^4 \rho^{a,4}+ \cdots \; ,\nonumber\\
&& \rho^{a_1\dots a_5}=\lambda^2 \rho^{a_1\dots a_5,2}+ \lambda^4
\rho^{a_1\dots a_5,4}+ \cdots \; ,
\nonumber \\
&&\nu^\alpha=\lambda\nu^{\alpha,1} +
\lambda^3\nu^{\alpha,3}+\cdots \; .
\end{eqnarray}
}Choosing the cutting orders $N_0=2$, $N_1=3$, $N_2=2$, as allowed
by the restriction (\ref{eq:tress2}), the MC equations of the
expanded algebra  $osp(1|32)(2,3,2)$ are obtained:
{\setlength\arraycolsep{0pt}
\begin{eqnarray}   \label{ospmaurerdDAF}
 &&      d\rho^{ab,0}=-\ft{1}{16}
 \rho^{ac,0}\wedge {\rho_c}^{b\,,0} \; , \nonumber \\
 &&      d\rho^{a\,,2}=-\ft{1}{16} \rho^{b,2}\wedge
{\rho_b}^{a,0} -i \nu^{\alpha,1}
\wedge \nu^{\beta,1} \Gamma^{a}_{\alpha\beta} \; , \nonumber\\
&&      d\rho^{ab,2}=-\ft{1}{16} \left( \rho^{ac,0}\wedge
{\rho_c}^{b,2} + \rho^{ac,2}\wedge {\rho_c}^{b,0} \right) -
\nu^{\alpha,1}
\wedge \nu^{\beta,1} \Gamma^{ab}_{\alpha\beta} \; , \nonumber\\
&&      d\rho^{a_1\dots a_5}{}^{,2}=\ft{5}{16} \rho^{b[a_1\dots
a_4|\,,2} \wedge \rho_b{}^{|a_5],0}\nonumber - i \nu^{\alpha,1}
\wedge \nu^{\beta,1} \Gamma^{a_1\dots
a_5}_{\alpha\beta} \; , \nonumber\\
&&      d\nu^{\alpha,1}=-\ft{1}{64} {\nu}^{\beta,1} \wedge
\rho^{ab,0}{\Gamma_{ab}}_\beta{}^\alpha \quad, \nonumber\\
&&     d\nu^{\alpha,3}=-\ft{1}{64} {\nu}^{\beta,3} \wedge
\rho^{ab,0}{\Gamma_{ab}}_\beta{}^\alpha \nonumber \\ && \qquad -
\ft{1}{32} \nu^{\beta,1} \wedge {\left(i\rho^{a,2}\Gamma_a +
\ft{1}{2}\rho^{ab,2}\Gamma_{ab} + \ft{i}{5!}\rho^{a_1\dots a_5,2}
\Gamma_{a_1\dots a_5}\right)_\beta}^\alpha  \, .\qquad\quad
\end{eqnarray}
}Now, setting $\rho^{ab,0} \equiv -16 \omega^{ab}$ and identifying
$\rho^{a,2} \equiv \Pi^a$, $\rho^{ab,2} \equiv \Pi^{ab}$,
$\rho^{a_1 \cdots a_5,2} \equiv \Pi^{a_1\cdots a_5}$,
$\nu^{\alpha,1} \equiv \pi^\alpha$ and $\nu^{\alpha,3} \equiv
\pi^{\prime \alpha}/64\gamma_1$ (notice that $\gamma_1 \neq 0$
just defines the scale of $Q^\prime_\alpha$), the set of equations
(\ref{ospmaurerdDAF}) coincides with the MC equations of
$\tilde{\mathfrak{E}}(0) \rtimes so(1,10)$ (obtained by restoring
the Lorentz part in equations (\ref{compacts0})). As a check,
notice that the dimensional counting is correct since, by equation
(\ref{eq:dimsuper}), {\setlength\arraycolsep{2pt}
\begin{eqnarray}
\textrm{dim}\, osp(1|32)(2,3,2) &=& 2\cdot55+2\cdot32+473=583+64 =
\nonumber \\ &=& \textrm{dim}\, \left( \tilde{\mathfrak{E}}(0)
\rtimes so(1,10) \right) \; .
\end{eqnarray}
}In conclusion \cite{Lett},
\begin{equation}
\tilde{\Sigma}(0) \rtimes SO(1,10)\approx OSp(1|32)(2,3,2) \; .
\end{equation}

The algebra $\tilde{\mathfrak{E}}(0) \rtimes sp(32)$ with its
enhanced automorphism symmetry $Sp(32)$ can also be obtained as an
expansion of $osp(1|32)$. Indeed, consider instead the splitting
$osp(1|32)=V_0\oplus V_1$ where $V_0$ is generated by all the
bosonic generators $\rho^{\alpha\beta}$ and $V_1$ by the fermionic
ones, $\nu^\alpha$. This splitting makes $V_1$ a symmetric coset
and, indeed, makes the algebra have the structure discussed in
section \ref{scoset} of chapter \ref{chapter5}. Cutting the
corresponding series at orders $N_0=2$ and $N_1=3$, in agreement
with  condition (\ref{eq:Nonez2}), the MC equations corresponding
to the expansion $osp(1|32)(2,3)$ are obtained:
{\setlength\arraycolsep{0pt}
\begin{eqnarray}\label{MC-sp}
&& d\rho^{\alpha\beta,0}=-i\rho^{\alpha\gamma,0}\wedge
\rho_\gamma{}^{\beta ,0} \; , \nonumber \\
&& d\rho^{\alpha\beta,2}=-i \left( \rho^{\alpha\gamma,0}\wedge
\rho_\gamma{}^{\beta ,  2} + \rho^{\alpha\gamma,2}\wedge
\rho_\gamma{}^{\beta ,  0} \right) -i\nu^{\alpha,1}\wedge
\nu^{\beta,1} \;,
\nonumber\\
&& d\nu^{\alpha,1}=- i\nu^{\beta,1} \wedge \rho_\beta{}^{\alpha ,
0}
 \; , \nonumber \\
&& d\nu^{\alpha,3}=-i \nu^{\beta,3} \wedge \rho_\beta{}^{\alpha ,
0} - i\nu^{\beta,1} \wedge \rho_\beta{}^{\alpha,2}\;.
\end{eqnarray}
}Identifying $\rho^{\alpha\beta,0}$ in (\ref{MC-sp}) with an
$sp(32)$ connection, equations (\ref{MC-sp}) are those of
${\tilde{\mathfrak{E}}(0) \rtimes sp(32)}$ (given by
(\ref{compacts0}) when $sp(32)$-automorphisms are included) with
$\rho^{\alpha\beta,2} \equiv \Pi^{\alpha\beta}$, $\nu^{\alpha,1}
\equiv \pi^\alpha$ and $\nu^{\alpha,3} \equiv \pi^{\prime
\alpha}/64\gamma_1$. Again, by equations (\ref{eq:dimsuper}), the
dimensions agree,
\begin{equation}
\textrm{dim}\, osp(1|32)(2,3) = 2 \cdot 528+64 = \textrm{dim}
\left( \tilde{\mathfrak{E}}(0) \rtimes sp(32) \right) \; ,
\end{equation}
and \cite{Lett}
\begin{equation}
\tilde{\Sigma}(0) \rtimes Sp(32)\approx OSp(1|32)(2,3) \; .
\end{equation}

\section{The composite nature of $A_3$} \label{nature}

We will now show how the set of one-forms (\ref{formsdualsoft}) of
the gauge FDA (\ref{trivializingFDA}) associated to
$\tilde{\mathfrak{E}}(s)$ allows for a composite structure of
$A_3$,
\begin{eqnarray}
\label{A3=A3def2} A_3 = A_3(e^a ,\, \psi^\alpha \; ; \; B^{ab}, \,
B^{abcde},\, \eta^\alpha )\;.
\end{eqnarray}
According to the discussion of section \ref{D=11FDA}, based on the
general arguments of section \ref{FDAandLie}, the problem is
equivalent to the trivialization of the superPoincar\'e algebra CE
four-cocycle $\omega_4$ of equation (\ref{a4=0}) in an extended
superalgebra. Thus, we are looking for a three-form
$\tilde{\omega}_3$ built up as an exterior polynomial of the
one-forms (\ref{formsdual}) of the family of extensions
$\tilde{\mathfrak{E}}(s)$ that fulfils equation (\ref{a4=0}),
namely,
\begin{equation} \label{a4=0bis}
d \tilde{\omega}_3 = \omega_4 \equiv - \ft14 \pi^\alpha \wedge
\pi^\beta \wedge \Pi^a \wedge \Pi^b \Gamma_{ab}{}_{\alpha\beta} \;
.
\end{equation}
In the process, it will be made apparent which of the
superalgebras in the family $\tilde{\mathfrak{E}}(s)$ allow for a
trivialization of $\omega_4$.

The most general expression for $\tilde{\omega}_3$ as an exterior
polynomial of the one-forms (\ref{formsdual}) of
$\tilde{\mathfrak{E}}(s)$ is {\setlength\arraycolsep{0pt}
\begin{eqnarray}
\label{A3=Ansflat} && 4 \ \tilde{\omega}_3 = \lambda \Pi^{ab}
\wedge \Pi_a \wedge \Pi_b \; - \alpha_1 \Pi_{ab} \wedge \Pi^b{}_c
\wedge \Pi^{ca} \nonumber  \\* && \quad - \alpha_2
\Pi_{b_1a_1\ldots a_4} \wedge \Pi^{b_1}{}_{b_2} \wedge
\Pi^{b_2a_1\ldots a_4}  \nonumber \\* && \quad - \alpha_3
\epsilon_{a_1\ldots a_5b_1\ldots b_5c} \Pi^{a_1\ldots a_5} \wedge
\Pi^{b_1\ldots b_5} \wedge \Pi^c \nonumber \\* && \quad - \alpha_4
\epsilon_{a_1\ldots a_6b_1\ldots b_5} \Pi^{a_1a_2 a_3}{}_{c_1c_2}
\wedge \Pi^{a_4a_5a_6c_1c_2} \wedge \Pi^{b_1\ldots b_5} \nonumber
\\* && \quad - 2i \pi^\beta \wedge \pi^{\prime \alpha} \wedge
\left( \beta_1 \, \Pi^a \Gamma_{a\alpha\beta} -i  \beta_2 \,
\Pi^{ab} \Gamma_{ab\; \alpha\beta} + \beta_3 \, \Pi^{a_1\ldots
a_5} \Gamma_{a_1\ldots a_5\; \alpha\beta} \right) \; , \nonumber
\\* &&
\end{eqnarray}
}where  $\alpha_1, \ldots, \alpha_4$, $\beta_1, \ldots, \beta_3$
\cite{D'A+F} {\it and} $\lambda$ \cite{Lett,AnnP04} are constants
to be determined by the requirement  that $\tilde{\omega}_3$ obeys
equation (\ref{a4=0bis}). The numerical factors  in the right hand
side of (\ref{A3=Ansflat}) have been introduced to make the
definition of the coefficients coincide with that in \cite{D'A+F}
while keeping our notation for the FDA. The only essential
difference with \cite{D'A+F} is the inclusion of the arbitrary
coefficient $\lambda$ in the first term; as we show below this
leads to a one-parametric family of solutions that includes  the
two D'Auria-Fr\'e ones.

Using the MC equations (\ref{Sigma(s)dual}) for
$\tilde{\mathfrak{E}}(s)$, the application in (\ref{A3=Ansflat})
of the differential $d$ leads to \cite{D'A+F}
{\setlength\arraycolsep{0pt}
\begin{eqnarray} \label{products}
&& 4 \ d \tilde{\omega}_3 = -(\lambda - 2 \delta\beta_1) \
\pi^\alpha \wedge \pi^\beta
\wedge \Pi^a \wedge \Pi^b \Gamma_{ab}{}_{\alpha\beta} \nonumber \\
&& +2(\beta_1 + 10\beta_2 - 6! \beta_3) \ \pi^\alpha \wedge
\pi^\beta \wedge \pi^\gamma \wedge  \pi^{\prime \delta}
\Gamma^a{}_{\alpha\beta} \Gamma_a{}_{\gamma\delta} \nonumber \\
&& +2i(\lambda- 2\gamma_1 \beta_1 - 2\delta\beta_2) \ \pi^\alpha
\wedge \pi^\beta \wedge \Pi^{ab} \wedge \Pi_b
\Gamma_a{}_{\alpha\beta} \nonumber \\
&& +(3\alpha_1 + 8 \gamma_1\beta_2) \ \pi^\alpha \wedge \pi^\beta
\wedge \Pi^a{}_c \wedge \Pi^{cb} \Gamma_{ab}{}_{\alpha\beta}
\nonumber  \\
&& +2i(\alpha_2 - 10\gamma_1 \beta_3 - 10\gamma_2 \beta_2) \
\pi^\alpha \wedge \pi^\beta \wedge \Pi^{a_1}{}_c \wedge \Pi^{ca_2
\cdots a_5}
\Gamma_{a_1 \cdots a_5}{}_{\alpha\beta} \nonumber \\
&& +\ft{2i}{5!} (5! \alpha_3 - \delta \beta_3 - \gamma_2 \beta_1)
\ \epsilon_{a_1 \cdots a_5 b_1 \cdots b_5 c} \pi^\alpha \wedge
\pi^\beta \wedge \Pi^{b_1 \cdots b_5} \wedge \Pi^c
\Gamma_{a_1 \cdots a_5}{}_{\alpha\beta} \nonumber \\
&& -(\alpha_2 - 5!\,  10\gamma_2 \beta_3) \ \pi^\alpha \wedge
\pi^\beta \wedge \Pi^{a_1}{}_{b_1 \cdots b_4} \wedge \Pi^{a_2 b_1
\cdots b_4}
\Gamma_{a_1 a_2}{}_{\alpha\beta} \nonumber \\
&& +i(\alpha_3 - 2\gamma_2 \beta_3) \ \epsilon_{a_1 \cdots a_5 b_1
\cdots b_5 c} \pi^\alpha \wedge \pi^\beta \wedge \Pi^{a_1 \cdots
a_5} \wedge  \Pi^{b_1 \cdots b_5}
\Gamma^c{}_{\alpha\beta} \nonumber \\
&&  +\ft{i}{3} (9 \alpha_4 + 10\gamma_2 \beta_3) \ \epsilon_{a_1
\cdots a_6 b_1 \cdots b_5} \pi^\alpha \wedge \pi^\beta \wedge
\Pi^{a_1 a_2 a_3}{}_{c_1c_2}  \wedge \Pi^{a_4a_5a_6 c_1c_2}
\Gamma^{b_1 \cdots b_5}{}_{\alpha\beta}  .   \nonumber \\* &&
\end{eqnarray}
}Finally, comparing the expressions for $d \tilde{\omega}_3$ in
(\ref{products}) and (\ref{a4=0bis}), and equating the
coefficients of the different, independent four-form terms, the
following non-homogeneous linear system of nine equations and
eight unknowns $\lambda$, $\alpha_1, \ldots, \alpha_4$, $\beta_1,
\ldots, \beta_3$, dependent on the parameters $\delta$,
$\gamma_1$, $\gamma_2$ is found\footnote{The factor $5!$ in the
equation $5! \alpha_3 - \delta \beta_3 - \gamma_2 \beta_1 =0$, and
the factor 9 in the equation $9 \alpha_4 + 10\gamma_2 \beta_3 =0$
were both missing in footnote 6 in \cite{Lett}, and the later
factor was missing in equation (4.39) of  \cite{AnnP04}.}
\cite{D'A+F,Lett,AnnP04} {\setlength\arraycolsep{0pt}
\begin{eqnarray}\label{systemofeqs}
&& \lambda - 2 \delta\beta_1=1 \; , \nonumber \\* && \beta_1 +
10\beta_2 - 6! \beta_3=0\; , \nonumber \\ && \lambda- 2\gamma_1
\beta_1 - 2\delta\beta_2 =0 \; , \nonumber \\* &&  3\alpha_1 + 8
\gamma_1\beta_2 =0 \; , \nonumber \\* && \alpha_2 - 10\gamma_1
\beta_3 - 10\gamma_2 \beta_2=0 \; , \nonumber \\* && 5! \alpha_3 -
\delta \beta_3 - \gamma_2 \beta_1 =0 \; , \nonumber \\* &&
\alpha_2 - 5!\,  10\gamma_2 \beta_3 =0 \; , \nonumber \\* &&
\alpha_3 - 2\gamma_2 \beta_3 =0 \; , \nonumber \\* && 9 \alpha_4 +
10\gamma_2 \beta_3 =0 \; .
\end{eqnarray}
}The existence of solutions to this system depends on the values
of the parameters $\delta$, $\gamma_1$, $\gamma_2$ that define it.
Actually, the system (\ref{systemofeqs}) depends effectively on
one parameter $s$ (defined in (\ref{s-def})) only, for the same
reason as the family of superalgebras $\tilde{\mathfrak{E}}(s)$
does: one parameter among $\delta$, $\gamma_1$, $\gamma_2$ can be
eliminated by means of the relation (\ref{idg}), and another one
by a redefinition of the extra fermionic generator
$Q^\prime_\alpha$ in (\ref{Sigma(s)}) (or $\pi^{\prime \alpha}$ in
(\ref{Sigma(s)dual})). The system (\ref{systemofeqs}) turns out to
be incompatible for $s=0$ (namely, for $\delta= 2\gamma_1$,
$\gamma_2 =2\gamma_1 /5!$) but has, otherwise, a unique solution
for each $s \neq 0$ given by \cite{Lett,AnnP04}
{\setlength\arraycolsep{0pt}
\begin{eqnarray}\label{sEq=g}
& \lambda =  {1 \over 5} \; \frac{s^2+2s+6}{s^2} \; , \;
\beta_1=-\frac{1}{10\gamma_1} \; \frac{2s-3}{s^2} \; , \;
\beta_2=\frac{1}{20\gamma_1} \; \frac{s+3}{s^2} \; , \;
\beta_3=\frac{3}{10 \cdot 6! \gamma_1} \; \frac{s+6}{s^2}  \; ,
\nonumber \\
& \alpha_1=-\frac{1}{15} \; \frac{2s+6}{s^2} \; ,  \;
\alpha_2=\frac{1}{6!} \; \frac{(s+6)^2}{s^2} \; , \;
\alpha_3=\frac{1}{5 \cdot 6! 5!} \; \frac{(s+6)^2}{s^2} \; , \;
\alpha_4=- \frac{1}{9 \cdot 6! 5!} \; \frac{(s+6)^2}{s^2} \; .
\nonumber \\ &
\end{eqnarray}

}

The two particular solutions in \cite{D'A+F} are recovered by
adjusting $s$ ({\it i.e.}, $\delta, \gamma_1$ in
equation~(\ref{s-def})) so that $\lambda=1$ in equation
(\ref{sEq=g}). This is achieved for $\delta =5\gamma_1$ ($\delta$
non vanishing but otherwise arbitrary), or for $\delta=0$ (with
$\gamma_1$ non vanishing but otherwise arbitrary). Thus, the two
D'Auria and Fr\'e decompositions of $A_3$ are characterized by
$s=3/2$, {\setlength\arraycolsep{0pt}
\begin{eqnarray}\label{D'A+F1}
&& \tilde {\mathfrak E}(3/2) \;  :  \; \delta=5\gamma_1  \not=0 \;
,
 \;  \gamma_2=\ft{\gamma_1}{2\cdot 4!} \; ,  \nonumber \\
 && \qquad \lambda=1 \; , \quad  \beta_1=0 \;
, \quad \beta_2=\ft{1}{10\gamma_1 } \; , \quad \beta_3=\ft{1}{6!
\, \gamma_1} \; ,  \nonumber \\
&& \qquad  \alpha_1=- \ft{4}{15} \; , \quad \alpha_2=\ft{25}{6!}
\; , \quad \alpha_3=\ft{1}{6!  4! } \; , \quad
\alpha_4=-\ft{1}{54\, (4!)^2} \; ,
\end{eqnarray}
}and by $s=-1$,
 {\setlength\arraycolsep{0pt}
\begin{eqnarray}\label{D'A+F2}
&& \tilde {\mathfrak E}(-1) \; : \;   \delta=0 \; , \; \gamma_1
\not=0 \; ,
 \;  \gamma_2=\ft{\gamma_1}{3\cdot 4!} \; , \nonumber
\\ && \qquad \lambda=1 \; , \quad \beta_1=\ft{1}{2\gamma_1} \; ,
\quad \beta_2=\ft{1}{10\gamma_1 } \; , \quad \beta_3=\ft{1}{4
\cdot 5! \, \gamma_1} \; , \nonumber \\
&& \qquad  \alpha_1=-\ft{4}{15} \; , \quad \alpha_2=\ft{25}{6!} \;
, \quad \alpha_3=\ft{1}{6!\, 4!} \; , \quad  \alpha_4=-\ft{1}{54\,
(4!)^2}\; .
\end{eqnarray}
}Here it has been shown that not only these two superalgebras
solve the problem, but that  {\it all the superalgebras in the
family $\tilde {\mathfrak E}(s)$, except $\tilde {\mathfrak
E}(0)$, allow for a trivialization of the four-cocycle $\omega_4$
in (\ref{a4=0bis})} \cite{Lett,AnnP04}. The three-form
$\tilde{\omega}_3$ that trivializes it in $\tilde {\mathfrak
E}(s)$, $s \neq 0$, is given by the expression (\ref{A3=Ansflat})
with the coefficients (\ref{sEq=g}). The composite expression
(\ref{A3=A3def2}) of the three-form $A_3$ in terms of the soft
one-form counterparts (\ref{formsdualsoft}) of the MC one-forms of
$\tilde {\mathfrak E}(s)$, $s \neq 0$, is thus given explicitly by
{\setlength\arraycolsep{0pt}
\begin{eqnarray} \label{A3=Ans} && 4A_3 =
\lambda B^{ab} \wedge e_a \wedge e_b \; - \alpha_1 B_{ab} \wedge
B^b{}_c \wedge B^{ca} \nonumber \\
&& \quad  - \alpha_2 B_{b_1a_1\ldots a_4} \wedge B^{b_1}{}_{b_2}
\wedge B^{b_2a_1\ldots
a_4}   \nonumber \\
&& \quad  - \alpha_3  \epsilon_{a_1\ldots a_5b_1\ldots b_5c}
B^{a_1\ldots
a_5} \wedge B^{b_1\ldots b_5} \wedge e^c \nonumber \\
&& \quad  -  \alpha_4 \epsilon_{a_1\ldots a_6b_1\ldots b_5}
B^{a_1a_2 a_3}{}_{c_1c_2} \wedge B^{a_4a_5a_6c_1c_2} \wedge
B^{b_1\ldots b_5}
\nonumber \\
&& \quad  - 2i \psi^\beta \wedge \eta^\alpha \wedge \left( \beta_1
\, e^a \Gamma_{a\alpha\beta} -i  \beta_2 \, B^{ab} \Gamma_{ab\;
\alpha\beta} + \beta_3 \, B^{a_1\ldots a_5} \Gamma_{a_1\ldots
a_5\; \alpha\beta} \right) \; , \nonumber \\
\end{eqnarray}
}where the coefficients are given by (\ref{sEq=g}).

It is worth stressing that allowing the coefficient $\lambda$ not
to be fixed from the onset, produces more possibilities for the
trivializing algebras than those in \cite{D'A+F} (equations
(\ref{D'A+F1}) and (\ref{D'A+F2})). A particularly interesting
superalgebra within the family $\tilde {\mathfrak E}(s)$ is
achieved for $s=-6$. The trivialization of the four-cocycle
$\omega_4$ (\ref{a4=0bis}) associated to $dA_3$ on $\tilde
{\mathfrak E}(-6)$ is obtained for the coefficients (\ref{sEq=g})
with $s=-6$, namely, {\setlength\arraycolsep{0pt}
\begin{eqnarray} \label{coeffminimal}
&& \tilde {\mathfrak E}(-6) \; : \;   \delta=-10 \gamma_1 \not=0
\; ,
 \;  \gamma_2= 0  \; , \nonumber
\\
&& \qquad  \lambda=\ft{1}{6} \; , \quad   \beta_1=\ft{1}{4!
\gamma_1 } \; , \quad \beta_2=-\ft{1}{ 2 \cdot 5! \gamma_1 } \; ,
\quad
\beta_3=0 \; , \nonumber \\
&& \qquad   \alpha_1=\ft{1}{90} \; , \quad   \alpha_2=0 \; , \quad
\alpha_3=0 \; , \quad  \alpha_4=0 \; .
\end{eqnarray}
}Interestingly enough, the vanishing coefficients $\alpha_2$,
$\alpha_3$, $\alpha_4$, $\beta_3$ are those in front of each term
involving the one-form $\Pi^{a_1 \ldots a_5}$ in the expression of
the trivializing three-form $\tilde{\omega}_3$ (\ref{A3=Ansflat}).
In consequence, the expression (\ref{A3=Ans}) for $A_3$ as a
composite of the gauge one-form fields of $\tilde {\mathfrak
E}(-6)$ becomes especially simple, {\setlength\arraycolsep{2pt}
\begin{eqnarray}\label{A3minimal}
A_3 & = & \ft{1}{4!} B^{ab} \wedge e_a \wedge e_b \; - \ft{1}{3
\cdot 5!} B_{ab} \wedge B^b{}_c \wedge B^{ca} \nonumber \\
&& - \ft{i}{4 \cdot 5! \, \gamma_1}  \psi^\beta \wedge \eta^\alpha
\wedge \left( 10 \, e^a \Gamma_{a\alpha\beta} + i \, B^{ab}
\Gamma_{ab\;\alpha\beta} \right) \; ,
\end{eqnarray}
}since it does not involve the gauge one-form field $B^{a_1 \ldots
a_5}$. This is, nevertheless, not surprising since, by the
definition (\ref{s-def}), the choice $s=-6$ fixes the parameter
$\gamma_2$ in the algebra (\ref{susyalg}) to $\gamma_2=0$,
rendering central the generator $Z_{a_1 \ldots a_5}$ associated to
the gauge field $B^{a_1 \ldots a_5}$.

For $s=-6$ the central generator $Z_{a_1 \ldots a_5}$ plays, in
short, no role in the trivialization of $\omega_4$. One can,
therefore, get rid of it and consider the smaller,
$(66+64)$-dimensional superalgebra $\tilde {\mathfrak E}_{min}$
\cite{Lett,AnnP04}, the extension of which by the central charge
$Z_{a_1\ldots a_5}$ gives the superalgebra $\tilde {\mathfrak
E}(-6)$. The `minimal' superalgebra $\tilde {\mathfrak E}_{min}$
is given explicitly by setting $\gamma_2=0$ (and, hence, $\delta
=-10 \gamma_1$) in (\ref{susyalg}) or, equivalently, by setting
$s=-6$ in (\ref{Sigma(s)}) and by removing the generator  $Z_{a_1
\ldots a_5}$:
{\setlength\arraycolsep{0pt}
\begin{eqnarray}\label{susymin} &&
\{Q_\alpha,Q_\beta\} = \Gamma^a_{\alpha\beta} P_a +
i\Gamma^{a_1a_2}_{\alpha\beta} Z_{a_1a_2} \; , \nonumber  \\
&& [ P_a , Q_\alpha ] =  -10 \gamma_1 \; \Gamma_{a \;
\alpha}{}^\beta Q^\prime_\beta \; , \nonumber  \\
&& [ Z_{a_1a_2} , Q_\alpha]=i \gamma_1 \Gamma_{a_1a_2 \;
\alpha}{}^\beta  Q^\prime_\beta\;,
\nonumber \\
&& [Q^\prime_\alpha , \textrm{all} \} =0 \; .
\end{eqnarray}
}It is worth noting that $\tilde {\mathfrak E}_{min}$ does not
belong to the family $\tilde {\mathfrak E}(s)$ and yet it provides
a composite structure of $A_3$ in terms of the soft field
counterparts of its MC one-forms. The expression of $A_3$ in terms
of the gauge field one-forms associated to $\tilde {\mathfrak
E}_{min}$ is precisely (\ref{A3minimal}), and it coincides with
the composite expression achieved for $\tilde {\mathfrak E}(-6)$.
In summary, the most economic extension of the standard
supertranslations algebra that allows for a composite structure of
$A_3$ is $\tilde {\mathfrak E}_{min}$, corresponding to the most
economical extended supergroup manifold
$\tilde{\Sigma}_{min}=\Sigma^{(66|32+32)}$ on which $\omega_4$
corresponding to $dA_3$ becomes trivial \cite{Lett,AnnP04}.

\setcounter{equation}0
\section{Dynamics with a composite $A_3$} \label{dyncomp}

In section  \ref{eom} of chapter \ref{chapter2}, the equations of
motion of ordinary $D=11$ CJS supergravity were derived (in a
first order formalism) from its action $S$, given by equations
(\ref{S11=}), (\ref{L11=}). It is now our aim to check for
possible dynamical consequences of a composite structure of $A_3$
(equation (\ref{A3=Ans})) in terms of the gauge one-forms
(\ref{formsdualsoft}) associated to any of the superalgebras
$\tilde {\mathfrak E}(s)$, $s \neq 0$. As noticed in \cite{D'A+F},
to perform the analysis in general, the expression (\ref{A3=Ans})
with the coefficients (\ref{sEq=g}) for a composed $A_3$ in terms
of gauge one-forms of $\tilde {\mathfrak E}(s)$, $s \neq 0$, would
have to be introduced in the first order action (\ref{S11=}),
(\ref{L11=}) of supergravity. We shall only consider here the case
in which $A_3$ is given by the expression (\ref{A3minimal}) in
terms of the soft one-forms of the minimal algebra $\tilde
{\mathfrak E}_{min}$ (equation (\ref{susymin})). The conclusions,
however, are general and can be translated to the general case in
which $A_3$ is given by equation (\ref{A3=Ans}).

Consider, thus, the expression (\ref{A3minimal}) for $A_3$ in
terms of the gauge one-forms of the superalgebra $\tilde
{\mathfrak E}_{min}$. Ignoring the variation of the vielbein $e^a$
and gravitino $\psi^\alpha$ fields, that would contribute,
respectively, to the Einstein (\ref{EqmE}) and the
Rarita-Schwinger (\ref{Eqmpsi}) equations with on-shell-vanishing
terms, the variation of $A_3$ as given by (\ref{A3minimal}) when
the fields $B^{a_1 a_2}$ and $\eta^\alpha$ are varied reads
{\setlength\arraycolsep{2pt}
\begin{eqnarray}
\label{varA3} \delta A_3  & = & \left( \ft{1}{4!} e_a \wedge e_b
\, - \ft{1}{5!} B_{a}{}^c \wedge B_{cb}  + \ft{1}{5! \, 4 }
\psi^\beta \wedge \eta^\alpha \Gamma_{ab\;\beta\alpha}  \right)
\wedge \delta B^{ab} \nonumber \\ && +
 \ft{i}{5! \, 4 }  \psi^\beta \wedge ( 10 \, e^a
\Gamma_{a\alpha\beta} + i B^{ab} \Gamma_{ab\;\alpha\beta} ) \wedge
\delta \eta^\alpha   \, .
\end{eqnarray}
}Once the composite $A_3$ given in (\ref{A3minimal}) has been
introduced into the supergravity action $S$ (equations
(\ref{S11=}), (\ref{L11=})), the variation of $S$ with respect to
the field $B^{ab}$ can be worked out, taking into account that
$B^{ab}$ enters the action $S$ only through $A_3$:
{\setlength\arraycolsep{2pt}
\begin{eqnarray}\label{varB}
{\delta S \over \delta B_{ab}} &=&  {\delta S \over \delta A_3}
\wedge {\delta A_3 \over \delta B_{ab} } \nonumber \\
&=& \ft{1}{4!} {\cal G}_8 \wedge \left(e^a\wedge e^b - \ft{1}{5}
B^{ac}\wedge
 B_{c}{}^b + \ft{1}{20}
\psi \wedge \eta \, \Gamma_{ab} \right) \; .
\end{eqnarray}
}In this expression, the eight-form ${\cal G}_8$ is the variation
of the action $S$ with respect to $A_3$, $\delta S / \delta A_3 =
{\cal G}_8$, and its explicit expression is given in equation
(\ref{defcG=}), namely, ${\cal G}_8 = d( \ast F_4 +b_7 - A_3\wedge
dA_3)$.

In the component approach we are dealing with, the action $S$ is
defined on eleven-dimensional spacetime $M^{11}$. Consequently,
all the forms involved in the action, including $B^{ab}$ and
$\eta^\alpha$ take arguments on $M^{11}$ and can, therefore, be
expressed in terms of the vielbein basis $e^a$; in particular,
$B^{ab} = e^c B_c{}^{ab}$, $\eta^\alpha = e^c \eta_c{}^\alpha$.
Thus, introducing the matrix
\begin{eqnarray}\label{K=()}
 {\cal K}_{cd}{}^{ab} =
\delta_{[c}{}^a \delta_{d]}^{b} + {1\over 5}
B_{[c}{}^{ae}B_{d]}{}^b{}_e + {1\over 20} \psi_{[c}{}^\beta \,
\eta_{d]}{}^\alpha \, \Gamma_{\alpha\beta}^{ab} \; ,
\end{eqnarray}
the variation (\ref{varB}) of the action with respect to $B^{ab}$
can be written as
\begin{eqnarray}\label{varBK}
{\delta S \over \delta B_{ab} }=
 {1\over 4!} {\cal G}_8 \wedge e    ^c\wedge e^d \; {\cal K}_{cd}{}^{ab}\; .
\end{eqnarray}
Now, as it can be seen {\it e.g.} at the linearized level, in
which the fields $B^{ab}$ are weak, the matrix $ {\cal
K}_{cd}{}^{ab}$ can be supposed to be invertible and the
requirement that the action be invariant under variations of
$B^{ab}$ leads to
\begin{eqnarray}\label{varBsim}
\textrm{det} ({\cal K}_{ab}{}^{cd} )\not=0 \ : \ {\delta S \over
\delta B^{ab}}=  0 \quad   \Rightarrow \quad {\cal G}_8 \wedge
e^c\wedge e^d =0 \; .
\end{eqnarray}
The last equation then implies the standard equations of motion
for $A_3$, equation (\ref{EqmA3}), but now for a composite, rather
than fundamental  $A_3$. Thus one may state, at least within the
$\textrm{det}({\cal K}_{ab}{}^{cd} )\not=0$ assumption, that the
variation with respect to the $B^{ab}$ field produces the same
equations as the variation with respect to the CJS three-form
$A_3$,
\begin{eqnarray}\label{varB=varA}
 \textrm{det} ({\cal K}_{ab}{}^{cd} )\not=0  \ : \ {\delta S
\over \delta B^{ab}}=  0 \quad  \Rightarrow \quad
 {\cal G}_8 := {\delta S \over \delta A_3} =0 \; .
\end{eqnarray}

Notice, however, that the $B^{ab}$ field carries more degrees of
freedom than $A_3$ does. In fact, the three index tensor
$B_c{}^{ab}=-B_c{}^{ba}$ has reducible symmetry properties
(product of two Young tableaux),
\begin{eqnarray}\label{B=Yt}
B_{c\; ab} \; \sim \; {\tiny \yng(1) }  \otimes {\tiny \yng(1,1) }
=  {\tiny \yng(2,1) }  \oplus {\tiny \yng(1,1,1) }
\end{eqnarray}
whereas the components $A_{abc}$ of  $A_3=\ft{1}{3!} e^c\wedge
e^b\wedge e^a A_{abc}$ are completely antisymmetric,
$A_{abc}=A_{[abc]}$,
\begin{eqnarray}\label{A=Yt}
A_{abc}\; \sim  \; {\tiny \yng(1,1,1) } \; .
\end{eqnarray}
Then, since a variation of the action with respect to $B^{ab}$
produces (for $\textrm{det} ({\cal K}_{[ab]}{}^{[cd]} )\not=0$)
the same equations as the variation with respect to $A_3$, one
concludes that the action for a composite $A_3$ must possess local
symmetries that make the {\it extra} degrees of freedom in
$B^{ab}$ ({\it i.e}, ${\tiny \yng(2,1) }$ but {\it not} ${\tiny
\yng(1,1,1) }$ ) pure gauge. Similarly, one may expect to have an
extra local fermionic symmetry under which the new fermionic
fields $\eta_a^\alpha$ in $\eta^\alpha =e^a\eta_a{}^\alpha$ are
also pure gauge.

This is indeed the case \cite{AnnP04}. Actually, the fact that the
above $\delta B^{ab}= e^c \delta B_{c}{}^{ab} $ variation produces
the same result as the variation with respect to $\delta A_{abc}=
\delta A_{[abc]}$  plays the role of Noether identities for all
these `extra' gauge symmetries. Let us show, for instance, that
the supergravity action with $A_3$ with the simple composite
structure of equation (\ref{A3minimal}) does possess extra
fermionic gauge symmetries with a spinorial one-form parameter.
Indeed, the equations of motion for $\eta^{\alpha}$,
\begin{eqnarray}\label{vareta}
{\delta S \over \delta \eta^{\alpha} } = 0\; \quad \Rightarrow
\quad {\cal G}_8 \wedge \psi^\beta \wedge \left( 10 \, e^a
\Gamma_{a\alpha\beta} + i \, B^{ab} \Gamma_{ab\;\alpha\beta}
\right) =0 \; ,
\end{eqnarray}
are satisfied identically on the $B^{ab}$  equations of motion
(${\cal G}_8=0$ for $\textrm{det} ({\cal K}_{[ab]}{}^{[cd]}
)\not=0$, equations (\ref{varBsim})). This is a Noether identity
that indicates the presence of a local fermionic symmetry  with
spinorial one-form parameter $\chi^{\alpha}$, $\chi^{\alpha}= e^a
\chi_{a}{}^{\alpha}$, such that {\setlength\arraycolsep{0pt}
\begin{eqnarray}\label{dchidchiB}
&& \delta_{\chi}\eta^{\alpha} = \chi^{\alpha} \; , \quad  \nonumber \\
&& \delta_{\chi}B^{ab} = {i\over 16} {\cal
K}^{-1}{}^{[ab]}{}^{[cd]} \, \psi_c{}^\alpha (10 \Gamma_d +
iB_d{}^{ef}\Gamma_{ef})_{\alpha\beta} \, \chi^\beta \; .
\end{eqnarray}
}We can see that the transformations (\ref{dchidchiB}), leave
invariant the composite three-form $A_3$ (\ref{A3minimal})
considered as a form on spacetime. In the same way, having in mind
that the contribution of {\it any} variation of the fundamental
fields in $\delta A_3$ on ${M}^{11}$ is always given by an {\it
antisymmetric} third-rank tensor contribution, one concludes that
{\it any} contribution to $\delta A_3$ from an arbitrary variation
of the ${\tiny \yng(2,1) }$ irreducible part of $\delta
B_c{}^{ab}$ (which carries also an antisymmetric contribution) can
always  be compensated by a contribution of a proper
transformation of its completely antisymmetric part $\delta
B_{[cba]}$, ${\tiny \yng(1,1,1) }$ \ .

When the more general form for $A_3$, (equations (\ref{A3=Ans}),
(\ref{sEq=g})) is considered, the same reasoning shows that any
transformations of the new form $B^{a_1\ldots a_5}$ can be
compensated by some properly chosen $B^{ab}$ transformations. The
key point is that the coefficient $\lambda$ in (\ref{sEq=g}) never
vanishes. Hence (omitting $\delta e^a$ and $\delta \psi^\alpha$),
{\setlength\arraycolsep{2pt}
\begin{eqnarray}
\label{varA3g} \delta A_3 &=& - \ft{\lambda}{4} e^c \wedge e^d
\wedge {\cal K}_{cd}{}^{ab} \delta B_{ab}  \; + {\cal S}_{2
a_1\ldots a_5} \wedge \delta B^{a_1\ldots a_5} +  {\cal
S}_{2}^{\alpha} \wedge \delta \eta_{\alpha} \; \nonumber
\\ \label{varA=l}
 & = & -   \ft{\lambda}{4} e^a \wedge e^b \wedge e^c \,
\delta B_{[c\; ab]}  + {\cal O}( B \wedge B) +  {\cal O}( \psi
\wedge \eta) \; ,
\end{eqnarray}
\begin{eqnarray} \label{K=g} {\cal K}_{cd}{}^{ab} &=&
\delta_{[c}{}^{a}\delta_{d]}{}^{b} + {\cal O}( B \wedge B) + {\cal
O}( \psi \wedge \eta)  \; , \qquad
\\
\nonumber && \lambda = {(20\gamma^2_1 +\delta^2)\over 5
 (2 \gamma_1-  \delta)^2}  \equiv {1 \over 5} \; \frac{s^2+2s+6}{s^2} \; \not=0
\end{eqnarray}
}and  the variation of the completely antisymmetric part
$B_{[abc]}$ of $B^{ab}=e^cB_c{}^{ab}$ always reproduces (for an
invertible ${\cal K}$ (\ref{K=g})) the same equation ${\cal
G}_8=0$ as it would an independent, fundamental  three-form $A_3$
\cite{AnnP04}.

One might also wonder whether the equations of motion of the first
order action with a composite $A_3$ produce any relations for the
curvatures ${\cal B}_2^{ab}$,  ${\cal B}_2^{a_1 \ldots a_5}$ and
${\cal B}_2^\alpha$ of the new fields $B^{ab}$,  $B^{a_1 \ldots
a_5}$ and $\eta^\alpha$, in the same way that they fix the
curvatures $\mathbf{R}^a$ and $\mathbf{R}_4$ of $e^a$ and $A_3$ to
be $\mathbf{R}^a=0$ and $\mathbf{R}_4=F_4$, where $F_4$ is the
auxiliary four-form of the first order supergravity action. An
expression for the curvature $\mathbf{R}_4$ of $A_3$ in terms of
the curvatures ${\cal B}_2^{ab}$,  ${\cal B}_2^{a_1 \ldots a_5}$,
${\cal B}_2^\alpha$ and $\mathbf{R}^\alpha$ may be obtained by
substituting the composite expression (\ref{A3=Ans}) for $A_3$ in
the expression (\ref{R4=}) for $\mathbf{R}_4$ \cite{AnnP04},
{\setlength\arraycolsep{2pt}
 \begin{eqnarray}
\label{R4=Ans}   \mathbf{R}_4 &=& \ft{\lambda}{4} {\cal B}_2^{ab}
\wedge e_a \wedge e_b \; - \ft{3 \alpha_1}{4} {\cal B}_{2 ab}
\wedge B^b{}_c \wedge B^{ca} \nonumber \\
&& - \ft{\alpha_2}{2} {\cal B}_{2\, a_1\ldots a_5} \wedge
B^{a_1}{}_{b} \wedge B^{ba_2\ldots a_5}
  + \ft{\alpha_2}{4} B_{a_1\ldots a_5} \wedge {\cal
B}_2^{a_1}{}_{b} \wedge B^{ba_2\ldots a_5} \nonumber \\
&& -  \ft{\alpha_3}{2}  \epsilon_{a_1\ldots a_5b_1\ldots b_5c} e^c
\wedge B^{a_1\ldots a_5} \wedge  {\cal B}_2^{b_1\ldots b_5}
 \nonumber \\
&& -  \ft{\alpha_4}{4} \epsilon_{a_1\ldots a_6b_1\ldots b_5}
B^{a_1a_2 a_3}{}_{c_1c_2} \wedge B^{a_4a_5a_6c_1c_2} \wedge {\cal
B}_2^{b_1\ldots b_5}  \nonumber \\
&& - \ft{\alpha_4}{2} \epsilon_{a_1\ldots a_6b_1\ldots b_5}
B^{a_4a_5a_6c_1c_2} \wedge B^{b_1\ldots b_5} \wedge {\cal
B}_2^{a_1a_2 a_3}{}_{c_1c_2}
\nonumber \\
&& - \ft{i}{2}  \psi^\beta \wedge \eta_1^\alpha \wedge \left(-i
\beta_2 \, {\cal B}_2^{ab} \Gamma_{ab\; \alpha\beta} + \beta_3 \,
{\cal
B}_2^{abcde} \Gamma_{abcde\; \alpha\beta} \right) \nonumber \\
&& + \ft{i}{2} \psi^\beta \wedge \left( \beta_1 \, e^a
\Gamma_{a\alpha\beta} -i  \beta_2 \, B^{ab} \Gamma_{ab\;
\alpha\beta} + \beta_3 \, B^{abcde} \Gamma_{abcde\; \alpha\beta}
\right) \wedge {\cal B}_2^\alpha  \nonumber \\
&&   + \ft{i}{2} \eta^\alpha \wedge \left( \beta_1 \, e^a
\Gamma_{a\alpha\beta} -i \beta_2 \, B^{ab} \Gamma_{ab\;
\alpha\beta} + \beta_3 \, B^{abcde} \Gamma_{abcde\; \alpha\beta}
\right) \wedge \mathbf{R}^\beta \; , \nonumber \\* &&
\end{eqnarray}
where $\mathbf{R}^a=0$ has been assumed by consistency with the
equations of motion for $e^a$, and the coefficients are given by
(\ref{sEq=g}).

}

If the condition $\mathbf{R}_4=F_4$, where $F_4=\ft{1}{4!} e^{a_4}
\wedge \ldots \wedge e^{a_1} F_{a_1 \ldots a_4}$, is now imposed,
equation (\ref{R4=Ans}) sets the value of $F_{a_1 \ldots a_4}$ in
terms of the curvatures ${\cal B}_2^{ab}$, ${\cal B}_2^{a_1 \ldots
a_5}$, ${\cal B}_2^\alpha$ and $\mathbf{R}^\alpha$. This reflects
the existence of the extra gauge symmetries that makes the theory
with a composite $A_3$ carry the same number of degrees of freedom
than the standard theory with a fundamental $A_3$, as discussed
previously in this section at the level of the equations of
motion. Indeed, equation (\ref{R4=Ans}) with $\mathbf{R}_4=F_4$ is
the only relation imposed on the new field strengths ${\cal
B}_2^{ab}$, $ {\cal B}_2^{a_1\ldots a_5}$, ${\cal B}_2^\alpha$ by
the first-order $D=11$ supergravity action (\ref{S11=}),
(\ref{L11=}) with a composite $A_3$. This makes the detailed
properties of the curvatures ${\cal B}^{ab}_2$, $ {\cal
B}_2^{a_1\ldots a_5}$, ${\cal B}_2^\alpha$  of the additional
gauge fields inessential: their only relevant properties are that
the field strength $F_4$ is constructed out of them in agreement
with equation (\ref{R4=Ans}), and that such a composite field
strength obeys the equation of motion (\ref{EqmA3}), ${\cal
G}_8=0$.

In summary, on the one hand,  the underlying gauge group structure
implied by the new one-form fields allows us to treat $D=11$
supergravity as a gauge theory of the supergroup
 $\tilde{\Sigma}(s) \rtimes SO(1,10)$, $s \neq 0$,
 that replaces  superPoincar\'e. On the other hand, the supergravity
 action (\ref{S11=}), (\ref{L11=}) with a composite $A_3$ also possesses
`extra' gauge symmetries  ({\it i.e.}, not in $\tilde{\Sigma}(s)
\rtimes SO(1,10)$, $s \neq 0$)
 that make the {\it additional} degrees of freedom in the
`new' fields $B^{ab}$, $B^{a_1\ldots a_5}$, $\eta^\alpha$ pure
gauge ({\it i.e.} $B^{ab}$, $B^{a_1\ldots a_5}$, $\eta^\alpha$
carry in all the same number of physical degrees of freedom as the
fundamental $A_3$ field). One might conjecture that the
superfluous degrees of freedom in the `new' one-form fields, which
are pure gauge in the pure supergravity action, could become
`alive' when supergravity is coupled to some M Theory objects.
These could not be the usual M-branes as they couple to the
standard fields and, hence, all the gauge symmetries preserving
the composite $A_3$ would remain preserved. Thus one might think
of some coupling of supergravity through some new action
containing explicitly the new one-form fields. A guide in the
search for such an action would be the preservation of the gauge
symmetries of the underlying
 $\tilde{\Sigma}(s) \rtimes SO(1,10)$, $s \neq 0$,
 gauge supergroup.


\section{Fields/extended superspace coordinates correspondence}
  \label{adboscoor}

In the previous section the additional one-forms $B^{a_1 a_2}$,
$B^{a_1 \ldots a_5}$ and $\eta^\alpha$ were introduced as forms on
conventional $D=11$ spacetime. In contrast, in section
\ref{nature}, the trivialization of the four-cocycle $\omega_4$
associated to $dA_3$ was carried out assuming that all those
forms, together with $e^a$ and $\psi^\alpha$, were independent.
This was explicitly used in the derivation of the linear system of
equations (\ref{systemofeqs}) for the coefficients of the
trivializing three-form $\tilde{\omega}_3$ from the expression
(\ref{products}) of $d \tilde{\omega}_3$. From this point of view,
for each value of the parameter $s$, the natural space on which
the MC one-forms $\Pi^a$, $\Pi^{a_1 a_2}$, $\Pi^{a_1 \ldots a_5}$,
$\pi^\alpha$ and $\pi^{\prime \alpha}$ of the superalgebra
$\tilde{\mathfrak{E}}(s)$ are defined, is the corresponding group
manifold $\tilde{\Sigma}(s)$ of $\tilde{\mathfrak{E}}(s)$, the
(rigid) enlarged superspace manifold.

The one-forms $\Pi^a$ and $\pi^\alpha$ are the usual MC one-forms
of the supertranslations algebra $\mathfrak{E} \equiv
\mathfrak{E}^{(11|32)}$, defined on the supertranslations group
manifold, that is, rigid superspace $\Sigma \equiv
\Sigma^{(11|32)}$. In eleven spacetime dimensions, a set of 11
bosonic coordinates $x^a$ and 32 fermionic coordinates
$\theta^\alpha$ can be introduced to parameterize the standard
rigid superspace,
\begin{equation} \label{eq:stancoor}
\Sigma \equiv \Sigma^{(11|32)} \; : \quad Z^M = (x^a ,
\theta^\alpha) \; .
\end{equation}
The MC equations of the supertranslations algebra (obtained from
the MC equations (\ref{eq:MCsuperP}) of the superPoincar\'e
algebra disregarding the Lo\-rentz part) can be solved,
accordingly, in terms of superspace coordinates as
{\setlength\arraycolsep{0pt}
\begin{eqnarray} \label{standardsuperspace}
&& \Pi^a = dx^a -id\theta^{\alpha } \Gamma^a_{\alpha\beta}
\theta^{\beta } \; , \nonumber \\
&& \pi^\alpha = d\theta^{\alpha } \; .
\end{eqnarray}

}

On standard superspace $\Sigma$, any (left-invariant) differential
form can be expressed in the basis provided by the MC one-forms
$\Pi^a$, $\pi^\alpha$ (with constant coefficients).  However, the
assumption that the one-forms $\Pi^{ab}$, $\Pi^{a_1 \ldots a_5}$
(or their `soft' counterparts $B^{ab}$, $B^{a_1 \ldots a_5}$) are
independent is equivalent to the assumption that the expressions
{\setlength\arraycolsep{0pt}
\begin{eqnarray} \label{solsupertrans}
&& d\Pi^{ab}= - d\theta^\alpha \wedge d\theta^\beta
\Gamma^{ab}_{\alpha\beta} \; , \nonumber \\
&& d\Pi^{a_1\ldots a_5}= - i d\theta^\alpha \wedge d\theta^\beta
\Gamma^{a_1\ldots a_5}_{\alpha\beta}
\end{eqnarray}
}(see equation (\ref{Sigma(s)dual})) cannot be solved in terms of
the left-invariant MC one-forms $\Pi^a$, $\pi^\alpha$ on standard
superspace  $\Sigma$. Although the forms $\Pi^{ab}$,  $\Pi^{a_1
\ldots a_5}$ are actually de Rham trivial (exact) and can indeed
be solved in terms of the coordinates $Z^M = (x^a ,
\theta^\alpha)$ of $\Sigma$, the resulting expressions $\Pi^{ab}=
- d\theta^\alpha \Gamma^{ab}_{\alpha\beta} \theta^\beta$,
$\Pi^{a_1\ldots a_5}= - i d\theta^\alpha
 \Gamma^{a_1\ldots a_5}_{\alpha\beta} \theta^\beta$ fail to be left
 invariant on $\Sigma$. In contrast, the introduction of new parameters $y^{ab}$,
$y^{a_1 \ldots a_5}$ does allow for a solution for $\Pi^{ab}$,
$\Pi^{a_1 \ldots a_5}$ in terms of them,
{\setlength\arraycolsep{0pt}
\begin{eqnarray} \label{solsupertransext}
&& \Pi^{ab} = dy^{ab} -d\theta^{\alpha } \Gamma^{ab}_{\alpha\beta}
\theta^{\beta } \; , \nonumber \\
&& \Pi^{a_1 \ldots a_5} = dy^{a_1 \ldots a_5} -id\theta^{\alpha }
\Gamma^{a_1 \ldots a_5}_{\alpha\beta} \theta^{\beta } \; ,
\end{eqnarray}
}such that, under suitable (and straightforward) transformation
rules for the new parameters, the forms $\Pi^{ab}$ and $\Pi^{a_1
\ldots a_5}$ become left invariant MC one-forms of an enlarged
algebra. The corresponding group manifold $\Sigma^{(528|32)}$ is
parameterized by the 11 bosonic $x^a$ and 32 fermionic
$\theta^\alpha$ coordinates of standard superspace, together with
the additional ${11 \choose 2} + {11 \choose 5} =517$ bosonic
coordinates $y^{ab}$, $y^{a_1 \ldots a_5}$; $\Sigma^{(528|32)}$
is, precisely, the group manifold associated to the M Theory
superalgebra\footnote{\label{footcahp6} The $\Sigma^{(528|32)}$
extended superspace group may be found in our spirit by searching
for a trivialization of the $\mathbb{R}^{528}$-valued two-cocycle
$d{\cal E}^{\alpha\beta}=-id\theta^{\alpha }\wedge
d\theta^{\beta}$, which leads to the one-form
 ${\cal E}^{\alpha\beta}=dX^{\alpha\beta} -id\theta^{(\alpha}\theta^{\beta)}$.
 This introduces in a natural way the 528 bosonic coordinates
$X^{\alpha \beta}$ including the coordinates $x^a$, $y^{ab}$,
$y^{a_1 \ldots a_5}$ in (\ref{MTcoor}) (see equation (\ref{11+})
of next chapter). The transformation law $\delta_\epsilon
X^{\alpha \beta} = i \theta^{(\alpha} \epsilon^{\beta)}$ makes
${\cal E}^{\alpha \beta}$ invariant, and hence leads to a central
extension structure for the extended superspace group
$\Sigma^{(528|32)}$. Thus, the (maximally extended in the bosonic
sector) superspace $\Sigma^{(528|32)}$ transformations make of
${\cal E}^{\alpha \beta}$ a MC form that trivializes, on the
extended superalgebra ${\mathfrak{E}}^{(528|32)}$, the non-trivial
CE two-cocycle on the original odd abelian algebra
$\Sigma^{(0|32)}$.} $\mathfrak{E}^{(528|32)}$:
\begin{equation} \label{MTcoor}
\Sigma^{(528|32)} \; : \quad (x^a , y^{ab}, y^{a_1 \ldots a_5} ,
\theta^\alpha) \; .
\end{equation}
When the curvatures are not zero, and in particular ${\cal
B}_2^{ab} \neq 0$, ${\cal B}_2^{a_1\ldots a_5} \neq 0$ {\it i.e.},
the invariant one-forms $\Pi^{a_1 a_ 2}$, $\Pi^{a_1\ldots a_5}$
become `soft', rendering  $\Sigma^{(528|32)}$ non-flat and no
longer a group manifold.

Likewise, if the additional fer\-mio\-nic one-form $\pi^{\prime
\alpha}$ is considered, 32 new  coordinates $\theta^{\prime
\alpha}$ must be introduced to solve for $\pi^{\prime \alpha}$ in
\begin{equation}
d \pi^{\prime \alpha} =- i d \theta^\beta \wedge  \left(\delta \,
\Pi^a\Gamma_a - i \gamma_1 \Pi^{ab} \Gamma_{ab} +  \gamma_2
\Pi^{a_1\ldots a_5} \Gamma_{a_1\ldots a_5} \right)_\beta{}^\alpha
\;
\end{equation}
(see the last equation of (\ref{Sigma(s)dual})), where $\Pi^a$,
$\Pi^{ab}$ and $\Pi^{a_1\ldots a_5}$ are given by
(\ref{standardsuperspace}), (\ref{solsupertransext}). In terms of
the new coordinates, $\pi^{\prime \alpha}$ reads
{\setlength\arraycolsep{0pt}
\begin{eqnarray}\label{eta=dt}
&& \pi^{\prime \alpha} = d\theta^{\prime \alpha} + i \theta^\beta
\left(\delta \, \Pi^a\Gamma_a - i \gamma_1  \Pi^{ab} \Gamma_{ab}
  + \gamma_2  \Pi^{a_1\ldots a_5} \Gamma_{a_1\ldots a_5} \right)_\beta{}^\alpha
\nonumber  \\* && \quad - \ft{2}{3}  \delta \,
d\theta\Gamma^a\theta \, (\Gamma_a\theta)^\alpha + \ft{2}{3}
\gamma_1 d\theta\Gamma^{ab}\theta \, (\Gamma_{ab}\theta)^\alpha -
\ft{2}{3} \gamma_2 d\theta\Gamma^{a_1\ldots a_5}\theta \,
(\Gamma_{a_1\ldots a_5}\theta)^\alpha \; . \nonumber \\*
\end{eqnarray}
}All these coordinates thus define the enlarged superspaces
$\tilde{\Sigma}(s)$  parameterized by the coordinates
\begin{eqnarray}\label{32+32}
\kern-.5em \tilde{\Sigma}(s) \equiv \Sigma^{(528|32+32)}(s) \ :
{\cal Z}^{{\cal N}}:= \left( x^a \; , \; y^{ab} \; , \;
y^{a_1\ldots a_5}\; ; \; \theta^\alpha \; , \; \theta^{\prime
\alpha} \right)   ,
\end{eqnarray}
which correspond to the group manifolds of the extended
superalgebras $\tilde{\mathfrak{E}}(s) \equiv
{\mathfrak{E}}^{(528|32+32)}(s)$. Again, when the curvatures are
not zero the invariant MC one-forms become `soft', and
$\tilde{\Sigma}(s)$  non-flat and no longer a group manifold.

The gauging of the superalgebra  $\tilde{\mathfrak{E}}(s)$
(\ref{Sigma(s)dual}) leads to the associated  FDA
(\ref{trivializingFDA}), including as many one-form gauge fields
(\ref{formsdualsoft}) as group parameters (\ref{32+32}) correspond
to the enlarged superspace $\tilde{\Sigma}(s)$. This points out to
the existence of a {\it fields/extended superspace coordinates
correspondence} \cite{JdA00,Azcarraga05}, according to which all
the (here, spacetime) fields entering the physical action are in
one-to-one correspondence with the coordinates (that is, group
parameters of their corresponding group manifolds) of suitably
enlarged superspaces. This one-to-one correspondence is further
supported by the fact that enlarged superalgebras also arise in
the description \cite{BESE,JdA00} of the strictly invariant
Wess-Zumino (WZ) terms of the scalar $p$-branes. These invariant
WZ terms trivialize their characterizing Chevalley-Eilenberg (CE)
$(p+2)$-cocycles \cite{AT89} on the standard supersymmetry
algebras $\mathfrak{E}^{(D|n)}$, including that of the $D=11$
supermem\-brane, since its WZ term is given by the pull-back to
${\cal W}$ of the three-form potential of the $dA_3$ superspace
four-cocycle. See \cite{Azcarraga05} for a review.

Enlarged superspaces can also be used in the case of D-branes
\cite{Saka-98,JdA00,Anguelova:2003sn}. Moreover, whereas the
coordinates corresponding to the enlarged superspaces enter
trivially (through a total derivative) the scalar $p$-brane
actions, that is not the case for D-branes or the M5-brane. The
Born-Infeld fields of D-branes and the antisymmetric tensor field
of the M5-brane are usually defined as `fundamental' gauge fields
{\it i.e.}, they are given, respectively, by one-forms $A_1(\xi)$
and a two-form $A_2(\xi)$ directly defined on the worldvolume
$\mathcal{W}$. It was shown in \cite{JdA00}  (see also
\cite{Saka-98}) that both $A_1(\xi)$ and $A_2(\xi)$ can be
expressed through pull-backs to $\mathcal{W}$ of forms defined on
superspaces $\Sigma^\prime$ suitably enlarged by additional
bosonic and fermionic coordinates, in agreement with the
worldvolume fields/extended superspace coordinates correspondence
for superbranes \cite{JdA00} (see also \cite{Az-Iz-Mi-04}). The
extra degrees of freedom that are introduced by considering
$A_1(\xi)$ and $A_2(\xi)$ to be the pull-backs to $\mathcal{W}$ of
forms given on $\Sigma^\prime$, and that produce the composite
structure of the Born-Infeld fields to be used in the superbrane
actions, are removed by the appearance of extra gauge symmetries
\cite{JdA00,Az-Iz-Mi-04}, as it is here the case for the composite
$A_3$ field of $D$=11 supergravity. Of course, these two problems
are not identical:
 for instance, in the case of $D$=11 supergravity with a composite
 $A_3$, the suitably enlarged {\it flat} superspace
 $\tilde{\Sigma}(s)=\Sigma^{(528|32+32)}(s)$
solves, for $s \neq 0$,  the associated problem of trivializing
the CE cocycle, but a dynamical $A_3$ field requires `softening'
the $\tilde{\mathfrak{E}} (s) = \mathfrak{E}^{(528|32+32)}(s)$, $s
\neq 0$,  MC equations by introducing nonvanishing  curvatures; in
contrast, the Born-Infeld worldvolume fields $A_1(\xi)$ and the
tensor field $A_2(\xi)$ are already dynamical in the flat
superspace situation considered in \cite{JdA00}. Nevertheless, in
both these seemingly different situations the fields/extended
superspace coordinates correspondence leads us to the convenience
of enlarging standard superspace $\Sigma \equiv \Sigma^{(11|32)}$.
In this way, all the fields in the theory under consideration (be
them on spacetime or on the worldvolume) correspond to coordinates
of a suitably enlarged superspace.

Superalgebras $\tilde{\mathfrak{E}}$ enlarged with additional
(bosonic and, possibly, fer\-mi\-o\-nic) generators have been
shown here to arise naturally when the underlying gauge structure
of $D=11$ supergravity is studied. The corresponding group
manifolds are, thus,  superspaces $\tilde{\Sigma}$ enlarged with
additional (bosonic and, possibly, fermionic) coordinates, that
generalize ordinary superspace $\Sigma \equiv \Sigma^{(11|32)}$.
The role in supergravity of these enlarged superspaces merits
further investigation. Just like ordinary supersymmetric objects
are formulated as dynamical systems in standard superspace
$\Sigma$, it is, thus, natural to pose actions describing the
dynamics of objects moving in the backgrounds provided by enlarged
superspaces $\tilde{\Sigma}$ and look for an interpretation for
these actions. As an example, the next chapter studies the
dynamics of a supersymmetric string moving in an enlarged
superspace (corresponding, in fact, to the group manifold
associated to the M Theory superalgebra). Such string can be
conveniently described in terms of supertwistors and,
interestingly enough, the model is argued to describe the
excitations of a system composed of two BPS preons.


\chapter[A 30/32-supersymmetric string in tensorial
superspace]{A 30/32-supersymmetric \\ string in tensorial
superspace}
 \label{chapter7}

In the early period of superstring theory, when it was found that
all  $D=10$ supergravities appear as low energy limits of
superstring models, a question arose: what is the origin of
maximally extended $D=11$ supergravity? Its relation with the
supermembrane \cite{BST} was established by studying the
supermembrane action in a supergravity background; however, the
quantization of the supermembrane was fraught with difficulties.
An indication was found \cite{Hoppe} that the quantum state
spectrum of the supermembrane is continuous, a problem now sorted
out by treating \cite{Nicolai} the supermembrane as an object
composed of D$0$-branes in the framework of the Matrix model
approach \cite{(M)atrix}. Another aspect of the same problem was
that the membrane was shown to develop string-like instabilities
\cite{Hoppe}. The Green-Schwarz superstring is free from these
problems, but it is a $D=10$ theory. Thus, it was tempting to
search for possible new  $D=11$ superstring models hoping that
their low energy limit would be eleven-dimensional supergravity.
Such a search requires going beyond the standard superspace
framework: in moving from $D=10$ to $D=11$ one has to add also
extra bosonic degrees of freedom, thus arriving to an {\it
enlarged} $D=11$ superspace rather than to the standard one.

In section \ref{models}, models in enlarged superspaces  are
argued to provide higher spin generalizations of their standard
superspace counterparts. A supersymmetric string action in
maximal, or tensorial superspace (the supergroup manifold
corresponding to the M Theory superalgebra, and its
generalizations containing $n$ fermionic and $\ft12 n(n+1)$
bosonic coordinates) is subsequently introduced in section
\ref{action}. The model does not use $D=11$ gamma-matrices, but
instead includes two auxiliary bosonic spinor
variables\footnote{\label{s-vector} Actually, the model possesses
$Sp(32)$ symmetry besides the $SO(1,10)$ one, so that
$\lambda_\alpha^\pm$ may be considered as symplectic vectors
(called `s-vectors' in \cite{V01s,V01c}) rather than Lorentz
spinors. See footnote \ref{fnote:spinor} of chapter
\ref{chapter4}. See \cite{W03} for a spacetime treatment of a
$\mathbb{CP}^{3}$ sigma model {\it i.e.}, of a string theory in
twistor space, and its relation to Yang-Mills amplitudes. See
\cite{BAME06} and references therein for very recent work in this
subject.}, $\lambda_\alpha^+$ and $\lambda_\alpha^-$. As a
consequence, the resulting supersymmetric string action possesses
$30$ local fermionic $\kappa$-symmetries, although it does not
include a Wess-Zumino term as that of \cite{Curtright}. In the
general formulation in terms of $n$ fermionic coordinates, the
number of preserved $\kappa$-symmetries is $n-2$, so that, for any
$n$, the models provide an extended object action for the
excitations of a state composed of two BPS preons.

Sections \ref{properties} and \ref{gauge}  describe, respectively,
the equations of motion and gauge symmetries of the model,
including the $(n-2)$ $\kappa$-symmetries and their
`superpartners', the $(n-1)(n-2)/2$ bosonic gauge $b$-symmetries.
The gauge symmetries are studied in section \ref{Hamiltonian} in
the Hamiltonian approach. We also discuss there the number of
degrees of freedom of our model. In Sec.~\ref{twistor} we show
that its action may be formulated in terms of a pair of
constrained $OSp(2n|1)$ supertwistors (see \cite{BL98}) which, by
definition, are invariant under both $\kappa$- and $b$-symmetries.
The hamiltonian analysis then simplifies considerably, as shown in
section \ref{twisthamilton}. Section \ref{unnormalized} contains
the hamiltonian analysis in terms of unconstrained supertwistors.
The generalization of the model to the super-$p$-brane case is
given in Sec.~\ref{pbrane}. Some details about the supertwistor
formulation of the model are included in Appendix
\ref{ap:breaking}. This chapter follows reference \cite{30/32}.

\section{Models in enlarged superspaces} \label{models}

A first example of a supersymmetric string action in an enlarged
$D=11$ superspace was found in \cite{Curtright}. The model,
possessing $32$ supersymmetries and $16$ $\kappa$-symmetries, was
constructed in the enlarged superspace $\Sigma^{(528|32)}$. This
contains $32$ fermionic coordinates $\theta^\alpha$ and $528$
bosonic coordinates $x^\mu , y^{\mu\nu}, y^{\mu_1\ldots \mu_5}$
($y^{\mu\nu}=- y^{\nu\mu}\equiv y^{[\mu\nu]}$, $y^{\mu_1\ldots
\mu_5}=y^{[\mu_1\ldots \mu_5]}$) which may be collected in a
symmetric spin-tensor $X^{\alpha\beta}= X^{\beta\alpha}$,
\begin{equation}\label{11+}
X^{\alpha\beta}= \ft{1}{32} \left( x^\mu \Gamma_\mu^{\alpha\beta}
- \ft{i}{ 2!} y^{\mu\nu} \Gamma_{\mu\nu}^{\alpha\beta} +
\ft{1}{5!} y^{\mu_1\ldots \mu_5} \Gamma_{\mu_1\ldots
\mu_5}^{\alpha\beta} \right) \; ,
\end{equation}
so that the coordinates of $\Sigma^{(528|32)}$ are (see equation
(\ref{MTcoor})):
\begin{eqnarray}
\label{528}  {\cal Z}^{{\cal M}}= (X^{\alpha\beta},
\theta^\alpha)\; , \quad X^{\alpha\beta}= X^{\beta\alpha}\; ,
\quad  \alpha, \beta = 1,2, \ldots , 32 \; .
\end{eqnarray}
Recall that the $\Sigma^{(528|32)}$ superspace has a special
interest because it is the supergroup manifold associated with the
maximal $D=11$ supersymmetry algebra $\mathfrak{E}^{(528|32)}$,
the M Theory superalgebra, defined in chapter \ref{chapter2} by
the (anti)commutators (\ref{QQP}) or the MC equations
(\ref{eq:MCMTsuperalg}). In fact, the coordinates $X^{\alpha
\beta}$ parameterizing the superspace $\Sigma^{(528|32)}$ can be
seen as canonically conjugate to the generalized momentum
$P_{\alpha \beta}$ entering the M--algebra and defined in equation
(\ref{n32}). Also, the MC one-forms $\Pi^a$, $\pi^\alpha$,
$\Pi^{ab}$ and $\Pi^{a_1 \ldots a_5}$ can be solved by the
coordinates $X^{\alpha \beta}$ (see equations
(\ref{standardsuperspace}), (\ref{solsupertransext})).

The model of \cite{Curtright} may also be restricted to the
superspaces\footnote{All these superspaces $\Sigma^{(528|32)}$,
$\Sigma^{(462|32)}$ and $\Sigma^{(66|32)}$, considered as
supergroup manifolds, may be seen as central extensions of an
abelian $32$-dimensional fermionic group  by tensorial
(equation~(\ref{11+})) bosonic groups \cite{JdA00}. See footnote
\ref{footcahp6} of chapter \ref{chapter6}.} $\Sigma^{(66|32)}$
($(x^\mu, y^{\mu\nu}, \theta^\alpha)$, with 66 bosonic
coordinates) and $\Sigma^{(462|32)}$ ($(x^\mu, y^{\mu_1\ldots
\mu_5}, \theta^\alpha)$, with $462$ bosonic coordinates). For the
sake of definiteness, we shall call here {\it maximal}, or {\it
tensorial superspaces},
 those with bosonic coordinates of symmetric
`spin-tensorial' type, like $\Sigma^{(528|32)}$ and its
counterparts $\Sigma^{({n(n+1)\over 2}|n)}$,
\begin{eqnarray} \label{Sn}
 \Sigma^{({n(n+1)\over 2}|n)} :  {\cal Z}^{\Sigma}=
(X^{\alpha\beta}, \theta^\alpha)  , \; X^{\alpha\beta}=
X^{\beta\alpha} , \;  \alpha, \beta = 1,2, \ldots n ,
\end{eqnarray}
where $n=2^l$ for suitable $l$, to allow for an interpretation of
$n$ as the number fermionic coordinates. This name distinguishes
the
 $\Sigma^{({n(n+1)\over 2}|n)}$
superspaces from other, not maximally extended (in the bosonic
sector) superspaces like $\Sigma^{(66|32)}$ and
$\Sigma^{(462|32)}$ whose bosonic coordinates may be described by
a spin-tensor $X^{\alpha\beta}$ only if it satisfies some
conditions.

The main problem of the approach in \cite{Curtright} is how to
treat the large number of additional bosonic degrees of
freedom\footnote{See \cite{DG} for a later related search based on
an attempt to replace the $\kappa$-symmetry requirement by a
dynamically generated projection constraint on the spinor
coordinate functions. This approach also suffers from the problem
of additional bosonic degrees of freedom.} in the coset(s) \\
$\Sigma^{(528|32)}/\Sigma^{(11|32)}$ (or
$\Sigma^{(462|32)}/\Sigma^{(11|32)}$,
$\Sigma^{(66|32)}/\Sigma^{(11|32)}$), where, as usual, $\Sigma
\equiv \Sigma^{(11|32)}$
is the standard $D=11$ superspace (see (\ref{eq:stancoor})).
Actually, the same problem arises in any approach dealing with
enlarged superspaces
\cite{JdA00,RS,BL98,BLS99,BPS01,V01s,V01c,Manvelyan,ZL,ZU,B02,30/32}.
Thus, one has to find a mechanism that either suppresses the
additional (with respect to the usual spacetime/superspace
$\Sigma^{(D|n)}$) degrees of freedom or provides a physical
interpretation for them. In this respect $\Sigma^{({n(n+1)\over
2}|n)}$, despite having a maximal bosonic part, has some
advantages with respect to non-maximally extended superspaces.
Indeed, the bosonic sector of the tensorial superspace (\ref{Sn}),
\begin{eqnarray}\label{Sn0}
& \Sigma^{({n(n+1)\over 2}|0)}\; :  \; X^{\alpha\beta}=
X^{\beta\alpha}\; , \quad
 \alpha, \beta = 1,2, \ldots , n  \;\; ,\quad
\end{eqnarray}
was proposed for $n=4$ \cite{Fr86} as a basis for the construction
of $D=4$ higher-spin theories \cite{V01,V01s,V01c}. Moreover, it
was shown in \cite{BLS99} that the quantization of a simple
superparticle model \cite{BL98} in $\Sigma^{({n(n+1)\over 2}|n)}$
for $n=2,4,8,16$ results in a wavefunction describing a tower of
massless fields of all possible spins (helicities). Such an
infinite tower of higher spin fields allows for a non-trivial
interaction in $AdS$ spacetimes \cite{Vasiliev89,V01}\footnote{A
relation between the generalized $n=4$ superparticle wavefunctions
\cite{BLS99} and Vasiliev's `unfolded' equations for higher spin
fields was noted in \cite{V01s}. This was elaborated in detail in
\cite{Dima}, where the quantization of an $AdS$ superspace
generalization of the $n=4$ model of \cite{BL98} was also carried
out (see also \cite{Misha} for a related study of higher spin
theories in the maximal generalized $AdS_4$ superspace).}.

To give an idea of the relation between higher spin theories and
maximally extended superspaces, let us consider the  free bosonic
massless higher-spin equations proposed in \cite{V01s} (for
$n=4$). These can be collected as the following set of equations
for a scalar function $b$ on $\Sigma^{({n(n+1)\over 2}|0)}$
\begin{eqnarray}\label{hsEqb0}
\partial_{\alpha[\beta}\partial_{\gamma]\delta} b(X)=0 \; ,
\end{eqnarray}
where $\partial_{\alpha\beta}=\partial/\partial X^{\alpha\beta}$.
Equation (\ref{hsEqb0}) states that
$\partial_{\alpha\beta}\partial_{\gamma\delta}$ is fully symmetric
on a non-trivial solution. In the generalized momentum
representation equation~(\ref{hsEqb0}) reads
\begin{eqnarray}\label{hsEqbk}
k_{\alpha[\beta} k_{\gamma]\delta} b(k) =0 \; .
\end{eqnarray}
This implies that $b(k)$ has support on the
$\frac{n(n+1)}2-\frac{n(n-1)}2=n$-dimensional surface in momentum
space $\Sigma^{({n(n+1)\over 2}|0)}$ (actually, in
$\Sigma^{({n(n+1)\over 2}|0)}\backslash \{ 0\}$) on which the rank
of the matrix $ k_{\gamma\delta}$ is equal to unity \cite{V01c}.
This is the surface defined by $k_{\alpha\beta}=\lambda_{\alpha}
\lambda_{\beta}$ (or $-\lambda_{\alpha} \lambda_{\beta}$)
characterized by the $n$ components of $\lambda_\alpha$. In a
`$GL(n,\mathbb{R})$-preferred' frame (an analogue of the standard
frame for lightlike ordinary momentum),
$\lambda_\alpha=(1,0,\ldots,0)$ and the surface is the
$GL(n,\mathbb{R})$-orbit of the point
$k_{\alpha\beta}=\delta_{\alpha 1} \delta_{\beta 1}$. Thus,
equation~(\ref{hsEqbk}) may also be written as
\begin{eqnarray}\label{hsEqbkl}
(k_{\alpha\beta} -\lambda_{\alpha} \lambda_{\beta}) b = 0 \; ,
\end{eqnarray}
which is equivalent to writing equation~(\ref{hsEqb0}) in the form
\begin{eqnarray}\label{hsEqbl}
(i\partial_{\alpha\beta} -\lambda_{\alpha} \lambda_{\beta}) b = 0
\; .
\end{eqnarray}
Equations~(\ref{hsEqbkl}) and (\ref{hsEqbl}) may be considered
\cite{30/32} as the generalized momentum ($k_{\alpha\beta}$) and
coordinate ($X^{\alpha\beta}$) representations of the definition
(\ref{BPSdef}) of a BPS preon \cite{BPS01} (see section
\ref{sec:BPSpreons} of chapter \ref{chapter4}). The solutions of
equations~(\ref{hsEqbkl}), (\ref{hsEqbl}) are the momentum and
coordinate `wavefunctions' corresponding to a BPS preon state
$|\lambda \rangle$, $b(X)= \langle X | \lambda \rangle$, $b(k)=
\langle k|\lambda \rangle $. These equations also appear as a
result of the quantization \cite{BLS99} of the superparticle model
in \cite{BL98}\footnote{In \cite{V01s,Misha+}
equation~(\ref{hsEqbl}) was written as $(\partial_{\alpha\beta} -
{\partial \over
\partial \mu^{\alpha}} {\partial \over \partial \mu^{\beta}})
b(X,\mu) = 0 $ which is an equivalent `momentum' representation
obtained by a Fourier transformation with respect to
$\lambda_\alpha$, see \cite{Dima}.}.

Thus, in contrast with other extended superspaces, the models in
the tensorial superspaces $\Sigma^{({n(n+1)\over 2}|n)}$ can be
regarded as higher spin generalizations of the models in standard
superspace\footnote{Formulations of higher spin theories are
currently known up to spacetime dimension $D=10$ \cite{BBAST05}.}
$\Sigma^{(D|n)}$. For a recent review of higher spin theory see
\cite{Sor05}.

\section{A supersymmetric string in
tensorial superspace } \label{action}

A superstring in $\Sigma^{({n(n+1)\over 2}|n)}$ is described by
worldsheet functions $X^{\alpha\beta}(\xi)$, $\theta^\alpha(\xi)$,
where $\xi=(\tau,\sigma)$ are the worldsheet $W^2$ coordinates. We
propose the following action \cite{30/32}:
\begin{eqnarray}\label{St}
\kern-2em S =  {1\over \alpha^\prime} \int_{W^2} [e^{++} \wedge
\Pi^{\alpha\beta}\, \lambda^-_{\alpha}\lambda^-_{\beta}  - e^{--}
\wedge \Pi^{\alpha\beta}\, \lambda^+_{\alpha}\lambda^+_{\beta} -
 e^{++} \wedge e^{--}] \; ,
\end{eqnarray}
where {\setlength\arraycolsep{0pt}
\begin{eqnarray}\label{Pi} &&
\Pi^{\alpha\beta}(\xi) = dX^{\alpha\beta}(\xi) - i
d\theta^{(\alpha}\,\theta^{\beta )}(\xi) = d\tau
\Pi_{\tau}^{\alpha\beta} + d\sigma \Pi_{\sigma}^{\alpha\beta} \; ;
\nonumber \\
&& \alpha, \beta =1,\ldots, n\; , \quad m=0,1 \; , \quad
\xi^m=(\tau, \sigma)\;  ,
\end{eqnarray}
}with dimensions $[1/\alpha^\prime]=\mathrm{ML}^{-1}\,$,
$[\Pi^{\alpha\beta}]=\mathrm{L}\,$, $[e^{\pm\pm}]=\mathrm{L}\;
(c=1)$. The two auxiliary worldvolume fields, the {\it bosonic}
spinors $\lambda^-_{\alpha}(\xi), \lambda^+_{\alpha}(\xi)$, are
{\it dimensionless} and constrained by
\begin{eqnarray}\label{l+l-}
C^{\alpha\beta} \lambda^+_{\alpha}\lambda^-_{\beta} = 1 \; ;
\end{eqnarray}
$e^{\pm\pm}(\xi) = d\xi^m e^{\pm\pm}_m(\xi) = d\tau
e^{\pm\pm}_\tau(\xi) + d\sigma e^{\pm\pm}_\sigma(\xi)$  are two
auxiliary worldvolume one-forms. The one-forms $e^{\pm\pm}$ are
assumed to be linearly independent and, hence, define an auxiliary
worldsheet zweibein
\begin{eqnarray}\label{vielbein}
e^a = (e^0, e^1) = d\xi^m e_m^a(\xi) =(\ft12 (e^{++}+ e^{--}),
\ft12 (e^{++}- e^{--})).
\end{eqnarray}
The $C^{\alpha\beta}$ in (\ref{l+l-}) is an invertible constant
antisymmetric  matrix
\begin{eqnarray}\label{Cab}
 C^{\alpha\beta}= -  C^{\beta\alpha}\; , \qquad d C^{\alpha\beta}=0\;
 ,
\end{eqnarray}
which can be used to rise and lower the spinor indices (as the
charge conjugation matrix in Minkowski spacetimes). The
invertibility of the matrix $C^{\alpha\beta}$ requires $n$ to be
even; this is not really a limitation since, after all, we are
interested in $n=2^l$ to allow for a spinor treatment of the
$\alpha, \beta$ indices.

We shall refer to this $n=32$ ($\Sigma^{(528|32)}$) model as a
$D=11$ superstring, which implies the decomposition of
equation~(\ref{n32}) for the generalized momentum. Nevertheless,
the $n=32$ case also admits a $D=10$, Type IIA treatment, which
uses the same $C^{\alpha\beta}$ of the $D=11$ case, and in which
the decomposition (\ref{n32}) is replaced by its $D=10$, IIA
counterpart obtained from (\ref{n32}) by separating the eleventh
value of the vector index.

The action (\ref{St}) is invariant under the supersymmetry
transformations
\begin{eqnarray} \label{susy} \label{susysinglets}
\delta_\epsilon X^{\alpha\beta}=i\theta^{(\alpha}\epsilon^{\beta)}
\; , \quad \delta_\epsilon\theta^\alpha=\epsilon^\alpha \; , \quad
\delta_\epsilon \lambda^\pm_\alpha=0 \; , \quad \delta_\epsilon
e^{\pm\pm}=0 \; ,
\end{eqnarray}
as well as under rigid $Sp(n,\mathbb{R})$ `rotations' acting on
the $\alpha, \beta$ indices. Note also that, although formally the
action (\ref{St}) possesses a manifest $GL(n,\mathbb{R})$
invariance, the constraint (\ref{l+l-}) breaks it down to
$Sp(n,\mathbb{R})$ $\subset$ $GL(n,\mathbb{R})$. Under the action
of $Sp(n,\mathbb{R})$, the Grassmann coordinate functions
$\theta^\alpha(\xi)$ and the auxiliary fields
$\lambda^{\pm}_\alpha(\xi)$ are transformed as symplectic vectors
and $X^{\alpha\beta}(\xi)$ as a symmetric symplectic tensor.
Nevertheless, we keep for them the `spinor' and `spin-tensor'
terminology having in mind their transformation properties under
the subgroup $Spin(t,D-t) \subset Sp(n,\mathbb{R})$, which would
appear in a `standard' $(t,D-t)$ spacetime treatment.

The above $\Sigma^{({n(n+1)\over 2}|n)}$ superstring model may
also be described by an action written in terms of {\it
dimensionful} unconstrained spinors $\Lambda^\pm_\alpha(\xi)$,
$[\Lambda^\pm_\alpha]=(\mathrm{ML}^{-1})^{1/2}$ \cite{30/32},
{\setlength\arraycolsep{0pt}
\begin{eqnarray}\label{St01}
 S  = \int_{W^2} [&& e^{++} \wedge \Pi^{\alpha\beta}\,
\Lambda^-_{\alpha}\Lambda^-_{\beta} - e^{--} \wedge
\Pi^{\alpha\beta}\, \Lambda^+_{\alpha}\Lambda^+_{\beta}  \nonumber
\\ &&  - \alpha^\prime e^{++} \wedge e^{--}
(C^{\alpha\beta}\Lambda^+_{\alpha}\Lambda^-_{\beta})^2] \; .
\end{eqnarray}
}Indeed, one can see that the action (\ref{St01}) possesses two
independent scaling gauge symmetries defined by the transformation
rules
\begin{eqnarray}\label{sc+}
e^{++}(\xi)\rightarrow e^{2\alpha (\xi)} e^{++}(\xi)\; , \quad
\Lambda^-_{\alpha}(\xi) \rightarrow e^{-\alpha (\xi)}
\Lambda^-_{\alpha}(\xi)
\end{eqnarray}
and
\begin{eqnarray}\label{sc-}
e^{--}(\xi)\rightarrow e^{2\beta (\xi)} e^{--}(\xi)\; , \quad
\Lambda^+_{\alpha}(\xi) \rightarrow e^{-\beta (\xi)}
\Lambda^+_{\alpha}(\xi)\; .
\end{eqnarray}
This allows one to obtain
$C^{\alpha\beta}\Lambda^+_{\alpha}\Lambda^-_{\beta}={1 /
\alpha^\prime }$ as a gauge fixing condition. Then the gauge fixed
version of the action (\ref{St01}) coincides with (\ref{St}) up to
the trivial redefinition $\Lambda^{\pm}_{\alpha}=
(\alpha^\prime)^{-1/2}\lambda^{\pm}_{\alpha}$. The gauge
$C^{\alpha\beta}\Lambda^+_{\alpha}\Lambda^-_{\beta}={1 /
\alpha^\prime }$ (equivalent to equation~(\ref{l+l-})) is
preserved by a one-parametric combination of (\ref{sc+}) and
(\ref{sc-}) with $\alpha=- \beta$, which is exactly the $SO(1,1)$
gauge symmetry (worldvolume Lorentz symmetry) of the action
(\ref{St}),
\begin{eqnarray}\label{SO(1,1)}
e^{\pm\pm}(\xi)\rightarrow e^{\pm 2\alpha (\xi)} e^{\pm\pm}(\xi)\;
, \;\; \lambda^{\pm}_{\alpha}(\xi) \rightarrow e^{\pm\alpha (\xi)}
\lambda^{\pm}_{\alpha}(\xi)\,.
\end{eqnarray}

The tension parameter $T=1/\alpha^\prime$ enters in the last
(`cosmological') term of the action (\ref{St01}) only. Setting in
it $\alpha^\prime=0$ one finds that the model is non-trivial only
for $e^{++} \propto e^{--}$ and $\Lambda^+ \propto \Lambda^-$ in
which case one arrives at the tensionless super-$p$-brane action
(with $p=1$) of reference \cite{B02}, $S=\int d^2\xi \rho^{++m}
\Pi^{\alpha\beta}_m \Lambda^-_\alpha \Lambda^-_\beta$. As we are
not interested in this case, we set $\alpha^\prime=1$ below since
the $\alpha^\prime$ factors can be restored by dimensional
considerations.

The most interesting feature of the model (\ref{St}), (\ref{St01})
is that, being formulated in the tensorial $\Sigma^{({n(n+1)\over
2}|n)}$ superspace with $n$ fermionic coordinates, it possesses
$(n-2)$ $\kappa$-symmetries \cite{30/32}; we will prove this in
section \ref{gauge}. For a supersymmetric extended object in
standard superspace, the $\kappa$-symmetry of its worldvolume
action determines the number $k$ of supersymmetries which are
preserved by the ground state, which is a $\nu=\frac{k}{n}$ BPS
state made out of $\n=n-k$ preons if at least one supersymmetry,
$k \ge 1$, is preserved (see section \ref{sec:BPSpreons} of
chapter \ref{chapter4}). In the present case, we may expect that
the ground state of our model should preserve ($n-2$) out of $n$
supersymmetries, {\it i.e.} that it is a $\nu={n-2\over n}$ BPS
state ($\n=2, {30\over 32}$ BPS state for the $D=11$ tensorial
superspace $\Sigma^{(528|32)}$).

For $n=2$, $X^{\alpha\beta}$ provides a representation of the
3-dimensional Min\-kow\-ski space coordinates, $X^{\alpha\beta}
\propto \Gamma_\mu^{\alpha\beta}x^\mu$ ($\alpha, \beta = 1,2 $;
$\mu=0,1,2$). Thus the $n=2$ model (\ref{St}) describes a string
in the $D=3$ standard $\Sigma^{(3|2)}$ superspace. However, in the
light of the above discussion, it does not possess any
$\kappa$-symmetry and, hence, its ground state is not a BPS state
since it does not preserve any supersymmetry. The situation
becomes different starting with the $n=4$ model (\ref{St}), which
possesses two $\kappa$-symmetries, the same number as the
Green-Schwarz superstring in the standard $D=4$ superspace. For $D
\ge 6$, $n \ge 8$ the number of $\kappa$-symmetries of our model
exceeds $n/2$ and thus the model describes the excitations of BPS
states with extra supersymmetries, a ${30\over 32}$ BPS state in
the $D=11$ $\Sigma^{(528|32)}$ superspace.

The number of {\it bosonic} degrees of freedom of our model is
$4n-6$ \cite{30/32} (see section \ref{Hamiltonian}). It is not as
large as it might look at first sight due to the `momentum space
dimensional reduction mechanism' \cite{BLS99} which occurs due to
the presence of auxiliary  spinor variables entering the
generalized Cartan-Penrose relation (see equation (\ref{PX})
below) generated by our model. However, it is larger than that of
the $(D=3,4,6,10)$ Green-Schwarz superstring (which has $D$
[$2n=4(D-2)$] bosonic [fermionic] configuration space real degrees
of freedom, which reduce to $D-2$ [$2(D-2)$] after taking into
account reparameterization invariance ($\kappa$-symmetry), thus
resulting in $2(D-2)$ bosonic and $2(D-2)$ fermionic phase space
degrees of freedom). Thus, the relation of models in tensorial
superspaces to higher spin theories mentioned in section
\ref{models}, allows us to consider our model as a higher spin
generalization of the Green-Schwarz superstring, containing
additional information about the non-perturbative states of the
String/M Theory.

The number of {\it fermionic} degrees of freedom of our model is
$2$ for any $n$, less than that of the $D=4,6,10$ $(N=2)$
Green-Schwarz superstring.

\section{Equations of motion} \label{properties}

Consider the variation of the action (\ref{St}). Allowing for
integration by parts one finds {\setlength\arraycolsep{2pt}
\begin{eqnarray}\label{variation}
\delta S  & = & \int_{W^2} d( e^{--}
 \lambda^+_{\alpha}\lambda^+_{\beta}-
e^{++}\lambda^-_{\alpha}\lambda^-_{\beta} ) \,  i_\delta
 \Pi^{\alpha\beta} \nonumber\\
&& -  2i \int_{W^2} e^{++} \wedge d \theta^\alpha \lambda^-_\alpha
\; \delta\theta^\beta \lambda^-_\beta +  2i \int_{W^2} e^{--}
\wedge d \theta^\alpha
\lambda^+_\alpha \; \delta\theta^\beta \lambda^+_\beta \nonumber\\
&& +  \int_{W^2} (\Pi^{\alpha\beta}\,
\lambda^+_{\alpha}\lambda^+_{\beta} -e^{++}) \wedge \delta e^{--}
- \int_{W^2} (\Pi^{\alpha\beta}
\lambda^-_{\alpha}\lambda^-_{\beta} - e^{--}) \wedge \delta e^{++}
 \nonumber
\\
&& +    \delta_{\lambda} S \; ,
\end{eqnarray}
}where $i_\delta \Pi^{\alpha\beta} \equiv \delta X^{\alpha\beta} -
i \delta \theta^{(\alpha}\theta^{\beta)}$ and the last term
\begin{equation}\label{varl}
 \delta_{\lambda} S = + \int_{W^2}
2e^{++}\wedge\Pi^{\alpha\beta} \lambda^-_{\beta} \delta
\lambda^-_{\alpha} - \int_{W^2} 2e^{--}\wedge\Pi^{\alpha\beta}
\lambda^+_{\beta} \delta \lambda^+_{\alpha} \; ,
\end{equation}
collects the variations of the bosonic spinors
$\lambda^\pm_{\alpha}(\xi)$.

The equations of motion for the bosonic coordinate functions,
$\delta S / \delta X^{\alpha\beta}$ $(= \delta S / i_\delta
\Pi^{\alpha\beta})=0$, turn out to restrict the auxiliary spinors
and auxiliary one-forms,
\begin{eqnarray}\label{vX}
d( e^{--}
 \lambda^+_{\alpha}\lambda^+_{\beta}-
e^{++}\lambda^-_{\alpha}\lambda^-_{\beta} ) =0  \, . \qquad
\end{eqnarray}
The equations for the fermionic coordinate functions, $\delta S /
\delta \theta^{\alpha}=0$, read
\begin{eqnarray}
\label{vTh0}  e^{++} \wedge d \theta^\alpha \lambda^-_\alpha
\lambda^-_\beta - e^{--} \wedge d \theta^\alpha \lambda^+_\alpha
\lambda^+_\beta =0 \; ,
\end{eqnarray}
which, due to the linear independence of the spinors
$\lambda^+_{\alpha}$ and $\lambda^-_{\alpha}$, imply
\begin{eqnarray}
\label{vTh} e^{++} \wedge d \theta^\alpha \lambda^-_\alpha =0\; ,
\qquad e^{--} \wedge d \theta^\alpha \lambda^+_\alpha =0 \; .
\end{eqnarray}
The equations for the one-forms $e^{\pm\pm}(\xi)$ express them
through the worldsheet covariant bosonic form (\ref{Pi}) of the
$\Sigma^{({n(n+1)\over 2}|n)}$ superspace and the spinors
$\lambda^\pm_{\alpha}(\xi)$,
\begin{eqnarray}
\label{ve-} e^{++} & = & \Pi^{\alpha\beta}\,
\lambda^+_{\alpha}\lambda^+_{\beta} \; , \quad
\\ \label{ve+}
e^{--} & = & \Pi^{\alpha\beta} \lambda^-_{\alpha}\lambda^-_{\beta}
\; . \qquad
\end{eqnarray}
This reflects the auxiliary nature of $e^{\pm\pm}$ and implies
that equations~(\ref{vX}) and (\ref{vTh}) actually restrict
$\Pi^{\alpha\beta}$ and $d\theta^\alpha$,
{\setlength\arraycolsep{0pt}
\begin{eqnarray}\label{vXs}
&& d( \Pi^{\gamma\delta} \lambda^-_{\gamma}\lambda^-_{\delta} \;
\lambda^+_{\alpha}\lambda^+_{\beta}- \Pi^{\gamma\delta}
\lambda^+_{\gamma}\lambda^+_{\delta} \;
\lambda^-_{\alpha}\lambda^-_{\beta} )  =  0  \, ,
\\ \label{vThs-}
&& \Pi^{\gamma\delta} \lambda^+_{\gamma}\lambda^+_{\delta} \wedge
d \theta^\alpha \lambda^-_\alpha  =  0\; , \qquad
\\ \label{vThs+}
&&  \Pi^{\gamma\delta} \lambda^-_{\gamma}\lambda^-_{\delta} \wedge
d \theta^\alpha \lambda^+_\alpha  =  0 \; . \; \qquad
\end{eqnarray}

}

The necessity of the constraints (\ref{l+l-}) on the bosonic
spinor variables can be seen to stem from the equations
(\ref{ve-}), (\ref{ve+}). Indeed, were the constraints
(\ref{l+l-}) ignored, the variation of the action (\ref{varl})
with respect to unconstrained $\lambda^\pm_\alpha$ would yield
$e^{++}\wedge \Pi^{\alpha\beta}\, \lambda^-_{\beta}=0$ and
$e^{--}\wedge \Pi^{\alpha\beta}\, \lambda^+_{\beta}=0$. By
(\ref{ve-}) (or (\ref{ve+})) this would imply, in particular,
$e^{++}\wedge e^{--}=0$, contradicting
 the original assumption of independence of the one-forms $e^{++}$ and
$e^{--}$ and, actually, reducing the present model to a $p=1$
version of the tensionless $p$-brane model \cite{B02}.

As  $\lambda^{\pm}_\alpha$ are restricted by the constraint
(\ref{l+l-}), this constraint has to be taken into account in the
variational problem. Instead of applying the Lagrange multiplier
technique, one may restrict the variations to those that preserve
(\ref{l+l-}), {\it i.e.} such that
\begin{equation}\label{vl+l-}
C^{\alpha\beta} \delta \lambda^+_{\alpha}\lambda^-_{\beta} +
C^{\alpha\beta} \lambda^+_{\alpha} \delta\lambda^-_{\beta}=0\; .
\end{equation}
One can solve (\ref{vl+l-}) by introducing a set of $n-2$
auxiliary spinors $u^I_{\alpha}$ `orthogonal' to the
$\lambda^{\pm}$ ({\it cf.}~\cite{BanZhel93,BZ95}),
\begin{equation} \label{cond1}
C^{\alpha\beta} u_\alpha^I \lambda^{\pm}_\beta = 0 \; , \qquad
I=1, \ldots, n-2 \quad ,
\end{equation}
and normalized by
\begin{equation} \label{cond2}
C^{\alpha\beta} u_\alpha^I u_\beta^J = C^{IJ} \, , \qquad
C^{IJ}=-C^{JI} \; ,
\end{equation}
where $C^{IJ}$ is an antisymmetric constant invertible $(n-2)
\times (n-2)$ matrix.

The $n$ spinors
\begin{equation} \label{basis}
\{\lambda^+_\alpha\; , \, \lambda^-_\alpha \; ,  u^I_\alpha\} \; ,
\; I=1, \ldots, n-2 \; ,
\end{equation}
provide a basis that can be used to decompose an arbitrary spinor
worldvolume function ({\it cf.} \cite{NP85}), and in particular
the variations $\delta \lambda^+$, $\delta\lambda^-$. Then one
finds that the only consequence of equation~(\ref{vl+l-}) is that
the sum of the coefficient for $\lambda^+$ in the decomposition of
$\delta \lambda^+$  and that of $\lambda^-$ in the decomposition
of $\delta \lambda^-$ vanishes . In other words, the general
solution of equation~(\ref{vl+l-}) reads
{\setlength\arraycolsep{0pt}
\begin{eqnarray}\label{vlpm}
&& \delta \lambda_\alpha^+ = \omega(\delta) \lambda_\alpha^+ +
\Omega^{++}(\delta)\lambda_\alpha^- +
\Omega^+_I(\delta)u_\alpha^I \; , \\
\label{vlpm2} && \delta \lambda_\alpha^- = - \omega(\delta)
\lambda_\alpha^- + \Omega^{--}(\delta)\lambda_\alpha^+ +
\Omega^-_I(\delta)u_\alpha^I \; ,
\end{eqnarray}
}where $\Omega^{\pm}_I(\delta)$, $\Omega^{\pm\pm}(\delta)$ and
$\omega(\delta)$ are arbitrary variational parameters.
Substituting equations~(\ref{vlpm}), (\ref{vlpm2}) into
(\ref{varl}), one finds {\setlength\arraycolsep{2pt}
\begin{eqnarray}\label{var2}
 \delta_{\lambda} S &=& -  \int_{W^2}
(2e^{++}\wedge\Pi^{\alpha\beta} \lambda^-_{\beta}
\lambda^-_{\alpha} + 2e^{--}\wedge\Pi^{\alpha\beta}
\lambda^+_{\beta}
\lambda^+_{\alpha}) \omega(\delta)  \nonumber \\
&& +  \int_{W^2} 2e^{++}\wedge\Pi^{\alpha\beta} \lambda^-_{\beta}
\lambda^+_{\alpha} \Omega^{--}(\delta)
  \nonumber \\ && +\int_{W^2} 2e^{--}\wedge\Pi^{\alpha\beta} \lambda^+_{\beta}
\lambda^-_{\alpha} \Omega^{++}(\delta)
\nonumber \\
&& +  \int_{W^2} 2e^{++}\wedge\Pi^{\alpha\beta} \lambda^-_{\beta}
u_{\alpha}^I \Omega^-_I(\delta) \nonumber \\
&& -   \int_{W^2} 2e^{--}\wedge\Pi^{\alpha\beta} \lambda^+_{\beta}
u^I_{\alpha} \Omega^+_I(\delta)\; .
\end{eqnarray}}

Now we can write the complete set of equations of motion which
include, in addition to equations~(\ref{vX}), (\ref{vTh}),
(\ref{ve-}), (\ref{ve+}), the set of equations for
$\lambda^{\pm}_\alpha$,  which follows from $\delta S /
\omega(\delta)=0$, $\delta S / \Omega^{++}(\delta)=0$, $\delta S /
\Omega^+_I(\delta)=0$, $\delta S / \Omega^{--}(\delta)=0$, and
$\delta S / \Omega^-_I(\delta)=0$, namely
{\setlength\arraycolsep{0pt}
\begin{eqnarray}
\label{vl0} && e^{++}\wedge \Pi^{\alpha\beta} \lambda^-_{\beta}
\lambda^-_{\alpha} + e^{--}\wedge\Pi^{\alpha\beta}
\lambda^+_{\beta}\lambda^+_{\alpha} =0 \; ,
\\ \label{vl--}
&& e^{++}\wedge\Pi^{\alpha\beta} \lambda^-_{\beta}
\lambda^+_{\alpha} =0 \; ,
\\ \label{vl++}
&& e^{--}\wedge\Pi^{\alpha\beta} \lambda^+_{\beta}
\lambda^-_{\alpha} =0 \; ,
\\ \label{vl-I}
&& e^{++}\wedge\Pi^{\alpha\beta} \lambda^-_{\beta} u_{\alpha}^I =0
\; ,
\\ \label{vl+I}
&& e^{--}\wedge\Pi^{\alpha\beta} \lambda^+_{\beta} u^I_{\alpha} =0
\; .
\end{eqnarray}
}Due to the linear independence of both one-forms $e^{++}=d\xi^m
e_m^{++}(\xi)$ and $e^{--}=d\xi^m e_m^{--}(\xi)$,
equations~(\ref{vl--}), (\ref{vl++}) imply
\begin{eqnarray}\label{P+-=0}
\Pi^{\alpha\beta} \lambda^-_{\beta} \lambda^+_{\alpha} =0 \; .
\end{eqnarray}
Decomposing the bosonic invariant one form $\Pi^{\alpha\beta}=
d\xi^m \Pi^{\alpha\beta}_m$ in the basis provided by $e^{\pm\pm}$,
{\setlength\arraycolsep{0pt}
 \begin{eqnarray}\label{P=eP}
&& \Pi^{\alpha\beta} = e^{++} \Pi^{\alpha\beta}_{++} + e^{--}
\Pi^{\alpha\beta}_{--} \; ,
\\ \label{Pe++}
&& \Pi^{\alpha\beta}_{\pm\pm} = \nabla_{\pm\pm} X^{\alpha\beta} -
i \nabla_{\pm\pm} \theta^{(\alpha } \; \theta^{\beta ) } \; ,
\end{eqnarray}
}where $\nabla_{\pm\pm}$ is defined by
\begin{equation} \label{d=en} d \equiv e^{\pm\pm} \nabla_{\pm\pm}
= e^{++} \nabla_{++} + e^{--} \nabla_{--}\; ,
\end{equation}
 one finds that equations~(\ref{vl-I}) and (\ref{vl+I}) restrict
only the derivatives $(\nabla_{++},\,\nabla_{--})$ of the bosonic
coordinate function $X^{\alpha\beta}(\xi)$, respectively,
{\setlength\arraycolsep{0pt}
\begin{eqnarray}
 \label{vl-I--}
&& \Pi_{--}^{\alpha\beta} \lambda^-_{\beta} u_{\alpha}^I \equiv
(\nabla_{--} X^{\alpha\beta} - i  \nabla_{--} \theta^{(\alpha } \;
\theta^{\beta ) }) \; \lambda^-_{\beta} u_{\alpha}^I =0 \; ,
\\ \label{vl+I++}
&& \Pi_{++}^{\alpha\beta} \lambda^+_{\beta} u^I_{\alpha} \equiv
(\nabla_{++} X^{\alpha\beta} - i  \nabla_{++} \theta^{(\alpha } \;
\theta^{\beta ) } ) \; \lambda^+_{\beta} u_{\alpha}^I =0 \; .
\end{eqnarray}
}In the same manner, equations~(\ref{vTh}) can be written as
\begin{eqnarray}
\label{vTh--} \nabla_{--} \theta^\alpha \, \lambda^-_\alpha =0\; ,
\qquad \nabla_{++}\theta^\alpha \, \lambda^+_\alpha =0 \; .
\end{eqnarray}

The analysis of the above set of equations in the tensorial
superspace, the search for solutions and their reinterpretation in
standard $D$-dimensional spacetime, possibly along the
fields/extended superspace coordinates correspondence of
\cite{JdA00} (see section \ref{adboscoor} of chapter
\ref{chapter6}) , or of the `two-time physics' approach of
\cite{Bars}, lies beyond the scope of this Thesis.

\section{Gauge symmetries} \label{gauge}

The expression (\ref{variation}), with (\ref{var2}), for the
general variation of the supersymmetric string action (\ref{St})
shows that the model possesses $n$ supersymmetries and $(n-2)$
$\kappa$-symmetries of the form \cite{30/32}
{\setlength\arraycolsep{0pt}
\begin{eqnarray}
\label{kappa1} && \delta_\kappa \theta^\alpha(\xi) =
C^{\alpha\beta}u^I_\beta (\xi) \kappa_I(\xi) \; ,\\
\label{kappa2} && \delta_\kappa X^{\alpha\beta}(\xi)
= i  \delta_\kappa \theta^{(\alpha}(\xi)\theta^{\beta)}(\xi) \; ,\\
\label{kappa3} && \delta_\kappa \lambda_\alpha^\pm(\xi)=0\; ,
\quad
 \delta_\kappa e^{\pm\pm}_m(\xi)=0 \; ,
\end{eqnarray}
}with $(n-2)$ fermionic gauge parameters $\kappa_I(\xi)$ (30 for
$\Sigma^{(528|32)})$. In the framework of the second Noether
theorem this $\kappa$-symmetry is reflected by the fact that only
2 of the $n$ fermionic equations (\ref{vTh0}) are independent. We
stress that the ($n-2$) $GL(n,\mathbb{R})$ vector fields
$u^I_\alpha$ defined by (\ref{cond1}) are auxiliary. They allow us
to write explicitly the general solution of the equations
\begin{equation}\label{kappa4}
\delta_\kappa \theta^\alpha(\xi)\lambda^{\pm}_\alpha(\xi)=0 \; ,
\end{equation}
which define implicitly the  $\kappa$-symmetry transformation
(\ref{kappa1}). Note that the dynamical system is
$\kappa$-symmetric despite it does not contain a Wess-Zumino term.
This property seems to be  specific of models defined on tensorial
superspaces.

Our model also possesses ${1 \over 2}(n-1)(n-2)$ $b$-symmetries,
which are the bosonic `superpartners' of the fermionic
$\kappa$-symmetries, defined by
\begin{eqnarray} \label{b4}
\kern-1em \delta_b X^{\alpha\beta}=b_{IJ}(\xi)u^{\alpha I}u^{\beta
J} \; ,
 \delta_b \theta^\alpha= 0\; ,\quad
\delta_b \lambda^\pm_\alpha=0 \;, \quad \delta_b e^{\pm\pm}=0 \; ,
\end{eqnarray}
where $b_{IJ}(\xi)$ is symmetric and $I,J=1,\ldots,n-2$. They are
reflected by  the $(n-1)(n-2)/2$ Noether identities stating that
the contractions of the bosonic equations (\ref{vX}) with the
$u^{\alpha I}u^{\beta J}$ bilinears of the $(n-2)$ auxiliary
bosonic spinors $u^{\alpha I}(=C^{\alpha\beta}u^I_\beta)$ vanish
\footnote{In the massless $\Sigma^{(\frac{n(n+1)}{2}|n)}$
superparticle and tensionless super-$p$-brane models the
$b$-symmetry \cite{BL98,BLS99,B02} is $n(n-1)/2$ parametric. This
comes from the fact that such models contain a single bosonic
spinor ${\lambda}_\alpha$ and the non-trivial $b$-symmetry
variation is the general solution of the spinorial equation
 $\delta_b X^{\alpha\beta} {\lambda}_\alpha=0$.
In our tensionful superstring model with two bosonic spinors
$\lambda^{\pm}_\alpha(\xi)$, the $(n-1)(n-2)/2$ parametric
$b$-symmetry transformations (equation (\ref{b4})) are the
solutions of two equations
 $\delta_b X^{\alpha\beta} {\lambda}^+_\alpha=0$ and
 $\delta_b X^{\alpha\beta} {\lambda}^-_\alpha=0$.}.

The remaining gauge symmetries of the action (\ref{St}) are the
$SO(1,1)$ worldsheet Lorentz invariance
\begin{eqnarray}\label{sc1}
\kern-2em \delta X^{\alpha\beta}=0 ,  \quad  \delta\theta^\alpha
=0 , \quad \delta\lambda_\alpha^\pm=
\pm\omega(\delta)\lambda_\alpha^\pm , \quad  \delta e^{\pm\pm}=\pm
2\omega(\delta) e^{\pm\pm} ,
\end{eqnarray}
which is reflected by the fact that equation~(\ref{vl0}) is
satisfied identically when equations~(\ref{ve-}), (\ref{ve+}) are
taken into account,
 and the symmetry under worldvolume general coordinate transformations.

As customary in  string models, the general coordinate invariance
and the $SO(1,1)$ gauge symmetry allows one to fix locally the
conformal gauge where $e_m{}^a(\xi) = e^{\phi(\xi)} \delta_m^a$
or, equivalently
{\setlength\arraycolsep{0pt}
\begin{eqnarray}\label{cg}
&& e^{++}=  e^{\phi(\xi)} (d\tau + d\sigma)\; , \qquad e^{--}=
e^{\phi(\xi)} (d\tau - d\sigma)\; , \quad
\\ \label{cgc}
&& \Leftrightarrow \quad e^{++}_\sigma = e^{++}_\tau =
e^{\phi(\xi)} \; , \quad e^{--}_\sigma = - e^{--}_\tau = -
e^{\phi(\xi)} \; .\quad
\end{eqnarray}
}This indicates that it makes sense to consider the fields
$e_\sigma^{\pm\pm}(\tau,\sigma)$ as nonsingular (${1\over
e_\sigma^{\pm\pm}}= \pm e^{-\phi(\xi)}$ in the conformal gauge), a
fact used in the Hamiltonian analysis below.

According to the  correspondence \cite{BKOP97,ST97} between the
$\kappa$-symmetry of the worldvolume action and the supersymmetry
preserved by a BPS state ({\it e.g.} by a solitonic solution of
the supergravity equations of motion), the action (\ref{St})
defines a dynamical model for the excitations of a BPS state
preserving {\it all but two} supersymmetries. Such a BPS state can
be treated as a composite of two BPS preons ($\n = 32-30$). This
will be proved after the Hamiltonian analysis of next section.

\section{Hamiltonian analysis} \label{Hamiltonian}

The gauge symmetry structure of the model has already been shown
in the Lagrangian framework. However, our dynamical system
possesses additional, second class, constraints \cite{Dirac}, one
of which is condition~(\ref{l+l-}). The Hamiltonian analysis of
our $\Sigma^{({n(n+1)\over 2}|n)}$ superstring model \cite{30/32},
that we perform in this section, will allow us to find the number
of field theoretical degrees of freedom of our model and to
establish its relation with the notion of BPS preons \cite{BPS01}
(see section \ref{sec:BPSpreons} of chapter \ref{chapter4}).

The Lagrangian density ${\cal L}$ for the action (\ref{St}),
\begin{equation}\label{StL}
S  =  \int_{W^2} d\tau d\sigma \; {\cal L} \; ,
\end{equation}
is given by {\setlength\arraycolsep{2pt}
\begin{eqnarray}\label{St2}
{\cal L} & = &  (e^{++}_{\tau} \Pi_{\sigma}^{\alpha\beta} -
e^{++}_{\sigma} \Pi_{\tau}^{\alpha\beta} )
\lambda^-_{\alpha}\lambda^-_{\beta} -  (e^{--}_{\tau}
\Pi_{\sigma}^{\alpha\beta} - e^{--}_{\sigma}
\Pi_{\tau}^{\alpha\beta} ) \lambda^+_{\alpha}\lambda^+_{\beta}
\nonumber  \\
&&  -  (e^{++}_{\tau} e^{--}_{\sigma} - e^{++}_{\sigma}
e^{--}_{\tau}) \; ,
\end{eqnarray}
}where
\begin{eqnarray} \label{Pits}
\Pi_{\tau}^{\alpha\beta} =
\partial_{\tau}X^{\alpha\beta} -i\partial_{\tau}
\theta^{(\alpha} \theta^{\beta)} \; , \quad
 \Pi_{\sigma}^{\alpha\beta} =
\partial_{\sigma}X^{\alpha\beta} -i\partial_{\sigma}
\theta^{(\alpha} \theta^{\beta)} \; ,
\end{eqnarray}
are the worldsheet components of the one-form (\ref{Pi}).

The momenta $P_{{\cal M}}$ canonically conjugate to the
configuration space variables
\begin{eqnarray}\label{cZcM}
{\cal Z}^{{\cal M}}  \equiv {\cal Z}^{{\cal M}}(\tau, \sigma) :=
\left( X^{\alpha\beta} \, , \, \theta^\alpha  , \lambda^\pm_\alpha
\, , \, e^{\pm\pm}_\tau \, , \, e^{\pm\pm}_\sigma \right)\;
\end{eqnarray}
are defined as usual:
\begin{eqnarray}\label{cPcM}
P_{{\cal M}} =(P_{\alpha\beta}\,,\, \pi_\alpha \, , \, P_{\pm
}^{\alpha (\lambda)} , \, P_{\pm\pm}^{\tau} \,, \,
P_{\pm\pm}^{\sigma}) = {\partial {\cal L} \over \partial
(\partial_\tau {\cal Z}^{{\cal M}})} \; .\quad
\end{eqnarray}

The canonical equal $\tau$ graded Poisson brackets,
\begin{equation}
[{\cal Z}^{{\cal N}}(\sigma) \, , \, P_{{\cal M}}
(\sigma^\prime)\}_{_P} =  - (-1)^{{\cal N}{\cal M}} [ P_{{\cal M}}
(\sigma^\prime) \, , \, {\cal Z}^{{\cal N}}(\sigma ) \}_{_P} \; ,
\end{equation}
 are defined by
\begin{eqnarray}
\label{canonical} [ {\cal Z}^{{\cal N}}(\sigma^{\prime}) \, , \,
P_{{\cal M}}(\sigma ) \}_{_P} &:=& (-1)^{\cal N} \delta^{{\cal
N}}_{{\cal M}} \delta(\sigma-\sigma^{\prime})\; ,
\end{eqnarray}
where $(-1)^{\cal N} \equiv (-1)^{\mathrm{deg}({\cal N})}$ and the
degree $\mathrm{deg}({\cal N})\equiv \mathrm{deg}({\cal Z}^{\cal
N})$ is $0$ for the bosonic fields, ${\cal Z}^{\cal
N}=X^{\alpha\beta}, \lambda^\pm_\alpha, e_m^{\pm\pm}$ (or for the
`bosonic indices' ${\cal N}=(\alpha\beta), (\alpha\pm ), (\pm\pm),
m$), and $1$ for the fermionic fields ${\cal Z}^{\cal
N}=\theta^\alpha$ (or for the `fermionic indices'  ${\cal
N}=\alpha $ and ${\cal N}= \pm$).

Since the action (\ref{St}) is of first order type, it is not
surprising that the expression of every momentum results in a
primary \cite{Dirac} constraint. Explicitly,
{\setlength\arraycolsep{0pt}
\begin{eqnarray} \label{PX}
&& {\cal P}_{\alpha\beta}  =   P_{\alpha\beta} + e_{\sigma}^{++}
\lambda^-_\alpha \lambda^-_\beta -
e_{\sigma}^{--} \lambda^+_\alpha \lambda^+_\beta \approx 0 \; , \\
\label{D} && {\cal D}_\alpha  =   \pi_\alpha + i
\theta^\beta P_{\alpha\beta} \approx 0 \; , \\
\label{Plam}
&& P_{\pm}^{\alpha (\lambda)}  \approx  0 \; , \\
\label{Psig}
&& P_{\pm \pm}^{\sigma}  \approx  0 \; , \\
\label{Ptau} &&  P_{\pm\pm}^{\tau} \approx  0 \; ,
\end{eqnarray}
}where only ${\cal D}_\alpha$ is fermionic. Condition
(\ref{l+l-}),
\begin{eqnarray}\label{cl+l-}
{\cal N} := C^{\alpha\beta} \lambda^+_{\alpha}\lambda^-_{\beta}
 - 1 \approx 0 \; ,
\end{eqnarray}
imposed on the bosonic spinors from the beginning, is also a
primary constraint and has to be treated on the same footing as
equations~(\ref{PX})-(\ref{Ptau}).

The {\it canonical} Hamiltonian density ${\cal H}_0$,
\begin{equation}\label{H0}
{\cal H}_0  =  \partial_\tau {\cal Z}^{\cal M} \, P_{\cal M} -
{\cal L} \quad ,
\end{equation}
calculated on the primary constraints (\ref{PX})--(\ref{Ptau})
hypersurface reads
\begin{eqnarray}\label{H1}
\kern-.5em{\cal H}_0 =  e^{--}_{\tau} \Pi^{\alpha\beta}_{\sigma}
\lambda^+_\alpha \lambda^+_\beta -e^{++}_{\tau}
\Pi^{\alpha\beta}_{\sigma} \lambda^-_\alpha \lambda^-_\beta +
(e^{++}_{\tau} e^{--}_{\sigma} - e^{++}_{\sigma} e^{--}_{\tau}) \;
.
\end{eqnarray}
The evolution of any functional $f({\cal Z}^{\cal M}, P_{\cal N})$
is then defined by
\begin{equation}
\partial_\tau f = [ f\; , \; \int
d\sigma {\cal H}^\prime ]_{_P} \; ,
\end{equation}
involving the total Hamiltonian, $\int d\sigma {\cal H}^\prime$,
where the Hamiltonian density ${\cal H}^\prime$ is the sum of
${\cal H}_0$ in equation~(\ref{H1}) and the terms given by
integrals of the primary constraints (\ref{PX})-(\ref{Ptau})
multiplied by arbitrary functions (Lagrange multipliers)
\cite{Dirac}. Then one has to check that the primary constraints
are preserved under the evolution, $\partial_\tau {\cal
P}_{\alpha\beta} \approx 0$, etc. At this stage additional,
secondary constraints may be obtained. This is the case for our
system.

Indeed, since the constraints (\ref{Ptau}) have zero Poisson
brackets with any other primary constraint, their time evolution
is just determined by the canonical Hamiltonian ${\cal H}_0$,
$\partial_\tau{\cal P}_{\pm \pm}^{\tau}= [{\cal P}_{\pm
\pm}^{\tau},\int d \sigma {\cal H}_0 ]_{_P}$. Then
$\partial_\tau{\cal P}_{\pm \pm}^{\tau} \approx 0$ can be seen to
produce a pair of secondary constraints,
\begin{eqnarray} \label{phipm1}
\Phi_{\pm \pm} &:=&  \Pi_{\sigma}^{\alpha \beta}
\lambda^\mp_\alpha \lambda^\mp_\beta - e^{\mp\mp}_\sigma
\nonumber \\&=& (\partial_{\sigma} X^{\alpha \beta}-i
\partial_{\sigma}\theta^{(\alpha}\theta^{\beta)})\lambda^\mp_\alpha
\lambda^\mp_\beta - e^{\mp\mp}_\sigma \approx 0\; . \quad
\end{eqnarray}
Slightly more complicated calculations with the total ${\cal
H}^\prime$ show that we also have the secondary constraint
\begin{eqnarray} \label{phi0}
\Phi^{(0)} :=  \Pi_{\sigma}^{\alpha \beta} \lambda^+_\alpha
\lambda^-_\beta = (\partial_{\sigma} X^{\alpha \beta}-i
\partial_{\sigma}\theta^{(\alpha}\theta^{\beta)})\lambda^+_\alpha
\lambda^-_\beta \approx 0
\end{eqnarray}
(details about its derivation can be found below
equation~(\ref{contractions})). The appearance of this secondary
constraint may be understood as well by comparing with the results
of the Lagrangian approach: it is just the $\sigma$ component of
the differential form equation (\ref{P+-=0}).

The secondary constraints (\ref{phipm1}) imply that the canonical
Hamiltonian ${\cal H}_0$, equation~(\ref{H0}), vanishes on the
surface of constraints (\ref{phipm1}),
\begin{eqnarray}\label{H0=0}
{\cal H}_0 \approx 0  \, ,
\end{eqnarray}
a characteristic property of theories with general coordinate
invariance. Hence the total Hamiltonian reduces to a linear
combination of the constraints (\ref{PX})--(\ref{Ptau}),
(\ref{phipm1}), (\ref{phi0}), {\setlength\arraycolsep{2pt}
\begin{eqnarray}\label{H=} {\cal H}
&=& - e_\tau^{++} \Phi_{++} + e_\tau^{--} \Phi_{--} + l^{(0)}
\Phi^{(0)} + L^{\alpha\beta} {\cal P}_{\alpha\beta} + \xi^\alpha
{\cal D}_\alpha \nonumber \\
&& +  l^\pm_\alpha P^{\alpha (\lambda )}_{\pm} + L^{\pm\pm}
P_{\pm\pm}^{\sigma} + h^{\pm\pm} P_{\pm\pm}^\tau  + L^{(n)} {\cal
N}
\end{eqnarray}
}where $l^{(0)}$, $ L^{\alpha\beta}$, $\xi^\alpha$,
$l^\pm_\alpha$, $L^{\pm\pm}$, $h^{\pm\pm}$,  $L^{(n)}$ {\it and}
$\pm e_\tau^{\pm\pm}$ are Lagrangian multipliers whose form should
be fixed from the preservation of all the primary and secondary
constraints under $\tau$-evolution.

Note  that the constraints (\ref{Ptau}) are trivially first class,
since their Poisson brackets with all the other constraints,
including (\ref{phipm1}) and (\ref{phi0}), vanish. This allows us
to state  that $e_\tau^{\pm\pm}(\xi)$ are not dynamical fields but
rather Lagrange multipliers (as the time component of
electromagnetic potential $A_0$ in electrodynamics). Nevertheless,
the appearance of these Lagrange multipliers from the $\tau$
components of the zweibein $e_m^{\pm\pm}$ puts a `topological'
restriction on a possible gauge fixing; in particular the gauge
$e_\tau^{\pm\pm}=0$ is not allowed. Indeed, the nondegeneracy of
the zweibein, assumed from the beginning, reads
\begin{eqnarray}\label{det(e)}
\mathrm{det}\,(e_m^a(\xi)) \equiv \ft12 (e_\tau^{--}e_\sigma^{++}-
e_\tau^{++} e_\sigma^{--} )\not= 0 \; .
\end{eqnarray}
Just due to this restriction, studying the $\tau$-preservation of
the primary constraints, one finds the secondary constraint
(\ref{phi0}).

If by checking the (primary and secondary) constraints
preservation under $\tau$-evolution one finds that some lagrangian
multipliers remain unfixed, then they correspond to {\it first
class constraints} \cite{Dirac} which generate gauge symmetries of
the system through the Poisson brackets. In other words, since the
canonical Hamiltonian vanishes in the weak sense, the total
Hamiltonian is a linear combination of all first class constraints
\cite{Dirac}. If some of the equations  resulting from the
$\tau$-evolution of the constraints (or their linear combinations)
do not restrict the Lagrangian multiplier, but imply the vanishing
of a combination of the canonical variables, they correspond to
new secondary constraints, which have to be added with new
Lagrange multipliers to obtain a new total Hamiltonian. In this
case the check that all the constraints are preserved under
$\tau$--evolution has to be repeated.

This does not happen for our dynamical system: a further check of
the constraints $\tau$--preservation does not result in the
appearance of new constraints. Indeed, it leads to the following
set of equations for the Lagrange multipliers
{\setlength\arraycolsep{0pt}
\begin{eqnarray}
\label{dtP} && \partial_\sigma ( e_\tau^{--}\lambda^+_{\alpha}
\lambda^+_{\beta} - e_\tau^{++} \lambda^-_{\alpha}
\lambda^-_{\beta} + l^{(0)} \lambda^+_{(\alpha} \lambda^-_{\beta
)})  \nonumber \\ &&
 -2 e_\sigma^{--} \lambda^+_{(\alpha} l^+_{\beta)}
+2 e_\sigma^{++} \lambda^-_{(\alpha} l^-_{\beta)} + L^{++}
\lambda^-_{\alpha} \lambda^-_{\beta} - L^{--} \lambda^+_{\alpha}
\lambda^+_{\beta} \approx  0 \; , \\ && \nonumber  \\[0.5pt]
\label{dtD} && \lambda^-_{\alpha} \, [2i  e^{++}_{\tau}
(\partial_\sigma \theta \lambda^-) - i l^{(0)} (\partial_\sigma
\theta \lambda^+) + 2i e^{++}_{\sigma} (\xi\lambda^-)]\,  \quad
\nonumber  \\ &&
 - \lambda^+_{\alpha} [2i  e^{--}_{\tau} (\partial_\sigma  \theta
 \lambda^+) + i l^{(0)} (\partial_\sigma  \theta \lambda^-)
  + 2i e^{--}_{\sigma} (\xi\lambda^+)]\,
 \approx  0 \; , \\ && \nonumber \\ &&
\label{dtPl+} -2 e^{--}_{\tau} \Pi^{\alpha\beta}_\sigma
\lambda^+_{\beta} - l^{(0)}\Pi^{\alpha\beta}_\sigma
\lambda^-_{\beta} + 2
e^{--}_{\sigma}L^{\alpha\beta}\lambda^+_{\beta} -
 L^{(n)} C^{\alpha\beta}\lambda^-_{\beta}
\approx  0 \qquad
\\ && \nonumber \\ &&
\label{dtPl-} 2 e^{++}_{\tau} \Pi^{\alpha\beta}_\sigma
\lambda^-_{\beta} - l^{(0)}\Pi^{\alpha\beta}_\sigma
\lambda^+_{\beta} - 2
e^{++}_{\sigma}L^{\alpha\beta}\lambda^-_{\beta} - L^{(n)}
C^{\alpha\beta}\lambda^+_{\beta} \approx  0 \; , \\ && \nonumber
\\ &&
\label{dtPs++} e^{--}_{\tau} -
L^{\alpha\beta}\lambda^-_{\alpha}\lambda^-_{\beta} \approx 0 \; ,
 \\ && \nonumber
\\ && \label{dtPs--} e^{++}_{\tau} -
L^{\alpha\beta}\lambda^+_{\alpha}\lambda^+_{\beta} \approx  0 \; ,
\\ && \nonumber
\\ && \label{dtCll}
l^+_{\alpha} C^{\alpha\beta}\lambda^-_{\beta} - l^-_{\alpha}
C^{\alpha\beta}\lambda^+_{\beta} \approx 0 \; , \\ && \nonumber
\\ &&
\label{dtP++}
\partial_\sigma L^{\alpha\beta}  \lambda^-_{\alpha}  \lambda^-_{\beta}
+ 2i (\xi\lambda^-) (\partial_\sigma  \theta \lambda^-) + 2
l^-\Pi_\sigma \lambda^- - L^{--}  \approx  0 \; ,
\\ && \nonumber
\\ &&
\label{dtP--}
\partial_\sigma L^{\alpha\beta}  \lambda^+_{\alpha}  \lambda^+_{\beta}
+ 2i (\xi\lambda^+) (\partial_\sigma  \theta \lambda^+) + 2
l^+\Pi_\sigma \lambda^+ - L^{++} \approx  0 \; ,
\\ && \nonumber
\\ &&
\label{dtP0}
\partial_\sigma L^{\alpha\beta}  \lambda^+_{\alpha}  \lambda^-_{\beta}
+ i (\xi\lambda^+) (\partial_\sigma  \theta \lambda^-)
- i (\xi\lambda^-) (\partial_\sigma \theta \lambda^+) + \nonumber
\\ && + l^+\Pi_\sigma \lambda^- + l^-\Pi_\sigma \lambda^+ \approx  0 \;
,
\end{eqnarray}
}where the weak equality sign is used to stress that one may use
the constraints in solving the above system of equations. For
brevity, in equations~(\ref{dtP})--(\ref{dtP0}) and below we often
omit spinor indices in the contractions
{\setlength\arraycolsep{0pt}
\begin{eqnarray}\label{contractions}
&& (\partial_\sigma  \theta \lambda^\pm)\equiv
\partial_\sigma  \theta^{\beta} \, \lambda^\pm_{\beta}\; , \quad
(\xi \lambda^\pm)\equiv  \xi^{\beta} \, \lambda^\pm_{\beta}\; ,
\nonumber \\
&& l^\pm\Pi_\sigma \lambda^\pm\equiv l^\pm_\alpha
\Pi^{\alpha\beta}_\sigma \lambda^\pm_{\beta}\; , \quad
 l^\pm L \lambda^\pm\equiv
l^\pm_\alpha L^{\alpha\beta}_\sigma \lambda^\pm_{\beta}\; .
\end{eqnarray}

}

Note that equations~(\ref{dtP})--(\ref{dtCll}) come from the
requirement of $\tau$--preservation of the primary constraints,
while that for the secondary constraints leads to
equations~(\ref{dtP++})--(\ref{dtP0}). Thus the above statement
about the appearance of the secondary constraint (\ref{phi0}) can
be checked by studying equations~(\ref{dtP})--(\ref{dtCll}) with
$l^{(0)}=0$. In this case the contraction of
equation~(\ref{dtPl+}) with ($-\lambda_\alpha^-$) and of
equation~(\ref{dtPl-}) with $\lambda_\alpha^+$ results,
respectively, in the equations $e_\tau^{--} \lambda^+\Pi_\sigma
\lambda^- - e_\sigma^{--} \lambda^+ L \lambda^- \approx 0$ and
$e_\tau^{++} \lambda^+\Pi_\sigma \lambda^- - e_\sigma^{++}
\lambda^+ L \lambda^- \approx 0$. Due to the nondegeneracy of the
zweibein, equation~(\ref{det(e)}), the solution to these two
equations is trivial, {\it i.e.} it implies $\lambda^+ L \lambda^-
\approx 0$ and $\lambda^+\Pi_\sigma \lambda^-  \approx 0$, the
last of which is just the secondary constraint (\ref{phi0}).

To solve this system of equations for the Lagrange multipliers and
thus to describe explicitly the first class constraints, we can
use the auxiliary spinor fields $u_\alpha^I(\xi)$ defined as in
(\ref{cond1}), (\ref{cond2}).
The general solution of equations~(\ref{dtP})--(\ref{dtP0})
obtained in such a framework reads {\setlength\arraycolsep{2pt}
\begin{eqnarray}
\label{Lab=A}  L^{\alpha\beta} &=& b_{IJ}u^{\alpha I} u^{\beta J}
 \nonumber \\ &&  + {e^{++}_{\tau}\over e^{++}_{\sigma}} \Big[
e^{++}_{\sigma} \lambda^{-\alpha}\lambda^{-\beta} + 2
\Big(\lambda^-_\gamma \Pi_{\sigma}^{\gamma(\alpha}
\lambda^{+\beta)}
-(\lambda^-\Pi_\sigma
\lambda^+)\lambda^{-(\alpha}\lambda^{+\beta)} \nonumber
\\ &&   \qquad \quad  +(\lambda^-\Pi_\sigma
\lambda^-)\lambda^{+(\alpha}\lambda^{+\beta)}
  \Big)\Big]\,
\nonumber \\ &&   + {e^{--}_{\tau} \over e^{--}_{\sigma}}
\Big[e^{--}_{\sigma}\lambda^{+\alpha}\lambda^{+\beta} -2
\Big(\lambda^+_\gamma \Pi_{\sigma}^{\gamma(\alpha}
\lambda^{-\beta)}
-(\lambda^+\Pi_\sigma
\lambda^+)\lambda^{-(\alpha}\lambda^{-\beta)} \nonumber
\\ &&  \qquad \quad
+(\lambda^+\Pi_\sigma
\lambda^-)\lambda^{+(\alpha}\lambda^{-\beta)}
  \Big)\Big]\, ,  \\ {} \nonumber \\
\label{xi=A} \xi^{\alpha} &=& \kappa_I \, u^{\alpha I} +
\frac{e^{++}_{\tau}}{e^{++}_{\sigma}} (\partial_{\sigma} \theta
\lambda^-) \lambda^{+ \alpha} -
\frac{e^{--}_{\tau}}{e^{--}_{\sigma}} (\partial_{\sigma} \theta
\lambda^+) \lambda^{- \alpha} \; , \\ {} \nonumber  \\
 \label{l+=A}
 l^+_{\alpha} &=&  \omega^{(0)}\lambda_{\alpha}^{+} +
{e^{--}_{\tau} \over e^{--}_{\sigma}}
\left(\partial_{\sigma}\lambda^+_\alpha - \Omega^{(0)}_\sigma
\lambda^+_\alpha\right)  \nonumber \\ &&  + { e^{--}_{\tau} \over
2e^{--}_{\sigma}e^{--}_{\sigma}} \Big[-e^{--}_{\sigma}
\Omega^{++}_{\sigma} - e^{++}_{\sigma} \Omega^{--}_{\sigma} +
i\partial_{\sigma}\theta \lambda^+\partial_{\sigma}\theta
\lambda^-
\nonumber \\ &&  \qquad \qquad \qquad -\Pi^{\alpha\beta}_\sigma
(\partial_{\sigma}\lambda^+_{\alpha} \lambda^-_{\beta} -
\lambda^+_{\alpha}
\partial_{\sigma}\lambda^-_{\beta}) \Big] \lambda^-_\alpha
\nonumber \\  &&  + { e^{++}_{\tau} \over
2e^{++}_{\sigma}e^{--}_{\sigma}} \Big[e^{--}_{\sigma}
\Omega^{++}_{\sigma} + e^{++}_{\sigma} \Omega^{--}_{\sigma} +
i\partial_{\sigma}\theta \lambda^+\partial_{\sigma}\theta
\lambda^-
\nonumber \\ &&  \qquad \qquad \qquad +\Pi^{\alpha\beta}_\sigma
(\partial_{\sigma}\lambda^+_{\alpha} \lambda^-_{\beta} -
\lambda^+_{\alpha}
\partial_{\sigma}\lambda^-_{\beta}) \Big] \lambda^-_\alpha \; , \\
\label{l-=A} l^-_{\alpha} &=& -\omega^{(0)}\lambda_{\alpha}^{-} +
{e^{++}_{\tau} \over e^{++}_{\sigma}}
\left(\partial_{\sigma}\lambda^-_\alpha + \Omega^{(0)}_\sigma
\lambda^-_\alpha\right) + \nonumber \\
&& \quad +  { e^{--}_{\tau} \over 2e^{++}_{\sigma}e^{--}_{\sigma}}
\Big[ -e^{--}_{\sigma} \Omega^{++}_{\sigma} - e^{++}_{\sigma}
\Omega^{--}_{\sigma} + i\partial_{\sigma}\theta
\lambda^+\partial_{\sigma}\theta
\lambda^- -  \nonumber \\
&&    \qquad \qquad \qquad \qquad - \Pi^{\alpha\beta}_\sigma
(\partial_{\sigma}\lambda^+_{\alpha} \lambda^-_{\beta} -
\lambda^+_{\alpha}
\partial_{\sigma}\lambda^-_{\beta}) \Big]\lambda^+_\alpha + \nonumber \\
&& \quad + {e^{++}_{\tau} \over 2e^{++}_{\sigma}e^{++}_{\sigma}}
\Big[ e^{--}_{\sigma} \Omega^{++}_{\sigma} + e^{++}_{\sigma}
\Omega^{--}_{\sigma} + i\partial_{\sigma}\theta
\lambda^+\partial_{\sigma}\theta \lambda^- +  \nonumber \\
&&  \qquad \qquad \qquad \qquad + \Pi^{\alpha\beta}_\sigma
(\partial_{\sigma}\lambda^+_{\alpha} \lambda^-_{\beta} -
\lambda^+_{\alpha} \partial_{\sigma}\lambda^-_{\beta}) \Big]
\lambda^+_\alpha \; ,
\\ && \nonumber \\
\label{Lpmpm=A}  L^{\pm\pm} &=&
\partial_{\sigma}e^{\pm \pm}_{\tau} + 2e^{\pm \pm}_\tau
\Omega^{(0)}_\sigma \pm 2e^{\pm \pm}_{\sigma} \omega^{(0)} \; , \\
\nonumber \\  \label{L(n)=A}  L^{(n)} &=& - 4 \mathrm{det}(e_m^a)
\equiv -2 (e_\tau^{--}e_\sigma^{++} - e_\tau^{++}e_\sigma^{--} )
\; ,\\ \nonumber \\ \label{l0=A} l^{(0)} &=& 0\; ,
\end{eqnarray}
}where, $\Omega_\sigma^{\pm\pm}$ and $\Omega_\sigma^{(0)}$ ({\it
cf.} equations (\ref{vlpm})) are given by
{\setlength\arraycolsep{0pt}
\begin{eqnarray}\label{Ompm=}
&& \Omega_\sigma^{++} := \partial_{\sigma} \lambda^+C \lambda^+\;
, \qquad \Omega_\sigma^{--} :=\partial_{\sigma} \lambda^-C
\lambda^- \; ,
\\ \label{Om0=} && \Omega_\sigma^{(0)} :=
\frac{1}{2}(\partial_{\sigma} \lambda^+C \lambda^- - \lambda^+C
\partial_{\sigma} \lambda^-) \; .
\end{eqnarray}
}In this solution the parameters {\setlength\arraycolsep{0pt}
\begin{eqnarray}
\label{paramb} && \mathrm{bosonic}\, : \quad   b^{IJ}=b^{JI}\, ,
\quad  \omega^{(0)} \, , \quad  e_\tau^{\pm\pm}\, , \quad
h^{\pm\pm}\, ,
\\ \label{paramf}
&& \mathrm{fermionic} \, :  \quad  \kappa_{I}\; ,
\end{eqnarray}
}are indefinite. They correspond to the first class constraints
{\setlength\arraycolsep{1pt}
\begin{eqnarray}\label{PIJ}
 {\cal P}^{IJ} & :=& {\cal P}_{\alpha\beta} u^{\alpha I} u^{\beta
J} \approx  0 \; ,
\\ \label{DI}
 {\cal D}^I & := & {\cal D}_\alpha u^{\alpha I} \approx 0 \; ,
\\ \label{G0}
 G^{(0)}& := & \lambda^+_\alpha P^{\alpha (\lambda)}_+ -
\lambda^-_\alpha P^{\alpha (\lambda)}_- + 2 e_\sigma^{++}
P^\sigma_{++} - 2 e_\sigma^{--} P^\sigma_{--} \approx 0 \; , \\
&& \nonumber \\ \label{phi11} \tilde{\Phi}_{++} & = & {\Phi}_{++}
+\partial_{\sigma} P_{++}^{\sigma}
 -2\Omega_{\sigma}^{(0)} P_{++}^{\sigma} -2e^{--}_{\sigma} {\cal
 N} \nonumber \\
 && -\frac{1}{e^{++}_{\sigma}}
 (\partial_{\sigma} \lambda^-_{\alpha} +
 \Omega_{\sigma}^{(0)}\lambda_{\alpha}^-)P_-^{\alpha(\lambda)} -
 \frac{1}{e^{++}_{\sigma}}(\partial_{\sigma}\theta
\lambda^-)(\lambda^{+\alpha} {\cal D_{\alpha}}) - \nonumber \\ &&
-\Big[ \lambda^{-\alpha}\lambda^{-\beta} +
\frac{2}{e^{++}_{\sigma}} \Big(\lambda^-_{\gamma}
\Pi_{\sigma}^{\gamma \alpha} \lambda^{+\beta}
 -(\lambda^-\Pi_{\sigma}\lambda^+)\lambda^{-\alpha}\lambda^{+\beta}
 \nonumber
\\ && \qquad   +(\lambda^-\Pi_{\sigma}\lambda^-)\lambda^{+\alpha}\lambda^{+\beta}
\Big) \Big] {\cal P}_{\alpha \beta}  \nonumber \\ &&
-\frac{1}{2e^{++}_{\sigma}} \Big[e^{--}_{\sigma}
\Omega^{++}_{\sigma} + e^{++}_{\sigma} \Omega^{--}_{\sigma} +
i\partial_{\sigma}\theta \lambda^+\partial_{\sigma}\theta
\lambda^-  \nonumber \\ &&  +\Pi^{\alpha\beta}_\sigma
(\partial_{\sigma}\lambda^+_{\alpha} \lambda^-_{\beta} -
\lambda^+_{\alpha}
\partial_{\sigma}\lambda^-_{\beta}) \Big] \left[ \frac{\lambda_{\alpha}^- P_+^{\alpha
(\lambda)}}{e^{--}_{\sigma}} + \frac{\lambda_{\alpha}^+
P_-^{\alpha (\lambda)}}{e^{++}_{\sigma}} \right]
\\ && \nonumber \\
\label{phi21} \tilde{\Phi}_{--} &:=& {\Phi}_{--}
-\partial_{\sigma} P_{--}^{\sigma}
 +2\Omega_{\sigma}^{(0)} P_{--}^{\sigma} -2e^{++}_{\sigma} {\cal N}
\nonumber \\ && -\frac{1}{e^{--}_{\sigma}}(\partial_{\sigma}\theta
\lambda^+)(\lambda^{-\alpha} {\cal D_{\alpha}})
+\frac{1}{e^{--}_{\sigma}}
 (\partial_{\sigma} \lambda^+_{\alpha} +
\Omega_{\sigma}^{(0)}\lambda_{\alpha}^+)P_+^{\alpha(\lambda)} +
\nonumber \\ && +\Big[ \lambda^{+\alpha}\lambda^{+\beta} -
\frac{2}{e^{--}_{\sigma}} \big(\lambda^+_{\gamma}
\Pi_{\sigma}^{\gamma \alpha} \lambda^{-\beta}
-(\lambda^+\Pi_{\sigma}\lambda^+)\lambda^{-\alpha}\lambda^{-\beta}
+ \nonumber
\\ && \qquad  +(\lambda^+\Pi_{\sigma}\lambda^-)\lambda^{+\alpha}\lambda^{-\beta}
\big) \Big] {\cal P}_{\alpha \beta}  \nonumber \\ &&
+\frac{1}{2e^{--}_{\sigma}} \Big[-e^{--}_{\sigma}
\Omega^{++}_{\sigma} - e^{++}_{\sigma} \Omega^{--}_{\sigma} +
i\partial_{\sigma}\theta \lambda^+\partial_{\sigma}\theta
\lambda^-  \nonumber \\ &&  -\Pi^{\alpha\beta}_\sigma
(\partial_{\sigma}\lambda^+_{\alpha} \lambda^-_{\beta} -
\lambda^+_{\alpha}
\partial_{\sigma}\lambda^-_{\beta}) \Big] \left[ \frac{\lambda_{\alpha}^-
P_+^{\alpha (\lambda)}}{e^{--}_{\sigma}} +
\frac{\lambda_{\alpha}^+ P_-^{\alpha (\lambda)}}{e^{++}_{\sigma}}
\right]
\end{eqnarray}
}
and
\begin{eqnarray}\label{Ptau1}
P_{\pm \pm}^{\tau} &\approx& 0 \; .
\end{eqnarray}
In equations~(\ref{phi11}), (\ref{phi21}) the relation
{\setlength\arraycolsep{0pt}
\begin{eqnarray}\label{I=}
&& \delta_\alpha{}^\beta \approx \lambda^+_\alpha\lambda^{-\beta}
- \lambda^-_\alpha\lambda^{+\beta} - u^I_\alpha u^{J\beta}C_{IJ}
\; ,
\\ \label{CL=}
&& \lambda^{\pm\beta}:= C^{\beta\alpha}\lambda^\pm_\alpha\; ,
\quad u^{I\beta}:= C^{\beta\alpha}u^I_\alpha\; ,
\end{eqnarray}
}has been used to remove the auxiliary variables $u^I_\alpha$ in
all places where it is possible. Note that (\ref{I=}) is a
consequence of the constraint (\ref{cl+l-}) and of the definition
of the $u^I_\alpha$ spinors, equations~(\ref{cond1}),
(\ref{cond2}) (see further discussion on the use of $u$ variables
below). Thus we are allowed to use them in the solution of the
equation for the Lagrange multipliers and, then, in the definition
of the first class constraints, as the product of any two
constraints is a first class one since its Poisson brackets with
any other constraint vanishes weakly.

Using the Poisson brackets (\ref{canonical}), the first class
constraints generate gauge symmetries. In our dynamical system the
fermionic first class constraints (\ref{DI}) are the generators of
the $(n-2)$--parametric $\kappa$--symmetry
(\ref{kappa1})--(\ref{kappa3}). The ${\cal P}^{IJ}$ in
equation~(\ref{PIJ}) are the ${1 \over 2}(n-1)(n-2)$ generators of
the $b$-symmetry (\ref{b4}). The constraint $G^{(0)}$ (\ref{G0})
generates the $SO(1,1)$ gauge symmetry (\ref{SO(1,1)}).  Finally,
the constraints $\tilde{\Phi}_{\pm\pm}$, equations~(\ref{phi11}),
(\ref{phi21}), generate worldvolume reparameterizations. They
provide a counterpart of the Virasoro constraints characteristic
of the Green--Schwarz superstring action. Thus, as it could be
expected, our $\Sigma^{(\frac{n(n+1)}2|n)}$ superstring is a
two-dimensional conformal field theory. As it was noted above, the
presence of the first class constraints
 (\ref{Ptau1}) indicates the pure gauge nature of the fields
$e_\tau^{\pm\pm}(\xi)$; the freedom of the gauge fixing is,
nevertheless, restricted by the `topological' conditions
(\ref{det(e)}).

Note that the $\kappa$--symmetry and $b$--symmetry generators, in
(\ref{DI}) and (\ref{PIJ}), are the $u_{\alpha}^I$ and
 $u_{\alpha}^I u_{\beta}^J$ components of
equation~(\ref{D})  and equation~(\ref{PX}), respectively, while
all other first class constraints can be defined without any
reference to auxiliary variables. The use of the auxiliary spinors
$u_\alpha^I(\xi)$ to define the first class constraints requires
some discussion. Any spinor can be decomposed in the basis
(\ref{basis}), but the use of $u_\alpha^I$ to define constraints
requires, to be rigorous, to consider them as (auxiliary)
dynamical variables, to introduce momenta, and to take into
account any additional constraints for them, including
equations~(\ref{cond2}) and the vanishing of the momenta conjugate
to $u_\alpha^I$ ({\it cf.}~\cite{BZstr}).

An alternative is to consider these auxiliary spinors as defined
by (\ref{cond1}), (\ref{cond2}) and by the gauge symmetries of
these constraints, {\it i.e.} to treat them as some implicit
functions of $\lambda^\pm_\alpha$ ({\it cf.}~\cite{Grassi}). Such
a description can be obtained rigorously by the successive gauge
fixings of all the additional gauge symmetries that act only on
$u_\alpha^I$ and by introducing Dirac brackets accounting for all
the second class constraints for the $u_\alpha^I$ variables.
Nevertheless, with some precautions, the above simpler alternative
can be used from the beginning. In this case, one has to keep in
mind, in particular, that the $u_\alpha^I$'s do not commute with
$P_{\pm}^{\alpha (\lambda)}$. Indeed, as conditions (\ref{cond1})
have to be treated in a strong sense, one has to assume
$[P_{\pm}^{\alpha (\lambda)}(\sigma), u^I_\beta (\sigma^\prime
)]_P \approx \pm \lambda^\pm_\beta C^{\alpha\gamma} u_\gamma^I
\delta (\sigma - \sigma^\prime)$. However, one notices that this
does not change the result of the analysis of the number of first
and second class constraints among
equations~(\ref{PX})--(\ref{cl+l-}), (\ref{phipm1}), (\ref{phi0}),
which do not involve $u_\gamma^I(\xi)$. The reason is that one
only uses $u_\gamma^I(\xi)$ as multipliers needed to extract the
first and second class constraints from the mixed ones (\ref{PX}),
(\ref{D}). Thus, the Poisson brackets of the projected constraints
${\cal P}_{\alpha\beta} u^{\alpha I} u^{\beta J}$, $u^{\alpha I}
{\cal D}_\alpha$ with other constraints ({\it e.g.}, $[{\cal
P}_{\alpha\beta} u^{\alpha I} u^{\beta J}\, , \ldots ]_P$) and the
projected Poisson brackets of the original constraints ${\cal
P}_{\alpha\beta}$, ${\cal D}_\alpha$ with the same ones ({\it
e.g.}, $u^{\alpha I} u^{\beta J} \,[{\cal P}_{\alpha\beta} \, ,
\ldots ]_P$) are equivalent in the sense that a non-zero
difference ($[{\cal P}_{\alpha\beta} u^{\alpha I} u^{\beta J}\, ,
\ldots ]_P\;-\, u^{\alpha I} u^{\beta J} \,[{\cal P}_{\alpha\beta}
\, , \ldots ]_P$) will be proportional to ${\cal P}_{\alpha\beta}$
or ${\cal D}_\alpha$ and, hence, will vanish weakly. This
observation allows us to use the basis (\ref{basis}) to solve the
equations (\ref{dtP})--(\ref{dtCll}), that is to say, to decompose
the constraints (\ref{PX})--(\ref{cl+l-}), (\ref{phipm1}),
(\ref{phi0}) into first and second class ones, without introducing
momenta for the $u_\gamma^I(\xi)$ and without studying the
constraints restricting these variables.

The remaining constraints are second class. In particular, these
are the $\lambda^{\pm}$ components of the fermionic constraints
(\ref{D}),
\begin{equation} \label{Dpm}
{\cal D}^\pm = {\cal D}_\alpha \lambda^{\pm\alpha} = \pi_\alpha
\lambda^{\pm\alpha} + i e_\sigma^{\pm\pm} \theta^\beta
\lambda_\beta^\mp \approx 0 \;
\end{equation}
with Poisson brackets {\setlength\arraycolsep{0pt}
\begin{eqnarray}\label{DpmDpm}
&& \{ {\cal D}^+(\sigma) , {\cal D}^+ (\sigma^\prime )\}_P \approx
+2i e_\sigma^{++}  \delta (\sigma -\sigma^\prime )\; ,
\nonumber \\
&& \{ {\cal D}^+(\sigma) , {\cal D}^+ (\sigma^\prime )\}_P \approx
-2i e_\sigma^{--}  \delta (\sigma -\sigma^\prime )\; ,
\nonumber \\
&& \{ {\cal D}^+(\sigma) , {\cal D}^+ (\sigma^\prime )\}_P \approx
0 \;
\end{eqnarray}
}(recall that, having in mind the possibility of fixing the
conformal gauge (\ref{cg}), we assume nondegeneracy of
$e_\sigma^{\pm\pm}(\sigma)$, {\it i.e.} that the expression
$1/e_\sigma^{\pm\pm}(\sigma)$ is well defined). The selection of
the basic second class constraints and the simplification of their
Poisson bracket algebra is a technically involved problem.

In the next section we show that the dynamical degrees of freedom
of our superstring in $\Sigma^{({n(n+1)\over 2}|n)}$, may be
presented in a more economic way in terms of constrained
$OSp(2n|1)$ supertwistors. The Hamiltonian mechanics also
simplifies in this symplectic supertwistor formulation. In
particular, all the first class constraints can be extracted
without using the auxiliary fields $u_\alpha^I$. The reason is
that the supertwistor variables are invariant under both
$\kappa$-- and $b$--symmetry. Thus, moving to the twistor form of
our action means rewriting it in terms of trivially $\kappa$-- and
$b$--invariant quantities, effectively removing all variables that
transform non-trivially under these gauge symmetries. Since the
description of $\kappa$-- and $b$--symmetries is the one requiring
the introduction of the $u_\alpha^I(\xi)$ fields, it is natural
that these are not needed in the supertwistor Hamiltonian
approach.

This consideration already allows us to calculate the number of
the (field theoretical worldsheet) degrees of freedom of our
superstring model \cite{30/32}. The dynamical system described by
the action (\ref{St}) possesses ${1 \over 2}(n-1)(n-2)+5$ bosonic
first class constraints (equations (\ref{PIJ}), (\ref{G0}),
(\ref{phi11}), (\ref{phi21}) and (\ref{Ptau1})) out of a total
number of $\frac12 n(n+1)+2n+8$ constraints (equations~(\ref{PX}),
(\ref{Plam}), (\ref{Psig}), (\ref{Ptau}), (\ref{cl+l-}),
(\ref{phipm1}) and (\ref{phi0})). This leaves $4n+2$ bosonic
second class constraints. Since the phase space dimension
corresponding to the worldvolume bosonic fields ${\cal Z}^{\cal
M}(\tau, \sigma)=(X^{\alpha \beta}, \lambda^{\pm}_{\alpha},$
$e^{\pm\pm}_{\sigma}, e^{\pm \pm}_{\tau})$ is $2({n(n+1)\over 2}
+2n+4)$, the action (\ref{St}) turns out to have $(4n-6)$ bosonic
degrees of freedom.

Likewise, the $(n-2)$ fermionic first class constraints (\ref{DI})
and the $2$ fermionic second class constraints,
equations~(\ref{Dpm}), reduce the original $2n$ phase space
fermionic degrees of freedom of the action (\ref{St}) down to $2$.

Thus our supersymmetric string model in $\Sigma^{({n(n+1)\over
2}|n)}$ superspace carries $(4n-6)$ bosonic and $2$ fermionic
worldvolume field theoretical degrees of freedom. Treating  the
number $n$ as the number of components of an irreducible spinor
representation of the $D$-dimensional Lorentz group $SO(1,D-1)$,
one finds \cite{30/32}

\bigskip
\begin{center}
\begin{tabular}
{|c|c|c|c|c|} \hline  $D$  &  $n$ & \, $\#_{bosonic \; d.o.f.} $ &
\, $\#_{fermionic \; d.o.f.}$ & \, BPS \; \cr
 &   & $ = 4n-6 $ &  $ =2$ & \,
states \; \cr \hline  3 & 2 & 2 & 2 & NO \cr \hline 4 & 4 & 10 & 2
& $1/2$ \cr \hline  6 & 8 & 26 & 2 & $6/8$ \cr \hline  10 & 16 &
58 & 2 & $14/16$ \cr \hline  11 & 32 & 122 & 2 & $30/32$ \cr
\hline
\end{tabular}
\\[6pt]
{\footnotesize Table 7.1. Bosonic and fermionic degrees  of
freedom \\ of the model in various dimensions}
\end{center}

\noindent Thus, the number of bosonic degrees of freedom of our
$\Sigma^{({n(n+1)\over 2}|n)}$ superstring model exceeds that of
the Green--Schwarz superstring (where it exists, $4n-6 > 2(D-2)$),
while  the number of fermionic dimensions, $2$, is smaller than
that of the Green--Schwarz superstring for $D=6,10$. Note that
here the $\#(\textrm{bosons}) = \#(\textrm{fermions})$ rule is not
valid. The additional bosonic degrees of freedom might be treated
as higher spin degrees of freedom and/or as corresponding to the
additional `brane' central charges in the maximal supersymmetry
algebra (\ref{QQP}). The smaller number of physical fermionic
degrees of freedom just reflects the presence of extra
$\kappa$--symmetries ($(n-2) > n/2$ for $n>4$) in our
 $\Sigma^{(528|32)}$ superstring model.
Our $\Sigma^{({n(n+1)\over 2}|n)}$ superstring model describes, as
argued, the excitations of a BPS state preserving $k=(n-2)$
supersymmetries (a ${30\over 32}$ BPS state for the $D=11$
superstring in $\Sigma^{(528|32)}$).

The search for solitonic solutions of the usual $D= 11$ and $D=10$
Type II  supergravities preserving exotic fractions of
supersymmetry is a subject of recent interest. If successful, it
would be interesting to study how the additional bosonic degrees
of freedom of our model are mapped into the moduli of these
solutions, presumably related to the gauge fields of the
supergravity multiplet ({\it cf.}~\cite{JdA00}). Nevertheless, if
it were shown that such solutions do not appear in the standard
$D=11$ supergravity, this could indicate that M Theory does
require an extension of the usual superspace for its adequate
description.

To conclude this section we comment on the BPS preon
interpretation of our model. In agreement with \cite{BPS01}, it
can be argued to describe a composite of $\n=n-k=2$ BPS preons. To
support this conclusion one can have a look at the constraint
(\ref{PX}). As we have shown, it is a mixture of first and second
class constraints. However, performing a `conversion' of the
second class constraints \cite{conversion} to obtain first class
constraints (in a way similar to the one carried out for a
point--like model in \cite{BLS99}), one arrives at the first class
constraint
\begin{eqnarray} \label{I-PX}
{\cal P}_{\alpha\beta}  =   P_{\alpha\beta} + e_{\sigma}^{++}
\tilde{\lambda}^-_\alpha \tilde{\lambda}^-_\beta - e_{\sigma}^{--}
\tilde{\lambda}^+_\alpha \tilde{\lambda}^+_\beta \approx 0 \; ,
\end{eqnarray}
where the $\tilde{\lambda}^\pm_\alpha$ are related to
$\lambda^\pm_\alpha$. In the quantum theory this first class
constraint can be imposed on quantum states giving rise to a
relation similar to equation~(\ref{npreon}) with $\n=n-k=2$.

\section{Supertwistor form of the action} \label{twistor}

Further analysis of the Hamiltonian mechanics of our
$\Sigma^{({n(n+1)\over 2}|n)}$ superstring model would become
quite involved. Instead, we present in this section a more
economic description of the system.

The action (\ref{St}) can be rewritten $(\alpha^\prime=1)$ in the
form \cite{30/32} {\setlength\arraycolsep{2pt}
\begin{eqnarray}\label{St1}
S &=&  \int_{W^2}  [e^{++} \wedge (d\mu^{-\alpha}
\lambda^-_{\alpha} - \mu^{-\alpha} d\lambda^-_{\alpha} - i d\eta^-
\eta^-)  \nonumber
\\
&& \qquad  - e^{--} \wedge (d\mu^{+\alpha}  \lambda^+_{\alpha} -
\mu^{+\alpha}  d\lambda^+_{\alpha} - i d\eta^+ \eta^+) \nonumber
\\ && \qquad - e^{++} \wedge e^{--}] \; ,
\end{eqnarray}
}where the bosonic $\mu^{\pm\alpha}$ and the fermionic
$\eta^{\pm}$ are defined by
\begin{equation}
\label{mu+}  \mu^{\pm\alpha} = X^{\alpha\beta}
\lambda^{\pm}_{\beta}- {i\over 2} \theta^{\alpha} \theta^{\beta}
\lambda^{\pm}_{\beta} \; , \quad \eta^{\pm} = \theta^{\beta}
\lambda^{\pm}_{\beta} \; .
\end{equation}
Equations~(\ref{mu+}) are reminiscent of the Ferber generalization
\cite{Ferber} of the Penrose correspondence relation \cite{Pen}
(see also \cite{BPS01,BL98}). The two sets of $2n+1$ variables
belonging to the same real one-dimensional (Majorana--Weyl spinor)
representation of the worldsheet Lorentz group $SO(1,1)$,
\begin{eqnarray} \label{Ypm}
(\mu^{+\alpha}, \lambda_{\alpha}^{+}, \eta^+) := Y^{+\Sigma} \;\;
, \;\;  (\mu^{-\alpha}, \lambda_{\alpha}^{-}, \eta^-) :=
Y^{-\Sigma}\; , \quad
\end{eqnarray}
 can be treated as the components
of two $OSp(2n|1)$ supertwistors, $Y^{+ \Sigma}$ and $Y^{-
\Sigma}$. However, equations~(\ref{mu+}) considered together imply
the following constraint:
\begin{eqnarray}\label{Con}
\lambda^+_{\alpha}\mu^{-\alpha} - \lambda^-_{\alpha} \mu^{+\alpha}
 - i \eta^- \eta^+ =0 \; .
\end{eqnarray}
One has to consider as well the `kinematic' constraint
(\ref{l+l-}), which breaks $GL(n,\mathbb{R})$ down to
$Sp(n,\mathbb{R})$. In terms of the two supertwistors $Y^{\pm
\Sigma}$ the action (\ref{St}) describing our tensionful string
model and the constraints (\ref{l+l-}), (\ref{Con}) can be written
as follows\footnote{See \cite{BarPi06} for a recent construction
of massive particle actions in terms of only one supertwistor.}
{\setlength\arraycolsep{2pt} \cite{30/32}
\begin{eqnarray}\label{St002}
S &=& \int_{W^2} [e^{++} \wedge
dY^{-\Sigma}\, \Omega_{\Sigma\Pi} Y^{-\Pi}\,   \nonumber \\
&& - e^{--} \wedge dY^{+\Sigma}\, \Omega_{\Sigma\Pi} Y^{+\Pi}\ -
\, e^{++} \wedge e^{--}] \; ; \qquad
\end{eqnarray}
\begin{eqnarray}
 \label{l+l-00}
 Y^{+\Sigma}\, C_{\Sigma\Pi} Y^{-\Pi} & = & 1 \; , \\
 \label{Con00}   Y^{+\Sigma}\, \Omega_{\Sigma\Pi} Y^{-\Pi} & = & 0
\; ,
\end{eqnarray}
}where the nondegenerate matrix $\Omega_{\Sigma\Pi} =
-(-1)^{\mathrm{deg}({\pm\Sigma}) \mathrm{deg}({\pm\Pi})}
\Omega_{\Pi \Sigma}$ is the orthosymplectic metric,
\begin{eqnarray}
\label{OmLP} \Omega_{\Sigma\Pi} = \left( \begin{matrix} 0 &
\delta_\alpha{}^\beta & 0 \cr - \delta_\beta{}^\alpha & 0 & 0 \cr
0 & 0 & -i \end{matrix} \right) \quad ,
\end{eqnarray}
preserved by $OSp(2n|1)$. The degenerate matrix $C_{\Sigma\Pi}$ in
equation~(\ref{l+l-00}) has the form
\begin{eqnarray} \label{CLP}
 C_{\Sigma\Pi} = \left( \begin{matrix} 0 & 0 & 0 \cr
0 & C^{\alpha\beta} & 0 \cr 0 & 0 & 0 \end{matrix} \right) \quad
\end{eqnarray}
with $C^{\alpha\beta}$ defined in (\ref{Cab}).

One can also find the orthosymplectic twistor form for the action
(\ref{St1}) with unconstrained spinors. It reads \cite{30/32}
{\setlength\arraycolsep{2pt}
\begin{eqnarray}\label{St3}
S &=& \int_{W^2}  [e^{++} \wedge (d{\cal M}^{-\alpha}
\Lambda^-_{\alpha} - {\cal M}^{-\alpha}  d\Lambda^-_{\alpha} - i
d\chi^- \chi^-)   \hspace{-1cm} \nonumber
\\
& & \qquad -e^{--} \wedge (d{\cal M}^{+\alpha}  \Lambda^+_{\alpha}
- {\cal M}^{+\alpha}  d\Lambda^+_{\alpha} - i d\chi^+ \chi^+)
\nonumber \hspace{-1cm}
\\
&& \qquad - e^{++} \wedge e^{--} (C^{\alpha\beta}
\Lambda^+_{\alpha}\Lambda^-_{\beta})^2] \; ,
\end{eqnarray}
}where
\begin{equation}\label{Mu+}
{\cal M}^{\pm\alpha} = X^{\alpha\beta} \Lambda^{\pm}_{\beta}-
{i\over 2} \theta^{\alpha} \theta^{\beta} \Lambda^{\pm}_{\beta} \;
, \quad \chi^{\pm} = \theta^{\beta} \Lambda^{\pm}_{\beta} \; .
\end{equation}
Equation~(\ref{Mu+}) differs from (\ref{mu+}) only by replacement
of the constrained dimensionless $\lambda^\pm$ by the
unconstrained dimensionful  $\Lambda^\pm$. But, as a result, the
$OSp(2n|1)$ supertwistors
\begin{eqnarray}
\label{Ups}  \Upsilon^{\pm \Sigma} := ({\cal M}^{\pm\alpha},
\Lambda^{\pm}_{\alpha}, \chi^{\pm})\; ,
\end{eqnarray}
 are restricted by only one condition similar to (\ref{Con00}),
\begin{eqnarray}
\label{Con001}  \Upsilon^{+\Sigma}\, \Omega_{\Sigma\Pi}
\Upsilon^{-\Pi} =0 \; .
\end{eqnarray}
The action in terms of $\Upsilon^{\pm\Sigma}$ includes the
degenerate matrix $C_{\Sigma\Pi}$, and reads \cite{30/32}
{\setlength\arraycolsep{2pt}
\begin{eqnarray}\label{St003}
S &=& \int_{W^2} [e^{++} \wedge d\Upsilon^{-\Sigma}\,
\Omega_{\Sigma\Pi} \Upsilon^{-\Pi}\,   - e^{--} \wedge
d\Upsilon^{+\Sigma}\, \Omega_{\Sigma\Pi} \Upsilon^{+\Pi}\
\nonumber \\ && \qquad  -\, e^{++} \wedge e^{--} \,
(\Upsilon^{+\Sigma}\, C_{\Sigma\Pi} \Upsilon^{-\Pi})^2] \; .
\end{eqnarray}}

The orthosymplectic supertwistors $\Upsilon^{\pm\Sigma}$ are both
in the fundamental representation of the $OSp(2n|1)$ supergroup.
The constraints (\ref{Con00}) (or (\ref{Con001})) are also
$OSp(2n|1)$ invariant. However, condition (\ref{l+l-00}) (or the
last term in the action (\ref{St003})) breaks the $OSp(2n|1)$
invariance down to the semidirect product ${\Sigma}^{({n(n+1)\over
2}|n)} \rtimes Sp(n,\mathbb{R})$, generalizing superPoincar\'e, of
$Sp(n,\mathbb{R})\subset Sp(2n,\mathbb{R})$ and the maximal
superspace group $\Sigma^{({n(n+1)\over 2}|n)}$ (see appendix
\ref{ap:breaking}). In contrast, both the point--like model in
\cite{BL98} and the tensionless superbrane model of \cite{B02}
possess full $OSp(2n|1)$ symmetry. This is in agreement with
treating $OSp(2n|1)$ as a generalized superconformal group, as the
standard conformal and superconformal symmetry is broken in any
model with mass, tension or another dimensionful parameter.

\section{Hamiltonian analysis in the supertwistor
formulation} \label{twisthamilton}

The Hamiltonian analysis simplifies in the supertwistor
formulation (\ref{St002}) of the action (\ref{St}) \cite{30/32}.
This is due to the fact that moving from (\ref{St}) to
(\ref{St002}) reduces the number of fields involved in the model.

The Lagrangian of the action (\ref{St002}) reads
{\setlength\arraycolsep{2pt}
\begin{eqnarray}
{\cal L} & = & (e^{++}_{\tau}
\partial_{\sigma} Y^{^{-\Sigma}}
 - e^{++}_{\sigma} \partial_{\tau}
Y^{^{-\Sigma}}) \Omega_{_{\Sigma \Pi}} Y^{^{-\Pi}} \nonumber \\
&&  - (e^{--}_{\tau} \partial_{\sigma} Y^{^{+\Sigma}}
 - e^{--}_{\sigma} \partial_{\tau}
Y^{^{+\Sigma}}) \Omega_{_{\Sigma \Pi}} Y^{^{+\Pi}} \nonumber
\\
&  &  - (e^{++}_{\tau} e^{--}_{\sigma} - e^{++}_{\sigma}
e^{--}_{\tau}) \; ,
\end{eqnarray}
}and involves the $2(2n+1+2)=4n+6$ configuration space worldvolume
fields
\begin{equation}\label{cZtwist}
\widetilde{\cal Z}^{\tilde{\cal M}}  \equiv \widetilde{\cal
Z}^{\tilde{\cal M}}(\tau, \sigma) =  \left( Y^{^{\pm \Sigma}} \, ,
\, e^{\pm\pm}_\tau \, , \, e^{\pm\pm}_\sigma \right)\; .
\end{equation}
The calculation of their canonical  momenta
\begin{eqnarray}\label{cPtwist}
\widetilde{P}_{\tilde{\cal M}} = (P_{_{\pm\Sigma}}\,,\,
P_{\pm\pm}^{\tau} \,, \, P_{\pm\pm}^{\sigma}) = {\partial {\cal L}
\over
\partial
(\partial_\tau \widetilde{\cal Z}^{\tilde{\cal M}})}
\end{eqnarray}
provides the following set of primary constraints:
{\setlength\arraycolsep{0pt}
\begin{eqnarray}
 \label{pY} &&  {\cal P}_{_{\pm\Sigma}}  =  P_{_{\pm \Sigma}} \mp
e_{\sigma}^{\mp \mp} \Omega_{_{\Sigma \Pi}}
Y^{^{\pm\Pi}} \approx  0 \; ,\\
\label{psigma3} &&  P_{\pm \pm}^{\sigma}  \approx  0 \;, \\
\label{tau3} && P_{\pm \pm}^{\tau}  \approx  0 \; .
\end{eqnarray}
}Conditions (\ref{Con00}), (\ref{l+l-00}) should also be taken
into account after all the Poisson brackets are calculated and,
hence, are also primary constraints, {\setlength\arraycolsep{0pt}
\begin{eqnarray}
\label{U1} && {\cal U} := Y^{+\Sigma}\, \Omega_{\Sigma\Pi}
Y^{-\Pi} \approx 0 \; , \\
\label{c2l+l-} && {\cal N}  :=
Y^{+\Sigma}\, C_{\Sigma\Pi} Y^{-\Pi} -1 \approx 0 \; .
\end{eqnarray}

}

The {\it canonical} Hamiltonian density ${\cal H}_0$ corresponding
to the action (\ref{St002}), reads {\setlength\arraycolsep{2pt}
\begin{eqnarray} \label{H2Y}
{\cal H}_0 & = & [-e^{++}_{\tau}
\partial_{\sigma} Y^{^{-\Sigma}} \Omega_{_{\Sigma \Pi}} Y^{^{-\Pi}} +
e^{--}_{\tau} \partial_{\sigma} Y^{^{+\Sigma}} \Omega_{_{\Sigma
\Pi}}
Y^{^{+\Pi}} \nonumber \\
&& + (e^{++}_{\tau} e^{--}_{\sigma} - e^{++}_{\sigma}
e^{--}_{\tau}) ] \; .
\end{eqnarray}
}The  preservation of the primary constraints under
$\tau$--evolution (see section \ref{Hamiltonian}) leads to the
secondary constraints {\setlength\arraycolsep{0pt}
\begin{eqnarray}
\label{Ph++} && \Phi_{++} = \partial_{\sigma} Y^{^{-\Sigma}}
\Omega_{_{\Sigma \Pi}} Y^{^{- \Pi}} - e^{--}_{\sigma} \approx 0 \;
,
\\
\label{Ph--} && \Phi_{--} = \partial_{\sigma} Y^{^{+ \Sigma}}
\Omega_{_{\Sigma \Pi}} Y^{^{+ \Pi}} - e^{++}_{\sigma} \approx 0 \;
,
\\
\label{Ph0} && \Phi^{(0)} = \partial_{\sigma} Y^{^{+ \Sigma}}
\Omega_{_{\Sigma \Pi}} Y^{^{- \Pi}} - Y^{^{+ \Sigma}}
\Omega_{_{\Sigma \Pi}}\partial_{\sigma}  Y^{^{- \Pi}} \approx 0 \;
.
\end{eqnarray}}

Again (see section \ref{Hamiltonian}) the canonical Hamiltonian
vanishes on the surface of constraints (\ref{Ph++}), (\ref{Ph--}),
and thus the $\tau$--evolution is defined by the Hamiltonian
density ({\it cf.} (\ref{H=})) {\setlength\arraycolsep{2pt}
\begin{eqnarray}\label{H(Y)=}
{\cal H}^\prime &=& - e_\tau^{++} \Phi_{++} + e_\tau^{--}
\Phi_{--}  + l^{(0)}  \Phi^{(0)} + L^{\pm \Sigma} {\cal P}_{\pm
\Sigma} +
\nonumber \\
&& + L^{(0)} {\cal U} +  L^{(n)} {\cal N} + L^{\pm\pm}
P_{\pm\pm}^{\sigma} + h^{\pm\pm} P_{\pm\pm}^\tau  \;
\end{eqnarray}
}and the canonical Poisson brackets {\setlength\arraycolsep{0pt}
\begin{eqnarray} \label{YPY}
&& [ P_{_{\pm\Lambda}}(\sigma)\, , \,  Y^{^{\pm\Sigma}}
(\sigma^{\prime}) \}_{_P} = - \delta^{^{\Sigma}}
_{_{\Lambda}} \;  \delta(\sigma-\sigma^{\prime}) , \quad  \\
\label{ePe1} && [ e_\sigma^{\pm\pm} (\sigma ) \, , \,
P_{{\pm\pm}}^{\sigma} (\sigma^{\prime}) ]_{_P} =
\delta(\sigma-\sigma^{\prime}) \; , \\
\label{ePe2} && [e_\tau^{\pm\pm} (\sigma ) \, , \,
P_{{\pm\pm}}^{\tau} (\sigma^{\prime}) ]_{_P} =
\delta(\sigma-\sigma^{\prime})
 \; , \quad
\end{eqnarray}}

Then the $\tau$--preservation requirement of the primary and
secondary constraints results in the following system of equations
for the Lagrange multipliers {\setlength\arraycolsep{0pt}
\begin{eqnarray}
\label{dtP+S} && L^{^{+\Sigma}} \approx  {e_\tau^{--} \over
e_\sigma^{--} }\partial_{\sigma}  Y^{^{+\Sigma}} +
{\partial_{\sigma} e_\tau^{--} - L^{--} \over 2e_\sigma^{--}
}Y^{^{+\Sigma}}  +  {l^{(0)} \over e_\sigma^{--}
}\partial_{\sigma} Y^{^{-\Sigma}}
 \nonumber \\ && \qquad  +
{\partial_{\sigma} l^{(0)} - L^{(0)} \over 2e_\sigma^{--}
}Y^{^{-\Sigma}} - {L^{(n)} \over 2e_\sigma^{--} }Y^{^{-\Pi}}
(C\Omega)_{_{\Pi}}{}^{^{\Sigma}} \; , \\ && \nonumber \\
 \label{dtP-S}
&& L^{^{-\Sigma}} \approx  {e_\tau^{++} \over e_\sigma^{++}
}\partial_{\sigma}  Y^{^{-\Sigma}} + {\partial_{\sigma}
e_\tau^{++} - L^{++} \over 2e_\sigma^{++} }Y^{^{-\Sigma}} -
{l^{(0)} \over e_\sigma^{++} }\partial_{\sigma} Y^{^{+\Sigma}}
\nonumber \\ && \qquad - {\partial_{\sigma} l^{(0)} + L^{(0)}
\over 2e_\sigma^{++} }Y^{^{+\Sigma}} - {L^{(n)} \over
2e_\sigma^{++} }Y^{^{+\Pi}} (C\Omega)_{_{\Pi}}{}^{^{\Sigma}} \; ,
\\ && \nonumber
\\
\label{dtU(Y)} && L^{^{+\Sigma}}  \Omega_{_{\Sigma\Pi}} Y
^{^{-\Pi}} \approx L^{^{-\Sigma}} \Omega_{_{\Sigma\Pi}} Y
^{^{+\Pi}}   \; ,
\\
\label{dtN(Y)} &&  L^{^{+\Sigma}}  C_{_{\Sigma\Pi}} Y ^{^{-\Pi}}
\approx  L^{^{-\Sigma}} C_{_{\Sigma\Pi}} Y ^{^{+\Pi}}   \; ,
\\
\label{dtP++Y} && L^{^{-\Sigma}}  \Omega_{_{\Sigma\Pi}} Y
^{^{-\Pi}}
 \approx  e_\tau^{--} \; ,
\\
\label{dtP--Y} && L^{^{+\Sigma}}  \Omega_{_{\Sigma\Pi}} Y
^{^{+\Pi}}  \approx e_\tau^{++} \; ,
\end{eqnarray}
}and {\setlength\arraycolsep{0pt}
\begin{eqnarray}
\label{dtPh++Y} && L^{--}  \approx \partial_\sigma L^{^{-\Sigma}}
\Omega_{_{\Sigma\Pi}} Y ^{^{-\Pi}} - L^{^{-\Sigma}}
\Omega_{_{\Sigma\Pi}} \partial_\sigma Y ^{^{-\Pi}} \; ,
\\
\label{dtPh--Y} && L^{++}  \approx  \partial_\sigma L^{^{+\Sigma}}
\Omega_{_{\Sigma\Pi}} Y ^{^{+\Pi}} - L^{^{+\Sigma}}
\Omega_{_{\Sigma\Pi}} \partial_\sigma Y ^{^{+\Pi}} \; ,
\\ \label{dtPh0Y}
&& \sum\limits_{\pm} \left(
\partial_\sigma
L^{^{\pm\Sigma}}  \Omega_{_{\Sigma\Pi}} Y ^{^{\mp\Sigma}} -
L^{^{\pm\Sigma}}  \Omega_{_{\Sigma\Pi}} \partial_\sigma Y
^{^{\mp\Sigma}} \right) \approx 0 \; . \qquad
\end{eqnarray}
}where $(C\Omega)_{_{\Pi}}{}^{^{\Sigma}}:= C_{_{\Pi\Lambda}}
\Omega^{^{\Lambda\Sigma}}$ and $\Omega^{^{\Sigma \Pi}} = -
\Omega_{_{\Sigma \Pi}}$ is the inverse of the orthosymplectic
metric (\ref{OmLP}),
\begin{equation}\label{Om-1}
\Omega_{_{\Sigma \Lambda}}  \Omega^{^{\Lambda \Pi}}=
\delta_{_{\Sigma}}^{^{\Pi}}\; , \quad \Omega^{^{\Sigma \Pi}} =
\left( \begin{matrix} 0 & - \delta_\beta{}^\alpha & 0 \cr
\delta_\alpha{}^\beta & 0 & 0 \cr 0 & 0 & i \end{matrix} \right)
\quad .
\end{equation}

Equations (\ref{dtP+S})--(\ref{dtP--Y}) come from the preservation
of the primary constraints, while
equations~(\ref{dtPh++Y})--(\ref{dtPh0Y}) from the preservation of
the secondary constraints. Again, as in section \ref{Hamiltonian},
one can follow the appearance of the secondary constraint
(\ref{Ph0}) by considering equations~(\ref{dtP+S})--(\ref{dtP--Y})
with $l^{(0)}=0$. Denoting {\setlength\arraycolsep{0pt}
\begin{eqnarray}
\label{A0=} && A_\sigma^{(0)} =  {1\over 2} \left(
\partial_\sigma Y^{^{+\Sigma}}
 C_{_{\Sigma\Pi}} Y^{^{-\Pi}}
- Y ^{^{+\Sigma}} C_{_{\Sigma\Pi}}
 \partial_\sigma Y^{^{-\Pi}} \right) \; ,
\qquad \\
\label{A++=} && A_\sigma^{++} =  \partial_\sigma Y^{^{+\Sigma}}
C_{_{\Sigma\Pi}} Y^{^{+\Pi}} \; ,
\\
\label{A--=} && A_\sigma^{--} =   \partial_\sigma Y^{^{-\Sigma}}
C_{_{\Sigma\Pi}} Y^{^{-\Pi}} \; ,
\\
\label{B0=} && B^{(0)} = {\cal S} -{\partial_\sigma
Y^{+}\Omega\partial_\sigma Y^{-} \over 2
e_\sigma^{++}e_\sigma^{--}} \; ,
\\ \label{calS=}
&& {\cal S} = {1\over 2} \left( {A_\sigma^{++}\over e_\sigma^{++}}
+ {A_\sigma^{--}\over e_\sigma^{--}} \right) \; ,
\end{eqnarray}
}one can write the general solution of
equations~(\ref{dtP+S})--(\ref{dtP--Y}) in the form
{\setlength\arraycolsep{2pt}
\begin{eqnarray}
\label{L+S} L^{^{+\Sigma}}& \approx & \omega^{(0)} Y^{^{+\Sigma}}
+ \nonumber \\ && + {e_\tau^{--}\over e_\sigma^{--} } \left(
\partial_{\sigma} Y^{^{+\Sigma}} - A_\sigma^{(0)}  Y^{^{+\Sigma}}
- e_\sigma^{++} B^{(0)} Y^{^{-\Sigma}} +
  e_\sigma^{++} (Y^-C\Omega)^{^{\Sigma}}
\right)  \nonumber \\
&& + e_\tau^{++} \left(  B^{(0)} Y^{^{-\Sigma}} -
(Y^-C\Omega)^{^{\Sigma}}\right)  \; ,
\\ \nonumber \\
\label{L-S} L^{^{-\Sigma}}& \approx & - \omega^{(0)}
Y^{^{-\Sigma}}  \nonumber \\ && +{e_\tau^{++}\over e_\sigma^{++} }
\left(
\partial_{\sigma}  Y^{^{-\Sigma}}
+ A_\sigma^{(0)}  Y^{^{-\Sigma}}
+ e_\sigma^{--} B^{(0)} Y^{^{+\Sigma}} -
 e_\sigma^{--} (Y^+C\Omega)^{^{\Sigma}}
\right)  \nonumber \\
&& - e_\tau^{--} \left(  B^{(0)} Y^{^{+\Sigma}} -
(Y^+C\Omega)^{^{\Sigma}}\right)  \; ,
\\ \nonumber \\ \label{L0(Y)}
L^{(0)} &=& 2(e_\tau^{--}e_\sigma^{++} - e_\tau^{++}e_\sigma^{--})
\, B^{(0)} \; ,
\\  \label{LpmpmY}
L^{\pm\pm}& = & \partial_{\sigma}e^{\pm \pm}_{\tau}\mp 2e^{\pm
\pm}_\tau  A^{(0)}_\sigma \pm 2e^{\pm \pm}_{\sigma} \omega^{(0)}
\; ,
\\  \label{Ln(Y)}
L^{(n)} &=& - 4 \mathrm{det}(e_m^a) \equiv -2
(e_\tau^{--}e_\sigma^{++} - e_\tau^{++}e_\sigma^{--} ) \; , \quad
\\ \label{l0(Y)}
l^{(0)} &=& 0 \; .
\end{eqnarray}
}Note that equations~(\ref{Ln(Y)}), (\ref{l0(Y)}) have the same
form as (\ref{L(n)=A}), (\ref{l0=A}) for the Lagrange multipliers
in the original formulation, and equations~(\ref{LpmpmY}) are
similar to equations~(\ref{Lpmpm=A}).

The above solution contains the indefinite worldsheet field
parameters $h^{\pm\pm}(\xi)$, $\omega^{(0)}(\xi)$ and
$e_\tau^{\pm\pm}(\xi)$ corresponding to the five first class
constraints which generate the gauge symmetries of the symplectic
twistor formulation of our $\Sigma^{({n(n+1)\over 2}|n)}$
superstring model. They are
\begin{eqnarray}\label{PtauY}
P_{\pm \pm}^{\tau} \approx 0 \;
\end{eqnarray}
and {\setlength\arraycolsep{2pt}
\begin{eqnarray}
\label{G0(Y)} G^{(0)} &:=& Y^{^{+\Sigma}}{\cal P}_{_{+\Sigma}} -
Y^{^{-\Sigma}} {\cal P}_{_{-\Sigma}} +
2e_\sigma^{++} P^\sigma_{++} - 2  e_\sigma^{--} P^\sigma_{--}
\approx 0 \, ,
\\
\label{IPh++Y} \tilde{\Phi}_{++} &:=& {\Phi}_{++} +
\partial_{\sigma} P_{++}^{\sigma} + 2A_{\sigma}^{(0)}
P_{++}^{\sigma} +
2 e^{--}_{\sigma} B^{(0)} {\cal U} \nonumber \\
&& -2e^{--}_{\sigma} {\cal N}    +{\cal F}_{++}^{^{\pm\Sigma}}
{\cal P}_{{\pm\Sigma}} \; , \\
\label{IPh--Y} \tilde{\Phi}_{--} &:=& {\Phi}_{--} -
\partial_{\sigma} P_{--}^{\sigma} +
2A_{\sigma}^{(0)} P_{--}^{\sigma} +
2 e^{++}_{\sigma} B^{(0)} {\cal U} \nonumber \\
&& - 2e^{++}_{\sigma} {\cal N}  + {\cal F}_{--}^{^{\pm\Sigma}}
{\cal P}_{{\pm\Sigma}}\; ,
\end{eqnarray}
}where {\setlength\arraycolsep{0pt}
\begin{eqnarray}
\label{cF+++S}  && \kern-2em {\cal F}_{++}^{^{+\Sigma}}  =  -
B^{(0)} Y^{^{-\Sigma}} + (Y^-C\Omega)^{^{\Sigma}} \; ,
\\ \label{cF++-S}
&& \kern-2em {\cal F}_{++}^{^{-\Sigma}}  =  - {1\over
e_\sigma^{++}} [\partial_{\sigma} Y^{^{-\Sigma}} +  A_\sigma^{(0)}
Y^{^{-\Sigma}} +
B^{(0)} e_\sigma^{--} Y^{^{+\Sigma}} - e_\sigma^{--}
(Y^+C\Omega)^{^{\Sigma}}] \; ,\nonumber \\* && \\
 \label{cF--+S} &&
\kern-2em {\cal F}_{--}^{^{+\Sigma}}  =  {1\over e_\sigma^{--}}
[\partial_{\sigma} Y^{^{+\Sigma}} - A_\sigma^{(0)} Y^{^{+\Sigma}}
-
B^{(0)} e_\sigma^{++} Y^{^{-\Sigma}} + e_\sigma^{++}
(Y^-C\Omega)^{^{\Sigma}}] ,
\\
\label{cF---S} && \kern-2em  {\cal F}_{--}^{^{-\Sigma}}  =  -
B^{(0)} Y^{^{+\Sigma}} + (Y^+C\Omega)^{^{\Sigma}} \; .
\end{eqnarray}
}Using Poisson brackets, the constraint (\ref{G0(Y)}) generates
the $SO(1,1)$ worldsheet Lorentz gauge symmetry, (\ref{IPh++Y})
and (\ref{IPh--Y})  are the reparameterization (Virasoro)
generators, and the symmetry generated by equations~(\ref{PtauY})
indicates the pure gauge nature of the $e_\tau^{\pm\pm}(\xi)$
fields (again, subject to the nondegeneracy condition
(\ref{det(e)}) that restricts the gauge choice freedom for them).

Note that both the $b$--symmetry and the $\kappa$--symmetry
generators, equations (\ref{PIJ}) and (\ref{DI}), are not present
in the symplectic supertwistor formulation. Actually, the number
of variables in this formulation minus the constraint among them,
equation~(\ref{Con00}), is $(4n+6)-1$ and equal to the number of
variables in the previous formulation $(\frac{n(n+1)}2+n+2n+4)$,
minus the number of $b$-- and $\kappa$--symmetry generators
$(\frac{(n-1)(n-2)}2+(n-2))$. This  indicates that the transition
to the supertwistor form of the action corresponds to an implicit
gauge fixing of these symmetries and the removal of the additional
variables, since the remaining supertwistor ones are invariant
under both $b$-- and $\kappa$--symmetry\footnote{This invariance
was known for the massless superparticle and the tensionless
superstring cases, see {\it e.g.}
\cite{BL98,B02,BLS99,Sokatchev}.}.

Other constraints are second class. Indeed, {\it e.g.} the algebra
of the constraints ${\cal P}_{\pm\Sigma}$, equation~(\ref{pY}),
{\setlength\arraycolsep{2pt}
\begin{eqnarray} \label{cPcP+}
 [ {\cal P}_{_{+\Sigma}} (\sigma ) \, , \, {\cal
P}_{_{+\Lambda}}(\sigma^{\prime}) \}_{_P} &=& 2 e_\sigma^{--}
\Omega_{_{\Lambda \Sigma}}
\delta(\sigma-\sigma^{\prime})\; ,  \\
\label{cPcP-}  [ {\cal P}_{_{-\Sigma}} (\sigma ) \, , \, {\cal
P}_{_{-\Lambda}}(\sigma^{\prime}) \}_{_P} &=& - 2 e_\sigma^{++}
\Omega_{_{\Lambda\Sigma}} \delta(\sigma-\sigma^{\prime})\; , \quad
\\
\label{cPcP+-}  [ {\cal P}_{_{+\Sigma}} (\sigma ) \, , \, {\cal
P}_{_{-\Lambda}}(\sigma^{\prime}) \}_{_P} &=& 0 \; ,
\end{eqnarray}
}shows their second class nature. As such, one can introduce the
graded Dirac (or starred \cite{Dirac}) brackets that allows one to
put them strongly equal to zero. For any arbitrary two (bosonic or
fermionic) functionals $f$ and $g$ of the canonical variables
(\ref{cZtwist}), (\ref{cPtwist}) they are defined by
{\setlength\arraycolsep{0pt}
\begin{eqnarray} \label{PBY}
&& [f(\sigma_1), g(\sigma_2)\}_{_D} = [f(\sigma_1) ,
g(\sigma_2)\}_{_P}  \nonumber \\* && \qquad  -{1 \over 2} \int
d\sigma \left( \frac{1}{e^{--}_{\sigma}(\sigma)} [f(\sigma_1) ,
{\cal P}_{_{+\Sigma}}(\sigma) \}_{_P}\Omega^{^{\Pi\Sigma}} [ {\cal
P}_{_{+\Pi}}(\sigma), g(\sigma_2)\}_{_P} \right. \nonumber
 \\*
 && \left. \qquad   -{1 \over e^{++}_{\sigma}(\sigma)} [f(\sigma_1) ,
 {\cal P}_{_{-\Sigma}}(\sigma) \}_{_P}\Omega^{^{\Pi\Sigma }} [ {\cal
P}_{_{-\Pi}}(\sigma), g(\sigma_2)\}_{_P} \right) \, . \qquad
\end{eqnarray}

}

\noindent Using these and reducing further the number of phase
space degrees of freedom by setting $P_{_{\pm\Sigma}}=0$ strongly,
the supertwistor becomes a self-conjugate variable,
\begin{equation} \label{DB2}
 [Y^{^{\pm \Sigma}}(\sigma), Y^{^{\pm
\Pi}}(\sigma^{\prime})\}_{_D} = \mp {1 \over 2e^{\mp
\mp}_{\sigma}} \, \Omega^{^{\Sigma \Pi}} \, \delta (\sigma -
\sigma^{\prime}) \; .
\end{equation}
For the `components' of the supertwistor, equation~(\ref{DB2})
implies {\setlength\arraycolsep{0pt}
\begin{eqnarray} \label{DBbose}
&& [\lambda^{\pm}_{\alpha}(\sigma), \mu^{\pm
\beta}(\sigma^\prime)]_{_D} = \mp {1 \over 2e^{\mp \mp}_{\sigma}}
\delta_{\alpha}^{\beta}  \delta(\sigma-\sigma^\prime) \ , \\
\label{DBfermi} && \{\eta^{\pm}(\sigma),
\eta^{\pm}(\sigma^\prime)\}_{_D} = \mp {i \over 2e^{\mp
\mp}_{\sigma}} \ \delta(\sigma-\sigma^\prime) \; .
\end{eqnarray}
}The Dirac brackets for $e_\sigma^{\pm\pm}$, $e_\tau^{\pm\pm}$ and
$P^\tau_{\pm\pm}$ coincide with the Poisson brackets, while for
$P^\sigma_{\pm\pm}$ one finds {\setlength\arraycolsep{0pt}
\begin{eqnarray}
\label{P++D} && [P^\sigma_{++}(\sigma), ... ]_{_D}  =
[P^\sigma_{++}(\sigma), ... ]_{_P} - {1\over 2e_\sigma^{++}}
Y^{^{-\Sigma}}(\sigma) [{\cal P}_{_{-\Sigma}}(\sigma), ... \}_{_P}
,
\\
\label{P--D} && [P^\sigma_{--}(\sigma), ... ]_{_D} =
[P^\sigma_{--}(\sigma), ... ]_{_P}  - {1\over 2e_\sigma^{--}}
Y^{^{+\Sigma}}(\sigma) [{\cal P}_{_{+\Sigma}}(\sigma), ... \}_{_P}
.
\end{eqnarray}
}However, $P^\sigma_{\pm\pm}(\sigma)$ still commute among
themselves, \begin{equation} [P^\sigma_{\pm\pm}(\sigma),
P^\sigma_{\pm\pm}(\sigma^\prime)]_{_D} =0=[P^\sigma_{++}(\sigma),
P^\sigma_{--}(\sigma^\prime)]_{_D} \; .
\end{equation}

When the constraints (\ref{pY}) are taken as strong equations, the
first class constraints (\ref{G0(Y)})--(\ref{IPh--Y}) simplify to
{\setlength\arraycolsep{0pt}
\begin{eqnarray}
 \label{G0(Y)D}  && \kern-2em G^{(0)} := 2e_\sigma^{++} P^\sigma_{++} - 2
e_\sigma^{--} P^\sigma_{--} \approx 0 \, ,
\\
\label{IPh++YD} && \kern-2em \tilde{\Phi}_{++} := {\Phi}_{++} +
\partial_{\sigma} P_{++}^{\sigma} + 2A_{\sigma}^{(0)}
P_{++}^{\sigma} +
2e^{--}_{\sigma} B^{(0)} {\cal U} - 2e^{--}_{\sigma} {\cal N}
\approx 0 ,
\\
\label{IPh--YD} && \kern-2em \tilde{\Phi}_{--} := {\Phi}_{--} -
\partial_{\sigma} P_{--}^{\sigma} +
2A_{\sigma}^{(0)} P_{--}^{\sigma} +
2 e^{++}_{\sigma} B^{(0)} {\cal U} - 2e^{++}_{\sigma} {\cal N}
\approx 0 ,
\end{eqnarray}
}and the remaining second class constraints can be taken in the
form {\setlength\arraycolsep{0pt}
\begin{eqnarray}
\label{K0} && K^{(0)} := e_\sigma^{++} P^\sigma_{++} +
e_\sigma^{--} P^\sigma_{--}
\approx 0 \, , \\
\label{N2} && {\cal N} =  Y^{+\Sigma}\, C_{\Sigma\Pi} Y^{-\Pi} -1
\approx 0
\; , \\
\label{U2} && {\cal U} =  Y^{+\Sigma}\, \Omega_{\Sigma\Pi}Y^{-\Pi}
\approx 0 \; ,
\\
\label{Ph0II} && \Phi^{(0)} = \partial_{\sigma} Y^{^{+ \Sigma}}
\Omega_{_{\Sigma \Pi}} Y^{^{- \Pi}} - Y^{^{+ \Sigma}}
\Omega_{_{\Sigma \Pi}}\partial_{\sigma}  Y^{^{- \Pi}} \approx 0 \;
.
\end{eqnarray}
}One has to take into account that, under the Dirac brackets,
$P^\sigma_{\pm\pm}$ and $Y^{\mp\Sigma}$ do not commute,
{\setlength\arraycolsep{0pt}
\begin{eqnarray}
\label{P++Y} && [P^\sigma_{++}(\sigma), Y^{-\Sigma}
(\sigma^\prime) ]_{_D}  = {1\over 2e_\sigma^{++}}
Y^{^{-\Sigma}}(\sigma) \delta(\sigma -  \sigma^\prime)\; ,  \qquad
\\
\label{P--Y} && [P^\sigma_{--}(\sigma), Y^{+\Sigma}
(\sigma^\prime) ]_{_D}  = {1\over 2e_\sigma^{--}}
Y^{^{+\Sigma}}(\sigma) \delta(\sigma -  \sigma^\prime)\; .  \qquad
\end{eqnarray}
}Then one checks that, under Dirac brackets, $G^{(0)}$ generates
the $SO(1,1)$ transformations of the supertwistors,
\begin{eqnarray}
\label{[G0Y]} [G^{(0)}(\sigma), Y^{\pm \Sigma} (\sigma^\prime)
]_{_D}  = \mp  Y^{^{\pm\Sigma}}(\sigma) \delta(\sigma -
\sigma^\prime)\; .
\end{eqnarray}
On the other hand, one finds that the second class constraint
${\cal U}$ interchanges the $Y^{+\Sigma}$ and $Y^{-\Sigma}$
supertwistors, {\setlength\arraycolsep{0pt}
\begin{eqnarray}
&& [\, {\cal U}(\sigma)\, , Y^{+\Sigma} (\sigma^\prime) ]_{_D} =
 \; {1\over 2e_\sigma^{--}} Y^{^{-\Sigma}}(\sigma)
\delta(\sigma -  \sigma^\prime)\; ,  \qquad  \nonumber
\\ \label{U0Y}
&& [\, {\cal U}(\sigma), Y^{-\Sigma} (\sigma^\prime) ]_{_D} =
  {1\over 2e_\sigma^{++}} Y^{^{+\Sigma}}(\sigma)
\delta(\sigma -  \sigma^\prime)\; .  \qquad
\end{eqnarray}
}
It is interesting to note that in the original supertwistor
formulation of the $D=4$, $N=1$ superparticle \cite{Ferber} there
exists a counterpart of the ${\cal U}$ constraint; however, there
it is the first class constraint generating the internal $U(1)$
symmetry\footnote{See \cite{SG89} for a detailed study of the
Hamiltonian mechanics in the twistor--like formulation of the
$D=4$ superparticle, where the possibility of constraint class
transmutation was noted.}.

The Dirac brackets of the above second class constraints
(\ref{K0})--(\ref{Ph0II}) are:
\begin{eqnarray} \label{DB1}
&& [\Phi^{(0)}(\sigma) , {\cal U}(\sigma^\prime)]_{_D} = \nonumber \\
&=& -\frac{1}{2} \left( \frac{\partial_\sigma Y^{+\Sigma}(\sigma)
\Omega_{\Sigma \Pi} Y^{+ \Pi}(\sigma)}{e^{++}_{\sigma}(\sigma)} +
\frac{\partial_\sigma Y^{-\Sigma}(\sigma) \Omega_{\Sigma \Pi} Y^{-
\Pi}(\sigma)}{e^{--}_\sigma(\sigma)} \right)
\delta(\sigma-\sigma^\prime)  \nonumber \\
&=& -\frac{1}{2} \left(
\frac{\Phi_{++}(\sigma)}{e^{++}_\sigma(\sigma)}
+\frac{\Phi_{--}(\sigma)}{e^{--}_\sigma(\sigma)} +2 \right)
\delta(\sigma-\sigma^\prime)
\nonumber \\
& \approx & -\delta(\sigma-\sigma^\prime) \; ,
\end{eqnarray}
\begin{eqnarray} \label{DBPh0N}
&& [\Phi^{(0)}(\sigma) , {\cal N}(\sigma^\prime)]_{_D} = \nonumber \\
&=& -\frac{1}{2} \left( \frac{\partial_\sigma Y^{+\Sigma}(\sigma)
C_{\Sigma \Pi} Y^{+ \Pi}(\sigma)}{e^{++}_{\sigma}(\sigma)} +
\frac{\partial_\sigma Y^{-\Sigma}(\sigma) C_{\Sigma \Pi} Y^{-
\Pi}(\sigma)}{e^{--}_\sigma(\sigma)} \right)
\delta(\sigma-\sigma^\prime) \nonumber \\
&=&  -\frac{1}{2} \left(
\frac{A_{\sigma}^{++}(\sigma)}{e^{++}_\sigma(\sigma)} +
\frac{A_{\sigma}^{--}(\sigma)}{e^{--}_\sigma(\sigma)} \right)
\delta(\sigma-\sigma^\prime) \equiv -{\cal S}(\sigma)
\delta(\sigma-\sigma^\prime) \; ,
\end{eqnarray}
{\setlength\arraycolsep{0pt}
\begin{eqnarray} \label{DB3}
&& [K^{(0)}(\sigma) , {\cal U}(\sigma^\prime)]_{_D} =
Y^{+\Sigma}(\sigma)\Omega_{\Sigma \Pi} Y^{-\Pi}(\sigma) \
\delta(\sigma-\sigma^\prime)  \nonumber  \\
&& \kern+7.6em =  {\cal U} \ \delta(\sigma-\sigma^\prime) \approx 0 \; , \\
&& \nonumber  \\
\label{DB4}  && [K^{(0)}(\sigma) , {\cal N}(\sigma^\prime)]_{_D} =
Y^{+\Sigma}(\sigma)C_{\Sigma \Pi} Y^{-\Pi}(\sigma) \
\delta(\sigma-\sigma^\prime)  \nonumber \\
&& \kern+8em = ({\cal N} +1) \ \delta(\sigma-\sigma^\prime)
\approx \delta(\sigma-\sigma^\prime)
\; , \\ 
\label{DB5} && [K^{(0)}(\sigma) , \Phi^{(0)}(\sigma^\prime)]_{_D}
= \nonumber \\ && \quad =\ft{1}{2} \left( \partial_\sigma
Y^{+\Sigma}(\sigma)\Omega_{\Sigma \Pi} Y^{-\Pi}(\sigma)
-Y^{+\Sigma}(\sigma)\Omega_{\Sigma\Pi}
\partial_\sigma Y^{-\Pi}(\sigma) \right)
\delta(\sigma-\sigma^\prime)  \nonumber \\ && \quad  = \ft{1}{2}
\Phi^{(0)} \ \delta(\sigma-\sigma^\prime) \approx 0 \; ,
\end{eqnarray}
}where, in (\ref{DBPh0N}), ${\cal S}(\sigma) \equiv {1\over 2}
\left( \frac{A_{\sigma}^{++}(\sigma)}{e^{++}_\sigma(\sigma)} +
\frac{A_{\sigma}^{--}(\sigma)}{e^{--}_\sigma(\sigma)} \right) $
(equation~(\ref{calS=})) and $\delta_{\sigma\,
\sigma^\prime}\equiv \delta(\sigma-\sigma^\prime)$.

These Dirac brackets (\ref{DB1})-(\ref{DB5}) can be summarized
schematically in the following table

\vspace{5pt}
\begin{center}
\begin{tabular}
{|c|c|c|c|c|} \hline \tt $[... \downarrow \, , ... \rightarrow
\}_{_D} \approx$ & $(\Phi^{(0)}(\sigma^\prime)+$ & ${\cal U}
(\sigma^\prime)$ &  $ K^{(0)} (\sigma^\prime)$ & ${\cal N}
(\sigma^\prime) $ \cr
 & $+{\cal S} K^{(0)} (\sigma^\prime))$   &  &    & \cr
\hline \tt $(\Phi^{(0)}+ {\cal S}K^{(0)}) (\sigma)$  & 0 & $-
\delta_{\sigma\,\sigma^\prime}$ & 0 &  0 \cr \hline \tt ${\cal U}
(\sigma)$  & $\delta_{\sigma\,\sigma^\prime}$ & 0 & 0 & 0 \cr
\hline \tt $K^{(0)} (\sigma)$ & 0 & 0 & 0 &
$\delta_{\sigma\,\sigma^\prime}$\cr \hline \tt ${\cal N} (\sigma)$
& 0 & 0 & $- \delta_{\sigma\,\sigma^\prime}$ & 0  \cr \hline
\end{tabular}
\\[6pt]
{\footnotesize Table 7.2. Schematic Dirac brackets of the second
class constraints \\ in the supertwistor formalism}
\end{center}
\vspace{5pt}
 \noindent  This table indicates that the
$K^{(0)}$ constraint is canonically conjugate to ${\cal N}$ while
the second class constraint $\Phi^{(0)} + {\cal S} K^{(0)}$ is
conjugate to ${\cal U}$. One may  pass to the (doubly starred)
Dirac brackets with respect to the above mentioned four second
class constraints. However, the new Dirac brackets for the
supertwistor variables would have a very complicated form, so that
it looks more practical either to apply the formalism using
(singly starred) Dirac brackets (equation~(\ref{PBY})) and simple
first and second class constraints,
equations~(\ref{G0(Y)D})--(\ref{IPh--YD}) and
(\ref{K0})--(\ref{Ph0II}), or to search for a conversion
\cite{conversion} of the remaining second  class constraints into
first class ones. Note that a phenomenon similar to conversion
occurs when one moves from (\ref{St002})  to the dynamical system
with unnormalized twistors described by the action (\ref{St003}).
We discuss on this in more detail in the next section.

As the simplest  application of the above Hamiltonian analysis let
us calculate the number of field theoretical degrees of freedom of
the dynamical system (\ref{St002}). In this supertwistor
formulation one finds from equations~(\ref{cZtwist}) and
(\ref{Ypm}) $(4n+4)$ bosonic and 2 fermionic configuration space
variables, which corresponds to a phase space with $2(4n+4)$ and 4
fermionic `dimensions'. The system has 5 bosonic first class
constraints, equations~(\ref{PtauY})-(\ref{IPh--Y}), out of a
total number of $4n+9$ bosonic constraints (the bosonic components
of (\ref{pY}) and (\ref{psigma3}), (\ref{tau3}),
(\ref{Ph++})--(\ref{Ph0})). Thus, in agreement with section
\ref{Hamiltonian}, one finds that the $\Sigma^{({n(n+1)\over
2}|n)}$ supersymmetric string described by the action
(\ref{St002}) possesses $4n-6$ bosonic degrees of freedom.
Likewise, the 2 fermionic constraints of the action (the fermionic
components of (\ref{pY})) reduce to 2 the fermionic degrees of
freedom \cite{30/32}.

\section{Hamiltonian analysis with
`unnormalized' supertwistors} \label{unnormalized}

As shown in Section \ref{twistor}, the action (\ref{St002}) may be
considered as a gauge fixed form of the action (\ref{St003})
written in terms of supertwistors (\ref{Ups}) restricted by only
one Lagrangian constraint (\ref{Con001}). The second constraint
(\ref{l+l-00}), the `normalization' condition that distinguishes
among the $Y^{\pm \Sigma}$ and $\Upsilon^{\pm \Sigma}$
supertwistors, may be obtained by gauge fixing the direct product
of the two scaling gauge symmetries (\ref{sc+}) and (\ref{sc-})
down to the $SO(1,1)$ worldsheet Lorentz symmetry (\ref{SO(1,1)})
of the action (\ref{St002}). As a result, one may expect that the
Hamiltonian structure of the model (\ref{St003}) will differ from
the one of the model  (\ref{St002}) by the absence of one second
class constraint  (\ref{N2}) and the presence of one additional
first class constraint replacing (\ref{K0}).

This is indeed the case \cite{30/32}. An analysis similar to the
one carried out in Section \ref{twisthamilton} allows one to find
the following set of primary {\setlength\arraycolsep{0pt}
\begin{eqnarray}
\label{pUY} && {\cal P}_{_{\pm\Sigma}}  =  P_{_{\pm \Sigma}} \mp
e_{\sigma}^{\mp \mp} \Omega_{_{\Sigma \Pi}}
\Upsilon^{^{\pm\Pi}} \approx  0 \; ,\\
\label{psU3}
&& P_{\pm \pm}^{\sigma}  \approx  0 \; , \\
\label{PtU3} &&  P_{\pm
\pm}^{\tau}  \approx  0 \; ,   \\
\label{UU1}  && {\cal U} \;  :=  \Upsilon^{+\Sigma}\,
\Omega_{\Sigma\Pi} \Upsilon^{-\Pi} \equiv \Upsilon^{+}\, \Omega
\Upsilon^{-} \approx 0 \; ,
\end{eqnarray}
}and secondary constraints {\setlength\arraycolsep{0pt}
\begin{eqnarray}
\label{Ph++U} && \Phi_{++} = \partial_{\sigma} \Upsilon^- \Omega
\Upsilon^- - e^{--}_{\sigma} (\Upsilon^+C\Upsilon^-)^2 \approx 0
\; ,
\\
\label{Ph--U} && \Phi_{--}  = \partial_{\sigma} \Upsilon^+ \Omega
\Upsilon^+ - e^{++}_{\sigma} (\Upsilon^+C\Upsilon^-)^2 \approx 0
\; ,
\\
\label{Ph0U} && \Phi^{(0)} = \partial_{\sigma} \Upsilon^+ \Omega
\Upsilon^- - \Upsilon^+ \Omega \partial_{\sigma} \Upsilon^-
\approx 0 \; ,
\end{eqnarray}
}that restrict the phase space variables
{\setlength\arraycolsep{0pt}
\begin{eqnarray}\label{cZU}
&& \widetilde{\cal Z}^{\tilde{\cal M}}  \equiv \widetilde{\cal
Z}^{\tilde{\cal M}}(\tau, \sigma) =  \left( \Upsilon^{^{\pm
\Sigma}} \, , \, e^{\pm\pm}_\tau \, , \, e^{\pm\pm}_\sigma
\right)\; , \\
\label{cPU} && \widetilde{P}_{\tilde{\cal M}} =
(P_{_{\pm\Sigma}}\,,\, P_{\pm\pm}^{\tau} \,, \,
P_{\pm\pm}^{\sigma}) = {\partial {\cal L} \over
\partial
(\partial_\tau \widetilde{\cal Z}^{\tilde{\cal M}})}\; .
\end{eqnarray}}

The set (\ref{pUY})--(\ref{Ph0U}) contains $6$ first class
constraints (versus five first class constraints
(\ref{PtauY})--(\ref{IPh--Y}) in the system (\ref{St002})), namely

{\setlength\arraycolsep{0pt}
\begin{eqnarray}\label{PtauU}
&& P_{\pm\pm}^\tau \approx 0 \; ,  \\
\label{P1++U} && 2e^{++}_\sigma P_{++}^\sigma-
\Upsilon^{-\Sigma}\mathcal{P}_{-\Sigma}
 \approx 0 \; , \\
\label{P1--U} && 2e^{--}_\sigma P_{--}^\sigma-
\Upsilon^{+\Sigma}\mathcal{P}_{+\Sigma}
 \approx 0 \; , \\ && \nonumber \\
&& \tilde{\Phi}_{++}  =   \Phi_{++}  \nonumber \\
&&  + \frac{2e^{--}_\sigma {\cal B}^{(0)}}
{(\Upsilon^+C\Upsilon^-)^2}\mathcal{U} - \frac{{\cal
B}^{(0)}}{(\Upsilon^+C\Upsilon^-)^2}\Upsilon^{-\Sigma}
 \mathcal{P}_{+\Sigma}
 - \partial_\sigma P_{++}^\sigma   +
(\Upsilon^+C\Upsilon^-)\Upsilon^- C \Omega \mathcal{P}_+
 \nonumber \\ && - \frac1{e_\sigma^{++}} \Big[\partial_\sigma \!
\Upsilon^{-\Sigma} \mathcal{P}_{-\Sigma} +
 \frac{e_\sigma^{--} {\cal B}^{(0)}}{(\Upsilon^+C\Upsilon^-)^2}
 \Upsilon^{+\Sigma} \mathcal{P}_{-\Sigma}
 - e^{--}_\sigma (\Upsilon^+C\Upsilon^-) \Upsilon^+
C \Omega \mathcal{P}_-\Big] \nonumber \\
&& \quad  \approx 0 \; , \\ \nonumber \\
&& \tilde{\Phi}_{--}  =  \Phi_{--} \nonumber \\
&& + \frac{2e^{++}_\sigma {\cal
B}^{(0)}}{(\Upsilon^+C\Upsilon^-)^2}
 \mathcal{U} -  \frac{{\cal
B}^{(0)}}{(\Upsilon^+C\Upsilon^-)^2}\Upsilon^{+\Sigma}
 \mathcal{P}_{-\Sigma} +  \partial_\sigma P_{--}^\sigma
 + (\Upsilon^+C\Upsilon^-)\Upsilon^+ C \Omega \mathcal{P}_-
 \nonumber \\
 &&  + \frac1{e_\sigma^{--}} \Big[\partial_\sigma \Upsilon^{+\Sigma}
 \mathcal{P}_{+\Sigma}
 -\frac{e_\sigma^{++} {\cal B}^{(0)}}{(\Upsilon^+C\Upsilon^-)^2}
 \Upsilon^{-\Sigma} \mathcal{P}_{+\Sigma}
 + e^{++}_\sigma (\Upsilon^+C\Upsilon^-)
 (\Upsilon^- C \Omega \mathcal{P}_+)\Big] \nonumber \\
 && \quad   \approx 0 \; ,
 \end{eqnarray}
}where ({\it cf.} (\ref{B0=})) {\setlength\arraycolsep{0pt}
\begin{eqnarray}
&& {\cal B}^{(0)} = \nonumber \\
&& = \frac{1}{2} \left({\partial_\sigma\Upsilon^+C\Upsilon^+\,
(\Upsilon^+C\Upsilon^-)\over e^{++}_\sigma} + {\partial_\sigma
\Upsilon^-C\Upsilon^- \, (\Upsilon^+C\Upsilon^-) \over
e^{--}_\sigma}
+\frac{\partial_\sigma\Upsilon^+\Omega\partial_\sigma\Upsilon^-}{e^{++}_\sigma
e^{--}_\sigma} \right) \; . \nonumber \\ &&
\end{eqnarray}

}

Using Dirac brackets to account for the second class constraints
(\ref{pUY}), where ({\it cf.} (\ref{DB2}))
\begin{equation} \label{DB2U}
 [\Upsilon^{^{\pm \Sigma}}(\sigma), \Upsilon^{^{\pm
\Pi}}(\sigma^{\prime})\}_{_D} = \mp {1 \over 2e^{\mp
\mp}_{\sigma}} \, \Omega^{^{\Sigma \Pi}} \, \delta (\sigma -
\sigma^{\prime}) \; ,
\end{equation}
the first class constraints simplify to
{\setlength\arraycolsep{0pt}
\begin{eqnarray}\label{PtauUU}
&& P_{\pm\pm}^\tau \approx 0 \; , \\
\label{P1++UU} && P_{++}^\sigma
 \approx 0 \; , \\
\label{P1--UU} && P_{--}^\sigma
 \approx 0 \; , \\
 && \tilde{\Phi}_{++}  = \Phi_{++}
 + \frac{2e^{--}_\sigma {\cal B}^{(0)}}
{(\Upsilon^+C\Upsilon^-)^2}\mathcal{U}  \approx 0 \; ,
\\
&&  \tilde{\Phi}_{--} = \Phi_{--} + \frac{2e^{++}_\sigma {\cal
B}^{(0)}}
 {(\Upsilon^+C\Upsilon^-)^2} \mathcal{U}
 \approx 0 \; ,
\end{eqnarray}
}which corresponds to the set of constraints
(\ref{G0(Y)D})--(\ref{IPh--YD}) of the description in terms of
`normalized' supertwistors with the addition of the constraint
(\ref{K0}), which is now `converted' into a first class one due to
the disappearance of the normalization constraint (\ref{N2}).

The remaining two bosonic constraints, equations (\ref{UU1}) and
(\ref{Ph0U}), are second class. Their Dirac bracket
{\setlength\arraycolsep{0pt}
\begin{eqnarray}\label{UPh0}
[{\cal U}(\sigma)\, , \, \Phi^{(0)} (\sigma^\prime)]_D &=&
(\Upsilon^+C\Upsilon^-)^2 \delta(\sigma - \sigma^\prime)  +
\left({\Phi_{++}\over 2e_\sigma^{++}} + {\Phi_{--}\over
2e_\sigma^{--}} \right) \delta(\sigma - \sigma^\prime)
 \nonumber \\ & \approx &
(\Upsilon^+C\Upsilon^-)^2 \delta(\sigma - \sigma^\prime)
  \end{eqnarray}
}is nonvanishing due to the linear independence of the
$\Upsilon^{^{+\Sigma}}$ and  $\Upsilon^{^{-\Sigma}}$ supertwistors
(\ref{Ups}) (coming from the linear independence of their
$\Lambda_\alpha^+$ and $\Lambda_\alpha^-$ components,
$\Lambda_\alpha^+C^{\alpha\beta}\Lambda_\alpha^-\not=0$). For a
further simplification of the Hamiltonian formalism it might be
convenient to make a conversion of this pair of second class
constraints into first class by adding a pair of canonically
conjugate variables, $q(\xi)$ and $P^{(q)}(\xi)$, ($[q(\sigma)\, ,
\, P^{(q)}(\sigma^\prime )]_P= \delta(\sigma -\sigma^\prime)$) to
our phase space.

The above Hamiltonian formalism and its further development can be
applied to quantize the $\Sigma^{(\frac{n(n+1)}2|n)}$ superstring
model. This should produce a quantum higher spin generalization of
the Green--Schwarz superstring for $n=4,8,16$ and, for $n=32$, an
exactly solvable quantum description of a conformal field theory
carrying, somehow, information about the non-perturbative brane
BPS states of M Theory.

\section{Supersymmetric $p$--branes in tensorial superspace} \label{pbrane}

The model may be generalized to describe higher-dimensional
extended objects (supersymmetric $p$--branes) in
$\Sigma^{({n(n+1)\over 2}|n)}$. The expression of the
supersymmetric $p$--brane action in terms of dimensionful
unconstrained bosonic spinors reads (cf.~(\ref{St01}))
\cite{30/32} {\setlength\arraycolsep{2pt}
\begin{eqnarray}\label{pbr}
S_p &=&  \int_{W^{p+1}} \; e^{\wedge p}_a \wedge \Pi^{\alpha\beta}
(\Lambda^r_\alpha \rho^a_{rs} \Lambda^s_\beta ) \nonumber \\* && -
(- \alpha^\prime )^p  \int_{W^{p+1}} \; e^{\wedge (p+1)} \;
\mathrm{det}(C^{\alpha\beta} \Lambda^r_\alpha \Lambda^s_\beta)\; ,
\qquad
\end{eqnarray}
}where $a=0,1, \ldots , p\;,\;
r=1,\ldots,\tilde{n}(p)\,,\;\alpha=1,\ldots,n\,,$
\begin{equation}\label{e10}
e^{\wedge p}_a \equiv
 \ft{1}{p!} \epsilon_{ab_1\ldots
b_{p}}e^{b_1} \wedge \ldots \wedge e^{b_p} \, ,
\end{equation}
(see equation (\ref{E11-n})) and $e^{\wedge (p+1)}$ is the
$W^{p+1}$ volume element
\begin{equation}
\label{e11} e^{\wedge (p+1)} \equiv  \ft{1}{(p+1)!}
\epsilon_{b_1\ldots b_{p+1}}e^{b_1} \wedge \ldots \wedge
e^{b_{p+1}}\;\;.
\end{equation}
In equation~(\ref{pbr}), the $(p+1)$ $e^a = d\xi^m e_m^a(\xi)$ are
auxiliary worldvolume vielbein fields, $\xi^m=(\tau, \sigma^1,
\ldots, \sigma^p)$ are the worldvolume  $W^{p+1}$ local
coordinates and $\Lambda^r_\alpha(\xi)$ is a set of $\tilde{n}=
\tilde{n}(p)$ unconstrained auxiliary real bosonic fields with a
`spacetime' spinorial (actually, a $Sp(n)$--vector) index $\alpha
=1,..., n$. The number $\tilde{n}(p)$ of real spinor fields
 $\Lambda^r_\alpha(\xi)$ as well as the meaning of the
 symmetric real matrices
$\rho^a_{rs}$ depend on the worldvolume dimension $d=p+1$. For
$d=2,3,4 \;$(mod $8$), where a Majorana spinor representation
exists, the $\rho^a_{rs}$ are $Spin(1,p)$ Dirac matrices
multiplied by the charge conjugation matrix or sigma matrices,
provided they are symmetric. If not, it is always possible to find
a real symmetric matrix by doubling the index $r$, $r^\prime= ri$
 $(i=1,2)$, as in the case of $d=6$ symplectic Majorana
spinors. For dimensions with only Dirac spinors (like $d=5$)
$\Lambda^r_\alpha \rho^a_{rs} \Lambda^s_\beta$ should be
understood as $\bar{\Lambda}_\alpha \rho^a \Lambda_\beta +
\bar{\Lambda}_\beta \rho^a \Lambda_\alpha$, etc.

For simplicity we present equation (\ref{pbr}) and other formulae
of this section for `Majorana dimensions' $d$ with symmetric $C
\rho$--matrices; the generalization to the other cases is
straightforward, although one should be careful determining the
value of $\tilde{n}(p)$ for a given $d=p+1$. For $p=1$, where the
irreducible Majorana--Weyl spinor is one-dimensional ($Spin(1,1)$
is abelian), one needs $\Lambda_\alpha^r$ to be in a reducible
Majorana representation in the worldsheet spinor index $r$, i.e.
$\Lambda_\alpha^r= (\Lambda_\alpha^+, \Lambda_\alpha^-)$;
otherwise the second term in (\ref{pbr}) would be zero and the
action would become that of a tensionless $\Sigma^{({n(n+1)\over
2}|n)}$ supersymmetric string. Then, the action (\ref{pbr})
reduces to (\ref{St01}) using (\ref{vielbein}).

The fermionic variation $\delta_f$ of the action (\ref{pbr}),
$\delta_{f} S_p$, comes only from the variation of
$\Pi^{\alpha\beta}$. Let us simplify it by taking
 $\delta_{f}X^{\alpha\beta}= i
\delta_{f}{\theta}^{(\alpha}\,{\theta}^{\beta)}$ ({\it cf.} below
equation~(\ref{variation})), so that $i_{\delta_{f}}
\Pi^{\alpha\beta}=0$ and $\delta_{f} \Pi^{\alpha\beta}= - 2i
d\theta^{(\alpha} \delta\theta^{\beta)}$. As $\Pi^{\alpha\beta}$
enters  the action (\ref{pbr}) only through its contraction with
$\Lambda^r_\alpha \rho^a_{rs} \Lambda^s_\beta $ we find
\begin{eqnarray}\label{vSp}
& \delta_{f}  S_p = -2i \int_{W^{p+1}} \, e^{\wedge p}_a \, \wedge
 \,d\theta^{\alpha} \Lambda^r_\alpha \, \rho^a_{rs} \,
\Lambda^s_\beta  \delta\theta^{\beta}\; .
\end{eqnarray}
Thus only $\tilde{n}(p)$ fermionic variations
 $\delta\theta^{\beta} \Lambda^s_\beta$ out of the $n$ variations
 $\delta\theta^{\beta}$ are effectively involved in
$\delta_{f}  S_p$.

This reflects the presence of $(n-\tilde{n}(p))$
$\kappa$--symmetries in the dynamical system described by the
supersymmetric $p$--brane action (\ref{pbr}). They are defined by
\begin{equation}
\label{kapp1} \delta_{\kappa}X^{\alpha\beta}= i
\delta_{\kappa}{\theta}^{(\alpha}\,{\theta}^{\beta)} \; , \quad
\delta_{\kappa} e^a =0\; ,
\end{equation}
and by the following condition on
$\delta_{\kappa}{\theta}^{\alpha}$,
\begin{equation}
\label{kapp2} \delta_{\kappa}{\theta}^{\alpha} \Lambda_{\alpha}^r
=0  \; , \qquad r=1,\ldots , \tilde{n}(p) \; . \qquad
\end{equation}
This can be solved, using the auxiliary spinor fields $u^{\alpha
J}$ [where now $J=1, \ldots, (n-\tilde{n}(p))$] orthogonal to
 $ \Lambda_{\alpha}^r$, as
{\setlength\arraycolsep{2pt}
\begin{eqnarray}
\label{kapp3} \delta_{\kappa}{\theta}^{\alpha}& = & \,
\kappa_{_J}(\xi) \;  u^{\alpha J}(\xi) \; , \qquad
 u^{\alpha J} (\xi)\; \Lambda_{\alpha}^r (\xi ) = 0  \; ,
 \nonumber
\\  && J=1, \ldots , (n-\tilde{n}(p)) \; ,  \quad
r=1,\ldots , \tilde{n}(p) \; . \qquad
\end{eqnarray}
}The $\kappa$--symmetry (\ref{kapp1}), (\ref{kapp3}) implies the
preservation of all but $\tilde{n}(p)$ supersymmetries by the
corresponding $\nu=\frac{n-\tilde{n}(p)}{n}$ BPS state.

For instance, for $p=2$, $n=32$, it is $\tilde{n}=2$. The action
(\ref{pbr}) then describes excitations of a membrane BPS state
preserving all but $2$ supersymmetries, a ${30\over 32}$ BPS
state. For $p=5$ and $\tilde{n}=8$ the action (\ref{pbr}) with
$n=32$ describes a ${24\over 32}$ supersymmetric 5--brane model in
$\Sigma^{(528|32)}$. Both the su\-per\-mem\-brane (M2--brane) and
the super--5--brane (M5--brane) are known in the standard $D=11$
superspace, where they correspond to ${16\over 32}$ BPS states.
The speculation could be made that the `usual' M2 and M5
superbranes are related to the generalized $\Sigma^{(528|32)}$
supersymmetric 2--brane and 5--brane described by the action
(\ref{vSp}) for $p=2$ and $5$. For instance, they might be related
with some particular solutions to the equations of motion of the
corresponding ${30\over 32}$ and ${24\over 32}$
$\Sigma^{(528|32)}$ models preserving $16$ supersymmetries and/or
with the result of a dimensional reduction of them. For the $p=5$
case a question of a special interest would be the role of the M5
selfdual worldvolume gauge field in the $\Sigma^{(528|32)}$
superspace description (see \cite{JdA00} for a related
discussion). For $p=3$ and $\tilde{n}=4$ we have a ${28\over 32}$
BPS state, a BPS 3--brane. Neither the Green--Schwarz superstring
nor the super--3--brane exist in the standard $D=11$ superspace,
but a super--D3--brane does exist in the $D=10$ Type IIB
superspace, as the superstring does. The possible relation of
these preonic branes with the usual Type II branes would require
further study.


\fancyhf{} \fancyhead[LE,RO]{\nouppercase{\it\thepage}}
\fancyhead[RE,LO]{\nouppercase{\it\leftmark}}

\chapter{Conclusions and outlook}
\label{chapter8}

A number of topics about eleven-dimensional supergravity, the low
energy limit of the hypothetical M Theory, have been covered in
this Thesis. The role of (generalized) holonomy in the description
of supersymmetric solutions of supergravity has been discussed and
applied, in particular, to the search of possible preonic
solutions. The notion of BPS preons leads naturally to that of
enlarged superspaces and supersymmetry algebras, the role of which
in supergravity has also been explored. In particular, enlarged
superspaces have been shown to allow for the construction of
supersymmetric objects with a manifest content of BPS preons, and
enlarged superalgebras have appeared in the discussion of the
underlying symmetry of $D=11$ supergravity.

After an introductory chapter \ref{chapter1}, the conventions and
notation used throughout the Thesis (with the exception of chapter
\ref{chapter3}), were set in chapter \ref{chapter2}. The later
contains a general discussion of topics about eleven-dimensional
supergravity relevant for the remainder of the Thesis. The M
Theory superalgebra is introduced, and the Lagrangian, equations
of motion and symmetries of $D=11$ supergravity discussed.
Especial attention has been payed to supersymmetry, in relation to
which the notions of generalized connection, curvature and
holonomy have been reviewed. The interplay between the generalized
curvature and the equations of motion has also been discussed. In
particular, we have shown \cite{BAPV05} that all the bosonic
equations of $D$=11 supergravity can be collected in a single
equation, (\ref{SfEq}), written in terms of the generalized
curvature (\ref{calR}) which takes values in the algebra of the
generalized holonomy group. The concise form (\ref{SfEq}) of all
the bosonic equations is obtained by factoring out  the fermionic
one--form $\psi^\beta$ in the selfconsistency (or integrability)
conditions ${\cal D}\Psi_{10\; \beta}=0$ [Eqs. (\ref{DEqmpsi})],
for  the gravitino equations $\Psi_{10\, \alpha}=0$, Eqs.
(\ref{Eqmpsi}). In this sense, one can say that in (the second
order formalism of) $D=11$ CJS supergravity all the equations of
motion and Bianchi identities are encoded in the fermionic
gravitino equation
 $\Psi_{10\; \beta}  := {\cal D}
\psi^\alpha \wedge \bar{\Gamma}^{(8)}_{\alpha\beta} =0$ (equation
(\ref{Eqmpsi})). Actually  this should be expected for a
supergravity theory including  only one fermionic field, the
gravitino,  and whose supersymmetry algebra closes on shell.  As
we have  discussed, the basis for such an expectation is provided
by the second Noether theorem.

Generalized holonomy has been further explored in chapter
\ref{chapter3}, where especial emphasis has been made in the role
of supercovariant derivatives of the curvature to characterize the
holonomy algebra. After a general review of how higher order
integrability conditions might be necessary to properly determine
the Killing spinors characterizing purely bosonic supersymmetric
solutions of supergravity, the generalized holonomy of some well
known solutions has been revised. Supercovariant derivatives of
the generalized curvatures corresponding to the M2 and M5 brane
solutions of supergravity only turn out to help to close the
generalized holonomy algebra obtained at first order. The
situation is different for other solutions, such as Freund-Rubin
compactifications. An example has been provided by the
compactification on the squashed $S^7$. Left squashing preserves
$N=1$ supersymmetry, while its right counterpart breaks all
supersymmetries. This situation cannot be described if the
generalized holonomy is $G_2$, as obtained in first order. Second
order integrability, namely, the supercovariant derivative of the
generalized curvature yields a holonomy algebra of $so(7)$
\cite{BLVW03}, which gives the correct decompositions of the
Killing spinor for both left and right compactifications.

In chapter \ref{chapter4}, the role of the BPS preon notion in the
analysis of supersymmetric solutions of $D=11$ supergravity is
studied. This notion suggests the moving $G$--frame method
\cite{BAIPV03}, which is proposed as a useful tool in the search
for supersymmetric solutions of $D=11$ and $D=10$ supergravity. We
used this method here to make a step towards answering whether the
standard CJS supergravity  possesses a solution preserving $31$
supersymmetries, a solution that would correspond to a BPS preon
state. Although this question has not been settled for the CJS
supergravity case, we have shown in our framework that preonic,
$\nu=31/32$ solutions do exist \cite{BAIPV03} in a Chern--Simons
type $D=11$ supergravity.

Although the main search for preonic solutions concerns the `free'
bosonic CJS supergravity equations, one should not exclude other
possibilities, both inside and outside the CJS standard
supergravity framework. When, {\it e.g.}, super--$p$--brane
sources are included, the Einstein equation, and possibly the
gauge field equations and even the Bianchi identities, acquire
r.h.s.'s and the situation would have to be reconsidered. Another
source of modification of the CJS supergravity equations might be
due to `radiative' corrections of higher order in curvature. Such
modified equations might also allow for preonic solutions not
present in the unmodified ones. If it were found that only the
inclusion of these higher--order curvature terms allows for
preonic BPS solutions, this would indicate that BPS preons cannot
be seen in a classical low energy approximation of M Theory and,
hence, that they are intrinsically quantum objects.

The special role of BPS preons in the algebraic classification of
all the M Theory BPS states allows us to conjecture that they are
elementary (`quark-like') necessary ingredients of any model
providing a more complete description of M Theory. In such a
framework, if the standard supergravity did not contain $\nu
=31/32$ solutions, neither in its `free' form, nor in the presence
of a super--$p$--brane source, this might just indicate the need
for a wider framework for an effective description of M Theory.
Such an approach could include Chern--Simons supergravities
\cite{CS} and/or the use of larger, extended superspaces (see
\cite{JdA00,JdAIMO} and refs.~therein), in particular with
additional tensorial coordinates (also relevant in the description
of massless higher--spin theories \cite{BLS99,V01s,Sor05}). In
this perspective our observation that the BPS preonic
configurations do solve the bosonic equations of Chern--Simons
supergravity models looks interesting. It might be also worthwhile
to look at the role of vectors and higher order tensors that may
be constructed from the preonic spinors $\lambda_\alpha{}^r$, in
analogy with the use of the Killing vectors $K_{IJ}^a=\epsilon_I
\Gamma^a \epsilon_J$ and higher order bilinears $\epsilon_I
\Gamma^{a_1\cdots a_s} \epsilon_J$ made in references
\cite{GMPW04,GP02,GaGuPa03,GaMaSpWa04,GauntMaSpWa04,MacConamhna:2004fb,GoNeWa03,H-JPaSm03}.

Chapter \ref{chapter5} contains the technical details of the
expansion method \cite{JdAIMO,JdA02}, a procedure of obtaining new
(super)algebras ${\cal G}(N_0,\ldots,N_p)$ from a given one ${\cal
G}$. It is based in the power expansion of the Maurer-Cartan
equations that results from rescaling some group parameters. These
expansions are in principle infinite, but some truncations are
consistent and define the Maurer-Cartan equations of new
(super)algebras, the structure constants of which are obtained
from those of the original (super)algebra ${\cal G}$. We have
considered the different possible  ${\cal G}(N_0,\ldots,N_p)$
algebras subordinated to various splittings of ${\cal G}$ and
discussed their structure. We have seen that in some cases (when
the splitting of ${\cal G}$ satisfies the Weimar-Woods conditions)
the resulting algebras include the simple or generalized
\.In\"on\"u-Wigner contractions of ${\cal G}$, but that this is
not always the case. In general, the new `expanded' algebras have
higher dimension than the original one. Since ${\cal G}$ is the
only ingredient of the expansion method, it is clear that the
extension procedure (which involves {\it two} algebras) is richer
when one is looking for new (super)algebras; the expansion method
is more constrained. Nevertheless, we have used it to obtain the M
Theory superalgebra, including its Lorentz part, from the
orthosymplectic superalgebra $osp(1|32)$ as the expansion
$osp(1|32)(2,1,2)$ \cite{JdAIMO}.

The expansion method is also applied, already in chapter
\ref{chapter6}, to discuss the relation of the gauge structure of
$D=11$ supergravity with $osp(1|32)$. In this chapter, the
consequences of a possible composite structure, {\it \`a la}
D'Auria-Fr\'e, of the three--form field $A_3$ of the standard CJS
$D=11$ supergravity are studied. In particular, we have shown that
$A_3$  may be expressed in terms of the one-form gauge fields
$B^{ab}$, $B^{a_1\ldots a_5}$, $\eta^{\alpha}$, $e^a$,
$\psi^\alpha$ associated to  a {\it family} of superalgebras
$\tilde {\mathfrak E}(s) \equiv
\tilde{\mathfrak{E}}^{(528|32+32)}(s)$, $s\not=0$, corresponding
to the supergroups $\tilde{\Sigma}(s) \equiv
\tilde{\Sigma}^{(528|32+32)}(s)$ \cite{Lett,AnnP04}. Two values of
the  parameter $s$ recover the two earlier D'Auria--Fr\'e
\cite{D'A+F} decompositions of $A_3$, while one value of $s$,
$s=-6$ leads to a simple expression for $A_3$ that does not
involve
 $B^{a_1\ldots a_5}$. Indeed, the generator $Z_{a_1\ldots
 a_5}$ associated to $B^{a_1\ldots  a_5}$ is central in
 $\tilde {\mathfrak E}(-6)$, so that the
 smaller supergroup $\Sigma_{min}$ obtained by removing
$Z_{a_ 1\ldots  a_5}$ from $\tilde {\Sigma}(-6)$ can be regarded
as the minimal underlying gauge supergroup of supergravity
\cite{Lett,AnnP04}. The supergroups
 $\tilde{\Sigma}(s) \rtimes SO(1,10)$ with $s\not=0$
 are non-trivial (non-isomorphic) deformations  of the
 $\tilde{\Sigma}(0) \rtimes SO(1,10)\subset
 \tilde{\Sigma}(0) \rtimes Sp(32)$ supergroup,
which is itself the expansion \cite{Lett,JdAIMO}
$OSp(1|32)(2,3,2)$ of $OSp(1|32)$. For any $s\not=0$,
$\tilde{\Sigma}(s) \rtimes SO(1,10)$ may be looked at as a hidden
gauge symmetry of the $D=11$ CJS supergravity generalizing the
$D$=11 superPoincar\'e group
 $\Sigma^{(11|32)} \rtimes SO(1,10)$.

To study the possible  dynamical consequences of the composite
structure of $A_3$ we have followed the original proposal
\cite{D'A+F} of substituting the composite $A_3$ for the
fundamental $A_3$ in the first order CJS supergravity action
\cite{D'A+F,J+S99} of chapter \ref{chapter2}. It has been seen
that such an action possesses the right number of `extra' gauge
symmetries to make the number of degrees of freedom the same as in
the standard CJS supergravity \cite{AnnP04}. These are symmetries
under the transformations of the new one--form fields that leave
the composite $A_3$ field invariant; their presence is related to
the fact that the new gauge fields enter the supergravity action
only inside the $A_3$ field. In other words, the extra degrees of
freedom carried by the new fields $B^{ab}$, $B^{a_1\ldots a_5}$,
$\eta^{\alpha}$  are pure gauge. One may conjecture that these
extra degrees of freedom might be important in M Theory and that,
correspondingly, the extra gauge symmetries that remove them would
be broken by including in the supergravity action some exotic
`matter' terms that couple to the `new' additional one--form gauge
fields. In constructing such an `M--theoretical matter' action,
the preservation of the $\tilde{\Sigma}(s) \rtimes SO(1,10)$ gauge
symmetry  would provide a guiding principle.

We have stressed the equivalence between the problem of searching
for a composite structure of the $A_3$ field and, hence, for a
hidden gauge symmetry of $D=11$ supergravity, and that  of
trivializing a four--cocycle of the standard $D=11$ supersymmetry
algebra $\mathfrak{E} \equiv \mathfrak{E}^{(11|32)}$ cohomology
 on the enlarged superalgebras
 $\tilde{\mathfrak{E}}(s)$, $s\neq 0$. The generators of
$\tilde{\mathfrak{E}}(s)$  are in one--to--one correspondence with
the one--form  fields $e^a$, $\psi^\alpha$, $B^{ab}$,
$B^{a_1\ldots a_5}$, $\eta^{\alpha}$. For zero curvatures these
fields satisfy the same equations as the
${\tilde{\Sigma}}(s)$--invariant Maurer-Cartan forms of
$\tilde{\mathfrak{E}}(s)$ which, before pulling them back to a
bosonic eleven--dimensional spacetime surface, are expressed
through the coordinates $(x^a, \theta^\alpha, y^{ab}$,
$y^{a_1\ldots a_5}, \theta^{\prime \alpha})$ of the
${\tilde{\Sigma}}(s)$ superspace. This is the content of the
fields/extended superspace coordinates correspondence principle,
that conjectures that for the relevant supergravity theories there
always exists an enlarged superspace whose coordinates are in
one-to-one correspondence with the fields of the theory
\cite{JdA00,Azcarraga05}. $D=11$ supergravity itself can be
conjectured to be embedded in a larger superspace (see
\cite{AnnP04}).

Several interesting questions arise concerning the composite
nature of $A_3$. The first one was already sketched in chapter
\ref{chapter7} and refers to the trivialization of the FDA
seven-cocycle $\omega_7$ related to the dual formulation of $D=11$
supergravity. It would be interesting to check if $\omega_7$ is
already trivial on the family of superalgebras
$\tilde{\mathfrak{E}}(s)$, or further extensions are needed
instead to trivialize it. Another issue that would be worth
studying is the trivialization of the FDAs corresponding to lower
dimensional supergravities. It would be interesting, in
particular, to study whether the FDA corresponding to IIA and IIB
supergravities can be trivialized by some superalgebra and, in
case it were possible, to study its relation with the family
$\tilde{\mathfrak{E}}(s)$ trivializing the $D=11$ supergravity
FDA. Another question that would be interesting to analyze would
be the implications of the composite nature of $A_3$ in the
problem of the cosmological constant of $D=11$ supergravity,
argued in \cite{BaDeHeSe97} to be forbidden on cohomological
grounds.

In chapter \ref{chapter7}, we have presented a supersymmetric
string model in the `maximal' or `tensorial' superspace
$\Sigma^{({n(n+1)\over 2}|n)}$ with additional tensorial central
charge coordinates (for $n>2$) \cite{30/32}. The model possesses
$n$ rigid supersymmetries and $n-2$ local fermionic
$\kappa$--symmetries. This implies that it provides the worldsheet
action for the excitations of a BPS state preserving $(n-2)$
supersymmetries. In particular, for $n=32$ our model describes a
supersymmetric string with $30$ $\kappa$--symmetries in
$\Sigma^{(528|32)}$, which corresponds to a BPS state preserving
$30$ out of $32$ supersymmetries. This model can be treated as a
composite of two BPS preons \cite{BPS01} and is the second (after
the $D=11$ Curtright model \cite{Curtright}) tensionful extended
object model in $\Sigma^{(528|32)}$. In contrast with the
Curtright model \cite{Curtright}, our supersymmetric string action
in the enlarged $D=11$ superspace $\Sigma^{(528|32)}$ does not
involve any gamma--matrices, but instead makes use of two
constrained bosonic spinor variables, $\lambda_\alpha^+$ and
$\lambda_\alpha^-$, corresponding to the two BPS preons from which
the superstring BPS state is composed. As a result, our model
preserves the $Sp(32)$ subgroup of the $GL(32,\mathbb{R})$
automorphism symmetry of the $D=11$ M--algebra. Our
$\Sigma^{({n(n+1)\over 2}|n)}$ supersymmetric string model can be
treated as a higher spin generalization of the classical
Green--Schwarz superstring. At the same time, the additional
bosonic tensorial coordinate fields of the $n=32$ case might
contain information about topological charges corresponding to the
higher branes of the superstring/M Theory [71].

The $\Sigma^{({n(n+1)\over 2}|n)}$ model may also be formulated in
terms of a pair of constrained worldvolume $OSp(2n|1)$
supertwistors. The transition to the supertwistor formulation is
similar to that for the massless superparticle and the tensionless
$\Sigma^{({n(n+1)\over 2}|n)}$ supersymmetric $p$--branes
\cite{BL98, B02}. In our case, however, the supertwistors are
restricted by a constraint that breaks the generalized
superconformal $OSp(64|1)$ symmetry down to a generalization of
the super--Poincar\'e group, ${\Sigma}^{(528|32)}\rtimes Sp(32)$.
Such a breaking is characteristic of tensionful models. We note
that this constrained $OSp(2n|1)$ supertwistor framework might
also be useful for massive higher spin theories.

We have developed the Hamiltonian formalism, both in the original
and in the symplectic supertwistor representation, and found that,
while the Hamiltonian analysis in the original formulation
requires the use of the additional auxiliary spinor variables
$u_\alpha^I$ ($I=1,...,(n-2)$) orthogonal to $\lambda_\alpha^\pm$,
the symplectic supertwistor Hamiltonian mechanics can be discussed
in terms of the original phase space variables. Moreover, under
Dirac brackets, supertwistors become selfconjugate variables and
the symplectic structure of the phase space simplifies
considerably. A natural application of the Hamiltonian approach
developed here would be the BRST quantization of the
$\Sigma^{({n(n+1)\over 2}|n)}$ superstring model, which might
provide a `higher spin' counterpart of the usual string field
theory.

We have also presented a generalization of our
$\Sigma^{({n(n+1)\over 2}|n)}$ supersymmetric string model for
supersymmetric $p$--branes in $\Sigma^{({n(n+1)\over 2}|n)}$. They
correspond to BPS states preserving all but $\tilde{n}(p)$ (see
below (\ref{pbr})) supersymmetries, composites of $\tilde{n}(p)$
BPS preons $(\tilde{n}(2)=2\,,\; \tilde{n}(3)=4\,,\;
\tilde{n}(5)=8)$. In particular, the ${\Sigma}^{(528|32)}$
supersymmetric membrane ($p=2$) also corresponds to ${30\over 32}$
a BPS state.

In this Thesis, preonic solutions have been shown to exist in
enlarged superspaces or in the context of Chern-Simons
supergravities. It would be very interesting to determine whether
preonic solutions also occur as solutions of standard  CJS $D=11$
supergravity. The definitive answer would be provided by a
complete classification of supergravity solutions, that would also
shed light into the structure of M Theory itself. As future work,
we aim to make some steps towards a complete classification of CJS
supergravity solutions. The study of the interplay between the
approaches to classify general supergravity solutions, is expected
to give us new insights towards that classification. In
particular,  the presence of Killing spinors implies the existence
of a $G$-structure \cite{GMPW04,GP02} on the tangent bundle, that
is, a sub-bundle of the frame bundle with structure group $G$. Its
existence can be seen from the fact that a set of covariantly
constant tensors exist on the tangent bundle, built as bilinears
of the Killing spinors. The differential and algebraic conditions
that these tensors satisfy turn out to constrain the geometry (the
metric) and the fluxes.

The generalized holonomy approach \cite{Duff03,Hull03}, on the
contrary, deals with the supergravity connection as a Clifford
algebra valued connection, as discussed in chapter \ref{chapter2},
without focussing on the tangent bundle or the spinor bundle of
the background. The difficulty in relating both approaches could
be put down to that fact. By dealing with an $sl(32,
\mathbb{R})$-valued connection, the Killing spinors are not any
longer regarded as spinors in the tangent bundle (transforming in
suitable representations of the tangent bundle structure group,
$SO(1,10)$), but are instead promoted to vectors of $SL(32,
\mathbb{R})$. It is not obvious that this step does not entail any
loss of the information contained in the spin bundle \cite{GMSW04}
so, if this were indeed the case, supplementary conditions should
be derived to account for the fact that the relevant $SL(32,
\mathbb{R})$-vectors are also tangent bundle spinors.

An interesting question that would shed light into the relation of
both approaches is what subgroups $H$ of the generalized structure
group can actually arise as generalized holonomy groups of
supergravity solutions. A refined definition of holonomy taking
into account covariant derivatives of the curvature (higher order
commutators of the covariant derivatives) \cite{BLVW03} could be
relevant with this regard. It could also be worth studying the
effects of the effect successive covariant derivatives of the
curvature when no fluxes are considered but higher order
gravitational corrections are taken into account \cite{LPS05}.
Alternatively, the study of the relevant $G$-structures of
solutions including higher order corrections could allow us both
to generate new examples, and help us to understand the origin of
this higher order gravitational terms in the fully-fledged M
Theory.

The classification of supergravity solutions is not only expected
to provide insights into the structure of M Theory, but also to
provide new backgrounds for Standard Model building and for the
AdS/CFT correspondence. In the later case, the $G$-structure
approach has provided solutions containing AdS factors
\cite{GMSW04,GaMaSpWa04,GauntMaSpWa04} suitable to test the
correspondence (see \cite{GMSW04,GMPW04} and references therein
for a review). We expect to be able to make progress also in this
direction.


\renewcommand{\theequation}{\thechapter.\arabic{equation}}

\appendix

\chapter[Second order integrability for the squashed $S^7$]
{Second order integrability \\ for the squashed $S^7$}
\label{holS7}

In this Appendix we present the details of the derivation of the
linearly independent generators (\ref{G2}) and (\ref{SO7}) of the
generalized holonomy algebra $\hol(\Omega_m)=so(7)$ of the
squashed $S^7$, associated to the second order integrability
condition (\ref{int2}). For convenience, we rewrite (\ref{int2})
with a modified normalization
\begin{equation} \label{second}
M_{abc} = 5\left( \sqrt{5}
     D_a C_{bcde} \Gamma^{de} \
     -m' C_{bcad} \Gamma^d \right)    \; ,
\end{equation}
where we have defined
\begin{equation}
m'=2\sqrt{5} i m ,
\end{equation}
and have chosen the $-$ sign in front of $m'$ for definiteness.

To obtain $M_{abc}$, we have computed both the Weyl tensor
$C_{bcad}$ (given in \cite{DNP}) and its (L\'evi-Civita) covariant
derivative $D_aC_{bcde}$. We obtain, for the non-vanishing
generators (\ref{second}) \cite{BLVW03}:
{\setlength\arraycolsep{1pt}
\begin{eqnarray}
\label{1} M_{00j} &=& 4 \Gamma_{0 \hat{j}} - \epsilon_{jkl}
\Gamma^{k \hat{l}} -2m' \Gamma_j \; , \\[4pt]
\label{2} M_{00\hat{j}} &=& 4 \Gamma_{0j} +
 \epsilon_{jkl}  \Gamma^{kl} +2m'\Gamma_{\hat{j}} \; , \\[4pt]
\label{3} M_{0ij} &=& 2 \epsilon_{ijk} \Gamma^{0 \hat{k}} +
\Gamma_{i \hat{j}}-\Gamma_{j \hat{i}} \; , \\[4pt]
\label{4} M_{0i\hat{j}} &=& -\epsilon_{ijk} \Gamma^{0k} +
\Gamma_{ij}-3\Gamma_{\hat{i} \hat{j}} + m' \epsilon_{ijk}
\Gamma^{\hat{k}} \; , \\[4pt]
\label{5} M_{0\hat{i}\hat{j}} &=& - 3\Gamma_{i \hat{j}}
+3\Gamma_{j\hat{i}} -2m' \epsilon_{ijk} \Gamma^{k} \; , \\[4pt]
\label{6} M_{h0j} &=& \epsilon_{hjk} \Gamma^{0 \hat{k}} +2
\Gamma_{h\hat{j}} + \delta_{hj} \delta^{kl} \Gamma_{k\hat{l}}
+\Gamma_{j\hat{h}} +2m' \delta_{hj} \Gamma_{0} \; ,  \\[4pt]
\label{7} M_{h0\hat{j}} &=& -\epsilon_{hjk} \Gamma^{0k} +
\Gamma_{hj} + 3 \Gamma_{\hat{h}\hat{j}} -m' \epsilon_{hjk}
\Gamma^{\hat{k}} \; , \\[4pt]
\label{8} M_{hij} &=&  \delta_{hi} \Gamma_{0\hat{j}} -\delta_{hj}
\Gamma_{0\hat{i}} + 4\epsilon_{ij}{}^k \Gamma_{h\hat{k}}
-\epsilon_{hij} \delta^{kl} \Gamma_{k\hat{l}} - \epsilon_{ij}{}^k
\Gamma_{k\hat{h}} \nonumber \\* && + 2m'(\delta_{hj} \Gamma_i -
\delta_{hi} \Gamma_j) \; , \\[6pt]
\label{9} M_{hi\hat{j}} &=& (2 \epsilon_{jkl} \delta_{hi} -\ft12
\epsilon_{hkl} \delta_{ij}  -\ft12 \epsilon_{ikl} \delta_{hj})
\Gamma^{kl}  -3\epsilon_{hi}{}^k \Gamma_{\hat{k} \hat{j}}
+2\delta_{hi} \Gamma_{0j}
\nonumber \\
&&  +\delta_{ij} \Gamma_{0h} +\delta_{hj} \Gamma_{0i}   +
m'(2\delta_{hi} \Gamma_{\hat{j}} - \delta_{ij} \Gamma_{\hat{h}}
+\delta_{hj} \Gamma_{\hat{i}}) \; ,  \\[6pt]
\label{10}
 M_{h\hat{i}\hat{j}} &=&  3\delta_{hi} \Gamma_{0\hat{j}}
-3\delta_{hj} \Gamma_{0\hat{i}} + 3 \epsilon_{hi}{}^k
\Gamma_{k\hat{j}} -3\epsilon_{hj}{}^k \Gamma_{k\hat{i}} \nonumber
\\ && + 2m'(\epsilon_{hij} \Gamma_0 + \delta_{hj} \Gamma_i
-\delta_{hi} \Gamma_j ) \; , \\[6pt]
\label{11} M_{\hat{h}0j} &=& -6 \Gamma_{\hat{h}\hat{j}} + 2m'
\epsilon_{hjk} \Gamma^{\hat{k}} \; , \\[4pt]
\label{12} M_{\hat{h}0\hat{j}} &=& 3\Gamma_{j \hat{h}} -3
\delta_{hj} \delta^{kl}\Gamma_{k\hat{l}} -m' (2\delta_{hj}
\Gamma_0 + \epsilon_{hjk} \Gamma^k) \; , \\[4pt]
\label{13} M_{\hat{h}ij} &=& 6 \epsilon_{ij}{}^k
\Gamma_{\hat{k}\hat{h}} +2m'( \delta_{hj}
\Gamma_{\hat{i}}-\delta_{hi} \Gamma_{\hat{j}}) \; , \\[4pt]
\label{14} M_{\hat{h}i\hat{j}} &=& 3\delta_{hj} \Gamma_{0 \hat{i}}
-3\delta_{ij} \Gamma_{0 \hat{h}}-
3\delta_{hi}\epsilon_{jkl}\Gamma^{k\hat{l}} -
3\epsilon_{ij}{}^l \Gamma_{h\hat{l}}\nonumber\\
&&+m'( -\epsilon_{hij} \Gamma_0 -2\delta_{hj} \Gamma_i
+\delta_{ij} \Gamma_{h} -\delta_{hi} \Gamma_j) \; , \\[8pt]
\label{15} M_{\hat{h}\hat{i}\hat{j}} &=&  6\delta_{hj} \Gamma_{0i}
 -6\delta_{hi} \Gamma_{0j}
-6 \epsilon_{ij}{}^k\Gamma_{kh} +4m'(\delta_{hj} \Gamma_{\hat{i}}
-\delta_{hi} \Gamma_{\hat{j}}) \; .
\end{eqnarray}
}Not all the generators included in (\ref{1})--(\ref{15}) are
linearly independent, however. After all, they are built up from
Dirac matrices $\{ \Gamma_{ab}, \Gamma_a \}$, that is, from
generators of $\SO(8)$, so at most 28 can be linearly independent.

In fact, only 21 linearly independent generators are contained in
(\ref{1})--(\ref{15}), as we will now show. Some redundant
generators are straightforward to detect, since the Bianchi
identities for the Weyl tensor, $D_{[a} C_{bc]de} =0$ and
$C_{[bca]d}=0$ place the restrictions
\begin{equation}
M_{[abc]}=0 \; .
\end{equation}
Further manipulations show that only the generators (\ref{9}) and
(\ref{14}) are relevant, the rest being linear combinations of
them. The generators  (\ref{2}), (\ref{4}), (\ref{7}), (\ref{11}),
(\ref{13}) and (\ref{15}) are obtained from (\ref{9}):
{\setlength\arraycolsep{1pt}
\begin{eqnarray}
M_{00\hat{j}} &=& \ft15 \delta^{kl}
( M_{kl\hat{j}} + M_{kj\hat{l}} + M_{jk\hat{l}} ) \; , \\[4pt]
M_{0i\hat{j}} &=& \ft15 \epsilon_{[i \mid}{}^{kl} (4 M_{k \mid j ]
\hat{l}} -M_{\mid j ] k\hat{l}})
-\ft15 \epsilon_{ij}{}^k \delta^{lm} M_{lm\hat{k}} \; , \\[4pt]
M_{i0\hat{j}} &=& -\ft15 \epsilon_{[i \mid}{}^{kl} (M_{k \mid j ]
\hat{l}} -4M_{\mid j ] k\hat{l}})
-\ft15 \epsilon_{ij}{}^k \delta^{lm} M_{lm\hat{k}} \; , \\[4pt]
M_{\hat{j}0i} &=& -\epsilon_{[i \mid}{}^{kl}
(M_{k \mid j ] \hat{l}} -M_{\mid j ] k\hat{l}}) \; , \\[4pt]
M_{\hat{h}ij} &=& M_{ji\hat{h}} - M_{ij\hat{h}} \; , \\[4pt]
M_{\hat{h}\hat{i}\hat{j}} &=& \ft15 ( M_{hi\hat{j}} -M_{hj\hat{i}}
+ M_{ih\hat{j}} - M_{jh\hat{i}} ) \nonumber \\* && -\ft45
\delta^{kl} (\delta_{hi} M_{kl\hat{j}} - \delta_{hj} M_{kl\hat{i}}
) \; ,
\end{eqnarray}
}while  (\ref{1}), (\ref{3}), (\ref{5}), (\ref{6}), (\ref{8}),
(\ref{10}) and (\ref{12}) are linear combinations of (\ref{14}):
{\setlength\arraycolsep{1pt}
\begin{eqnarray}
M_{00j} &=& \ft13 \delta^{kl}
(M_{\hat{k}j\hat{l}}-M_{\hat{j}k\hat{l}} ) \; , \\[4pt]
M_{0hj} &=& -\ft13 \epsilon_h{}^{kl} (M_{\hat{k}l\hat{j}} +3
M_{\hat{j}k\hat{l}} ) +\ft13 \epsilon_j{}^{kl}
 \left(M_{\hat{k}l\hat{h}} +3M_{\hat{h}k\hat{l}} \right) \; , \\[4pt]
M_{0\hat{h}\hat{j}} &=& \epsilon_h{}^{kl} (M_{\hat{k}l\hat{j}} +2
M_{\hat{j}k\hat{l}} ) - \epsilon_j{}^{kl}
 (M_{\hat{k}l\hat{h}} +2M_{\hat{h}k\hat{l}} ) \; , \\[4pt]
M_{h0j} &=& -\ft16 \epsilon_h{}^{kl} (2 M_{\hat{k}l\hat{j}} +5
M_{\hat{j}k\hat{l}} ) +\ft16 \epsilon_j{}^{kl}
M_{\hat{h}k\hat{l}} \; , \\[4pt]
M_{hij} &=& \ft12 \delta^{kl} \left( \delta_{hi} (
M_{\hat{k}l\hat{j}} - 2M_{\hat{j}k\hat{l}} ) -\delta_{hj} (
M_{\hat{k}l\hat{i}} - 2M_{\hat{i}k\hat{l}} ) \right) +
M_{\hat{i}j\hat{h}} \nonumber \\
&& \kern-1.5em - M_{\hat{j}i\hat{h}} + \ft73 ( M_{\hat{h}i\hat{j}}
- M_{\hat{h}j\hat{i}} )  - \ft23 \epsilon_h{}^{kl}
\epsilon_{ij}{}^m ( M_{\hat{k}l\hat{m}}+4M_{\hat{m}k\hat{l}}
)\;,\\[6pt]
M_{h\hat{i}\hat{j}} &=&  M_{\hat{i}h\hat{j}} -  M_{\hat{j}h\hat{i}} \; , \\[4pt]
M_{\hat{h}0\hat{j}}&=&\epsilon_h{}^{kl} (M_{\hat{k}l\hat{j}} +
M_{\hat{j}k\hat{l}} ) -\epsilon_j{}^{kl} M_{\hat{h}k\hat{l}} ,
\end{eqnarray}
}

 Moreover, both (\ref{9}) and (\ref{14}) contain redundant
generators. The following combinations obtained from (\ref{9}):
{\setlength\arraycolsep{0pt}
\begin{eqnarray}
&& \Cm_{0i} = \ft16 \delta^{kl} M_{ik\hat{l}} \; , \\[4pt]
&& \Cm_{ij} = -\ft1{30} \epsilon_{[i \mid}{}^{kl} (M_{k \mid j ]
\hat{l}} -9M_{\mid j ] k\hat{l}})
-\ft1{30} \epsilon_{ij}{}^k \delta^{lm} M_{lm\hat{k}} \; , \\[4pt]
&& M_{ij} = \ft16 M_{\hat{j}0i}= -\ft16 \epsilon_{[i \mid}{}^{kl}
(M_{k \mid j ] \hat{l}} -M_{\mid j ] k\hat{l}}) \;
\end{eqnarray}
}(the expressions of which in terms of Dirac matrices are the
first two equations in (\ref{G2}) and the first equation in
(\ref{SO7}), respectively) are linearly independent.  Thus
(\ref{9}) [and so (\ref{2}), (\ref{4}), (\ref{7}), (\ref{11}),
(\ref{13}) and (\ref{15})] can be uniquely written in terms of
them:
{\setlength\arraycolsep{2pt}
\begin{eqnarray}
M_{hi\hat{j}} &=& 2\delta_{hi} \Cm_{0j}+ \delta_{ij} \Cm_{0h} +
\delta_{hj} \Cm_{0i} - 3\delta_{hi} \epsilon_j{}^{kl} M_{kl} -3
\epsilon_{hi}{}^k M_{kj}
\nonumber \\
&& + (2 \epsilon_j{}^{kl} \delta_{hi} -\ft12 \epsilon_h{}^{kl}
\delta_{ij}  -\ft12 \epsilon_i{}^{kl} \delta_{hj}) \Cm_{kl} \; .
\end{eqnarray}

}

 Similarly, the following combinations contained in (\ref{14}):
{\setlength\arraycolsep{0pt}
\begin{eqnarray}
&& \Cm_{i \hat{j}} = \ft13 \epsilon_i{}^{kl} M_{\hat{j}k\hat{l}} -
\ft16  \epsilon_j{}^{kl} (M_{\hat{k}l\hat{i}} +
M_{\hat{i}k\hat{l}} ) \; , \\[4pt]
&& M_i = \ft1{12} \delta^{kl} (M_{\hat{k}l\hat{i}} - 2
M_{\hat{i}k\hat{l}}) \; , \\[4pt]
&& M = -\ft16 \epsilon^{hij} M_{\hat{h}i\hat{j}} \;
\end{eqnarray}
}(which can be written in terms of Dirac matrices as in the final
equation of (\ref{G2}) and the last two equations of (\ref{SO7}),
respectively) are linearly independent.  Hence (\ref{14}) [and so
(\ref{1}), (\ref{3}), (\ref{5}), (\ref{6}), (\ref{8}), (\ref{10})
and (\ref{12})] can be uniquely written in terms of them:
{\setlength\arraycolsep{2pt}
\begin{eqnarray}
M_{\hat{h}i\hat{j}} &=& 6 \delta_{hi} \epsilon_j{}^{kl}
\Cm_{k\hat{l}}
-2\epsilon_{ij}{}^k( \Cm_{k\hat{h}}-2 \Cm_{h\hat{k}}) \nonumber \\
&&+ 6 \delta_{hj} M_i  + 3\delta_{ih} M_j - 3\delta_{ij} M_h
-\epsilon_{hij} M \; .
\end{eqnarray}
}

 In summary, the linearly independent generators associated to
the second order integrability condition (\ref{int2}) are the 21
linearly independent generators (\ref{G2}) and (\ref{SO7}), namely
$\{ \Cm_{0i}$, $\Cm_{ij}$, $\Cm_{i\hat{j}}$, $M_{ij}$, $M_i, M \}$
(notice that $\Cm_{i\hat{j}}$ contains 8 generators, since it is
traceless), which close into an algebra whenever $m^2$ takes the
value required by the equations of motion,  $m^2 = \frac{9}{20}$.
Since the only condition for the generators to close the algebra
is placed on $m^2$, they will close regardless of the orientation
({\it i.e.}, of the sign of $m$). In fact, they generate the
21-dimensional algebra $so(7)$, for both orientations
\cite{BLVW03}.

Note that, by further choosing linear combinations of (\ref{G2}),
the 14 generators $\{\Cm_{0i}$, $\Cm_{ij}$, $\Cm_{i\hat j}\}$ of
$G_2$ may be re-expressed in symmetric form
\begin{eqnarray}
&&\Gamma_{1\hat1}-\Gamma_{2\hat2},\qquad\Gamma_{1\hat1}-\Gamma_{3\hat3},
\nonumber\\
&&\Gamma_{0\hat i}+\Gamma_{j\hat k},\qquad \Gamma_{0\hat
i}+\Gamma_{\hat jk},\qquad(i,j,k=123,231,312)
\nonumber\\
&&\Gamma_{0i}+\Gamma_{\hat j\hat k},\qquad
\Gamma_{0i}-\Gamma_{jk},\qquad(i,j,k=123,231,312).
\label{eq:g2can}
\end{eqnarray}
The 7 additional generators $\{M_{ij}$, $M_i$, $M\}$ of
(\ref{SO7}) extending (\ref{eq:g2can}) to $so(7)$ may also be
simplified in appropriate linear combinations.  One possible set
of generators is given by \cite{BLVW03}:
\begin{eqnarray}
&&\Gamma_{1\hat1}\pm i\Gamma_0,\nonumber\\
&&\Gamma_{0\hat i}\mp i\Gamma_{i},\nonumber\\
&&\Gamma_{\hat j\hat k}\mp i\Gamma_{\hat
i},\qquad(i,j,k=123,231,312).
\end{eqnarray}

\chapter{Expansion of $d\omega^{k_s,\alpha_s}$}
\label{ap:expansion}

This appendix contains the details of the derivation of the
results summarized in table 5.3, about what one-form coefficients
$\omega^{i_p, \beta_p}$ are needed to express $d \omega^{k_s,
\alpha_s}$ when the original algebra ${\cal G}$ is split as in
(\ref{eq:splitn}), with the structure constants satisfying
(\ref{eq:cont}).

Inserting (\ref{eq:szero})-(\ref{eq:sn}) into (\ref{eq:MC}) where
now $p,q,s=0,1,\ldots,n$, and using
\begin{equation} \label{eq:sumatorio}
\left( \sum_{\alpha=p}^{\infty} \lambda^\alpha \omega^{i_{ p},
\alpha} \right) \wedge \left( \sum_{\alpha=q}^{\infty}
\lambda^\alpha \omega^{j_{ q}, \alpha} \right) =
      \sum_{\alpha=p+q}^{\infty}
\lambda^\alpha \sum_{\beta=p}^{\alpha-q} \omega^{i_{ p}, \beta}
\wedge \omega^{j_{ q}, \alpha-\beta} \; ,
\end{equation}
we obtain the expansion of the MC equations for ${\cal G}$,
\begin{equation} \label{eq:MCap}
\sum_{\alpha=s}^{\infty} \lambda^\alpha d\omega^{k_{ s}, \alpha}=
\sum_{\alpha=s}^{\infty} \lambda^\alpha \left[ -\frac{1}{2} c_{i_{
p}j_{ q}}^{k_{ s}} \sum_{\beta=0}^{\alpha} \omega^{i_{ p}, \beta}
\wedge \omega^{j_{ q}, \alpha-\beta} \right] \; ,
\end{equation}
since the W-W conditions (\ref{eq:cont}) will give zero in the
r.h.s.~unless $\alpha=p+q \geq s$, in agreement with the
l.h.s.~equation~(\ref{eq:MCap}) can be made explicit for
$p,q,s=0,1,\ldots,n$ as follows \cite{JdAIMO}:
{\setlength\arraycolsep{2pt}
\begin{eqnarray}
\sum_{\alpha=s}^{\infty} \lambda^\alpha d \omega^{k_s , \alpha} &
= & -\frac{1}{2} \left[c_{i_0 j_0}^{k_s} \sum_{\alpha=0}^{\infty}
\lambda^\alpha \sum_{\beta=0}^\alpha \omega^{i_0 , \beta} \wedge
\omega^{j_0 ,\alpha-\beta} \right. +
      \nonumber \\
&& \left. \qquad  + 2 c_{i_0 j_1}^{k_s} \sum_{\alpha=1}^{\infty}
\lambda^{\alpha} \sum_{\beta=0}^{\alpha-1} \omega^{i_0 , \beta}
\wedge \omega^{j_1 ,\alpha-\beta} + \ldots + \right.
      \nonumber  \\
& & \left. \qquad  +2 c_{i_0 j_n}^{k_s} \sum_{\alpha=n}^{\infty}
\lambda^{\alpha} \sum_{\beta=0}^{\alpha-n} \omega^{i_0 , \beta}
\wedge \omega^{j_n ,\alpha-\beta} + \right.
      \nonumber  \\
& & \left. \qquad + c_{i_1 j_1}^{k_s} \sum_{\alpha=2}^{\infty}
\lambda^{\alpha} \sum_{\beta=1}^{\alpha-1} \omega^{i_1 , \beta}
\wedge \omega^{j_1 ,\alpha-\beta} + \ldots + \right.
\nonumber  \\
& & \left. +2 c_{i_1 j_n}^{k_s} \sum_{\alpha=1+n}^{\infty}
\lambda^{\alpha} \sum_{\beta=1}^{\alpha-n} \omega^{i_1 , \beta}
\wedge \omega^{j_n ,\alpha-\beta} + \ldots + \right.
\nonumber  \\
& & \left. +c_{i_{n-1} j_{n-1}}^{k_s} \sum_{\alpha=2n-2}^{\infty}
\lambda^{\alpha} \sum_{\beta=n-1}^{\alpha-n+1} \omega^{i_{n-1} ,
\beta} \wedge \omega^{j_{n-1} ,\alpha-\beta} + \right.
\nonumber  \\
& & \left. +2 c_{i_{n-1} j_{n}}^{k_s} \sum_{\alpha=2n-1}^{\infty}
\lambda^{\alpha} \sum_{\beta=n-1}^{\alpha-n} \omega^{i_{n-1} ,
\beta} \wedge \omega^{j_{n} ,\alpha-\beta} + \right.
\nonumber  \\
& & \left. + c_{i_{n} j_{n}}^{k_s} \sum_{\alpha=2n}^{\infty}
\lambda^{\alpha} \sum_{\beta=n}^{\alpha-n} \omega^{i_{n} , \beta}
\wedge \omega^{j_{n} ,\alpha-\beta} \right] \quad .
\end{eqnarray}}

\noindent Rearranging powers we get {\setlength\arraycolsep{0pt}
\begin{eqnarray} \label{eq:calc}
&& \sum_{\alpha=s}^{\infty} \lambda^\alpha d \omega^{k_s , \alpha}
=
    -\frac{1}{2} \Bigg[ c_{i_0 j_0}^{k_s} \omega^{i_0, 0} \wedge
\omega^{j_0 , 0} + \nonumber \\
&& \quad  + \lambda \Big( c_{i_0 j_0}^{k_s} \sum_{\beta=0}^{1}
\omega^{i_{0} , \beta} \wedge \omega^{j_{0} ,1-\beta} + 2 c_{i_0
j_1}^{k_s} \omega^{i_0, 0} \wedge \omega^{j_1, 1} \Big) +
\nonumber  \\
& & \quad  + \lambda^2 \Big( c_{i_0 j_0}^{k_s} \sum_{\beta=0}^{2}
\omega^{i_{0} , \beta} \wedge \omega^{j_{0} ,2-\beta} + \nonumber
\\
&& \quad \qquad \quad +2 c_{i_0 j_1}^{k_s} \sum_{\beta=0}^{1}
\omega^{i_0, \beta} \wedge \omega^{j_1 , 2-\beta} + 2 c_{i_0
j_2}^{k_s} \omega^{i_{0} , 0}
\wedge \omega^{j_{2} ,2}+ \nonumber  \\
 & & \quad \qquad \quad   +c_{i_1 j_1}^{k_s} \omega^{i_{1} , 1} \wedge
\omega^{j_{1} ,1}  \Big) + \ldots \Bigg] \; .
\end{eqnarray}}

\noindent Equation~(\ref{eq:calc}) now gives
{\setlength\arraycolsep{2pt}
\begin{eqnarray}
\sum_{\alpha=s}^{\infty} \lambda^\alpha d \omega^{k_s , \alpha} &
= &
      -\frac{1}{2} c_{i_0 j_0}^{k_s} \omega^{i_0, 0} \wedge \omega^{j_0 ,
0} -
\nonumber \\
&& - \sum_{\alpha=1}^{n-1} \lambda^\alpha \left[
      \frac{1}{2} \sum_{p=0}^{[\frac{\alpha}{2}]}
c_{i_p j_p}^{k_s} \sum_{\beta=p}^{\alpha-p} \omega^{i_p , \beta}
\wedge \omega^{j_p , \alpha-\beta} + \right. \nonumber \\
&& \left. \kern+4em + \sum_{p=0}^{[\frac{\alpha-1}{2}]}
\sum_{q=p+1}^{\alpha-p} c_{i_p j_q}^{k_s}
\sum_{\beta=p}^{\alpha-q} \omega^{i_p , \beta} \wedge
\omega^{j_q , \alpha-\beta} \right] - \nonumber \\
&& - \sum_{\alpha=n}^{2n-1} \lambda^\alpha \left[
      \frac{1}{2} \sum_{p=0}^{[\frac{\alpha}{2}]}
c_{i_p j_p}^{k_s} \sum_{\beta=p}^{\alpha-p} \omega^{i_p , \beta}
\wedge \omega^{j_p , \alpha-\beta} + \right. \nonumber \\
&& \left. \kern+4em  + \sum_{p=0}^{[\frac{\alpha-1}{2}]}
\sum_{q=p+1}^{\textrm{min} \left\{ \alpha-p \, , n \right\}}
c_{i_p j_q}^{k_s} \sum_{\beta=p}^{\alpha-q} \omega^{i_p , \beta}
\wedge
\omega^{j_q , \alpha-\beta} \right] - \nonumber  \\
&& - \sum_{\alpha=2n}^{\infty} \lambda^\alpha \left[
      \frac{1}{2} \sum_{p=0}^{n}
c_{i_p j_p}^{k_s} \sum_{\beta=p}^{\alpha-p} \omega^{i_p , \beta}
\wedge \omega^{j_p , \alpha-\beta} + \right. \nonumber \\
&& \left. \kern+4em + \sum_{p=0}^{n-1} \sum_{q=p+1}^{n} c_{i_p
j_q}^{k_s} \sum_{\beta=p}^{\alpha-q} \omega^{i_p , \beta} \wedge
\omega^{j_q , \alpha-\beta} \right] ,
\end{eqnarray}}

\noindent that is {\setlength\arraycolsep{0pt}
\begin{eqnarray}
&& \sum_{\alpha=s}^{\infty} \lambda^\alpha d \omega^{k_s , \alpha}
=      -\frac{1}{2} c_{i_0 j_0}^{k_s} \omega^{i_0, 0} \wedge
\omega^{j_0 ,
0} - \nonumber \\
&& \quad  - \sum_{\alpha=1}^{\infty} \lambda^\alpha \left[
\frac{1}{2} \sum_{p=0}^ {\textrm{min} \left\{ [\frac{\alpha}{2}]
\, , n \right\}} c_{i_p j_p}^{k_s} \sum_{\beta=p}^{\alpha-p}
\omega^{i_p , \beta} \wedge
\omega^{j_p , \alpha-\beta} + \right . \nonumber \\
& & \qquad + \left . \sum_{p=0}^{\textrm{min} \left\{
[\frac{\alpha-1}{2}] \, , n-1 \right\}} \sum_{q=p+1}^{\textrm{min}
\left\{ \alpha-p \, , n \right\}} c_{i_p j_q}^{k_s}
\sum_{\beta=p}^{\alpha-q}
      \omega^{i_p , \beta} \wedge \omega^{j_q , \alpha-\beta} \right] \; ,
\end{eqnarray}}

\noindent from which we obtain, upon explicit imposition of the
contraction condition (\ref{eq:cont}) on the structure constants
$c$'s \cite{JdAIMO}:

\vspace*{2em}

\noindent $\alpha=s=0$:
\begin{equation}\label{facil}
d \omega^{k_0 , 0} = - \frac{1}{2} c_{i_0 j_0}^{k_0} \omega^{i_0 ,
0} \wedge \omega^{j_0 , 0} \; ;
\end{equation}
$\alpha =s \geq 1$, $s$ odd:
\begin{equation}
d \omega^{k_s , s} = -\sum_{p=0}^{\frac{s-1}{2}} c_{i_p
j_{s-p}}^{k_s} \omega^{i_p , p} \wedge \omega^{j_{s-p} , s-p} \; ;
\end{equation}
$\alpha=s \geq 1$, $s$ even:
\begin{equation}
d \omega^{k_s , s} = - \frac{1}{2} c_{i_{\frac{s}{2}}
j_{\frac{s}{2}}}^{k_s} \omega^{i_{\frac{s}{2}} , \frac{s}{2}}
\wedge \omega^{j_{\frac{s}{2}} , \frac{s}{2}}
-\sum_{p=0}^{\frac{s-2}{2}} c_{i_p j_{s-p}}^{k_s} \omega^{i_p , p}
\wedge \omega^{j_{s-p} , s-p} \; ;
\end{equation}
$\alpha > s \geq 0$: {\setlength\arraycolsep{0pt}
\begin{eqnarray} \label{eq:MCexp2}
&& d \omega^{k_s , \alpha} = - \frac{1}{2} \sum_{p= \left[
\frac{s+1}{2} \right]}^{\textrm{min} \left\{ [\frac{\alpha}{2}] \,
, n \right\}} c_{i_p j_p}^{k_s} \sum_{\beta=p}^{\alpha-p}
\omega^{i_p , \beta} \wedge \omega^{j_p , \alpha-\beta} -
\nonumber \\
&& \quad -  \sum_{p=0}^{\textrm{min} \left\{ [\frac{\alpha-1}{2}]
\, , n-1 \right\}} \sum_{q=\textrm{max} \left\{ s-p, p+1
\right\}}^ {\textrm{min} \left\{ \alpha-p \, , n \right\}} c_{i_p
j_q}^{k_s} \sum_{\beta=p}^{\alpha-q}
      \omega^{i_p , \beta} \wedge \omega^{j_q , \alpha-\beta} \; .
\end{eqnarray}}

\chapter[Symmetry breaking of the  supertwistor
string]{Symmetry breaking of \\ the  supertwistor string}
\label{ap:breaking}



This appendix contains the details of the breaking of the
$OSp(2n|1)$ symmetry down to the supergroup
${\Sigma}^{({n(n+1)\over 2}|n)} \rtimes Sp(n)$, generalizing
superPoincar\'e, in the supertwistor formulation (section
\ref{twistor}) of the supersymmetric string model in tensorial
superspace of chapter \ref{chapter7}.

The supergroup $OSp(2n|1)$ is characterized by the $(2n+1) \times
(2n+1)$ supermatrices ${\cal G}_{\Sigma}{}^{\Pi}$ that preserve
the graded-antisymmetric matrix $\Omega_{\Sigma\Pi} =
-(-1)^{\mathrm{deg}({\Sigma}) \mathrm{deg}({\Pi})} \Omega_{\Pi
\Sigma}$, `orthosymplectic metric',
\begin{eqnarray}
\label{GOmG} {\cal G}_{\Sigma}{}^{\Sigma^\prime}
\Omega_{\Sigma^\prime\Pi^\prime} {\cal G}_{\Pi}{}^{\Pi^\prime}
(-1)^{\mathrm{deg}(\Pi)(\mathrm{deg}(\Pi^\prime) + 1)} =
\Omega_{\Sigma\Pi}\; ,
\end{eqnarray}
the canonical form of which is given by equation (\ref{OmLP}). The
grading is defined by
\begin{equation}
(-1)^{\mathrm{deg}(\Sigma)}
=
\begin{cases} \, 1 & \mathrm{for} \; \Sigma=1, \dots , 2n \cr -1 &
\mathrm{for} \; \Sigma=2n+1 \;
 \end{cases}
\end{equation}
and coincides with $\mathrm{deg}(\pm\Sigma)$ for $Y^{\pm\Sigma}$
(see below equation (\ref{canonical})). The fundamental
representation of  $OSp(2n|1)$ acts on supertwistors
\begin{eqnarray} \label{Y}
Y^{\Sigma} = (\mu^{\alpha}, \lambda_{\alpha}, \eta ) \; ,
\end{eqnarray}
with even $\mu^{\alpha}, \lambda_{\alpha}$ and odd $\eta$. Near
the unity,
\begin{eqnarray}
\label{GIXi} {\cal G}_{\Sigma}{}^{\Pi} \sim
{\delta}_{\Sigma}{}^{\Pi} + {\Xi}_{\Sigma}{}^{\Pi}\; ,
\end{eqnarray}
where ${\Xi_{\Sigma}}^\Pi$ is an element of the $osp(2n|1)$
superalgebra. It has the form
\begin{eqnarray}
\label{SA} \Xi_{\Sigma}{}^{\Pi} = \left( \begin{matrix}
G_{\alpha}{}^\beta & {K}_{\alpha\beta} & \zeta_\alpha \cr
A^{\alpha\beta} & -  G_{\beta}{}^\alpha & \epsilon^\alpha \cr i
\epsilon^\beta & -i \zeta_\beta & 0 \end{matrix} \right) \; \quad
,
\end{eqnarray}
where the even $n\times n$ matrix $G_{\alpha}{}^\beta$ is
arbitrary and the even $n\times n$ ${K}_{\alpha\beta}=
{K}_{\beta\alpha}$ and ${A}^{\alpha\beta}= {A}^{\beta\alpha}$
matrices  are symmetric. They define a $gl(n)$ and two $sp(n)$
subalgebras  of  $osp(2n|1)$,
\begin{eqnarray}
\label{Agl} G_{\alpha}{}^\beta \, \in \, gl(n)\; , \quad
A^{\alpha\beta} \, \in \, sp(n)\; , \quad \;{K}_{\alpha\beta} \,
\in \, sp(n)\; .
\end{eqnarray}

Exploiting the analogy with the matrix representation of the
standard $4$-dimensional conformal algebra $su(2,2|N)$ and the
$4$-dimensional super-Poincar\'e algebra, one can look at the
$gl(n)$ boxes $G$ as a generalization of the $spin(1,D-1)$ and
dilatation algebras ($L_\alpha{}^\beta + \delta_\alpha{}^\beta
D$), at the elements $A^{\alpha\beta}\in sp(n)$ as a
generalization of the translation one, and at
${K}_{\alpha\beta}\in sp(n)$ as a generalization of the special
conformal transformations. Equation (\ref{SA}) also contains two
fermionic parameters, $\epsilon^\alpha$ and $\zeta_\alpha$, which
can be identified as those of the of `usual' and special conformal
supersymmetries. A specific check is provided by the $n=2$ case,
where $SL(2,\mathbb{R})=Spin(1,2)$, the symmetric spin-tensor
provides an equivalent representation for a $SO(1,2)$ vector, and
the superconformal group is $OSp(2|1)$.

If we now demand in addition that the degenerate matrix
$C_{\Sigma\Pi}$ (equation (\ref{CLP})) is preserved,
\begin{eqnarray}
\label{GCG} {\cal G}_{\Sigma}{}^{\Sigma^\prime}
C_{\Sigma^\prime\Pi^\prime} {\cal G}_{\Pi}{}^{\Pi^\prime}
(-1)^{\mathrm{deg}(\Pi) (\mathrm{deg}(\Pi^\prime) +1)} =
C_{\Sigma\Pi}\; ,
\end{eqnarray}
we see that this is satisfied by the $osp(2n|1)$ elements of the
form
\begin{eqnarray}
\label{SAC} \Xi_{\Sigma}{}^{\Pi} = \left( \begin{matrix}
S_{\alpha}{}^\beta & 0 & 0 \cr A^{\alpha\beta} & -
S_{\beta}{}^\alpha & \epsilon^\alpha \cr i  \epsilon^\beta & 0 & 0
\end{matrix} \right) \; \equiv  \Xi_{\Sigma}{}^{\Pi} (S,A,
\epsilon) \quad ,
\end{eqnarray}
where $S_{\alpha}{}^\beta \in sp(n)$,
\begin{eqnarray}
 S^{\alpha\beta}\equiv C^{\alpha\gamma} S_\gamma{}^{\beta}
= S^{\beta \alpha} \; ,
\end{eqnarray}
{\it i.e.} by those of (\ref{SA}) with $K_{\alpha\beta}=0$,
$\zeta_\alpha=0$ and $G_{\alpha}{}^\beta =S_{\alpha}{}^\beta  \in
sp(n)$.
 Thus the condition (\ref{GCG}) not only reduces $GL(n)$ symmetry down to
$Sp(n)$, but also breaks the generalized special conformal
transformations and the superconformal supersymmetry.

The right action of ${\cal G}_{\Sigma}{}^{\Pi}(S,A,\epsilon)$
(Eqs. (\ref{GIXi}), (\ref{SAC})) on the supertwistor (\ref{Y}),
$\delta Y^\Sigma= Y^\Pi \Xi_\Pi{}^\Sigma$, defines the generalized
super-Poincar\'e transformation of the supertwistor components,
{\setlength\arraycolsep{0pt}
\begin{eqnarray}
\label{YS-P} && \delta \mu^\alpha = \mu^\beta S_\beta{}^\alpha +
\lambda_\beta A^{\beta\alpha} + i \epsilon^\alpha \eta \; , \qquad
\nonumber \\
&& \delta \lambda_\alpha = - S_\alpha {}^\beta \lambda_\beta \; ,
\qquad \delta \eta = \epsilon^\alpha \lambda_\alpha \; .
\end{eqnarray}
}These can be reproduced from the following transformations of the
coordinates of $\Sigma^{({n(n+1)\over 2}|n)}$,
\begin{eqnarray}
\label{dZ-P} \delta X^{\alpha\beta} = A^{\alpha\beta} + i
\theta^{(\alpha}\epsilon^{\beta )} + 2 X^{(\alpha|\gamma}
S_\gamma{}^{|\beta)} \; , \quad
 \delta \theta^{\alpha} = \epsilon^{\alpha} +  \theta^{\beta} A_\beta{}^\alpha
\; ,
\end{eqnarray}
using the generalization \cite{BL98} of the Penrose correspondence
relation \cite{Ferber,Pen}  given in equation (\ref{mu+}),
\begin{equation}
\label{mu}  \mu^{\alpha} = X^{\alpha\beta} \lambda_{\beta}-
{i\over 2} \theta^{\alpha} \theta^{\beta} \lambda_{\beta} \; ,
\quad \eta = \theta^{\alpha} \lambda_{\alpha} \; .
\end{equation}
The transformations (\ref{dZ-P}) of the $\Sigma^{({n(n+1)\over
2}|n)}$ variables are a straightforward generalization of the
super-Poincar\'e transformations of the standard superspace
coordinates. This justifies calling  the resulting supergroup
${\Sigma}^{({n(n+1)\over 2}|n)} \rtimes Sp(n)$ a generalization of
the  super-Poincar\'e group.

Going back to $osp(2n|1)$, let us note that the generalized
special superconformal transformations $(K_{\alpha\beta},
\zeta_\alpha)$ act on the supertwistor components by
\begin{eqnarray}
\label{YS-K} \delta \mu^\alpha = 0 \; , \quad \delta
\lambda_\alpha = \mu^\beta K_{\beta\alpha} - i \eta \zeta_\alpha
\; , \quad \delta \eta =  \mu^\beta  \zeta_\beta \; . \quad
\end{eqnarray}
Using equation (\ref{mu}) one may find from (\ref{YS-K}) the
generalized special superconformal transformations of the
${\Sigma}^{({n(n+1)\over 2}|n)}$ coordinates
{\setlength\arraycolsep{0pt}
\begin{eqnarray}
\label{dZ-K} &&  \delta X^{\alpha\beta} = i \theta^{(\alpha}
X^{\beta)\gamma}\zeta_\gamma - (XKX)^{\alpha\beta}\; ,
\nonumber \\
&&  \delta \theta^{\alpha}  = X^{\alpha\beta}\zeta_\beta - {i\over
2} (\theta\zeta)\, \theta^{\alpha} - (\theta KX)^\alpha \; .
\end{eqnarray}}

Note that (\ref{dZ-P}) follows as well  from a nonlinear
realization of the generalized super-Poincar\'e group
${\Sigma}^{({n(n+1)\over 2}|n)} \rtimes Sp(n)$ on the
${\Sigma}^{({n(n+1)\over 2}|n)}$ coset, {\it i.e.} from the left
action of ${\cal G}_{\Sigma}{}^{\Pi}(S, A, \epsilon)  \sim
{\delta}_{\Sigma}{}^{\Pi} + {\Sigma}_{\Sigma}{}^{\Pi}(S, A,
\epsilon) $ (\ref{SAC}) on ${\cal K}_{\Sigma}{}^{\Pi}(X, \theta )
\sim {\delta}_{\Sigma}{}^{\Pi} + K_{\Sigma}{}^{\Pi}(X, \theta)$
with
\begin{eqnarray}
\label{SAX} K_{\Sigma}{}^{\Pi}(X, \theta)= \left( \begin{matrix} 0
& 0 & 0 \cr X^{\alpha\beta} &   0 & \theta^\alpha \cr i
\theta^\beta & 0 & 0 \end{matrix} \right) \quad \; .
\end{eqnarray}
Indeed, the infinitesimal form of
\begin{equation}\label{GK=}
{\cal G}_{\Sigma}{}^{\Pi}(S, A, \epsilon) {\cal
K}_{\Sigma}{}^{\Pi}(X, \theta)= {\cal K}_{\Sigma}{}^{\Pi}(X^\prime
, \theta^\prime) {\cal G}_{\Sigma}{}^{\Pi}(A, 0, 0)\;
\end{equation}
reads {\setlength\arraycolsep{2pt}
\begin{eqnarray}\label{GKi=}
{K}(\delta X, \delta \theta) &=& {\Xi}(0, A, \epsilon) + {\Xi}(0,
A, \epsilon) {K}(X, \theta)
\nonumber \\
&& + [ {\Xi}(S, 0, 0)\, , \, K(X, \theta)]\;
\end{eqnarray}
}and reproduces the generalized super-Poincar\'e transformations
(\ref{dZ-P}) \cite{30/32}.


\chapter*{Note added}

\addcontentsline{toc}{chapter}{Note added}

\vskip 1.4cm

After this Thesis was submitted, bosonic $k=31$-supersymmetric
({\it i.e.}, preonic) solutions were shown to be ruled out in IIB
\cite{Gran:2006ec} and in IIA \cite{Bandos:2006xz} supergravity,
at least when no $\alpha^\prime$ corrections are taken into
account. This conclusion can be drawn from an analysis of the
algebraic equation for the Killing spinors arising from the
requirement that the dilatino transformation under supersymmetry
vanishes (see footnote 1, page 36 above). The question of whether
(bosonic) $k=31$-supersymmetric solutions of $D=11$ supergravity
do exist remains open, although the impossibility of such a
solution to be dimensionally-reduced to a $k=31$-supersymmetric
IIA solution places particular restrictions \cite{Bandos:2006xz}.
See \cite{Duff:2002rw} for a discussion on the possible numbers $0
\leq k \leq 32$ of supersymmetries allowed by supergravity
solutions.

A possible way out \cite{Bandos:2006xz}, both in Type II and in
$D=11$ supergravity, for the existence of preonic solutions, would
be the consideration of stringy (in Type II) and M-theoretical (in
$D=11$) corrections \cite{W00,Howe03,LPST,Zw85,LPS05}.




\begin{thebibliography}{990}

\addcontentsline{toc}{chapter}{Bibliography}

\vskip 1.4cm


\bibitem{BAPV05}
I.A.  Bandos, J.A. de Azc\'arraga, M. Pic\'on and O. Varela, {\it
Generalized curvature and the equations of $D=11$ supergravity},
Phys. Lett.  {\bf B615}, 127--133 (2005) [arXiv:hep-th/0501007].

\bibitem{BLVW03}
A. Batrachenko, J.T. Liu, O. Varela and W.Y. Wen, {\it Higher
order integrability in generalized holonomy},
arXiv:hep-th/0412154.

\bibitem{BAIPV03}
I.A. Bandos, J.A. de Azc\'arraga, J.M. Izquierdo, M. Pic\'on and
O. Varela, {\it BPS preons, generalized holonomies and $D=11$
supergravities}, Phys. Rev. {\bf D69}, 105010 (2004)
 [arXiv:hep-th/0312266].

\bibitem{JdAIMO}
J.A. de Azc\'arraga, J.M. Izquierdo, M. Pic\'on and O. Varela,
{\it Generating Lie and gauge free differential (super)algebras by
expanding Maurer-Cartan forms and Chern-Simons supergravity},
Nucl. Phys. {\bf B662}, 185--219 (2003) [arXiv:hep-th/0212347].

\bibitem{JdA02}
J.A. de Azc\'arraga, J.M. Izquierdo, M. Pic\'on and O. Varela,
{\it Extensions, expansions, Lie algebra cohomology and enlarged
superspaces}, Class. Quantum Grav. {\bf 21}, S1375--S1384 (2004)
[arXiv:hep-th/0401033].


\bibitem{Lett}
I.A. Bandos, J.A. de Azc\'arraga, J.M. Izquierdo, M. Pic\'on and
O. Varela, {\it On the underlying gauge group structure of $D=11$
supergravity}, Phys.~Lett.~{\bf B596}, 145--155 (2004)
[arXiv:hep-th/0406020].


\bibitem{AnnP04}
I.A. Bandos, J.A. de Azc\'arraga, M. Pic\'on and O. Varela, {\it
On the formulation of $D = 11$ supergravity and the composite
nature of its three-form gauge field}, Ann. Phys. {\bf 317},
238--279 (2005) [arXiv:hep-th/0409100].

\bibitem{30/32}
I.A. Bandos, J.A. de Azc\'arraga, M. Pic\'on and O. Varela, {\it
Supersymmetric string model with 30 $\kappa$-syymetries in an
extended $D= 11$ superspace and 30/32 BPS states}, Phys. Rev. {\bf
D69}, 085007 (2004) [arXiv:hep-th/0307106].





\bibitem{GolLik72} Yu.A.~Gol'fand and E.P.~Likhtman, {\it
Extensions of the algebra of Poicar\'e group generators and
violation of P-invariance}, JETP Lett.~{\bf 13}, 323--326 (1972).

\bibitem{VolAk72} D.V.~Volkov and V.P.~Akulov,
{\it Possible universal neutrino interaction}, JETP Lett.~{\bf
16}, 438--440 (1972).

\bibitem{WeZu74} J.~Wess and B.~Zumino, {\it Supergauge
transformations in four dimensions}, Nucl.~Phys.~{\bf B70}, 39--50
(1974).

\bibitem{FayFer77} P.~Fayet and S.~Ferrara, {\it Supersymmetry},
Phys.~Rept.~{\bf 32}, 249--334 (1977).

\bibitem{Sohn85} M.F.~Sohnius, {\it Introducing supersymmetry},
Phys.~Rept.~{\bf 128}, 39--204 (1985).

\bibitem{collsusy}
S.~Ferrara (ed.), {\it Supersymmetry} (vols. I,II), World
Sci.~(1987).

\bibitem{vNPhysRep} P.~van Nieuwenhuizen, {\it Supergravity},
Phys.~Rept.~{\bf 68}, 189--398 (1981).

\bibitem{vProeyen03} A.~Van Proeyen, {\it Structure of
supergravity theories}, Proceedings of the XI Fall Workshop on
Geometry and Physics, Oviedo, (2002) [arXiv:hep-th/0301005].

\bibitem{Ortin04} T.~Ort\'in, {\it Gravity and strings}, Cambridge
University Press (2004).

\bibitem{Cd'AF91} L. Castellani, R. D'Auria and P. Fr\'e,
{\it Supergravity and superstrings: a geometric perspective},
vols. I, II, III, World Scientific (1991).


\bibitem{FvNF76} D.Z.~ Freedman, P.~ van Nieuwenhuizen and S.~
Ferrara, {\it Progress towards a theory of supergravity}, Phys.~
Rev.~ {\bf D13}, 3214--3218 (1976).

\bibitem{DZ76} S.~ Deser, B.~ Zumino, {\it Consistent supergravity},
Phys.~ Lett.~ {\bf B62}, 335--337 (1976).


\bibitem{Grisaru77} M.T. Grisaru, {\it Two-loop renormalizability
of supergravity}, Phys. Lett. {\bf B66}, 75--76 (1977).

\bibitem{vNV77} P. van Nieuwenhuizen and J.A.M. Vermaseren, {\it
Finiteness of $SO(3)$ and $SO(4)$ supergravity and
nonrenormalizability of the scalar multiplet in explicit one-loop
examples}, Phys. Rev. {\bf D16}, 298--303 (1977).

\bibitem{CJ78} E. Cremmer and B. Julia, {\it The $N=8$ supergravity
theory. I. The lagrangian}, Phys. Lett. {\bf B80}, 48--51 (1978);
{\it The $SO(8)$ supergravity}, Nucl. Phys. {\bf B159}, 141--212
(1979).

\bibitem{dWitNic82} B. de Wit and H. Nicolai, {\it $N=8$
supergravity with local $SO(8) \times SU(8)$ invariance}, Phys.
Lett. {\bf B108}, 285--290 (1982); {\it $N=8$ supergravity}, Nucl.
Phys. {\bf B208}, 323--364 (1982).




\bibitem{SalSez89} A. Salam and E. Sezgin, {\it Supergravities in
diverse dimensions}, Elsevier and World Scientific (1989).


\bibitem{Nahm78} W. Nahm, {\it Supersymmetries and their
representations}, Nucl. Phys. {\bf B135}, 149--166 (1978).


\bibitem{CJS}
E. Cremmer, B. Julia and J. Scherk, {\it Supergravity theory in
eleven dimensions}, Phys.~Lett.~{\bf B76}, 409--412 (1978).

\bibitem{dWitFreed77} B. de Wit and D. Freedman, {\it On $SO(8)$
extended supergravity}, Nucl. Phys. {\bf B130}, 105--113 (1977).

\bibitem{DNP}
M.J.~Duff, B.E.W.~Nilsson and C.N.~Pope, {\it Kaluza-Klein
supergravity}, Phys.~Rept.~{\bf 130}, 1--142 (1986).


\bibitem{ADP}
M.A.~Awada, M.J.~Duff and C.N.~Pope, {\it $N=8$ supergravity
breaks down to $N=1$}, Phys.~Rev.~Lett.~{\bf 50}, 294 (1983).


\bibitem{Witten81} E. Witten, {\it Search for a realistic
Kaluza-Klein theory}, Nucl. Phys. {\bf B186}, 412 (1981).

\bibitem{FR}
P.G.O.~Freund and M.A.~Rubin, {\it Dynamics of dimensional
reduction}, Phys.~Lett.~{\bf B97}, 233 (1980).

\bibitem{Witten85} E. Witten, {\it Fermion quantum numbers in
Kaluza-Klein theory}, in {\it Shelter Island II}, R. Jackiw, N.
Khuri, S. Weinberg and E. Witten eds., M.I.T. Press (1985).

\bibitem{AchWit01} B.~Acharya and E.~Witten, {\it Chiral fermions
from manifolds of $G_2$ holonomy}, arXiv:hep-th/0109152.


\bibitem{Veneziano68} G.~ Veneziano, {\it Construction of a
crossing-symmetric, Regge behaved amplitude for linearly rising
trajectories}, Nuovo Cim.~ {\bf A57}, 190--197 (1968).

\bibitem{GSW87} M.B. Green, J.H. Schwarz and E. Witten, {\it
Superstring Theory}, vols. I, II, Cambridge University Press
(1987).

\bibitem{Pol98} J. Polchinski, {\it String Theory}, vols. I, II,
Cambridge University Press (1998).

\bibitem{SS74} J. Scherk and J.H. Schwarz, {\it Dual models for
non-hadrons}, Nucl. Phys. {\bf B81}, 118--144 (1974).



\bibitem{GrSchvocsup84} M.B. Green and J.H. Schwarz,
{\it Covariant description of superstrings}, Phys. Lett. {\bf
B136}, 367 (1984).

\bibitem{AI95} J.A. de Azc\'arraga and J.M. Izquierdo,
{\it Lie groups, Lie algebras, cohomology and some applications in
physics}, Cambridge University Press (1995).


\bibitem{AL82}
J.A.~de Azc\'arraga and J.~Lukierski, {\it Supersymmetric
particles with internal symmetries and central charges},
Phys.~Lett.~{\bf B113}, 170 (1982); {\it Supersymmetric particles
in $N=2$ superespace: phase space variables and hamiltonian
dynamics}, Phys. Rev. {\bf D28}, 1337 (1983).

\bibitem{Siegel83} W. Siegel, {\it Hidden local supersymmetry in
the supersymmetric particle action}, Phys. Lett. {\bf B128},
397-399 (1983); {\it Spacetime-supersymmetric quantum mechanics},
Class. Quantum Grav. {\bf 2}, L95--L97 (1985).


\bibitem{GSchw84} M.B. Green and J.H. Schwarz, {\it Anomaly
cancellation in supersymmetric $D=10$ gauge theory and superstring
theory}, Phys. Lett. {\bf B149}, 117--122 (1984).


\bibitem{GHMR84} D.J. Gross, J.A. Harvey, E.J.
Martinec and R. Rohm, {\it Heterotic string}, Phys. Rev. Lett.
{\bf 54}, 502--505 (1985).

\bibitem{VST83} D.V.~Volkov, D.P.~Sorokin and V.I.~Tkach, {\it
Mechanisms of spontaneous compactification of $N=2$, $d=10$
supergravitation}, JETP Lett.~{\bf 38}, 481--485 (1983).

\bibitem{CHSW85} P.~Candelas, G.T. Horowitz, A. Strominger and E.
Witten, {\it Vacuum configurations for superstrings}, Nucl. Phys.
{\bf B258}, 46--74 (1985).

\bibitem{Bog76} E.B. Bogomol'nyi,
{\it The stability of classical solutions}, Sov. J. Nucl. Phys.
{\bf 24} 449 (1976).

\bibitem{PradSom75} M.K. Prasad and
C.M. Sommerfield, {\it An exact classical solution for the 't
Hooft monopole and the Julia-Zee dyon}, Phys. Rev. Lett. {\bf 35},
760--762 (1975).

\bibitem{GiPoRa94} A. Giveon, M. Porrati and E. Rabinovici,
{\it Target space duality in string theory}, Phys. Rept. {\bf
244}, 77--202 (1994) [arXiv:hep-th/9401139].

\bibitem{FILQ} A. Font, L. Ib\'a\~nez, D. L\"ust and F. Quevedo,
{\it Strong--Weak coupling duality and nonperturbative effects in
string theory}, Phys. Lett. {\bf B249}, 35--43 (1990).

\bibitem{MonOl77} C. Montonen, D.I. Olive, {\it Magnetic monopoles
as gauge particles?}, Phys. Lett. {\bf B72}, 117 (1977).

\bibitem{HT95} C. Hull and P. Townsend, {\it Unity of superstring
dualities}, Nucl. Phys. {\bf B450}, 69 (1995)
[arXiv:hep-th/9410167].

\bibitem{SchwMTh}
J.H.~Schwarz, {\it The second superstring revolution}, in Proc. of
the {\it Second Int. Sakharov conference on physics, Moscow,
1996}, I.M. Dremin and A.M. Semikhatov eds., World Sci.~1997, pp.
562-569 [arXiv:hep-th/9607067]; {\it Lectures on superstring and M
theory dualities}, Nucl.~Phys.~Proc.~Suppl.~{\bf 55B}, 1-32 (1997)
[arXiv:hep-th/9607201].

M.J. Duff, {\it M Theory (the theory formerly known as Strings)},
Int. J. Mod. Phys. {\bf A11}, 5623--5642 (1996)
[arXiv:hep-th/9608117].

\bibitem{Achucarro:1987nc}
  A. Ach\'ucarro, J.M. Evans, P.K. Townsend and D.L. Wiltshire,
  {\it Super p-branes},
  Phys.\ Lett.\ {\bf B198}, 441 (1987).

\bibitem{Evans88} J.M. Evans, {\it Super p-brane Wess-Zumino terms}, Class.
Quantum Grav. {\bf 5}, L87--L90 (1988).


\bibitem{Topbrane} P.K.~Townsend, {\it The eleven-dimensional
supermembrane revisited}, Phys.~Lett.~{\bf B350}, 184--188 (1995)
[arXiv:hep-th/9501068].


\bibitem{JdAT}
J.A. de Azc\'arraga, J.P. Gauntlett, J.M. Izquierdo and P.K.
Townsend, {\it Topological extensions of the supersymmetry algebra
for extended objects}, Phys.~Rev.~Lett. {\bf 63}, 2443 (1989).


\bibitem{ST97}
 D.P. Sorokin and P.K. Townsend,
 {\it M Theory superalgebra from the M five-brane},
 Phys.~Lett.~{\bf B412}, 265 (1997)  [arXiv:hep-th/9708003].


\bibitem{M-alg}
P.K. Townsend, {\it $p$-brane democracy}, in {\it Particles,
strings and cosmology}, J. Bagger {\it et al.} eds., World Sci.,
271--285 (1996) [arXiv:hep-th/9507048]; {\it M-theory from its
superalgebra}, in {\it Strings, branes and dualities}, NATO ASI
Ser. C, Math. and Phys.~ Sci., {\bf 520}, 141--177 (1999)
[arXiv:hep-th/9712004], and refs. therein.


\bibitem{Witten95M} E. Witten, {\it String theory dynamics in
various dimensions}, Nucl. Phys. {\bf B443}, 85--126 (1995)
[arXiv:hep-th/9503124].


\bibitem{HW96}
P.~Ho\v{r}ava and E.~Witten, {\it Eleven-Dimensional Supergravity
on a Manifold with Boundary}, Nucl.~ Phys.~  {\bf B475}, 94 (1996)
[arXiv:hep-th/9603142].

\bibitem{Duff11dim} M.J. Duff, {\it The world in eleven dimensions:
supergravity, supermembranes and M-theory}, Institute of Physics
Publishing (1999).

\bibitem{DHIS87} M.J. Duff, P.S. Howe, T. Inami and
K.S. Stelle, {\it Superstrings in D=10 from supermembranes in
D=11}, Phys.~Lett.~{\bf B191}, 70 (1987).

\bibitem{Duff94}
M.J. Duff, R.R. Khuri and J.X. Lu, {\it String solitons}, Phys.
Rept. {\bf 259}, 213 (1995) [arXiv:hep-th/9412184].


\bibitem{Pol95} J. Polchinski, {\it Dirichlet branes and
Ramond-Ramond charges}, Phys. Rev. Lett. {\bf 75}, 4724--4727
(1995) [arXiv:hep-th/9510017].

\bibitem{CJDbranes} C.V. Johnson, {\it D-Branes}, Cambridge
University Press (2003).

\bibitem{Kir03} E. Kiritsis,
{\it D-branes in standard model building, gravity and cosmology},
Fortsch. Phys. {\bf 52}, 200--263 (2004) and Phys. Rept. {\bf
421}, 105--190 (2005) [arXiv:hep-th/0310001].

\bibitem{Maldacena97} J. Maldacena, {\it The large N limit of
superconformal field theories and supergravity}, Adv. Theor. Math.
Phys. {\bf 2}, 231--252 (1998) and Int. J. Theor. Phys. {\bf 38},
1113--1133 (1999) [arXiv:hep-th/9711200].

\bibitem{AGMOO00} O. Aharony, S.S. Gubser, J.M. Maldacena,
H. Ooguri, Y. Oz, {\it Large N field theories, string theory and
gravity}, Phys. Rept. {\bf 323}, 183--386 (2000).

\bibitem{AFFHS99} B.S. Acharya, J.M. Figueroa-O'Farrill, C.M. Hull and
 B.J. Spence, {\it Branes at conical singularities and
 holography}, Adv. Theor. Math. Phys. {\bf 2}, 1249--1286 (1999)
 [arXiv:hep-th/9808014].


\bibitem{BST}
E.~Bergshoeff, E.~Sezgin and P.K.~Townsend, {\it Supermembranes
and eleven-dimensional supergravity}, Phys.~ Lett.~ {\bf B189},
75--78 (1987); {\it Properties of the eleven dimensional super
membrane theory}, Ann.~ Phys.~ (NY)  {\bf 185}, 330 (1988);  \,


\bibitem{blnpst}
I.~Bandos, K.~Lechner, A.~Nurmagambetov, P.~Pasti, D.P.~Sorokin
and M.~Tonin, {\it Covariant action for the super-five-brane of
M-theory}, Phys.~Rev.~Lett.~{\bf 78}, 4332 (1997)
[arXiv:hep-th/9701149].

 M.~Aganagic, J.~Park, C.~Popescu and
J.H.~Schwarz, {\it World-volume action of the M-theory
five-brane}, Nucl.~Phys.~{\bf B 496}, 191 (1997)
[arXiv:hep-th/9701166].


\bibitem{DuffStelle91} M.J.~Duff and K.S.~Stelle, {\it Multi--membrane solutions of D
= 11 Supergravity}, Phys.~Lett.~{\bf B253}, 113--118 (1991).


\bibitem{Gu92}
R. G\"uven, {\it Black $p$-brane solutions of $D = 11$
supergravity theory}, Phys. Lett.  {\bf B276}, 49--55 (1992).



\bibitem{Stelle98}
K.S.~Stelle, {\it BPS branes in supergravity}, in {\it High energy
physics and cosmology 1997: Proceedings of the ICTP Summer school
in high energy physics and cosmology} (E. Gava et al eds.),
29--127 (1997) [arXiv:hep-th/9803116].

\bibitem{interscbranes}

  G. Papadopoulos and P.K. Townsend,
  {\it Intersecting M-branes},
  Phys. Lett. {\bf B380}, 273 (1996)
  [arXiv:hep-th/9603087].

  A.A. Tseytlin, {\it Harmonic superpositions of M-branes},
  Nucl. Phys. {\bf B475}, 149 (1996)
  [arXiv:hep-th/9604035].

  J.P. Gauntlett, D.A. Kastor and J.H. Traschen, {\it
  Overlapping branes in M-Theory},
  Nucl. Phys. {\bf B478}, 544 (1996)
  [arXiv:hep-th/9604179].

  E. Bergshoeff, M. de Roo, E. Eyras, B. Janssen and J.P. van der Schaar,
  {\it Multiple intersections of D-branes and M-branes},
  Nucl. Phys. {\bf B494}, 119 (1997)
  [arXiv:hep-th/9612095].



\bibitem{Hull84}
C.M. Hull, {\it Exact $pp$ wave solutions of 11--dimensional
supergravity}, Phys. Lett. {\bf B139}, 39--41 (1984).

\bibitem{Duff03}
M.J.~Duff and J.T.~Liu, {\it Hidden spacetime symmetries and
generalized holonomy in M-theory}, Nucl.~Phys.~{\bf B674},
217--230 (2003) [arXiv:hep-th/0303140].


\bibitem{GMPW04} J.P. Gauntlett, D. Martelli, S. Pakis and D.
Waldram, {\it $G$-structures and wrapped NS5-branes}, Comm. Math.
Phys. {\bf 247}, 421--445 (2004) [arXiv:hep-th/0205050].

\bibitem{GP02}
J.P. Gauntlett and S. Pakis, {\it The geometry of $D=11$ Killing
spinors}, JHEP {\bf 0304}, 039 (2003) [arXiv:hep-th/0212008].

\bibitem{Gillard:2004xq}
  J. Gillard, U. Gran and G. Papadopoulos,
  {\it The spinorial geometry of supersymmetric backgrounds},
  Class. Quant. Grav.  {\bf 22} (2005) 1033
  [arXiv:hep-th/0410155].


\bibitem{vHvP82} J.W. van Holten and A. Van Proeyen,
{\it N=1 supersymmetry algebras in d=2,3,4 mod 8}, J. Phys. {\bf
A15}, 3763--3783 (1982).


\bibitem{BPS01}
I.A. Bandos, J.A. de Azc\'arraga, J.M. Izquierdo and J. Lukierski,
{\it BPS states in M theory and twistorial constituents}, Phys.
Rev. Lett. {\bf 86}, 4451 (2001) [arXiv:hep-th/0101113].



\bibitem{Bandos05} I.A. Bandos, {\it BPS preons in supergravity and higher
spin theories. An overview from the hill of twistor approach}, in
{\it Fundamental interactions and twistor-like methods}, J.~
Lukierski and D.~ Sorokin eds., AIP Conf.~ Proc.~ {\bf 767},
141--171 (2005) [arXiv:hep-th/0501115].


\bibitem{JdA00}
C. Chryssomalakos, J.A. de Azc\'{a}rraga, J.M. Izquierdo and J.C.
P\'{e}rez Bueno, {\it The geometry of branes and extended
superspaces}, Nucl. Phys. {\bf B567}, 293--330 (2000)
[arXiv:hep-th/9904137].

J.A.~de Azc\'arraga and J.M.~Izquierdo, {\it Superalgebra
cohomology, the geometry of extended superspaces and superbranes},
in {\it New developments in fundamental interactions theories.
Procs. of  37th Karpacz Winter School of Theoretical Physics}, AIP
Conf. Proc. {\bf 589}, 3--17 (2001) [arXiv:hep-th/0105125].


\bibitem{Azcarraga05} J.A. de Azc\'arraga, {\it Superbranes,
$D=11$ CJS supergravity and enlarged superspace coordinates/fields
correspondence}, in {\it Fundamental interactions and twistor-like
methods}, J.~ Lukierski and D.~ Sorokin eds., AIP Conf.~
Proc.~{\bf 767}, 243--267 (2005) [arXiv:hep-th/0501198].


\bibitem{Hull03}
C.~Hull, {\it Holonomy and symmetry in M-theory},
arXiv:hep-th/0305039.

\bibitem{WestE11} P. West, {\it $E_{11}$ and M theory}, Class.
Quantum Grav. {\bf 18}, 4443-4460 (2001) [arXiv:hep-th/0104081].

\bibitem{BDM99}
I. Bars, C. Deliduman and D. Minic, {\it Lifting M-theory to
two-time physics},
 Phys. Lett. {\bf B457}, 275-284 (1999) [arXiv:hep-th/9904063].

\bibitem{West00} P. West,
{\it Hidden superconformal symmetry in M theory}, JHEP {\bf 0008},
007 (2000) [arXiv:hep-th/0005270].


\bibitem{CS}
R.~Troncoso and J.~Zanelli, {\it New gauge supergravity in seven
and eleven dimensions}, Phys.~Rev.~{\bf D58}, 101703 (1998)
[arXiv:hep-th/9710180]; \,

J.~Zanelli, {\it Chern-Simons gravity: from $2+1$ to $2n+1$
dimensions}, Braz.~J.~Phys.~{\bf30},251--267 (2000)
[arXiv:hep-th/0010049]; {\it (Super)-gravities beyond 4
dimensions}, arXiv:hep-th/0206199.

P.~Ho\v rava, {\it M-theory as a holographic field theory},
Phys.~Rev.~{\bf D59}, 046004 (1999) [arXiv:hep-th/9712130];\,

H.~Nastase, {\it Towards a Chern-Simons M theory of $OSp(1|32)
\times OSp(1|32)$}, arXiv:hep-th/0306269;

M.~Ba\~nados, {\it The linear spectrum of OSp(32$|$1) Chern-Simons
supergravity in eleven dimensions}, Phys.~Rev.~Lett.~{\bf 88},
031301 (2002) [arXiv:hep-th/0107214].


\bibitem{D'A+F}
R.~D'Auria and P.~Fr\'e, {\it Geometric supergravity in D = 11 and
its hidden supergroup}, Nucl.~Phys.~{\bf B201}, 101--140 (1982)
[{\it Erratum-ibid.}\ {\bf B206}, 496 (1982)].

\bibitem{Ha-Tr-Za-04}
M. Hassaine, R. Troncoso and J. Zanelli, {\it Poincar\'e invariant
gravity with local supersymmetry as a gauge theory for the
M-algebra}, Phys.~Lett.~{\bf B596}, 132--137 (2004)
[arXiv:hep-th/0306258].


\bibitem{BDLW03} A. Batrachenko, M.J. Duff, J.T. Liu and W.Y. Wen,
{\it Generalized holonomy of M-theory vacua}, Nucl. Phys. {\bf
B726}, 275--293 (2005) [arXiv:hep-th/0312165].


\bibitem{Duff:2002rw}
M.J.~Duff, {\it M-theory on manifolds of G$_2$ holonomy: The first
twenty years}, arXiv:hep-th/0201062.


\bibitem{Su77} D.~ Sullivan, {\it Infinitesimal computations in
topology}, Inst. des Haut. \'Etud. Sci., Pub.~ Math.~ {\bf 47},
269--331 (1977).

\bibitem{Ni83} P. van Nieuwenhuizen, {\it Free graded differential
superalgebras}, in {\it Group theoretical methods in physics}, M.
Serdaro\v{g}lu and E. \.In\"on\"u Eds., Lect. Notes in Phys. {\bf
180}, 228--247 (1983).






\bibitem{BaDeHeSe97}
K. Bautier, S. Deser, M. Henneaux and D. Seminara, {\it No
cosmological $D=11$ supergravity}, Phys. Lett. {\bf B406}, 49-53
(1997) [arXiv:hep-th/9704131].


\bibitem{Se97} E. Sezgin, {\it The M-algebra}, Phys. Lett.
{\bf 392}, 323--331 (1997) [arXiv:hep-th/9609086].


\bibitem{Hull97}
C.M. Hull, {\it Gravitational Duality, Branes and Charges}, Nucl.
Phys. {\bf B509}, 216 (1998) [arXiv:hep-th/9705162].


\bibitem{BergshoeffKK}
E.~Bergshoeff, E.~Eyras and Y.~Lozano, {\it  The massive
Kaluza-Klein monopole}, Phys.~Lett.~{\bf B430}, 77 (1998)
[arXiv:hep-th/9802199]; \,

E.~Eyras, Y.~Lozano, {\it  The Kaluza-Klein monopole in a massive
IIA background}, Nucl.~Phys.~ {\bf B546}, 197--218 (1999)
[arXiv:hep-th/9812188].

\bibitem{BergshoeffM9}
E. Bergshoeff and J.P. van der Schaar, {\it On M-9-branes}, Class.
Quantum Grav. {\bf 16}, 23 (1999) [arXiv:hep-th/9806069].

\bibitem{H-JPaSm03} E.J. Hackett-Jones, D.C. Page and D.J. Smith,
{\it Topological charges for branes in M-theory}, JHEP {\bf 0310},
005 (2003) [arXiv:hep-th/0306267].


\bibitem{Bars97}
I. Bars, {\it Duality covariant type IIB supersymmetry and
nonperturbative consequences}, Phys. Rev. {\bf D56}, 7954 (1997)
[arXiv:hep-th/9706185].


\bibitem{J+S99}
B. Julia and S. Silva, {\it On first order formulations of
supergravities}, JHEP {\bf 0001}, 026 (2000)
[arXiv:hep-th/9911035].


\bibitem{BEWN83}
B. Biran, F. Englert, B. de Wit and H. Nicolai, {\it Gauged $N=8$
supergravity and its breaking from spontaneous compactification},
Phys.~ Lett.~ {\bf B124}, 45 (1983) [E.:{\it
 ibid}. {\bf B 128}, 461 (1983)].


\bibitem{P+T03}
G. Papadopoulos and D. Tsimpis, {\it The holonomy of the
supercovariant connection and Killing spinors}, JHEP {\bf 0307},
018 (2003) [arXiv:hep-th/0306117].


\bibitem{P+T031}
G. Papadopoulos and D. Tsimpis, {\it The holonomy of IIB
supercovariant connection}, Class. Quantum  Grav. {\bf 20}, L253
(2003) [arXiv:hep-th/0307127].

\bibitem{FFP02}
J. Figueroa O'Farrill and G. Papadopoulos, {\it Maximally
supersymmetric solutions of ten and eleven-dimensional
supergravities}, JHEP  {\bf 0303}, 048 (2003)
[arXiv:hep-th/0211089].


\bibitem{BW}
A. Batrachenko, W.Y. Wen, {\it Generalized holonomy of
supergravities with 8 real supercharges}, Nucl. Phys. {\bf B690},
331--340 (2004) [arXiv:hep-th/0402141].

\bibitem{LPS05} H. Lu, C.N. Pope, K.S. Stelle, {\it Generalized
holonomy for higher order corrections to supersymmetric
backgrounds in String and M-Theory}, Nucl. Phys. {\bf B741},
17--23 (2006) [arXiv:hep-th/0509057].

\bibitem{Besse} A.L.~Besse, {\it Einstein manifolds},
Springer-Verlag (1987).


\bibitem{Gross:1983hb}
  D.J. Gross and M.J. Perry, {\it
  Magnetic Monopoles In Kaluza-Klein Theories},
  Nucl. Phys. {\bf B226}, 29 (1983).

\bibitem{Sorkin:1983ns}
  R.D. Sorkin, {\it Kaluza-Klein Monopole},
  Phys. Rev. Lett.  {\bf 51}, 87 (1983).


\bibitem{Duff91}
M.J. Duff and J.X. Lu, {\it A Duality between strings and
five-branes}, Class. Quantum Grav.  {\bf 9}, 1 (1992).

\bibitem{Pilch:1984xy}
  K.~Pilch, P.~van Nieuwenhuizen and P.~K.~Townsend,
  {\it Compactification of $D = 11$ supergravity on S(4) (or $11 = 7 + 4$, too)},
  Nucl. Phys. {\bf B242}, 377 (1984).

\bibitem{KowGlikwave}
  J. Kowalski-Glikman, {\it Vacuum states in supersymmetric Kaluza-Klein theory},
  Phys. Lett. {\bf B134}, 194 (1984).

  P.T. Chrusciel and J. Kowalski-Glikman, {\it
  The isometry group and killing spinors for the p p wave space-time in $D =
  11$ supergravity},
  Phys.  Lett. {\bf B149}, 107 (1984).

\bibitem{GauntHull00} J.P. Gauntlett and C.M. Hull,
{\it  BPS states with extra supersymmetry}, JHEP 0001 {\bf 004}
(2000) [arXiv:hep-th/9909098].

\bibitem{CLP02} M. Cveti\v c, H. L\"u and C.N. Pope, {\it M-theory pp-waves,
Penrose limits and supernumerary supersymmetries}, Nucl. Phys.
{\bf B644}, 65 (2002) [arXiv:hep-th/0203229].


\bibitem{GH02} J.P. Gauntlett and C.M. Hull, {\it PP-waves in 11-dimensions
with extra supersymmetry}, JHEP {\bf 0206}, 013 (2002)
[arXiv:hep-th/0203255].

\bibitem{BJ02} I. Bena and R. Roiban, {\it A supergravity pp-wave solutions
with 28 and 24 supercharges}, Phys. Rev. {\bf D67}, 125014 (2003)
[arXiv:hep-th/0206195].

\bibitem{Michelson} J. Michelson, {\it (Twisted) toroidal compactification of
pp-waves}, Phys.  Rev.  D {\bf 66}, 066002 (2002)
[arXiv:hep-th/0203140]; {\it A pp-Wave With 26 Supercharges},
Class. Quantum Grav. {\bf 19}, 5935 (2002) [arXiv:hep-th/0206204].


\bibitem{Gauntlett:2002nw}
  J.P. Gauntlett, J.B. Gutowski, C.M. Hull, S. Pakis and
  H.S. Reall, {\it All supersymmetric solutions of minimal
  supergravity in five dimensions},
  Class. Quant. Grav.  {\bf 20} (2003) 4587
  [arXiv:hep-th/0209114].


\bibitem{Harmark:2003ud}
  T. Harmark and T. Takayanagi, {\it
  Supersymmetric G\"odel universes in string theory},
  Nucl. Phys. {\bf B662}, 3 (2003)
  [arXiv:hep-th/0301206].


\bibitem{W00}
K. Peeters, P. Vanhove and A. Westerberg, {\it Supersymmetric
higher-derivative actions in ten and eleven dimensions, the
associated superalgebras and their formulation in superspace},
Class. Quantum  Grav.  {\bf 18}, 843 (2001)
[arXiv:hep-th/0010167]; {\it Supersymmetric R$^4$ actions and
quantum corrections to superspace  torsion constraints}, in {\it
Kiev 2000, Noncommutative structures in mathematics and physics}
[arXiv:hep-th/0010182].

\bibitem{Howe03}
P.S.~Howe and D.~Tsimpis, {\it On higher-order corrections in M
theory}, JHEP {\bf 0309}, 038 (2003) [arXiv:hep-th/0305129].

\bibitem{LPST} H. Lu, C.N. Pope, K.S. Stelle, P.K. Townsend,
{\it Supersymmetric deformations of $G_2$ manifolds from higher
order corrections to string and M Theory}, JHEP {\bf 0410}, 019
(2004) [arXiv:hep-th/0312002]; {\it String and M Theory
deformations of manifolds with special holonomy}, JHEP {\bf 0507},
075 (2005) [arXiv:hep-th/0410176].


\bibitem{Zw85}
B. Zwiebach, {\it Curvature squared terms and string theories},
Phys. Lett.  {\bf B156}, 315 (1985).





\bibitem{ambrose}
W. Ambrose and I.M. Singer, {\it A theorem on holonomy}, Trans.
Amer. Math. Soc. {\bf 75}, 428 (1953).

\bibitem{KobNom63} S.~Kobayashi and K.~Nomizu, {\it Foundations of
differential geometry}, vols. I, II, John Wiley \& Sons (1963).

\bibitem{Berger}
M. Berger, {\it Sur les groupes d'holonomie homogene des varietes
a connexion affine et des varietes riemanniennes}, Bull. Soc.
Math. France {\bf 83}, 225 (1955).

\bibitem{Bryant}
R.L. Bryant, {\it Pseudo-Riemannian metrics with parallel spinor
fields and non-vanishing Ricci tensor}, S\'emin. Congr., {\bf 4},
53--94, Soc. Math. France (2000) [arXiv:math.DG/0004073].


\bibitem{FoFG05} J.M. Figueroa-O'Farrill and S. Gadhia,
{\it Supersymmetry and spin structures}, Class. Quantum Grav. {\bf
22}, L121--L126 [arXiv:hep-th/0506229].

\bibitem{GMSW04} J.P.  Gauntlett, D.  Martelli, J.  Sparks and D.
Waldram, {\it Supersymmetric  $AdS$ backgrounds in string and M
Theory}, published in {\it Strasbourg 2003, AdS/CFT
correspondence}, 217--252 [arXiv:hep-th/0411194].


\bibitem{GaGuPa03} J.P. Gauntlett, J.B. Gutowski and S. Pakis,
{\it The geometry of $D=11$ null Killing spinors}, JHEP {\bf
0312}, 049 (2003) [arXiv:hep-th/0311112].

\bibitem{GaMaSpWa04} J.P.  Gauntlett, D.  Martelli, J.  Sparks and D.
Waldram, {\it Supersymmetric  $AdS_5$ solutions of M Theory},
Class. Quantum Grav. {\bf 21}, 4335--4366 (2004)
[arXiv:hep-th/0402153].

\bibitem{GauntMaSpWa04} J.P.  Gauntlett, D.  Martelli, J.  Sparks and D.
Waldram, {\it Supersymmetric  $AdS_5$ solutions of IIB
supergravity}, arXiv:hep-th/0510125.

%
\bibitem{MacConamhna:2004fb}
  O.A.P. Mac Conamhna, {\it Refining G-structure classifications},
  Phys. Rev. {\bf D70} (2004) 105024
  [arXiv:hep-th/0408203].

\bibitem{GoNeWa03} C.N. Gowdigere, D. Nemeschansky and
N.P. Warner, {\it Supersymmetric solutions with fluxes from
algebraic spinors}, Adv.  Theor.  Math.  Phys.  {\bf 7}, 787--806
(2004) [arXiv:hep-th/0306097].


\bibitem{Gauntlett05} J.P.  Gauntlett, {\it Classifying
supergravity solutions}, Fortsch. Phys. {\bf 53}, 468--479 (2005)
[arXiv:hep-th/0501229].



\bibitem{Grana06} M. Gra\~na, {\it Flux compactifications
in string theory: A comprehensive review}, Phys. Rept. {\bf 423},
91--158 (2006) [arXiv:hep-th/0509003].

\bibitem{BeCvL05} K. Behrndt, M. Cveti\v c and T. Liu, {\it
Classification of supersymmetric flux vacua in M Theory},
arXiv:hep-th/0512032.



\bibitem{DNP83}
M.J. Duff, B.E.W. Nilsson and C.N. Pope, {\it Spontaneous
supersymmetry breaking by the squashed seven-sphere}, Phys. Rev.
Lett. {\bf 50}, 2043 (1983).


\bibitem{vNW}
P. van Nieuwenhuizen and N. Warner, {\it Integrability conditions
for Killing spinors}, Commun. Math. Phys. {\bf 93}, 277 (1984).


\bibitem{joyce}
D.D. Joyce, {\it Compact Manifolds with Special Holonomy}, Oxford
University Press, Oxford, (2000).

\bibitem{Bar93}
C. B\"ar, {\it Real Killing spinors and holonomy}, Comm. Math.
Phys. {\bf 154}, 509--521 (1993).





\bibitem{BKOP97}
 E.  Bergshoeff, R.  Kallosh, T.  Ort\'{\i}n and G.  Papadopoulos,
{\it Kappa-Symmetry, Supersymmetry and Intersecting Branes}, Nucl.
Phys. {\bf B502}, 149--169 (1997) [arXiv:hep-th/9705040].

\bibitem{BL98}
I.  Bandos and J.  Lukierski, {\it Tensorial central charges and
new superparticle models with fundamental spinor coordinates},
Mod. Phys. Lett. {\bf 14}, 1257 (1999) [arXiv:hep-th/9811022].


\bibitem{B02}
I.A.  Bandos, {\it BPS preons and tensionless super--p--branes in
generalized superspace}, Phys.  Lett.  {\bf B558}, 197 (2003)
[arXiv:hep-th/0208110].

\bibitem{ZU}
 A.A.  Zheltukhin and D.V.  Uvarov,
{\it An Inverse Penrose Limit and Supersymmetry Enhancement in the
Presence of Tensor Central Charges}, JHEP {\bf 0208}, 008 (2002)
[arXiv:hep-th/0206214].

\bibitem{BdAIL}
I.A.  Bandos, J.A.  de Azc\'arraga, J.M.  Izquierdo and J.
Lukierski, {\it $D=4$ supergravity dynamically coupled to a
massless superparticle in a superfield Lagrangian approach},
Phys.  Rev.  {\bf D67}, 065003 (2003)[arXiv:hep-th/0207139]; {\it
On dynamical supergravity interacting with super--$p$--brane
sources}, Invited talk at the 3rd Sakharov International
Conference, Moscow, June 24-29, 2002, arXiv:hep-th/0211065;

I.A. Bandos and J.M. Isidro, {\it D = 4 supergravity dynamically
coupled to superstring in a superfield Lagrangian approach}, Phys.
Rev. {\bf D69}, 085009 (2004) [arXiv:hep-th/0308102].



\bibitem{BdAI}
I.A.  Bandos, {{J.A. de Azc\'arraga}} and J.M.  Izquierdo, {\it
Supergravity interacting with bosonic $p$--branes and local
supersymmetry}, Phys.  Rev.  {\bf D65}, 105010  (2002)
[arXiv:hep-th/0112207].


\bibitem{Bansuper90}
I. Bandos, {\it Superparticle in Lorentz harmonic superspace},
Sov. J. Nucl. Phys. {\bf 51}, 906--914 (1990).

\bibitem{BanZhel93}
I.A.  Bandos  and  A.A.   Zheltukhin, {\it Null Super-p-branes in
4-dimensional Space-time}, Fortschr.  Phys.  {\bf 41}, 619--676
(1993).

\bibitem{BLS99}
I.A. Bandos, J. Lukierski and D. Sorokin, {\it Superparticle
models with tensorial central charges}, Phys. Rev.  {\bf D61},
045002 (2000) [arXiv:hep-th/9904109].


\bibitem{HaSa01}
M. Hatsuda and M. Sakaguchi, {\it Wess-Zumino term for the AdS
superstring and generalized \.In\"on\"u-Wigner contraction},
Prog.  Theor.  Phys.  {\bf 109}, 853-869 (2003)
[arXiv:hep-th/0106114].

\bibitem{Seg51} I.E.  Segal, {\it A class of operator algebras which are
determined by groups}, Duke Math.  J.  {\bf 18}, 221--265 (1951).

\bibitem{IW53} E.  \.In\"on\"u and E.P.  Wigner, {\it On the contraction of
groups and their representations}, Proc. Nat. Acad. Sci. U.S.A.
{\bf 39}, 510--524 (1953); E. \.In\"on\"u, {\it contractions of
Lie groups and their representations}, in
      {\it Group theoretical concepts in elementary particle physics},
F.G\"ursey ed., Gordon and Breach, 391--402 (1964).

\bibitem{Sal61} E.J.  Saletan, {\it Contractions of Lie groups},
J. Math.  Phys.  {\bf 2}, 1--21 (1961).

\bibitem{AC79} D.  Arnal and J.C.  Cortet, {\it Contractions and group
representations}, J.  Math.  Phys.  {\bf 20}, 556--563 (1979).

\bibitem{CelTar} E.  Celeghini and M.  Tarlini,  {\it Contractions of group
representations}, Nuov.  Cim.  {\bf 61B}, 265--277, 172--180
(1981), {\it ibid.} {\bf 68B}, 133-141 (1982).

\bibitem{Lord85} E.A.  Lord, {\it Geometrical interpretation of
       \.In\"on\"u-Wigner contractions},
Int.  J.  Theor.  Phys.  {\bf 24}, 723--730 (1985).

\bibitem{MonPat91} M.  de Montigny and J.  Patera, {\it Discrete and
continuous graded contractions of Lie algebras and superalgebras},
J.  Phys.  {\bf A24}, 525-547 (1991); R.V.  Moody and J.  Patera,
{\it Discrete and continuous graded contractions of
representations of Lie algebras}, J.  Phys.  {\bf A24}, 2227--2257
(1991).

\bibitem{HMOS94}F.  Herranz, M.  De Montigny, M.A.  del Olmo
and M.  Santander, {\it Cayley-Klein algebras as graded
contractions of $so(N+1)$}, J.  Phys.  {\bf A27}, 2515--2526
(1994) [arXiv:hep-th/9312126].

\bibitem{Wei:00} E.  Weimar-Woods,{\it Contractions of Lie algebras:
generalized \.In\"on\"u-Wigner contractions versus graded
contractions}, J.  Math.  Phys.  {\bf 36}, 4519--4548 (1995); {\it
Contractions, generalized \.In\"on\"u and Wigner contractions and
deformations of finite-dimensional Lie algebras}, Rev.  Math.
Phys.  {\bf 12}, 1505--1529 (2000).



\bibitem{Gerst64} M.  Gerstenhaber,
{\it On the deformations of rings and algebras}, Ann.  Math.  {\bf
79}, 59--103 (1964) .

\bibitem{NijRich66} A.  Nijenhuis and R.W.  Richardson Jr.,
{\it Cohomology and deformations in graded Lie algebras}, Bull. A,
Math.  Soc.  {\bf 72}, 1--29 (1966).

\bibitem{NijRich67b} A.  Nijenhuis and R.W.  Richardson Jr.,
{\it Deformations of Lie algebra structures}, J.  Math.  Mech.
{\bf 171}, 89--105 (1967).

\bibitem{Rich67} R.W.  Richardson, {\it On the rigidity of semi-direct
products of Lie algebras}, Pac.  J.  Math.  {\bf 22}, 339--344
(1967).

\bibitem{Le67} M.  L\'evy-Nahas, J.  Math.  Phys., {\it Deformation and
contraction of Lie algebras}, J.  Math.  Phys.  {\bf 8},
1211--1222 (1967).

\bibitem{Her70} R.  Hermann, {\it Analytic continuation of group
representations III}, Commun.  Math.  Phys.  {\bf 3}, 75-97
(1996); {\it Vector bundles in mathematical physics} (vol. II),
W.A. Benjamin (N.Y.), 107 (1970).



\bibitem{Kallosh84}
R.E.  Kallosh, {\it Geometry of eleven-dimensional supergravity},
Phys.  Lett.   {\bf B143}, 373--387 (1984).

I. Bars and S.W. MacDowell, {\it Gravity with extra gauge
symmetry}, Phys.  Lett.  {\bf B129}, 182--191 (1983).

I. Bars and A. Higuchi, {\it First order formulation and
geometrical interpretation of $D=11$ supergravity}, Phys.  Lett.
{\bf 145B}, 329--332 (1984).


\bibitem{DoMa77}
S.W. MacDowell and F. Mansouri, {\it Unified geometric theory of
gravity and supergravity}, Phys. Rev. Lett. {\bf 38}, 739--742
(1977).

F. Mansouri, {\it Superunified theories based on the geometry of
local \mbox{(super-)}gauge invariance}, Phys. Rev. {\bf D16},
2456--2467 (1977).


\bibitem{Fre05} P.  Fr\'e, {\it M Theory FDA, twisted tori and
Chevalley cohomology}, arXiv:hep-th/0510068.

\bibitem{CE48} C.  Chevalley and S.  Eilenberg, {\it Cohomology
theory of Lie groups and Lie algebras}, Trans.  Am.  Math.  Soc.
{\bf 63}, 85--124 (1948).


\bibitem{CremmerFerrara80}
E. Cremmer and S. Ferrara, {\it Formulation of eleven-dimensional
supergravity in superspace}, Phys.  Lett.  {\bf B91}, 61 (1980).

\bibitem{BrinkHowe80}
L. Brink and P.S. Howe, {\it Eleven-dimensional supergravity on
the mass-shell in superspace}, Phys.  Lett.  {\bf B91}, 384
(1980).

\bibitem{AT89} J.A.  de Azc\'arraga and P.K.  Townsend,
{\it Superspace geometry and the formulation of supersymmetric
extended objects}, Phys.  Rev.  Lett.  {\bf 62}, 2579--2582 (1989)

\bibitem{BESE} E.  Bergshoeff and E.  Sezgin, {\it Super
$p$-brane  theories and new space-time superalgebras}, Phys.
Lett.  {\bf B354}, 256--263 (1995) [arXiv:hep-th/9504140].

\bibitem{Anguelova:2003sn}
  L. Anguelova and P.A. Grassi,
{\it Super D-branes from BRST symmetry},
  JHEP {\bf 0311} (2003) 010
  [arXiv:hep-th/0307260].

\bibitem{BNS98} I.A.  Bandos, N.  Berkovits, D.P. Sorokin, {\it
Duality-symmetric eleven-dimensional supergravity and its coupling
to M-branes}, Nucl. Phys. {\bf B522}, 214--233 (1998)
[arXiv:hep-th/9711055].




\bibitem{Saka-98}
M. Sakaguchi, {\it Type-II superstrings and new spacetime
superalgebras}, Phys.  Rev.  {\bf D59}, 046007 (1999)
[arXiv:hep-th/9809113].

\bibitem{Az-Iz-Mi-04}
J.A. de Azc\'arraga and J.M. Izquierdo and C.  Miquel-Espanya,
{\it Spacetime scale-invariant super-$p$-brane actions on enlarged
superspaces and the geometry of $\kappa$-symmetry}, Nucl. Phys.
{\bf B706}, 181--203 (2005) [arXiv:hep-th/0407238].


\bibitem{Hoppe} B. de Wit, J. Hoppe and H. Nicolai, {\it On the
quantum mechanics of supermembranes}, Nucl. Phys. {\bf B305}, 545
(1988).


\bibitem{Nicolai}
H. Nicolai and J. Plefka, {\it Supersymmetric effective action of
matrix theory}, Phys. Lett. {\bf B477}, 309 (2000)
[arXiv:hep-th/0001106].

 A. Dasgupta, H. Nicolai and J. Plefka, {\it Vertex operators for the
supermembrane}, JHEP {\bf 0005}, 007 (2000)
[arXiv:hep-th/0003280]; {\it An introduction to the quantum
supermembrane}, Grav. Cosmol.  {\bf 8}, 1 (2002)
[arXiv:hep-th/0201182], and refs. therein.


\bibitem{(M)atrix}
 T. Banks, W. Fischler, S.H. Shenker and L. Susskind, {\it
 M--theory as a matrix model: a conjecture},
 Phys. Rev. {\bf D55}, 5112  (1997) [arXiv:hep-th/9610043].

\bibitem{V01s}
M.A.  Vasiliev, {\it  Conformal higher spin symmetries of 4d
massless supermultiplets and $osp(L,2M)$ invariant equations in
generalized  (super)space}, Phys.  Rev.  {\bf D66}, 066006 (2002)
[arXiv:hep-th/0106149].

\bibitem{V01c}
M.A.  Vasiliev, {\it Relativity, causality, locality, quantization
and duality in the $Sp(2M)$ invariant generalized space-Time}, in
{\it Multiple facets of quantization and supersymmetry} ({\it
Marinov's memorial volume}, M.Olshanetsky and A.Vainshtein eds.),
826 [arXiv:hep-th/0111119].

\bibitem{W03}
E.  Witten, {\it Perturbative gauge theory and a string theory in
twistor space}, Commun.  Math.  Phys.  {\bf252}, 189--258 (2004)
[arXiv:hep-th/0312171].

\bibitem{BAME06} I.A. Bandos, J.A. de Azc\'arraga
and C. Miquel-Espanya, {\it Superspace formulations of the
(super)twistor string}, arXiv:hep-th/0604037.

\bibitem{Curtright}
T. Curtright, {\it Are there any superstrings in eleven
dimensions?}, Phys. Rev. Lett. {\bf 60}, 393 (1988).

\bibitem{DG}
A.A.  Deriglazov and A.V.  Galajinsky, {\it A covariant action for
the eleven dimensional superstring}, Mod.  Phys.  Lett.  {\bf
A12}, 2993 (1997) [arXiv:hep-th/9711196].

N.  Berkovits, {\it A Problem with the superstring action of
 Deriglazov and Galajinsky},
Phys.  Rev.  {\bf D59}, 048901 (1999) [arXiv:hep-th/9712056].

\bibitem{RS}
I.  Rudychev and E.  Sezgin, {\it Superparticles in $D=11$},
Phys.  Lett.  {\bf B415}, 363 (1997); {\bf B424}, 411 (1998)
[arXiv:hep-th/9704057].

\bibitem{Manvelyan}
 R.  Manvelyan and R.  Mkrtchyan,
{\it Towards $SO(2,10)$-invariant M-Theory: multilagrangian
fields} Mod.  Phys.  Lett.  {\bf A15}, 747 (2000)
[arXiv:hep-th/9907011]; {\it Free fields equations for space--time
algebras with
 tensorial momentum},
Mod.  Phys.  Lett.  {\bf A17}, 1393 (2002) [arXiv:hep-th/0112233].

 R.  Mkrtchyan, {\it Little groups and statistics of branes},
arXiv:hep-th/0209040; {\it On an (interacting) field theories with
tensorial momentum}, arXiv:hep-th/0209175; {\it Brain content of
branes' states}, Phys. Lett.  {\bf B558}, 205--212 (2003)
[arXiv:hep-th/0212174].

\bibitem{ZL}
 A.A.  Zheltukhin and U.  Lindstr\"{o}m,
{\it Strings in a space with tensor central charge coordinates},
Nucl. Phys. Proc. Suppl. {\bf 102}, 126 (2001)
[arXiv:hep-th/0103101]; {\it Hamiltonian of tensionless strings
with tensor central charge coordinates}, JHEP {\bf 0201}, 034
(2002) [arXiv:hep-th/0112206].

\bibitem{Fr86}
C.  Fronsdal, {\it Massless particles, orthosymplectic symmetry
and another type of Kaluza--Klein theory}, in {\it Essays on
Supersymmetry} (C. Fronsdal ed.), Mathematical Physics Studies
{\bf 8}, D. Reidel Pub. Co., 163 (1986).

\bibitem{V01}
M.A.  Vasiliev, {\it Progress in higher spin gauge theories}, in
Moscow 2000, {\it Quantization, gauge theory and strings}, {\bf
vol I} 452 (2001) [arXiv:hep-th/0104246], and refs. therein.

\bibitem{Vasiliev89}
M.A. Vasiliev, {\it Consistent equations for interacting masless
fields of all spins in the first order in curvatures}, Ann. Phys.
{\bf 190}, 59 (1989).

\bibitem{Dima}
 M.  Plyushchay, D.  Sorokin and M.  Tsulaia, {\it
Higher spins from tensorial charges and $OSp(N|2n)$ symmetry},
JHEP {\bf 0304}, 013 (2003) [arXiv:hep-th/0301067].

\bibitem{Misha}
V.E.  Didenko and M.A.  Vasiliev, {\it Free field dynamics in the
generalized $AdS$ (super)space}, J.  Math.  Phys.  {\bf 45}
197--215 [arXiv:hep-th/0301054].

 M.A.
Vasiliev, {\it Higher-spin theories and $Sp(2M)$ invariant
space--time}, arXiv:hep-th/0301235.


\bibitem{Misha+}
 O.A. Gelfond and M.A. Vasiliev, {\it
Higher rank conformal fields in the $Sp(2M)$ symmetric generalized
space-time}, Theor.  Math.  Phys.  {\bf 145}, 1400--1424 (2005)
[arXiv:hep-th/0304020].

\bibitem{BBAST05} I.Bandos, X. Bekaert, J.A. de Azc\'arraga,
D. Sorokin, M. Tsulaia, {\it Dynamics of higher spin fields and
tensorial space}, JHEP 0505 {\bf 031} (2005)
[arXiv:hep-th/0501113].

\bibitem{Sor05} D. Sorokin, {\it Introduction to the classical theory
of higher spins}, AIP Conf. Proc. {\bf 767}, 172--202 (2005)
[arXiv:hep-th/0405069].

\bibitem{BZ95}
I.A.  Bandos and A.A.  Zheltukhin, {\it $N=1$ Super-p-branes in
twistor--like Lorentz harmonic formulation}, Class.  Quantum
Grav.  {\bf 12}, 609 (1995) [arXiv:hep-th/9405113].

\bibitem{NP85}
E.R. Nissimov and S.J. Pacheva, {\it Quantization of the $N=1$,
$N=2$ superparticle with irreducible contraints}, Phys. Lett. {\bf
B189}, 57 (1986).

\bibitem{Bars}
 I.  Bars and C.  Deliduman,
{\it High Spin Gauge Fields and Two-Time Physics}, Phys.  Rev.
{\bf D64}, 045004 (2001) [arXiv:hep-th/0103042], and refs.
therein.


\bibitem{Dirac}
P.A.M.  Dirac, {\it Lectures on quantum mechanics}, Academic
Press, NY (1967).

\bibitem{BZstr}
I.A. Bandos and A.A. Zheltukhin, {\it Spinor Cartan N hedron,
Lorentz harmonic formulations of superstrings and kappa symmetry},
 JETP Lett. {\bf 54}, 421 (1991); {\it Green-Schwarz superstrings
 in spinor moving frame formalism},
Phys. Lett. {\bf B288}, 77 (1992).

\bibitem{Grassi} P.A. Grassi, G. Policastro and M. Porrati, {\it
Covariant quantization of the Brink--Schwarz superparticle}, Nucl.
Phys. {\bf B606}, 380 (2001) [arXiv:hep-th/0009239].

\bibitem{conversion}
I.A. Batalin and E.S. Fradkin, {\it Operatorial quantization of
dynamical systems subject to second class constraints}, Nucl.
Phys. {\bf B279}, 514 (1987).

L.D. Faddeev and S.L. Shatashvili, {\it Realization of the
Schwinger term in the Gauss law and the possibility of correct
quantization of a theory with anomalies}, Phys. Lett. {\bf B167},
225 (1986).

E.S. Egorian and R.P. Manvelyan, {\it BRST quantization of
Hamiltonian systems with second class constraints}, {\it Preprint}
YERPHI-1056-19-88 [unpublished]; Theor. Math. Phys. {\bf 94}
173--181 (1993).

I.A. Batalin, E.S. Fradkin and T.E. Fradkina, {\it Another version
for operatorial quantization of dynamical systems with irreducible
constraints}, Nucl. Phys. {\bf B314}, 158 (1989).

\bibitem{Ferber}
A. Ferber, {\it Supertwistor and conformal supersymmetry}, Nucl.
Phys. {\bf B132}, 55 (1978).

T. Shirafuji, {\it Lagrangian mechanics of massless particles with
spin}, Prog.  Theor.  Phys.  {\bf 70}, 18 (1983).

\bibitem{Pen}
R. Penrose and M.A.H. MacCallum, {\it Twistor theory: an approach
to the quantization of fields and space-time}, Phys. Rept. {\bf
6}, 241 (1972).

R. Penrose, {\it The twistor program}, Rep. Math. Phys. {\bf 12},
65 (1977).

\bibitem{BarPi06} I. Bars and M. Pic\'on,
{\it Twistor transform in d dimensions and a unifying role for
twistors}, Phys. Rev. {\bf D73}, 064033 (2006)
[arXiv:hep-th/hep-th/0512348]; {\it  Single twistor description of
massless, massive, AdS, and other interacting particles}, Phys.
Rev. {\bf D73}, 064002 (2006) [arXiv:hep-th/0512091].

\bibitem{Sokatchev}
E. Sokatchev, {\it Light cone harmonic superspace and its
applications}, Phys. Lett. {\bf B169}, 209 (1986); {\it Harmonic
superparticle}, Class. Quantum Grav. {\bf 4}, 237 (1987).

\bibitem{SG89}
A.I. Gumenchuk and D.P. Sorokin, {\it Relativistic superparticle
dynamics and twistor correspondence}, Sov. J. Nucl. Phys. {\bf
51}, 350 (1990).







\bibitem{Gran:2006ec}
  U.~Gran, J.~Gutowski, G.~Papadopoulos and D.~Roest,
  {\it N = 31 is not IIB}, arXiv:hep-th/0606049.


\bibitem{Bandos:2006xz}
  I.A.~Bandos, J.A.~de Azc\'arraga and O.~Varela,
  {\it On the absence of BPS preonic solutions in IIA and IIB supergravities},
  arXiv:hep-th/0607060.




\newpage{\pagestyle{empty}\cleardoublepage} 

\newpage{\pagestyle{empty}\cleardoublepage}

\thispagestyle{empty}



\vspace*{2cm}

\begin{center}

\includegraphics[scale=0.04]{Escudo01}

\end{center}


\end{thebibliography}
\end{document}